\documentclass[11pt,a4paper,onecolumn]{IEEEtran}

\usepackage{graphics}
\usepackage{graphicx}
\usepackage{dsfont}
\usepackage{fancybox}
\usepackage{colordvi}
\usepackage{color}
\usepackage{amssymb}
\usepackage{amsmath}
\usepackage{latexsym}
\usepackage{amsfonts}
\usepackage{acronym}
\usepackage[ruled]{algorithm2e}
\usepackage{subfigure}
\usepackage{hyperref}
\usepackage{setspace}
\usepackage{titlepic}
\usepackage{times}
\usepackage{blindtext}
\usepackage[inner=3cm, outer=3cm, vmargin=3cm, a4paper]{geometry}
\usepackage{epsfig,psfrag,color,graphics,epsf}
\usepackage{theorem}
\usepackage{epsfig,psfrag,color,graphics,epsf}

\usepackage{amsmath, amssymb, amsopn}
\usepackage{mathrsfs}
\usepackage{graphicx}
\usepackage{psfrag}
\usepackage{epsfig}
\usepackage{makeidx}
\usepackage{fancyhdr}

\newtheorem{theorem}{Theorem}
\newtheorem{lemma}{Lemma}
\newtheorem{corollary}{Corollary}

\newtheorem{proposition}{Proposition}
\newtheorem{remark}{Remark}

\newcommand{\Var}[1]{\textnormal{Var}\left(#1\right)}

\newcommand{\Markov}{\mathrel{\multimap}\joinrel\mathrel{-}\joinrel\mathrel{\mkern-6mu}\joinrel\mathrel{-}}
\newcommand{\Reals}{\mathbb R}

\newcommand{\qed}{\hfill\ensuremath{\IEEEQEDopen}}

\begin{document}

\IEEEoverridecommandlockouts
\interdisplaylinepenalty=1000

\title{Sending a Bivariate Gaussian Source Over a Gaussian MAC with Unidirectional Conferencing
Encoders }

\author{Shraga~I.~Bross  and Yaron Laufer
\thanks{
S. Bross is with the Engineering Department, Bar-Ilan University,
 Ramat Gan 52900, Israel.
 Email:brosss@biu.ac.il. The work of S.~Bross was supported by the Israel Science Foundation
under Grant 497/09 and Grant 455/14.}\thanks{Y. Laufer is with the Engineering Department,
Bar-Ilan University, Ramat Gan 52900, Israel. Email:yaron\textunderscore laufer@walla.com. }

}

\maketitle

\begin{abstract}

We consider the transmission of a memoryless bivariate Gaussian
 source over a two-user additive Gaussian multiple-access
channel with unidirectional conferencing encoders. Here, prior to
each transmission block, Encoder~1, which observes the first source component,
is allowed to communicate with Encoder~2, which observes the second source
component, via a unidirectional noise-free bit-pipe of given
capacity.
The main results of this work are sufficient conditions and a necessary condition for
the achievability of a distortion pair expressed as a function of the channel ${\sf SNR}$
and of the source correlation.
The main sufficient condition is obtained by an extension of the vector-quantizer scheme
suggested by Lapidoth-Tinguely, for the case without conferencing, to the case
with unidirectional conference.
In the high-${\sf SNR}$ regime, and when the capacity
of the conference channel is unlimited, these necessary and sufficient conditions are
shown to agree.
We evaluate the precise high-${\sf SNR}$ asymptotics for a subset of
distortion pairs when the capacity of the conference channel is unlimited in which case we show
that a separation based scheme attains these optimal distortion pairs.
However, with symmetric average-power constraints and fixed conferencing
capacity, at high-${\sf SNR}$ the latter separation based scheme is shown to be suboptimal.

\end{abstract}



\section{Introduction and Problem Statement}
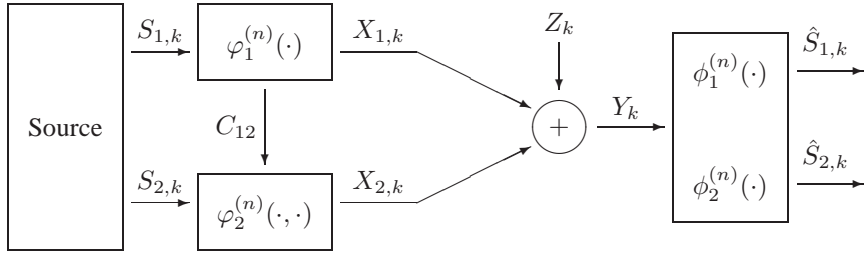
\begin{figure}
\centering\setlength{\unitlength}{0.25mm}
\begin{picture}(500,150)(10,10)
 \linethickness{0.1mm}


\put(290,80){\circle{30}} \put(290,80){\makebox(0,0){\small $+$}}
\put(290,125){\vector(0,-1){25}} \put(290,135){\makebox(0,0){\small
$Z_k$}} \put(325,90){\makebox(0,0){\small $Y_k$}}
\put(310,80){\vector(1,0){35}}

\put(100,105){\framebox(70,40){}} \put(135,125){\makebox(0,0){\small
$\varphi_1^{(n)}(\cdot)$}} \put(100,15){\framebox(70,40){}}
\put(135,35){\makebox(0,0){\small $\varphi_2^{(n)}(\cdot,\cdot)$}}
\put(215,120){\vector(2,-1){60}} \put(215,40){\vector(2,1){60}}

\put(135,100){\vector(0,-1){40}}
\put(120,80){\makebox(0,0){\small $C_{12}$}}
\put(175,120){\line(1,0){40}} \put(195,130){\makebox(0,0){\small
$X_{1,k}$}} \put(175,40){\line(1,0){40}}
\put(195,50){\makebox(0,0){\small $X_{2,k}$}}

\put(350,30){\framebox(60,100){}}
\put(380,110){\makebox(0,0){\small $\phi_1^{(n)}(\cdot)$}}
\put(380,50){\makebox(0,0){\small $\phi_2^{(n)}(\cdot)$}}

\put(0,15){\framebox(60,130){}} \put(30,80){\makebox(0,0){\small
Source}}

\put(80,130){\makebox(0,0){\small $S_{1,k}$}}
\put(65,120){\vector(1,0){30}} \put(80,50){\makebox(0,0){\small
$S_{2,k}$}} \put(65,40){\vector(1,0){30}}

\put(415,110){\vector(1,0){35}} \put(430,125){\makebox(0,0){\small
$\hat{S}_{1,k}$}}

\put(415,50){\vector(1,0){35}} \put(430,65){\makebox(0,0){\small
$\hat{S}_{2,k}$}}
\end{picture}

\caption{Transmission of bivariate Gaussian source over
a Gaussian multiple-access channel with unidirectional conferencing
encoders.}\label{fig:BivGau source over Gaussian MAC}
\end{figure}

We consider a communication scenario where two encoders
transmit a memoryless bivariate Gaussian source to a single receiver
over a two-user additive white Gaussian multiple-access channel (MAC).
The source is observed separately by the two encoders; Encoder~1 observes the first source
component and Encoder~2 observes the second source component.
The encoders are allowed to partially cooperate in the sense that prior to each transmission block,
Encoder~1 is allowed to communicate with Encoder~2 via a unidirectional noise-free bit-pipe of given
capacity, as shown in Fig.~\ref{fig:BivGau source over Gaussian MAC}.
Both encoders then cooperate in describing the source components to a common receiver,
via an average-power constrained Gaussian MAC.
From the output of the multiple-access channel, the receiver wishes to reconstruct
each source component with the least possible expected squared-error distortion.
Our interest is in characterizing the distortion pairs that are simultaneously
achievable on the two source components.
Special cases are the classical MAC considered by Lapidoth-Tinguely in \cite{Stephan},
where the encoders are ignorant of each others inputs (the bit-pipe
is of strictly zero capacity --i.e. no connection at all) and the asymmetric setting, where Encoder~2 is
fully cognizant of the source input at Encoder~1 (the bit-pipe is of
infinite capacity).

\medskip

Henceforth, we adopt the following notation conventions. Random
variables will be denoted by capital letters, while their
realizations will be denoted by the respective lower case letters.
Whenever the dimension of a random vector is clear from the context
the random vector will be denoted by a bold face letter, that is,
$\textbf{X}$ denotes the random vector
$(X_1,X_2,\ldots,X_n)$, and $\textbf{x}=(x_1,x_2,\ldots,x_n)$
will designate a specific sample value of
$\textbf{X}$. 
The alphabet of a scalar random variable $X$ will be designated by a
calligraphic letter ${\mathcal X}$.  The $n$-fold Cartesian power of
a generic alphabet ${\mathcal V}$, that is, the set of all
$n$-vectors over ${\mathcal V}$, will be denoted ${\mathcal V}^n$.
An estimator of a random variable $X$ is denoted by $\hat{X}$.
For a real-valued parameter $0\leq \beta\leq 1$ we define $\bar{\beta}\triangleq 1-\beta$,
and for a nonnegative distortion constraint $D$ the corresponding normalized distortion
is defined by $d\triangleq D/\sigma^2$ where $\sigma^2$ is the source variance.

\medskip

Formally, the time-$k$ output of the Gaussian MAC is given by
\begin{equation}
Y_k=x_{1,k}+x_{2,k}+Z_k,
\end{equation}
where $(x_{1,k},x_{2,k})\in \mathbb{R}^2$ are the symbols sent by
the transmitters, and $Z_k$ is the time-$k$ additive noise term. The
sequence $\{Z_k\}$ consists of independent identically distributed
(IID) zero-mean variance $N$ Gaussian random variables that are
independent of the source sequence.

The input source sequence $\{(S_{1,k},S_{2,k})\}$ consists of
zero-mean Gaussians of covariance
\begin{equation}
{\sf K}_{SS}=\left( \begin{array}{cc} \sigma^2 & \rho\sigma^2 \\
\rho\sigma^2 & \sigma^2 \end{array} \right)
\label{eq:covariancemat1}
\end{equation}
with $\rho\in[0,1]$, and $0<\sigma^2<\infty$ (for a justification for the restriction
to $\rho\in[0,1]$ and $\sigma_1^2=\sigma_2^2=\sigma^2$ see \cite[Section~II.C]{Stephan}).

\medskip

{\it Note:}
There are just two exceptions to the notation conventions defined above.
Throughout this work we define several scalings of the
source correlation coefficient. Specifically, we define $\tilde{\rho}$ and $\bar{\rho}$ as per \eqref{eq:constants}
(in which case $\bar{\rho} \neq 1-\rho$), and similarly $\hat{\rho}$ as per \eqref{eq:constants11} (in which case $\hat{\rho}$
does not refer to an estimator of $\rho$).

\medskip

The sequence
$\{S_{1,k}\}$ is observed by Encoder~1 and the sequence
$\{S_{2,k}\}$ is observed by Encoder~2. Prior to each block of $n$
channel uses, the encoders may exchange information via the use of
the unidirectional bit-pipe which is assumed to be:
\begin{itemize}
\item perfect in the sense that any input symbol is available
immediately and error-free at the output of the pipe; and
\item of limited capacity $C_{12}$, in the sense that when the
input to the pipe from Encoder~1 to Encoder~2 takes values in the
set ${\cal W}$, such that
$W=f^{(n)}(\textbf{S}_1)$ for some encoding
function $f^{(n)}\colon \mathbb{R}^n\mapsto {\cal W}$, then
\begin{equation}
 \log |{\cal W}|\leq nC_{12}.
\label{eq:confcap1}
\end{equation}
\end{itemize}
We define an $(n,C_{12})$-{\it conference} to be a collection of an
input alphabet ${\cal W}$,
and an encoding function $f^{(n)}(\cdot)$ as above, where
$n,C_{12}$ and the alphabet set satisfy \eqref{eq:confcap1}.

After the conference, Encoder~2 is cognizant of the random variable
$W$ so the channel inputs $\textbf{X}_1=(X_{1,1},\ldots,X_{1,n})$
and $\textbf{X}_2=(X_{2,1},\ldots,X_{2,n})$ can be described via
encoding functions $\varphi^{(n)}_1$ and $\varphi^{(n)}_2$ as
\begin{IEEEeqnarray}{rCl}
\textbf{X}_1 & = & \varphi^{(n)}_1(\textbf{S}_1), \nonumber \\
\textbf{X}_2 & = &
\varphi^{(n)}_2(\textbf{S}_2,W)=\varphi^{(n)}_2(\textbf{S}_2,f^{(n)}(\textbf{S}_1)),
\label{eq:encmappingsdef1}
\end{IEEEeqnarray}
where
\begin{IEEEeqnarray}{rCl}
\varphi^{(n)}_1 & \colon & \mathbb{R}^n\mapsto \mathbb{R}^n, \nonumber \\
\varphi^{(n)}_2 & \colon & \mathbb{R}^n\times {\cal W} \mapsto \mathbb{R}^n . \label{eq:encoder1}
\end{IEEEeqnarray}
The channel input sequences are average-power limited to $P_1$ and $P_2$ respectively, i.e.
\begin{equation}
\frac{1}{n} {\sf E}\left[\sum_{k=1}^n\left(X_{\nu,k}\right)^2\right]\leq P_{\nu},
\qquad \nu=1,2 \label{eq:powerconst1}
\end{equation}
where ${\sf E}$ denotes the expectation operator.
Based on the channel output $\textbf{Y}=(Y_1,\ldots,Y_n)$ the
receiver forms its estimates
$\hat{\textbf{S}}_1=\phi^{(n)}_1(\textbf{Y})$ and
$\hat{\textbf{S}}_2=\phi^{(n)}_2(\textbf{Y})$ for the source
sequences respectively, where
\begin{equation}
\phi^{(n)}_{\nu} \colon \mathbb{R}^n\mapsto \mathbb{R}^n, \qquad
\nu=1,2. \label{eq:reconstruction1}
\end{equation}
We are interested in the minimal expected squared-error distortions
at which the receiver can reconstruct each of the source sequences.

\medskip

{\it Definition 1:} Given $\sigma^2>0,\rho\in[0,1],P_1,P_2,N>0$ and
$C_{12}>0$ we say that the distortion pair
$(D_1,D_2)$ is {\it achievable} if
there exists a sequence of block-lengths $n$, encoding functions
$f^{(n)}$ that belong to an $(n,C_{12})$-{\it conference}, encoders
$(\varphi_1^{(n)},\varphi_2^{(n)})$ as in \eqref{eq:encoder1}
satisfying the average-power constraints \eqref{eq:powerconst1}, and
reconstruction functions $(\phi_1^{(n)},\phi_2^{(n)})$ as in
\eqref{eq:reconstruction1} resulting in average distortions that
fulfill
\begin{equation}
\varlimsup_{n\to\infty} \ \frac{1}{n}\sum_{k=1}^n{\sf
E}\left[(S_{\nu,k}-\hat{S}_{\nu,k})^2\right]\leq D_{\nu}, \qquad
\nu=1,2,
\end{equation}
whenever $Y_k=\varphi_{1,k}^{(n)}(\textbf{S}_1)+\varphi_{2,k}^{(n)}(\textbf{S}_2,f^{(n)}(\textbf{S}_1))+Z_k,
 k=1,\ldots,n$,
and $\{(S_{1,k},S_{2,k})\}$ are IID zero-mean bivariate Gaussian
vectors with covariance matrix ${\sf K}_{SS}$ as in
\eqref{eq:covariancemat1} and $\{Z_k\}$ are IID zero-mean
variance-$N$ Gaussian random variables that are independent of
$\{(S_{1,k},S_{2,k})\}$.


\medskip

\medskip


In \cite{CovEl} the authors provided sufficient conditions for
reliable transmission of correlated sources over a regular MAC and
demonstrated that, in general, the separation approach is not
optimal.
For the regular MAC, the separation approach is known to be optimal when
the channel is lossless (Slepian-Wolf source coding theorem
\cite{SpWolf}), or when the sources are independent. In the special
case of transmitting correlated sources losslessly over an asymmetric MAC it is
shown in \cite{DeBru} that necessary and sufficient conditions for
reliable transmission do exist and, moreover, these conditions can
be established by applying the separation approach.
In \cite{Deniz} the authors consider the model \cite{DeBru} with
a single distortion constraint namely,
when $D_1=0$ (i.e. $\textbf{S}_1$ is recovered losslessly at the receiver),
and show that source-channel separation is optimal.

A lossy Gaussian version of the problem addressed by Cover-El
Gamal-Salehi \cite{CovEl} has been considered in \cite{Stephan},
wherein the power-versus-distortion tradeoff for the distributed
transmission of a memoryless bivariate Gaussian source over a
two-to-one average-power limited Gaussian MAC is considered.
Necessary and sufficient conditions for the achievability of a
distortion pair are presented and it is shown that
if the channel signal-to-noise ratio (${\sf SNR}$) is below a certain threshold uncoded
transmission is optimal.
Furthermore, the authors derive the high-${\sf SNR}$ asymptotics for a subset of distortion pairs
and show that the source-channel vector-quantizer, by means of which they derive their sufficient condition,
is optimal at  high-${\sf SNR}$.
In the symmetric case of equal average-power constraints and equal distortions this vector-quantizer
outperforms source-channel separation at all ${\sf SNR}$'s.

Our problem is also related to the correlated sources with partially separated encoders source-coding
problem \cite{Kaspi}, and to the Gaussian MAC with conferencing encoders channel-coding problem \cite{Michelle}
(see also \cite{Willems}).
However, the above two problems are source/channel coding problems, whereas ours is one of the combined
source-channel coding.

\medskip

We present four sufficient conditions and one necessary condition for the achievability of a
distortion pair $(D_1,D_2)$.
These conditions are expressed as a function of the channel signal-to-noise ratio (${\sf SNR}$)
and of the source correlation.

 Our contribution is in the following aspects:
\begin{itemize}
\item We suggest an extension for the Lapidoth-Tinguely vector-quantizer \cite{Stephan} to the case
with unidirectional conferencing and derive the corresponding achievable rate-distortion region.
\item We derive an achievable rate-distortion region when the capacity of the conference channel
is unlimited.
\item We derive a necessary condition for the achievability of a distortion pair $(D_1,D_2)$.
This condition is obtained by some arguments reducing the multiple-access problem to a point-to-point
problem. The key step therein is to upper-bound the maximal correlation between two simultaneous
channel inputs, subject to conditional rate-distortion constraints,
by using a result from maximum correlation theory.
\item We derive the high-${\sf SNR}$ asymptotics of an optimal scheme when the capacity
of the conference channel is unlimited.
In particular, we show that in this case a source-channel separation scheme is optimal.
\item For a fixed conferencing capacity, high-${\sf SNR}$, and symmetric average-power constraints,
we show that the latter source-channel separation scheme, which is optimal for unlimited conferencing capacity,
is suboptimal compared to the vector-quantizer.
\end{itemize}

The paper is organized as follows. Section~II presents our main results, while
in Section~III we prove the necessary condition. In Section~IV and the Appendix we present our code construction and
analyze its performance. The analysis for the rest of our main results appears in Sections~V-IX.


\section{Main Results}

In this section we present one necessary condition and four sufficient conditions
for the achievability of a distortion pair $(D_1,D_2)$; the sufficient conditions are stated in Theorem~\ref{th:vector-quantizer},
Corollary~\ref{cor:VQ with unlimited capacity},
and via the two source-channel separation schemes considered in Section~II.C. The necessary condition also
establishes the asymptotic behavior of an optimal scheme for a subset of distortion pairs, when the capacity of the
conference channel is unlimited.

\medskip

\subsection{Necessary condition for the achievability of $(D_1,D_2)$}

\begin{theorem}\label{th:necessary1}
A necessary condition for the achievability of a distortion pair
$(D_1,D_2)$ over the Gaussian MAC with unidirectional conferencing
is that for some $0\leq \beta\leq 1$
\begin{IEEEeqnarray}{rCl}
R_{S_1,S_2}(D_1,D_2) & \leq &
\frac{1}{2}\log_2\left(1+\frac{P_1+P_2+2\sqrt{({\rho}^2\bar{\beta}+\beta)}\sqrt{P_1P_2}}{N}\right)
\label{eq:necessaryup1} \\
R_{S_2|S_1}(D_2) &  \leq &
\frac{1}{2}\log_2\left(1+\frac{\bar{\beta}P_2(1-\rho^2)}{N}\right),
\label{eq:necessaryup10}
\end{IEEEeqnarray}
where $R_{S_1,S_2}(D_1,D_2)$ denotes the rate-distortion function of a bivariate Gaussian source
$\{(S_{1,k},S_{2,k})\}$, which is derived first in \cite{Xiao} and then in \cite[Theorem~III.1]{Stephan}, and $R_{S_2|S_1}(D_2)$
denotes the rate-distortion function for $\{S_{2,k}\}$ when $\{S_{1,k}\}$ is given as side-information
to both the encoder and the decoder.
\end{theorem}
\begin{IEEEproof}
See Section \ref{Proof of Necessary Condition}.
\end{IEEEproof}

\begin{remark}
The necessary condition \eqref{eq:necessaryup1} is of the same flavor as the necessary
condition in \cite[Theorem~IV.1]{Stephan}. Specifically, Condition  \eqref{eq:necessaryup1}
corresponds to the necessary and sufficient condition for the achievability of a distortion
pair $(D_1,D_2)$ when the source $\{(S_{1,k},S_{2,k})\}$ is transmitted over a point to point
additive white Gaussian noise (AWGN) channel of input power constraint
$P_1+P_2+2\sqrt{({\rho}^2\bar{\beta}+\beta)}\sqrt{P_1P_2}$.
\end{remark}

\begin{remark}\label{th:highsnr00}
The necessary condition \eqref{eq:necessaryup1}--\eqref{eq:necessaryup10} is not a function of $C_{12}$.
Therefore, we expect that it will be tight when the conferencing capacity is unlimited.
\end{remark}

\subsection{Vector-quantizer scheme}
Our achievability result is based on an extension of the
vector-quantizer  scheme presented in \cite{Stephan}, which benefits
from the presence of the unidirectional conference channel. The
encoding steps of our scheme are presented in Fig.~\ref{fig:1vq}.

The source sequence $\textbf{S}_1$ is quantized by Encoder~1 in two
steps; first it is quantized by a rate-$R_1$ vector-quantizer where
the quantized sequence is denoted by $\textbf{U}_1^{*}$, then the
quantization error of the first step is quantized by a rate-$R_\textnormal{c}$
vector-quantizer, where
\begin{IEEEeqnarray}{l}
R_\textnormal{c}+1/2\log\left(1-\rho^22^{-2R_1}(1-2^{-2R_\textnormal{c}})\right) \leq C_{12},
\label{eq:vecquanconfconstraint}
\end{IEEEeqnarray}
and the quantized sequence is denoted by $\textbf{V}^{*}$. The source
sequence $\textbf{S}_2$ is quantized by Encoder~2 via a rate-$R_2$
vector-quantizer where the quantized sequence is denoted by
$\textbf{U}_2^{*}$. Encoder~1 informs Encoder~2 via the conference
channel on the index of $\textbf{V}^{*}$, taking into account that Encoder~2 has side-information
$\textbf{S}_2$, and consequently both encoders can cooperate in
transmitting this sequence.\\
The channel input $\textbf{X}_1$ is now given by
\begin{equation}
\textbf{X}_1=a_{1,1}\textbf{U}_1^{*}+a_{1,2}
\textbf{V}^{*}, \label{eq:inputX1}
\end{equation}
where for $0\leq \beta_1\leq 1$ the gains $a_{1,1}$ and $a_{1,2}$ are chosen as
\begin{equation*}
a_{1,1} = \sqrt{\frac{\bar{\beta_1}P_1}{\sigma^2(1-2^{-2R_1})}}
\quad , \quad
a_{1,2} = \sqrt{\frac{\beta_1
P_1}{\sigma^22^{-2R_1}(1-2^{-2R_\textnormal{c}})}}.
\end{equation*}
This ensures that the input $\textbf{X}_1$ satisfies the
average-power constraint $P_1$.\\
The channel input $\textbf{X}_2$ is now given by
\begin{equation}
\textbf{X}_2=a_{2,1}\textbf{U}_2^{*}+a_{2,2}
\textbf{V}^{*}, \label{eq:inputX2}
\end{equation}
where for $0\leq \beta_2\leq 1$ and $\sigma_v^2\triangleq \sigma^2 2^{-2R_1}(1-2^{-2R_\textnormal{c}})$, the gains $a_{2,1}$ and $a_{2,2}$ are chosen as
\begin{IEEEeqnarray*}{rCl}
a_{2,1} & = &
\sqrt{\frac{\bar{\beta_2}P_2}{\sigma^2(1-2^{-2R_2})}}, \nonumber \\
a_{2,2} 
 & = & \sqrt{\frac{P_2}{\sigma^2}}\left(
 \sqrt{\rho^2\bar{\beta}_2(1-2^{-2R_2})+\frac{\sigma^2\beta_2}{\sigma_v^2}}
 -\sqrt{\rho^2\bar{\beta}_2(1-2^{-2R_2})}\right).
\end{IEEEeqnarray*}
This ensures that the input $\textbf{X}_2$ satisfies the
average-power constraint $P_2$.

Based on the channel output $\textbf{Y}$, the decoder first estimates
the triplet $(\textbf{U}_1^{*},\textbf{V}^{*},\textbf{U}_2^{*})$ by
performing joint decoding which takes into account the correlation
between the sequences. The resulting decoded triplet is denoted by
$(\hat{\textbf{U}}_1,\hat{\textbf{V}},\hat{\textbf{U}}_2)$. The
decoder then treats $(\textbf{S}_1,\textbf{S}_2,\hat{\textbf{U}}_1,\hat{\textbf{V}},\hat{\textbf{U}}_2)$ as a jointly Gaussian tuple
and forms its estimates of the source sequences
$\textbf{S}_{\nu}, \ \nu=1,2$ using minimum-mean-square-error (MMSE) estimates of
${\textbf{S}}_{\nu}$ based on
$(\hat{\textbf{U}}_1,\hat{\textbf{V}},\hat{\textbf{U}}_2)$, i.e.,
\begin{IEEEeqnarray}{rCl}
\hat{\textbf{S}}_1 & = &
\gamma_{1,1}\hat{\textbf{U}}_1+\gamma_{1,2}\hat{\textbf{U}}_2+\gamma_{1,3}\hat{\textbf{V}}
\thickapprox {\sf{E}}\left[\textbf{S}_1\bigl|\hat{\textbf{U}}_1,\hat{\textbf{V}},\hat{\textbf{U}}_2\right] \nonumber \\
\hat{\textbf{S}}_2 & = &
\gamma_{2,1}\hat{\textbf{U}}_1+\gamma_{2,2}\hat{\textbf{U}}_2+\gamma_{2,3}\hat{\textbf{V}}
\thickapprox {\sf{E}}\left[\textbf{S}_2 \bigl|\hat{\textbf{U}}_1,\hat{\textbf{V}},\hat{\textbf{U}}_2\right],
\label{eq:MMSEestimation1}
\end{IEEEeqnarray}
where the approximate sign is due to the assumption that $(\textbf{S}_1,\textbf{S}_2,\hat{\textbf{U}}_1,\hat{\textbf{V}},\hat{\textbf{U}}_2)$ are jointly Gaussian.
Here
\begin{IEEEeqnarray}{rCl}\label{eq:MMSE_coefficients}
\gamma_{1,1}=\gamma_{1,3}
& = & \frac{1-\rho^2(1-2^{-2R_2})}{1-\rho^2(1-2^{-2R_2})(1-2^{-2(R_1+R_\textnormal{c})})}
\nonumber \\
\gamma_{1,2}
& = & \frac{\rho2^{-2(R_1+R_\textnormal{c})}}{1-\rho^2(1-2^{-2R_2})(1-2^{-2(R_1+R_\textnormal{c})})}
\nonumber \\
\gamma_{2,1}=\gamma_{2,3}
& = & \frac{\rho2^{-2R_2}}{1-\rho^2(1-2^{-2R_2})(1-2^{-2(R_1+R_\textnormal{c})})}
\nonumber \\
\gamma_{2,2}
& = & \frac{1-\rho^2(1-2^{-2(R_1+R_\textnormal{c})})}{1-\rho^2(1-2^{-2R_2})(1-2^{-2(R_1+R_\textnormal{c})})}
\end{IEEEeqnarray}
are the coefficients of the linear MMSE estimators of ${\textbf{S}}_{\nu}$ given $(\hat{\textbf{U}}_1,\hat{\textbf{V}},\hat{\textbf{U}}_2)$.
In Lemma~\ref{MMSE_coefficients_bounds} (in the Appendix) we prove that
\begin{equation}\label{eq:MMSE_coefficients2}
0<\gamma_{1,1},\gamma_{1,3},\gamma_{2,2}\leq 1 \ \ \mbox{  and  }\ \ 0<\gamma_{1,2},\gamma_{2,1},\gamma_{2,3}\leq \rho.
\end{equation}
A detailed description of the scheme is given in Section \ref{Proof of VQ scheme}.

{\bf
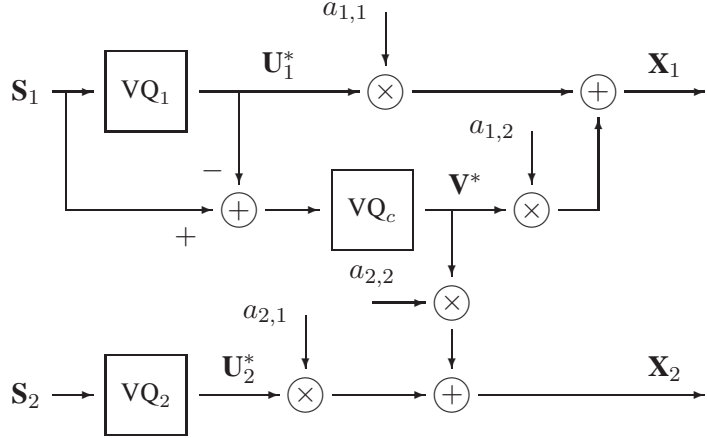
\begin{figure}
\centering \setlength{\unitlength}{0.35mm}
\begin{picture}(360,175)(10,10)



\put(40,130){\makebox(0,0){$\textbf{S}_1$}}
\put(50,130){\vector(1,0){15}} \put(70,115){\framebox(30,30){}}
\put(85,130){\makebox(0,0){\small $\mbox{VQ}_1$}}
\put(135,140){\makebox(0,0){$\textbf{U}_1^{*}$}}
\put(105,130){\vector(1,0){60}}

\put(185,130){\vector(1,0){60}}

\put(175,130){\circle{13}} \put(175,130){\makebox(0,0){$\times$}}
\put(160,160){\makebox(0,0){$a_{1,1}$}}
\put(175,160){\vector(0,-1){20}}

\put(255,130){\circle{13}} \put(255,130){\makebox(0,0){$+$}}
\put(265,130){\vector(1,0){30}}
\put(280,140){\makebox(0,0){$\textbf{X}_1$}}

\put(55,130){\line(0,-1){45}} \put(120,130){\vector(0,-1){35}}
\put(120,85){\circle{13}} \put(120,85){\makebox(0,0){$+$}}

\put(110,100){\makebox(0,0){$-$}} \put(100,75){\makebox(0,0){$+$}}

\put(55,85){\vector(1,0){55}} \put(130,85){\vector(1,0){20}}

\put(155,70){\framebox(30,30){}} \put(170,85){\makebox(0,0){\small $\mbox{VQ}_c$}}
\put(190,85){\vector(1,0){30}}
\put(205,95){\makebox(0,0){$\textbf{V}^{*}$}}

\put(230,85){\circle{13}} \put(230,85){\makebox(0,0){$\times$}}

\put(240,85){\line(1,0){15}} \put(255,85){\vector(0,1){35}}
\put(230,115){\vector(0,-1){20}}
\put(215,115){\makebox(0,0){$a_{1,2}$}}

\put(200,85){\vector(0,-1){25}} \put(200,50){\circle{13}}
\put(200,50){\makebox(0,0){$\times$}}
\put(200,40){\vector(0,-1){15}}

\put(170,50){\vector(1,0){20}}
\put(170,60){\makebox(0,0){$a_{2,2}$}}

\put(200,15){\circle{13}} \put(200,15){\makebox(0,0){$+$}}

\put(70,0){\framebox(30,30){}} \put(85,15){\makebox(0,0){\small $\mbox{VQ}_2$}}
\put(40,15){\makebox(0,0){$\textbf{S}_2$}}
\put(50,15){\vector(1,0){15}} \put(105,15){\vector(1,0){30}}
\put(155,15){\vector(1,0){35}}
\put(120,25){\makebox(0,0){$\textbf{U}_2^{*}$}}

\put(145,15){\circle{13}} \put(145,15){\makebox(0,0){$\times$}}
\put(145,45){\vector(0,-1){20}}
\put(130,45){\makebox(0,0){$a_{2,1}$}}

\put(210,15){\vector(1,0){85}}
\put(280,25){\makebox(0,0){$\textbf{X}_2$}}


\end{picture}
\caption{The vector-quantizer flow} \label{fig:1vq}
\end{figure}
}

\medskip

The distortion pairs achieved by this vector-quantizer (VQ) scheme are
described in the next theorem.


\begin{theorem}\label{th:vector-quantizer}
The distortions achieved by the vector-quantizer scheme are all
pairs $(D_1,D_2)$ satisfying
\begin{IEEEeqnarray}{rCl}
D_1 & > & \sigma^22^{-2(R_1+R_\textnormal{c})} \frac{1-\rho^2(1-2^{-2R_2})}{1-\rho^2( 1-2^{-2R_2})(1-2^{-2(R_1+R_\textnormal{c})})}  \nonumber \\
D_2 & > & \sigma^22^{-2R_2}\frac{1-\rho^2(1-2^{-2(R_1+R_\textnormal{c})})}
{1-\rho^2(1-2^{-2R_2})(1-2^{-2(R_1+R_\textnormal{c})})}
\label{eq:distortions1}
\end{IEEEeqnarray}
where, for some $0\leq\beta_1,\beta_2\leq 1$, the rate-triple
$(R_1,R_2,R_\textnormal{c})$ satisfies
\begin{IEEEeqnarray}{rCl}
R_{1} &<&  \frac{1}{2}\log
\left(\frac{\bar{\beta}_1P_1(1-\tilde{\rho}^2-\bar{\rho}^2)+N(1-\bar{\rho}^2)}
 {N(1-\tilde{\rho}^2-\bar{\rho}^2)} \right) \nonumber \\
R_{2}  &<&  \frac{1}{2}\log
\left(\frac{\bar{\beta}_2P_2(1-\tilde{\rho}^2-\bar{\rho}^2)+N}
 {N(1-\tilde{\rho}^2-\bar{\rho}^2)+\lambda_2} \right) \nonumber \\
R_\textnormal{c}  &<&  \frac{1}{2}\log
\left(\frac{\eta^2(1-\tilde{\rho}^2-\bar{\rho}^2)+N(1-\tilde{\rho}^2)}
 {N(1-\tilde{\rho}^2-\bar{\rho}^2)+\lambda_c} \right) \nonumber \\
R_1+R_2  &<&  \frac{1}{2}\log \left(
\frac{\lambda_{12}-\bar{\beta}_2P_2\bar{\rho}^2+N}
{(1-\bar{\beta}_2P_2\bar{\rho}^2\lambda_{12}^{-1})N(1-\tilde{\rho}^2)}
\right) \nonumber \\
R_1+R_\textnormal{c}  &<&  \frac{1}{2}\log \left( \frac{(\lambda_{1c}+N)(\bar{\beta}_1P_1+\eta^2)} {\lambda_{1c}N} \right)
\nonumber \\
R_2+R_\textnormal{c}  &<&  \frac{1}{2}\log \left(
\frac{\lambda_{2c}-\bar{\beta}_2P_2\tilde{\rho}^2+N}
{(1-\bar{\beta}_2P_2\tilde{\rho}^2\lambda_{2c}^{-1})N\left( 1-{{{\bar{\rho }}}^{2}}\  \right)} \right)
\nonumber \\
R_1+R_2+R_\textnormal{c}  &<&  \frac{1}{2}\log \left( \frac{\lambda_{12}+2\eta
\bar{\rho}\sqrt{\bar{\beta}_2P_2}
 +\eta^2+N}{N(1-\tilde{\rho }^2)(1-\bar{\rho}^2)} \right) \nonumber \\
C_{12} & > & R_\textnormal{c}+\frac{1}{2}\log \left( 1-{{\rho
}^{2}}{{2}^{-2{{R}_{1}}}}\left( 1-{{2}^{-2{{R}_\textnormal{c}}}} \right)\right)
\label{eq:rateconstraints1}
\end{IEEEeqnarray}
and where
\begin{IEEEeqnarray}{rCl}
\tilde{\rho} & \triangleq & \rho\sqrt{(1-2^{-2R_1})(1-2^{-2R_2})} \nonumber \\
\bar{\rho} & \triangleq & \rho\sqrt{2^{-2R_1}(1-2^{-2R_2})(1-2^{-2R_\textnormal{c}})} \nonumber \\
\lambda_2 & \triangleq &
\frac{N^2\bar{\rho}^2\tilde{\rho}^2(2+\tilde{\rho}^2)}{\beta_2P_2(1-\tilde{\rho}^2-\bar{\rho}^2)+N}
\nonumber \\
\eta & \triangleq & \sigma_v(\sqrt{\beta_1
P_1}\sigma^{-1}_v+a_{2,2})
\nonumber \\
\lambda_c & \triangleq &
\frac{N^2\bar{\rho}^2(\bar{\rho}^2\bar{\beta}_1P_1-\tilde{\rho}^2\sigma^2_v)}
{\sigma^2_v(\eta^2(1-\tilde{\rho}^2-\bar{\rho}^2)+N(1-\tilde{\rho}^2))}
\nonumber \\
\lambda_{12} & \triangleq & \bar{\beta}_1P_1
 +2\tilde{\rho}\sqrt{\bar{\beta}_1\bar{\beta}_2P_1P_2}+\bar{\beta}_2P_2
\nonumber \\
\lambda_{1c} & \triangleq &
\bar{\beta}_1P_1(1-\tilde{\rho}^2)+\eta^2(1-\bar{\rho}^2)
 -2\eta\sigma^{-1}_v\bar{\rho}^2\sqrt{\bar{\beta}_1P_1\sigma^2(1-2^{-2R_1})} \nonumber \\
\lambda_{2c} & \triangleq & \bar{\beta}_2P_2+2\eta\bar{\rho}
\sqrt{\bar{\beta}_2P_2}+\eta^2 . \label{eq:constants}
\end{IEEEeqnarray}
\end{theorem}
\begin{IEEEproof}
See Section \ref{Proof of VQ scheme}.
\end{IEEEproof}

\medskip

\begin{remark}
The substitution of $C_{12}=0$ in Theorem~\ref{th:vector-quantizer} (which then implies
$R_\textnormal{c}=0$, as well as $\bar{\beta}_1=\bar{\beta}_2=1$ as per \eqref{eq:inputX1} and \eqref{eq:inputX2}
based on the code-construction in Section~\ref{Proof of VQ scheme}.A) recovers the
Lapidoth-Tinguely achievable rate-distortion region
\cite[Theorem~IV.4]{Stephan}.
\end{remark}

\medskip

Based on Theorem~\ref{th:vector-quantizer} we now present sufficient conditions for the
achievability of  $(D_1,D_2)$ when $C_{12}=\infty$.

\begin{corollary}\label{cor:VQ with unlimited capacity}
When $C_{12}$ is unlimited, the distortions achieved by the
vector-quantizer scheme are all pairs $(D_1,D_2)$ satisfying
\begin{IEEEeqnarray*}{rCl}
D_1 & > & \sigma^22^{-2R_\textnormal{c}} \frac{1-\rho^2(1-2^{-2R_2})}{1-\hat{\rho}^2} \nonumber \\
D_2 & > & \sigma^22^{-2R_2}\frac{1-\rho^2(1-2^{-2R_\textnormal{c}})}
{1-\hat{\rho}^2}
\end{IEEEeqnarray*}
where, for some $0\leq\beta\leq 1$, the rate-pair $(R_2,R_\textnormal{c})$
satisfies
\begin{IEEEeqnarray}{rCl}
R_{2} & < & \frac{1}{2}\log
\left(\frac{\bar{\beta}P_2(1-\hat{\rho}^2)+N}
 {N(1-\hat{\rho}^2)} \right) \nonumber \\
R_\textnormal{c} & < & \frac{1}{2}\log
\left(\frac{\delta_1^2(1-\hat{\rho}^2)+N}
 {N(1-\hat{\rho}^2)} \right) \nonumber \\
R_2+R_\textnormal{c} & < & \frac{1}{2}\log \left( \frac{\delta_2+N}
{N(1-\hat{\rho }^2)} \right),
\end{IEEEeqnarray}
and where
\begin{IEEEeqnarray}{rCl}
\hat{\rho} & \triangleq & \rho\sqrt{(1-2^{-2R_2})(1-2^{-2R_\textnormal{c}})} \nonumber \\
\delta_1 & \triangleq &
\sqrt{P_1}+\sqrt{P_2}\left(\sqrt{\bar{\beta}\hat{\rho}^2+\beta}-\sqrt{\bar{\beta}\hat{\rho}^2}\right)
\nonumber \\
\delta_2 & \triangleq & 
 P_1+P_2+2\sqrt{(\bar{\beta}\hat{\rho}^2+\beta)P_1P_2}.
\label{eq:constants11}
\end{IEEEeqnarray}
\end{corollary}

\begin{remark}
For the achievability of the distortion pairs in Corollary~\ref{cor:VQ with unlimited capacity},
it suffices that
$R_\textnormal{c} +\frac{1}{2}\log \left( 1-\rho^2\left( 1-2^{-2R_\textnormal{c}} \right)
\right)  \leq C_{12} $.
\end{remark}

\medskip

\begin{figure}
\centering

\includegraphics[width=0.50\textwidth]{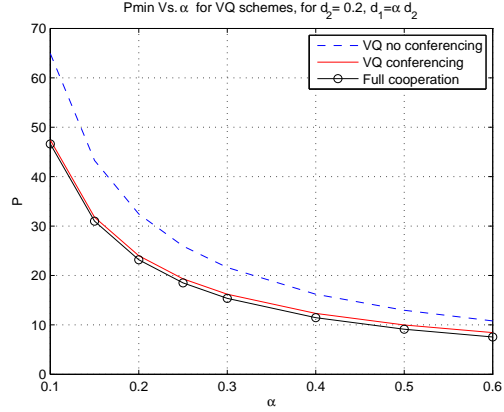}
\caption{$P_{min}$ for VQ when $d_2=0.2,d_1=\alpha d_2,\rho=0.5,N=1$}\label{fig:VQ-LT}
\end{figure}

To demonstrate the benefit of conferencing for the VQ scheme we compare the performance of the VQ with unlimited conferencing capacity
to the performance of the VQ without conferencing (i.e. the VQ in the Lapidoth-Tinguely MAC model).
We fix $d_2$ and let $d_1=\alpha d_2$ and assume that the encoders are subject to symmetric average-power constraints.
Fig.~\ref{fig:VQ-LT} compares the required average-power for the VQ with unlimited conferencing capacity and without
conferencing, for attaining a desired distortion pair $(\alpha d_2,d_2)$.
The figure displays also the minimum required power for attaining the desired distortions
when $(\textbf{S}_1,\textbf{S}_2)$ is available at both encoders hence they can fully cooperate
in the source description and therefore $R_{\textbf{S}_1,\textbf{S}_2}(\alpha d_2,d_2)=\frac{1}{2}\log{(1+\frac{4P}{N})}$.

\subsection{Source-Channel Separation}

Next, we compare the performance of our vector-quantizer scheme with the performance of
two optional source-channel separation schemes, for the case of unlimited conferencing capacity.

\medskip

\subsubsection{Source-Channel Separation Scheme~1}

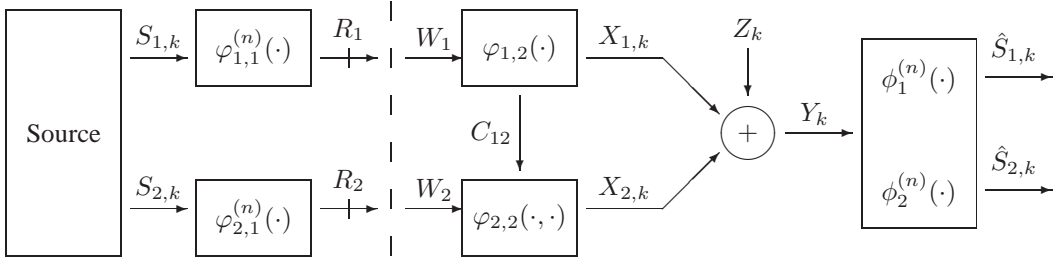
\begin{figure}
\centering\setlength{\unitlength}{0.25mm}
\begin{picture}(550,150)(10,10)
 \linethickness{0.1mm}


\put(390,80){\circle{30}} \put(390,80){\makebox(0,0){\small $+$}}
\put(390,125){\vector(0,-1){25}} \put(390,135){\makebox(0,0){\small
$Z_k$}} \put(425,90){\makebox(0,0){\small $Y_k$}}
\put(410,80){\vector(1,0){35}}

\put(100,105){\framebox(60,40){}} \put(130,125){\makebox(0,0){\small
$\varphi_{1,1}^{(n)}(\cdot)$}} \put(100,15){\framebox(60,40){}}
\put(130,35){\makebox(0,0){\small $\varphi_{2,1}^{(n)}(\cdot)$}}

\put(345,120){\vector(1,-1){30}} \put(345,40){\vector(1,1){30}}

\put(270,100){\vector(0,-1){40}}
\put(255,80){\makebox(0,0){\small $C_{12}$}}

\put(240,105){\framebox(60,40){}} \put(270,125){\makebox(0,0){\small
$\varphi_{1,2}(\cdot)$}} \put(240,15){\framebox(60,40){}}
\put(270,35){\makebox(0,0){\small $\varphi_{2,2}(\cdot,\cdot)$}}

\put(305,120){\line(1,0){40}} \put(325,130){\makebox(0,0){\small
$X_{1,k}$}} \put(305,40){\line(1,0){40}}
\put(325,50){\makebox(0,0){\small $X_{2,k}$}}

\put(450,30){\framebox(60,100){}}
\put(480,110){\makebox(0,0){\small $\phi_1^{(n)}(\cdot)$}}
\put(480,50){\makebox(0,0){\small $\phi_2^{(n)}(\cdot)$}}

\put(0,15){\framebox(60,130){}} \put(30,80){\makebox(0,0){\small
Source}}

\put(80,130){\makebox(0,0){\small $S_{1,k}$}}
\put(65,120){\vector(1,0){30}} \put(80,50){\makebox(0,0){\small
$S_{2,k}$}} \put(65,40){\vector(1,0){30}}

\put(180,135){\makebox(0,0){\small $R_1$}}
\put(165,120){\vector(1,0){30}} \put(180,55){\makebox(0,0){\small
$R_2$}} \put(165,40){\vector(1,0){30}} \put(180,115){\line(0,1){10}}
\put(180,35){\line(0,1){10}}

\put(225,130){\makebox(0,0){\small $W_1$}}
\put(210,120){\vector(1,0){30}} \put(225,50){\makebox(0,0){\small
$W_2$}} \put(210,40){\vector(1,0){30}}

\put(515,110){\vector(1,0){35}} \put(530,125){\makebox(0,0){\small
$\hat{S}_{1,k}$}}

\put(515,50){\vector(1,0){35}} \put(530,65){\makebox(0,0){\small
$\hat{S}_{2,k}$}}

\multiput(202,15)(0,25){6}{\line(0,1){10}}
\end{picture}

\caption{Separation scheme~1; Gaussian two-encoder
source-coding combined with Gaussian MAC with unidirectional
conferencing channel-coding}\label{fig:SC1}
\end{figure}

We consider the set of distortion pairs that are achieved by
combining the optimal scheme for the source-coding problem without
conferencing with the optimal scheme for the
channel-coding problem with unidirectional conferencing, as shown
in Fig.~\ref{fig:SC1}.

The rate-distortion region associated with the
source-coding problem can be found in \cite{Oohama,Wagner} and is
described as follows.

\begin{proposition}\cite{Oohama,Wagner}\label{proposition_SC separation1}
A distortion pair $(d_1,d_2)$ is achievable for the Gaussian two-terminal source-coding problem
if, and only if, $(R_1,R_2)\in {\cal R}(d_1,d_2)$ where
\begin{IEEEeqnarray*}{rCl,rCl}
{\cal R}(d_1,d_2) =  \Biggl\{(R_1,R_2)\colon
&&& R_1 & \ge & \frac{1}{2}\log_2^{+}\left[\frac{1-\rho^2\bigl(1-2^{-2R_2}\bigr)}{d_1}\right]
\nonumber \\
&&& R_2 & \ge & \frac{1}{2}\log_2^{+}\left[\frac{1-\rho^2\bigl(1-2^{-2R_1}\bigr)}{d_2}\right]
\nonumber \\
&&& R_1+R_2 & \ge & \frac{1}{2}\log_2^{+}\left[\frac{\bigl(1-\rho^2\bigr) \> \gamma(d_1,d_2)}{2d_1d_2}\right]\Biggr\},
\end{IEEEeqnarray*}
with $\gamma(d_1,d_2)=1+\sqrt{1+\frac{4\rho^2d_1d_2}{(1-\rho^2)^2}}$ and $\log_2^{+}[x]=\max \{0,\log_2(x)\}$.
\end{proposition}

\medskip

\medskip

The distortion pairs achievable by source-channel separation
follow now by combining the latter set of rate pairs with the
capacity region of the Gaussian MAC with unidirectional conference
link reported in \cite{Michelle}, which for $C_{12}=\infty$, is
expressed by
\begin{IEEEeqnarray}{rCl,rCl}\label{eq:unlimited conf capacity}
{\cal C}  =  \bigcup_{0\leq \beta\leq 1}
\biggl\{(R_1,R_2)\colon 
&&& R_2 & \le & \frac{1}{2}\log_2\left(1+\bar{\beta} P_2/N\right)
\nonumber \\
&&& R_1+R_2 & \le & \frac{1}{2}\log_2\left(1+(P_1+P_2+2\sqrt{\beta P_1P_2})/N\right)
\biggr\}.
\end{IEEEeqnarray}
Note that, by \cite[Theorem~1]{DeBru}, when $C_{12}=\infty$ source-channel separation is optimal
for lossless transmission of both sources and by \cite{Deniz} source-channel separation is optimal
also when $d_1=0$ and $d_2>0$.


%
%
%

\medskip

\begin{figure}
\centering

\includegraphics[width=0.450\textwidth]{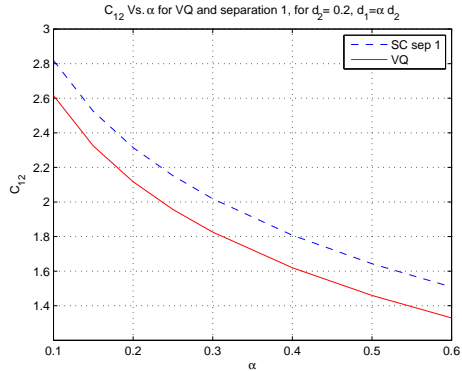}
\caption{$C_{12}$ for VQ and SC sep~1 when $d_2=0.2,d_1=\alpha d_2,\rho=0.5,N=1$}\label{fig:SC1 graph}
\end{figure}

Next, we compare the performance of the vector-quantizer scheme, with that of source-channel separation scheme~1,
for lossy trasmission.
We fix $d_2=0.2$ and let $d_1=\alpha d_2$.
In addition, we assume that the encoders are subject to symmetric average-power constraints.
Fig.~\ref{fig:SC1 graph} shows the required conferencing capacity for the VQ and for separation scheme~1,
for attaining a desired distortion pair $(\alpha d_2,d_2)$ (The figure uses the shorthand notation SC for source-channel).
While both schemes require the same average-power, the VQ requires a smaller conferencing capacity.

\medskip

For the set of distortion pairs $(d_1<1,d_2=1)$ we can show analytically that the
VQ scheme outperforms separation scheme~1 in the required conferencing rate.

For separation scheme~1, by choosing $R_2=0$ we obtain the following bounds on $R_1$,
\begin{itemize}
\item Source coding: $R_1\geq \frac{1}{2}\log \frac{1}{d_1}$.
\item Channel coding: $R_1\leq \frac{1}{2}\log\left(1+ \frac{4P}{N}\right)$,
\end{itemize}
where
$R_1\leq C_{12}$.

On the other hand, for the VQ scheme by choosing $R_2=0$ we obtain the following bounds on $R_\textnormal{c}$
(which plays the role of $R_1$ in separation scheme~1),
\begin{itemize}
\item $R_\textnormal{c}\geq \frac{1}{2}\log \frac{1}{d_1}$.
\item $R_\textnormal{c}\leq \frac{1}{2}\log\left(1+ \frac{4P}{N}\right)$,
\end{itemize}
where
$R_\textnormal{c}+1/2\log\left[1-\rho^2(1-2^{-2R_\textnormal{c}})\right]\leq C_{12}$.

\medskip

\medskip

Moreover, in this special case, the $C_{12}$ versus $(P,d_1)$ tradeoff of the VQ is optimal as can be argued as follows:
\begin{itemize}
\item Over a point-to-point channel with average power $4P$ quantizing the source at
the channel capacity rate attains the minimal distortion $d_1=\frac{N}{4P+N}$.
\item The Wyner-Ziv  (WZ) \cite{WZ} rate for the Gaussian WZ problem, coincides with our lower bound on $C_{12}$:
\begin{IEEEeqnarray*}{rCl}
R_{\textnormal{WZ}}(d_1) & = & 1/2\log\left[(1-\rho^2)/d_1+\rho^2\right]  \nonumber \\
  & = & R_\textnormal{c}+1/2\log\left[1-\rho^2(1-2^{-2R_\textnormal{c}})\right].
\end{IEEEeqnarray*}
\end{itemize}

\subsubsection{Source-Channel Separation Scheme~2}

\medskip

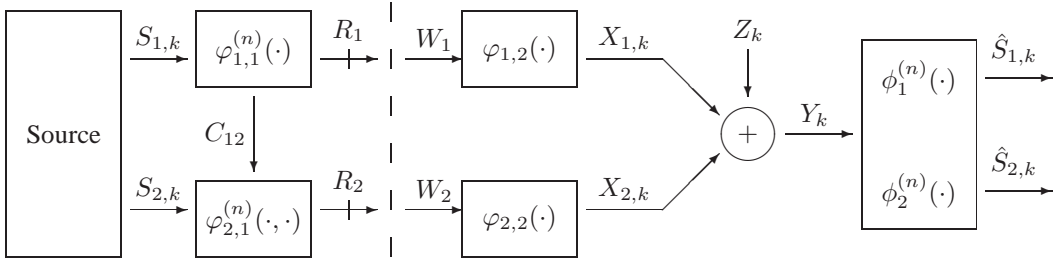
\begin{figure}
\centering\setlength{\unitlength}{0.25mm}
\begin{picture}(550,150)(10,10)
 \linethickness{0.1mm}


\put(390,80){\circle{30}} \put(390,80){\makebox(0,0){\small $+$}}
\put(390,125){\vector(0,-1){25}} \put(390,135){\makebox(0,0){\small
$Z_k$}} \put(425,90){\makebox(0,0){\small $Y_k$}}
\put(410,80){\vector(1,0){35}}

\put(130,100){\vector(0,-1){40}}
\put(115,80){\makebox(0,0){\small $C_{12}$}}

\put(100,105){\framebox(60,40){}} \put(130,125){\makebox(0,0){\small
$\varphi_{1,1}^{(n)}(\cdot)$}} \put(100,15){\framebox(60,40){}}
\put(130,35){\makebox(0,0){\small
$\varphi_{2,1}^{(n)}(\cdot,\cdot)$}}

\put(345,120){\vector(1,-1){30}} \put(345,40){\vector(1,1){30}}

\put(240,105){\framebox(60,40){}} \put(270,125){\makebox(0,0){\small
$\varphi_{1,2}(\cdot)$}} \put(240,15){\framebox(60,40){}}
\put(270,35){\makebox(0,0){\small $\varphi_{2,2}(\cdot)$}}

\put(305,120){\line(1,0){40}} \put(325,130){\makebox(0,0){\small
$X_{1,k}$}} \put(305,40){\line(1,0){40}}
\put(325,50){\makebox(0,0){\small $X_{2,k}$}}

\put(450,30){\framebox(60,100){}}
\put(480,110){\makebox(0,0){\small $\phi_1^{(n)}(\cdot)$}}
\put(480,50){\makebox(0,0){\small $\phi_2^{(n)}(\cdot)$}}

\put(0,15){\framebox(60,130){}} \put(30,80){\makebox(0,0){\small
Source}}

\put(80,130){\makebox(0,0){\small $S_{1,k}$}}
\put(65,120){\vector(1,0){30}} \put(80,50){\makebox(0,0){\small
$S_{2,k}$}} \put(65,40){\vector(1,0){30}}

\put(180,135){\makebox(0,0){\small $R_1$}}
\put(165,120){\vector(1,0){30}} \put(180,55){\makebox(0,0){\small
$R_2$}} \put(165,40){\vector(1,0){30}} \put(180,115){\line(0,1){10}}
\put(180,35){\line(0,1){10}}

\put(225,130){\makebox(0,0){\small $W_1$}}
\put(210,120){\vector(1,0){30}} \put(225,50){\makebox(0,0){\small
$W_2$}} \put(210,40){\vector(1,0){30}}

\put(515,110){\vector(1,0){35}} \put(530,125){\makebox(0,0){\small
$\hat{S}_{1,k}$}}

\put(515,50){\vector(1,0){35}} \put(530,65){\makebox(0,0){\small
$\hat{S}_{2,k}$}}

\multiput(202,15)(0,25){6}{\line(0,1){10}}
\end{picture}

\caption{Separation scheme~2; Gaussian two-encoder with
unidirectional conferencing source-coding combined with Gaussian MAC
channel-coding}\label{fig:SC2}
\end{figure}

We consider next the set of distortion pairs that are achieved by
combining an achievable rate-distortion scheme for the source-coding
problem with unidirectional conference link, with the optimal scheme
for the channel-coding problem without conferencing,  as shown in Fig.~\ref{fig:SC2}.
An achievable rate-distortion region for the
source-coding problem with unidirectional conference link can be
found in \cite[Theorem 5.1]{Kaspi} \ (for the open switch problem) and
is described as follows. Let ${\cal P}(D_1,D_2)$ be the set of all
triples of random variables $(U,V,W)$ jointly distributed with
$(S_1,S_2)$ such that
\begin{enumerate}
\item $U \Markov (S_2,W)\Markov (S_1,W)\Markov V$ and $W \Markov S_1\Markov S_2$
are Markov chains,
\item $\sigma^{2}_{S_1|U,V,W}\leq D_1$ , $\sigma^{2}_{S_2|U,V,W}\leq D_2$.
\end{enumerate}
Furthermore, define
\begin{IEEEeqnarray*}{rCl,rCl}
{\cal R}^{(\textnormal{in})}{(D_1,D_2)}  =  \bigcup_{(U,V,W)\in {\cal P}(D_1,D_2)}
\bigl\{(R_1,R_2)\colon
&&& C_{12} & \ge & I(S_1;W|S_2)
\nonumber \\
&&& R_1 &\ge & I(S_1;V|U,W)
\nonumber \\
&&& R_2 &\ge & I(S_2;U|V,W)
\nonumber \\
&&& R_1+R_2 & \ge & I(S_1,S_2;U,V,W)\bigr\}.
\end{IEEEeqnarray*}

\begin{proposition}\cite[Theorem 5.1]{Kaspi}
${\cal R}^{(\textnormal{in})}{(D_1,D_2)}$ is contained within the rate-distortion region
${\cal R}{(D_1,D_2)}$ for source-coding of correlated sources with
unidirectional conference link of capacity $C_{12}$.
The inner bound is tight when $S_1$ is reconstructed almost perfectly.
\end{proposition}

\medskip

\medskip

The Gaussian achievable rate-distortion region
associated with ${\cal R}^{(\textnormal{in})}{(D_1,D_2)}$ is characterized as follows.

\begin{proposition}\label{Gaussian RD with conferencing}
For a nonnegative pair  $(D_1, D_2)$,
the rate-distortion region ${\cal R}{(D_1,D_2)}$ contains the region
${\cal R}_{\textnormal{G}}{(D_1,D_2)}$ defined by

\begin{IEEEeqnarray*}{rCl,rCl}
{\cal R}_{\textnormal{G}}(D_1,D_2)  =  \bigcup_{\sigma^2_{w},\sigma^2_{u},\sigma^2_{v}}
\Biggl\{(R_1,R_2)\colon
&&& C_{12} & \ge & \frac{1}{2}\log_2\left[1+\frac{\sigma^2(1-\rho^2)}{\sigma^2_{w}}\right]
\nonumber \\
&&& R_1 & \ge & \frac{1}{2}\log_2\left[ 1+\frac{{{\sigma }^{2}}}
{\sigma^{2}_{v}}\frac{\left( 1-\rho^{2} \right)
+\frac{\sigma^{2}_{u}}{{{\sigma}^{2}}}}
{1+ \frac{{{\sigma }^{2}}\left( 1-{{\rho }^{2}} \right)}{\sigma^{2}_{w}}
+\left( \frac{\sigma^{2}_{u}}{\sigma^{2}_{w}}+\frac{\sigma^{2}_{u}}{\sigma^{2}} \right)} \right]
\nonumber \\
&&& R_2 & \ge & \frac{1}{2}\log_2 \left[ 1+\frac{{{\sigma }^{2}}}{\sigma^2_{u}}
\frac{ 1+\left( \frac{\sigma^{2}}{\sigma^2_{w}}+\frac{\sigma^{2}}{\sigma^2_{v}} \right)\left( 1-{{\rho }^{2}} \right) }
{1+\left( \frac{\sigma^2}{\sigma^2_{w}}+\frac{\sigma^2}{\sigma^2_{v}} \right)} \right]
\nonumber \\
&&& R_1+R_2 & \ge & \frac{1}{2}\log_2\left[\Delta \right] \nonumber \\
&&& D_1  & \leq & {{\sigma }^{2}}\frac{1+\frac{{{\sigma }^{2}}}{\sigma^2_{u}}\left( 1-{{\rho }^{2}} \right)}{\Delta}\nonumber \\
&&& D_2  & \leq &  {{\sigma }^{2}}\frac{1+{{\sigma }^{2}}\left( 1-{{\rho }^{2}} \right)\left( \frac{1}{\sigma^2_{v}}+\frac{1}{\sigma^2_{w}} \right)}{\Delta}\Biggr\},
\end{IEEEeqnarray*}
\end{proposition}
with
$\Delta=1+\frac{{{\sigma }^{2}}}{\sigma^2_{u}}+{{\sigma }^{2}}\left( 1+\frac{{{\sigma }^{2}}}{\sigma^2_{u}}\left( 1-{{\rho }^{2}} \right) \right)\left( \frac{1}{\sigma^2_{v}}+\frac{1}{\sigma^2_{w}} \right)$.

\medskip

%

\begin{IEEEproof}
See Section \ref{Proof of Gaussian RD with conferencing}.
\end{IEEEproof}

\medskip

The distortion pairs achievable by this source-channel separation scheme
follow now by combining the latter set of rate pairs with the
capacity region of the Gaussian MAC without unidirectional conference
link, which is expressed by
\begin{IEEEeqnarray*}{rCl,rCl}
{\cal C}_{\textnormal{G}}(P_1,P_2)=\biggl\{(R_1,R_2)\colon
&&& R_1 & \le & \frac{1}{2}\log_2\left(1+\frac{P_1}{N}\right)
\nonumber \\
&&& R_2 & \le & \frac{1}{2}\log_2\left(1+\frac{P_2}{N}\right)
\nonumber \\
&&& R_1+R_2 & \le & \frac{1}{2}\log_2\left(1+\frac{P_1+P_2}{N}\right)
\biggr\}.
\end{IEEEeqnarray*}

\medskip

\begin{figure}
\centering

\includegraphics[width=0.450\textwidth]{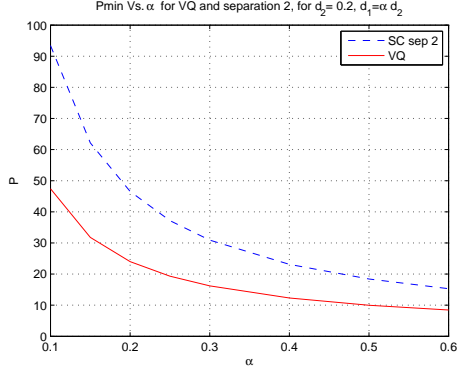}
\caption{$P_{\mbox{min}}$ for VQ and SC sep~2 when $d_2=0.2,d_1=\alpha d_2,\rho=0.5,N=1$}\label{fig:SC2 graph}
\end{figure}

We compare the performance of the source-channel separation scheme~2 inner bound with that of the vector-quantizer,
for unlimited conferencing capacity.
We fix $d_2=0.2$ and let $d_1=\alpha d_2$.
In addition, we assume that the encoders are subject to symmetric average-power constraints.
Fig.~\ref{fig:SC2 graph} compares the required average-power for attaining a desired distortion pair $(\alpha d_2,d_2)$.
We see that the VQ scheme requires less average-power than source-channel separation scheme~2 while both schemes require
the same conferencing capacity.

\medskip

\subsection{High-{\sf{SNR}} asymptotics with unlimited conferencing capacity}

We consider next the high-${\sf SNR}$ asymptotics of an optimal
scheme when the conferencing capacity is unlimited. To this end, let $(d^{*}_1,d^{*}_2)$ denote an arbitrary normalized distortion
pair resulting from an optimal scheme. For a subset of those distortion
pairs --i.e. distortion pairs satisfying $d_1d_2=O(\frac{N}{P_1+P_2})$ where $\frac{N}{P_1+P_2}\ll 1$,
the high-${\sf SNR}$ behavior is described in the following theorem.

\medskip

\begin{theorem}\label{th:highsnr1}
The high-${\sf SNR}$ asymptotics for the Gaussian MAC with unlimited
$C_{12}$ satisfies 
\begin{IEEEeqnarray}{l}
\lim_{N\to 0}
\frac{P_1+P_2+2\varrho_{\infty}^{*}\sqrt{P_1P_2}}{N}
d^{*}_1d^{*}_2=(1-\rho^2),
\label{eq:highsnrvq11}
\end{IEEEeqnarray}
provided that $d^{*}_1\leq 1,d^{*}_2\leq 1$, and that
\begin{IEEEeqnarray}{l}
\lim_{N\to 0}\frac{N}{d^{*}_1P_1}=0 \ , \quad
\mbox{and} \quad \lim_{N\to
0}\frac{N}{d^{*}_2P_2}=0 ,
\label{eq:highsnrvq11condition00}
\end{IEEEeqnarray}
where
\begin{IEEEeqnarray}{l}
\varrho_{\infty}^{*} = \sqrt{1-\frac{N(1-\rho^2)}{d^{*}_2P_2}}.
\end{IEEEeqnarray}
\end{theorem}

\medskip

\begin{IEEEproof}
See Section \ref{Proof of highsnr1}.
\end{IEEEproof}

\medskip

\begin{corollary}\label{th:highsnr2}
The high-${\sf SNR}$ asymptotics for separation scheme 1 for the Gaussian MAC with unlimited
$C_{12}$  satisfies  
\begin{IEEEeqnarray}{l}
\lim_{N\to 0}
\frac{P_1+P_2+2\varrho_{\textnormal{sep1}}^{*} \sqrt{P_1P_2}}{N}
d_1d_2 \geq (1-\rho^2),
\label{eq:highsnrsep11}
\end{IEEEeqnarray}
provided that $d_1\leq 1,d_2\leq 1$, and that 
\begin{IEEEeqnarray}{l}
\lim_{N\to 0}\frac{N}{d_1P_1}=0 \ , \quad
\mbox{and} \quad \lim_{N\to 0}\frac{N}{d_2P_2}=0 ,
\label{eq:highsnrvq11condition}
\end{IEEEeqnarray}
where
\begin{IEEEeqnarray}{l}
\varrho_{\textnormal{sep1}}^{*} = \sqrt{1-\frac{N(1-\rho^2)}{d_2P_2}}.
\end{IEEEeqnarray}
\end{corollary}

\medskip

\begin{IEEEproof}
See Section \ref{Proof of highsnr2}.
\end{IEEEproof}

\medskip


We conclude with the following extension to \cite{Deniz} which asserts that:
\begin{corollary}\label{th:highsnr3}
For high-${\sf SNR}$ with $\rho>0$, $C_{12}=\infty$,
and $(d_1,d_2)$ such that
$\lim_{N\to 0}\frac{N}{d_1P_1}=0$, and
$\lim_{N\to 0}\frac{N}{d_2P_2}=0$,
source-channel separation scheme~1 is optimal in the sense of attaining the optimal $d_1d_2$
given the system parameters $(\rho,P_1,P_2,N)$.
\end{corollary}

\medskip

We restate Theorem~\ref{th:highsnr1} more specifically for the "semi-symmetric" case
where $P_1=P_2=P$ while $(d_1,d_2)$ satisfies \eqref{eq:highsnrvq11condition}.

\begin{corollary}\label{th:semi-symmetric}
In the "semi-symmetric" case, when $(d_1,d_2)$ satisfies \eqref{eq:highsnrvq11condition}
\begin{IEEEeqnarray}{rCl}
\lim_{P/N\gg 1}  d_1d_2
 & = & \frac{N}{2P}\cdot \frac{1-\rho^2}{1+\sqrt{1-\frac{N(1-\rho^2)}{d_2P}}} \nonumber \\
 & \approx &  \frac{N}{2P}\cdot \frac{1-\rho^2}{2-\frac{N(1-\rho^2)}{2d_2P}} .
\label{eq:semi-symmetric}
\end{IEEEeqnarray}
\end{corollary}

\medskip

{\it Discussion:}  The asymptotic correlation can be explained as follows.
Separation scheme~1, when generating the channel inputs $(\textbf{X}_{1},\textbf{X}_{2})$,
 ignores the source correlation and transmits two independent messages via Willems's
code construction for the MAC with conferencing \cite{Willems}.
As a result, the correlation between the channel inputs is $\sqrt{\beta}$, where $\beta$ is the fraction of power
that Encoder~2 transmits in coherence with Encoder~1.
In contrast, the vector-quantizer does exploit the source correlation
and exhibits additional gain due to the correlation between $\textbf{V}^{*}$ and $\textbf{U}^{*}_2$
which is reflected by the larger correlation coefficient $\sqrt{\rho^2\bar{\beta}+\beta}$.
Nevertheless, when $C_{12}=\infty$, the final maximization of both
correlation expressions, each over its admissible domain of $\beta$, yields an identical result.
This is explained by the fact that
the asymptotic product \eqref{eq:semi-symmetric} is attained via the separation scheme by the rate pair
\begin{IEEEeqnarray*}{l}
R_1 = \frac{1}{2}\log\frac{2d_2P\left(1+\sqrt{1-\frac{N(1-\rho^2)}{d_2P}}\right)}{N(1-\rho^2)}
\nonumber \\
R_2  =  \frac{1}{2}\log\frac{1-\rho^2}{d_2},
\end{IEEEeqnarray*}
and by the vector-quantizer via the rate pair
\begin{IEEEeqnarray*}{rCl}
R_\textnormal{c} & = & \frac{1}{2}\log\frac{2d_2P\left(1+\sqrt{1-\frac{N(1-\rho^2)}{d_2P}}\right)}{N(1-\rho^2)}
\nonumber \\
R_2 & = & \frac{1}{2}\log\frac{1}{d_2}.
\end{IEEEeqnarray*}
Consequently, the separation strategy which exploits the source correlation at
the source-coding part sends at a lower $R_2$ rate which in turn increases the admissible domain of $\beta$.
For separation scheme~1, $C_{12}=R_1$, while for the vector-quantizer $C_{12}=R_\textnormal{c}+\frac{1}{2}\log(1-\rho^2)$.

\medskip



Next, let us compare this asymptotic behavior to the asymptotic behavior of the vector-quantizer without conferencing,
when $(d_1,d_2)$ satisfies \eqref{eq:highsnrvq11condition}, as reported in \cite[Section IV.D]{Stephan}
\begin{equation}
\lim_{P/N\gg 1}  d_1d_2
 \approx \frac{N}{2P}\cdot \frac{1-\rho^2}{1+\rho}  .
\label{eq:semi-symmetricnoc}
\end{equation}
As noted in \cite{Stephan}, the gain $1-\rho^2$ in the numerator on the r.h.s. of  \eqref{eq:semi-symmetricnoc} is due to the fact that the
receiver exploits the source correlation in joint-typicality decoding, while the gain $1+\rho$ in the denominator is due to
the correlation $\rho$ that the encoders build on the channel inputs $(\textbf{X}_{1},\textbf{X}_{2})$.
The asymptotic expression \eqref{eq:semi-symmetric} demonstrates that with unlimited unidirectional conferencing capacity,
both the vector quantizer and separation scheme~1, exploit the source correlation---each in its own way--- and increase the
correlation on the channel inputs to $2-\frac{N(1-\rho^2)}{2d_2P}\to 2$ for $\frac{N}{d_2P}\ll 1$.

\subsection{High-{\sf{SNR}} asymptotics with fixed conferencing capacity}

We consider first the high-${\sf SNR}$ asymptotics of source-channel separation scheme~1
when the conferencing capacity $C_{12}$ is fixed.

\medskip

\begin{corollary}\label{th:highsnr31}
The high-${\sf SNR}$ asymptotics for separation scheme 1 for the Gaussian MAC with fixed
unidirectional conferencing capacity $C_{12}=C$ satisfies
\begin{IEEEeqnarray}{l}
\lim_{N\to 0}
d_1d_2 \geq \frac{N(1-{\rho}^2)}
{P_1+P_2+2\varrho^{*}_{\textnormal{sep1}}\sqrt{P_1P_2}}
\label{eq:highsnrvecquanfixed3}
\end{IEEEeqnarray}
provided that $d_1\leq 1,d_2\leq 1$ and that $(d_1,d_2)$ satisfy \eqref{eq:highsnrvq11condition}, where
\begin{IEEEeqnarray}{l}
\varrho^{*}_{\textnormal{sep1}}= \sqrt{1-\frac{N(1-\rho^2)}{d_1P_1}2^{-2C}}\sqrt{1-\frac{N(1-\rho^2)}{d_2P_2}}.
\end{IEEEeqnarray}
\end{corollary}

\medskip

\begin{IEEEproof}
See Section \ref{Proof of highsnr2 fixed}.
\end{IEEEproof}

\medskip

We consider next the high-${\sf SNR}$ asymptotics of the vector-quantizer
scheme when the conferencing capacity $C_{12}$ is fixed.

\medskip

\begin{corollary}\label{th:highsnr22}
The high-${\sf SNR}$ asymptotics for the vector-quantizer scheme for the Gaussian MAC with fixed
unidirectional conferencing capacity $C_{12}=C$ satisfies
\begin{IEEEeqnarray}{l}
\lim_{N\to 0}
d_1d_2 \geq \frac{N(1-{\rho}^2)(1-\check{\rho}^2)}
{P_1+P_2+2\varrho^{*}_{\textnormal{VQ}}\sqrt{P_1P_2}}
\label{eq:highsnrvecquanfixed1}
\end{IEEEeqnarray}
provided that $d_1\leq 1,d_2\leq 1$, and that $(d_1,d_2)$ satisfy \eqref{eq:highsnrvq11condition},
where
\begin{IEEEeqnarray}{rCl}
\check{\rho} & \triangleq & \rho\sqrt{2^{-2R_1}(1-2^{-2R_\textnormal{c}})}
\nonumber \\
R_\textnormal{c} & \leq & C_{12} - 1/2\log\left(1-\check{\rho}^2\right),
\label{eq:highsnrvecquanfixed2}
\end{IEEEeqnarray}
and
\begin{IEEEeqnarray}{l}
\varrho^{*}_{\textnormal{VQ}}  \geq \rho 2^{-C}\sqrt{\frac{N}{d_1P_1}}\sqrt{\frac{N}{d_2P_2}}+ \sqrt{1-\frac{N}{d_1P_1}2^{-2C}}\sqrt{1-\frac{N}{d_2P_2}}.
\end{IEEEeqnarray}

\end{corollary}

\medskip

\begin{IEEEproof}
See Section \ref{Proof of highsnr1 fixed}.
\end{IEEEproof}



\medskip

Next, we compare the maximum correlation that can be achieved by the two schemes when $P_1=P_2=P$.
For separation scheme~1 we obtain
\begin{IEEEeqnarray}{rCl}
\varrho^{*}_\textnormal{sep1}
& = & \sqrt{1-\frac{N(1-\rho^2)}{d_1P}2^{-2C}}\sqrt{1-\frac{N(1-\rho^2)}{d_2P}}
\nonumber \\
& \approx & \left[1-\frac{N(1-\rho^2)}{2d_1P}2^{-2C}\right]\left[1-\frac{N(1-\rho^2)}{2d_2P}\right]
\nonumber \\
& \approx & 1-\frac{N}{2P} \left(\frac{2^{-2C}}{d_1}+\frac{1}{d_2}\right) (1-\rho^2),
\label{eq:optrholimitedsep1}
\end{IEEEeqnarray}
where in both approximation steps we use that $\frac{N}{d_{\nu}P}\ll 1, \ \nu=1,2$.


For the vector-quantizer we obtain
\begin{IEEEeqnarray}{rCl}
\varrho^{*}_{\textnormal{VQ}} & \geq &  \rho 2^{-C}\sqrt{\frac{N}{d_1P}}\sqrt{\frac{N}{d_2P}}+ \sqrt{1-\frac{N}{d_1P}2^{-2C}}\sqrt{1-\frac{N}{d_2P}}
\nonumber \\
& \approx & \rho 2^{-C}\sqrt{\frac{N}{d_1P}}\sqrt{\frac{N}{d_2P}}+ \left[1-\frac{N}{2d_1P}2^{-2C}\right]
       \left[1-\frac{N}{2d_2P}\right]
\nonumber \\
& \approx & \rho 2^{-C}\sqrt{\frac{N}{d_1P}}\sqrt{\frac{N}{d_2P}}+ 1-\frac{N}{2P}  \left(\frac{2^{-2C}}{d_1}+\frac{1}{d_2}\right),
\label{eq:optrholimitedvq}
\end{IEEEeqnarray}
where in both approximation steps we use that $\frac{N}{d_{\nu}P}\ll 1, \ \nu=1,2$.

Next, let $d_2=d$ and $d_1=\alpha d$ in which case the r.h.s.  of \eqref{eq:optrholimitedsep1} yields
\begin{IEEEeqnarray}{rCl}
\varrho^{*}_\textnormal{sep1}
& = & 1-\frac{N}{2Pd} \left(\frac{2^{-2C}}{\alpha}+1\right) (1-\rho^2),
\label{eq:optrholimitedsep11}
\end{IEEEeqnarray}
while  the r.h.s. of \eqref{eq:optrholimitedvq} yields
\begin{IEEEeqnarray}{rCl}
\varrho^{*}_\textnormal{VQ}  & \geq & \rho\frac{2^{-C}}{\sqrt{\alpha}}\frac{N}{dP}+1
 -\frac{N}{2Pd}  \left(\frac{2^{-2C}}{\alpha}+1\right).
\label{eq:optrholimitedvq1}
\end{IEEEeqnarray}
The r.h.s. of \eqref{eq:optrholimitedvq1} is strictly larger than the r.h.s. of \eqref{eq:optrholimitedsep11} as long as
\begin{IEEEeqnarray}{l}
\rho   <  2\cdot \frac{2^{-C}\sqrt{\alpha}}{2^{-2C}+\alpha}.
\label{eq:compare1}
\end{IEEEeqnarray}
It is easy to verify that  $\alpha=2^{-2C}$ satisfies \eqref{eq:compare1}.

We conclude that:

\begin{corollary}\label{th:highsnr7}
With fixed conferencing capacity and symmetric average-power constraints,
for high-${\sf SNR}$, and $(d_1,d_2)$ such that
$\lim_{N\to 0}\frac{N}{d_1P}=0$ and
$\lim_{N\to 0}\frac{N}{d_2P}=0$,
separation scheme~1 is suboptimal in the sense of attaining the optimal $d_1d_2$
given the system parameters $(\rho,P,N)$.
\end{corollary}

\section{Proof of Theorem~\ref{th:necessary1}}
\label{Proof of Necessary Condition}

\begin{lemma}
For a multiple-access channel with unidirectional conferencing, let
$\{X_{1,k}\}$, $\{X_{2,k}\}$ and $\{Y_{k}\}$ be the channel inputs
and channel outputs of a coding scheme achieving a distortion pair
$(D_1,D_2)$. Then, for every $\delta>0$ there exists an
$n_0(\delta)>0$ such that for all $n>n_0(\delta)$
\begin{IEEEeqnarray}{rCl}
nR_{S_1,S_2}(D_1+\delta,D_2+\delta) &  \leq &  \sum_{k=1}^n
I(X_{1,k},X_{2,k};Y_k) \label{eq:macconf1}  \\
nR_{S_2|S_1}(D_2+\delta) &  \leq &  \sum_{k=1}^n
I(X_{2,k};Y_k|X_{1,k},U_k), \label{eq:macconf10}
\end{IEEEeqnarray}
for $\ p_{X_{1,k}X_{2,k}U_k}=p_{U_k}p_{X_{1,k}|U_k}p_{X_{2,k}|U_k}$.
\end{lemma}

\medskip

\begin{IEEEproof}
By the definition of an achievable distortion pair $(D_1,D_2)$ and
the monotonicity of $R_{S_1,S_2}(\Delta_1,\Delta_2)$ in
$(\Delta_1,\Delta_2)$, for any $\delta>0$ there exists an
$n_0(\delta)>0$ such that for every $n>n_0(\delta)$
\begin{IEEEeqnarray}{l}
nR_{S_1,S_2}(D_1+\delta,D_2+\delta)
{\leq}
I(\textbf{S}_1,\textbf{S}_2;\textbf{Y}) ,  \label{eq:macrdpf1}
\end{IEEEeqnarray}
as reported in \cite[Appendix I]{Stephan feedback}.
Next,
\begin{IEEEeqnarray}{rCl}
I(\textbf{S}_1,\textbf{S}_2;\textbf{Y}) & = &
h(\textbf{Y})-h(\textbf{Y}|\textbf{S}_1,\textbf{S}_2)
\nonumber \\
 & = & h(\textbf{Y})- \sum_{k=1}^n h(Y_k|\textbf{S}_1,\textbf{S}_2,Y^{k-1})
\nonumber \\
 & \stackrel{(a)}{=} & h(\textbf{Y})- \sum_{k=1}^n h(Y_k|\textbf{S}_1,\textbf{S}_2,Y^{k-1},W,X_{1,k},X_{2,k})
\nonumber \\
 & \stackrel{(b)}{=} & h(\textbf{Y})- \sum_{k=1}^n h(Y_k|X_{1,k},X_{2,k})
\nonumber \\
 & \leq & \sum_{k=1}^n h(Y_k)- \sum_{k=1}^n h(Y_k|X_{1,k},X_{2,k})
\nonumber \\
 & = &  \sum_{k=1}^n I(X_{1,k},X_{2,k};Y_k).   \label{eq:macrdpf2}
\end{IEEEeqnarray}
Here, $(a)$ follows since $W$ is a deterministic function of
$\textbf{S}_1$ and by the encoding relations \eqref{eq:encmappingsdef1},
while $(b)$ follows since $Y_k\Markov (X_{1,k},X_{2,k})\Markov
(\textbf{S}_1,\textbf{S}_2,Y^{k-1},W)$ is a Markov chain. The
combination of  \eqref{eq:macrdpf1} and \eqref{eq:macrdpf2}
establishes \eqref{eq:macconf1}.

\medskip

In a similar way
\begin{IEEEeqnarray}{l}
nR_{S_2|S_1}(D_2+\delta)
\leq I(\textbf{S}_2;\textbf{Y}|\textbf{S}_1),  \label{eq:macrdpf3}
\end{IEEEeqnarray}
as reported in \cite[Appendix I]{Stephan feedback}.
Next,
\begin{IEEEeqnarray}{rCl}
I(\textbf{S}_2;\textbf{Y}|\textbf{S}_1) & = &
h(\textbf{Y}|\textbf{S}_1)-h(\textbf{Y}|\textbf{S}_1,\textbf{S}_2)
\nonumber \\
 & \stackrel{(c)}{=} & h(\textbf{Y}|\textbf{S}_1,W)- \sum_{k=1}^n h(Y_k|\textbf{S}_1,\textbf{S}_2,Y^{k-1},W)
\nonumber \\
 & \stackrel{(d)}{=} & \sum_{k=1}^n h(Y_k|\textbf{S}_1,W,Y^{k-1},X_{1,k})
  - \sum_{k=1}^n h(Y_k|\textbf{S}_1,\textbf{S}_2,W,Y^{k-1},X_{1,k},X_{2,k})
\nonumber \\
 & \leq & \sum_{k=1}^n h(Y_k|\textbf{S}_1,W,X_{1,k})
  - \sum_{k=1}^n h(Y_k|\textbf{S}_1,\textbf{S}_2,W,Y^{k-1},X_{1,k},X_{2,k})
\nonumber \\
 & \stackrel{(e)}{=} & \sum_{k=1}^n h(Y_k|\textbf{S}_1,W,X_{1,k})
  - \sum_{k=1}^n h(Y_k|\textbf{S}_1,W,X_{1,k},X_{2,k})
\nonumber \\
 & = &  \sum_{k=1}^n I(X_{2,k};Y_k|X_{1,k},\textbf{S}_1,W)   \nonumber \\
 & = &  \sum_{k=1}^n I(X_{2,k};Y_k|X_{1,k},U_k).   \label{eq:macrdpf4}
\end{IEEEeqnarray}
Here, $(c)$ follows since $W$ is a deterministic function of
$\textbf{S}_1$; $(d)$ follows by the encoding relations
\eqref{eq:encmappingsdef1}; and $(e)$ follows since $Y_k\Markov
(X_{1,k},X_{2,k})\Markov (\textbf{S}_1,\textbf{S}_2,Y^{k-1},W)$ is a
Markov chain. Furthermore, in the last step we've defined
$U_k=(\textbf{S}_1,W)$ in which case, by the definition of the
encoding relation $\textbf{X}_1 = \varphi_1^{(n)}(\textbf{S}_1)$,
$U_k$ satisfies the Markov chain $X_{1,k}\Markov U_k\Markov
X_{2,k}$. The combination of \eqref{eq:macrdpf4} and
\eqref{eq:macrdpf3} establishes \eqref{eq:macconf10}.
\end{IEEEproof}

\medskip

\begin{lemma} \label{th:lem1}
For a multiple-access channel with unidirectional conferencing, let
the sequences $\{X_{1,k}\}$ and $\{X_{2,k}\}$ satisfy $\frac{1}{n}\sum_{k=1}^n
{\sf{E}}\bigl[X_{\nu,k}^2\bigr]\leq P_{\nu}, \ \nu=1,2$. Let
$Y_{k}=X_{1,k}+X_{2,k}+Z_k$, where $\{Z_{k}\}$ are IID zero-mean
variance-$N$ Gaussian, and $Z_k$ is independent of
$(X_{1,k},X_{2,k})$ for every $k$. Define
$\varrho(\mathbf{X}_1,\mathbf{X}_2)\in [0,1]$ by
\begin{IEEEeqnarray}{l}
\varrho(\mathbf{X}_1,\mathbf{X}_2) =
\frac{|\frac{1}{n}\sum_{k=1}^n{\sf{E}}\left[X_{1,k}X_{2,k}\right]|}
{\sqrt{(\frac{1}{n}\sum_{k=1}^n{\sf{E}}\bigl[X_{1,k}^2\bigr])(\frac{1}{n}\sum_{k=1}^n{\sf{E}}\bigl[X_{2,k}^2\bigr])}}.
\label{eq:macconf2}
\end{IEEEeqnarray}
Then,
\begin{IEEEeqnarray}{rCl}
 \sum_{k=1}^n I(X_{1,k},X_{2,k};Y_k) & \leq &
 \frac{n}{2}\log_2\left(1+\frac{P_1+P_2+2\varrho(\mathbf{X}_1,\mathbf{X}_2)\sqrt{P_1P_2}}{N}\right)
 \label{eq:macconf3} \\
 \sum_{k=1}^n I(X_{2,k};Y_k|X_{1,k},U_k) & \leq &
 \frac{n}{2}\log_2\left(1+\frac{\Var{X_{2}|U}}{N}\right)
\label{eq:macconf31}
\end{IEEEeqnarray}
for a Gaussian Markov triple $X_{1}\Markov U\Markov X_{2}$.
\end{lemma}

\medskip

\begin{IEEEproof}
By the Max-Entropy Theorem \cite[Theorem~11.1.1]{cov} and the fact
that the variance is always smaller than or equal to the second moment:
\begin{IEEEeqnarray*}{rCl}
I(X_1,X_2;Y) & \leq &
\frac{1}{2}\log_2\left(1+\frac{\Var{X_1+X_2}}{N}\right) \nonumber \\
 & \leq & \frac{1}{2}\log_2\left(1+\frac{{\sf{E}}\left[(X_1+X_2)^2\right]}{N}\right)  \nonumber \\
 & = & \frac{1}{2}\log_2\left(1+\frac{{\sf{E}}\left[X_1^2\right]+{\sf{E}}\left[X_2^2\right]+2{\sf{E}}\left[X_1X_2\right]}{N}\right).
\end{IEEEeqnarray*}
This step reduces the multiple access problem to the problem of transmitting the source $(\textbf{S}_1,\textbf{S}_2)$
over a point to point AWGN channel of input power constraint ${\sf{E}}\left[(X_1+X_2)^2\right]$.
The first inequality \eqref{eq:macconf3} follows now from the proof of the converse in \cite[Section~III]{Ozar} using Jensen's inequality.

\medskip

For the second inequality \eqref{eq:macconf31}, again apply the Max-Entropy Theorem
conditioned on $U=u$, and then use Jensen's inequality
\begin{IEEEeqnarray*}{rCl}
I(X_2;Y|X_1,U) & \leq & \int
\frac{1}{2}\log_2\left(1+\frac{\Var{X_2|U=u}}{N}\right){\sf d}P_U(u) \nonumber \\
 & \leq & \frac{1}{2}\log_2\left(1+\frac{\Var{X_2|U}}{N}\right).
\end{IEEEeqnarray*}
It remains to show that for evaluating the upper bound it is
sufficient to consider only Gaussian distributions. The proof
follows the same lines as the proof of the main result in
\cite{Michelle} and is omitted (see also \cite[Lemma~3.15, and Appendix
B.2]{Michelle2}).
\end{IEEEproof}

\medskip

The last step in evaluating the upper bound, follows from two results from
Maximum Correlation Theory, which are stated now.
First, we recall Witsenhausen's lemma.

\medskip

\begin{lemma} \cite[Theorem~1, p. 105]{WIT}
\label{th:witsen1}
Consider a sequence of independent (across the time) pairs of random variables $\{(X_k,Y_k)\}$
and two Borel measurable arbitrary functions $g_{1,k},g_{2,k}\colon \mathbb{R}\to \mathbb{R}$
satisfying
\begin{IEEEeqnarray*}{rCl}
{\sf{E}}\left[g_{1,k}(X_k)\right] & = & {\sf{E}}\left[g_{2,k}(Y_k)\right]=0, \nonumber \\
{\sf{E}}\Bigl[\bigl(g_{1,k}(X_k)\bigr)^2\Bigr] & = & {\sf{E}}\Bigl[\bigl(g_{2,k}(Y_k)\bigr)^2\Bigr]=1 .
\end{IEEEeqnarray*}
Define
\begin{IEEEeqnarray}{l}\label{eq:sup}
\varrho^{*} \triangleq
\sup_{\substack{g_{1,k},g_{2,k}  \\
1\le k \le n}}
{\sf{E}}\left[g_{1,k}(X_k)g_{2,k}(Y_k)\right].
\end{IEEEeqnarray}
Then, for any two Borel measurable arbitrary functions
$g^{(n)}_1,g^{(n)}_2\colon \mathbb{R}^n\to \mathbb{R}$ satisfying
\begin{IEEEeqnarray*}{rCl}
{\sf{E}}[g^{(n)}_1(\mathbf{X})] & = & {\sf{E}}[g^{(n)}_2(\mathbf{Y})]=0, \nonumber \\
{\sf{E}}\Bigl[\bigl(g^{(n)}_1(\mathbf{X})\bigr)^2\Bigr] & = & {\sf{E}}\Bigl[\bigl(g^{(n)}_2(\mathbf{Y})\bigr)^2\Bigr]=1,
\end{IEEEeqnarray*}
and for length-$n$ sequences $\mathbf{X}$ and $\mathbf{Y}$, we have
\begin{IEEEeqnarray*}{l}
 \sup_{g^{(n)}_1,g^{(n)}_2} {\sf{E}}\bigl[g^{(n)}_1(\mathbf{X})g^{(n)}_2(\mathbf{Y})\bigr]
\leq \varrho^{*}.
\end{IEEEeqnarray*}
\end{lemma}
When $\{(X_k,Y_k)\}$ is IID, we define $\varrho^{*}=\varrho(X,Y)$ where
\begin{IEEEeqnarray*}{l}
\varrho(X,Y) \triangleq \sup_{g_1,g_2} {\sf{E}}\bigl[g_1(X)g_2(Y)\bigr].
\end{IEEEeqnarray*}

The second result states that when $(X_k,Y_k)$ is a bivariate Gaussian, the supremum in \eqref{eq:sup}
is obtained by linear mappings, as stated in the following lemma.

\medskip

\begin{lemma}\cite[Lemma~10.2, p. 182]{ROZ}
\label{th:roz}
Consider two jointly Gaussian random variables $W_{1,k}$ and
$W_{2,k}$ with correlation coefficient $\rho_k$. Then,
\begin{equation*}
\sup_{g_{1,k},g_{2,k}}
{\sf{E}}\left[g_{1,k}({W}_{1,k})g_{2,k}({W}_{2,k})\right]
=|\rho_{k}|,
\end{equation*}
where the supremum is taken over all functions $g_{i,k}\colon \mathbb{R}\to \mathbb{R}$,
satisfying ${\sf{E}}\left[g_{i,k}(W_{i,k})\right]  =0$ and ${\sf{E}}\left[(g_{i,k}(W_{i,k}))^2\right] =1 , i\in \{1,2\}$.
\end{lemma}

\vskip.2truein

Using Witsenhausen's lemma we may upper-bound
$\varrho(\textbf{X}_1,\textbf{X}_2)$ as follows
\begin{IEEEeqnarray}{rCl}
\varrho(\textbf{X}_1,\textbf{X}_2) & \leq &
\sup_{\substack{\varphi_1^{(n)},\varphi_2^{(n)} \\
1\le k\le n}}
{\sf{E}}\left[\varphi_{1,k}^{(n)}(\textbf{S}_1)\varphi_{2,k}^{(n)}(\textbf{S}_2,W)\right] \nonumber \\
& = &
\sup_{\substack{\varphi_1^{(n)},\varphi_2^{(n)} \\
1\le k\le n}}
{\sf{E}}\left[\varphi_{1,k}^{(n)}(\textbf{S}_1)\varphi_{2,k}^{(n)}(\textbf{S}_2,f^{(n)}(\textbf{S}_1))\right] \nonumber \\
& \stackrel{(a)}{\leq} &
\sup_{\substack{\varphi_1^{(n)},\varphi_2^{(n)} \\
1\le k\le n}}
{\sf{E}}\left[\varphi_{1,k}^{(n)}(\textbf{S}_1)\varphi_{2,k}^{(n)}(\textbf{S}_2,\textbf{S}_1)\right] \nonumber \\
& \stackrel{(b)}{\leq} &
\sup_{\varphi_{1,k},\varphi_{2,k}}
{\sf{E}}[\varphi_{1,k}({S}_{1,k})\varphi_{2,k}({S}_{2,k},{S}_{1,k})].
\label{eq:upboundrho1}
\end{IEEEeqnarray}
Here, $(a)$ follows since $f^{(n)}\colon \mathbb{R}^n\to {\cal W}$
is a deterministic function of $\textbf{S}_1$, and $(b)$ follows
since $(\textbf{S}_1,\textbf{S}_2)$ is IID generated, hence
Lemma~\ref{th:witsen1} applies.

Next, define
\begin{IEEEeqnarray*}{l}
 \sup_{\varphi_1,\varphi_2}
{\sf{E}}\left[\varphi_{1}(S_1)\varphi_{2}(S_2,S_1)\right]\triangleq \varrho(S_1,(S_1,S_2)).
\end{IEEEeqnarray*}
Then
\begin{IEEEeqnarray}{rCll}
\varrho(S_1,(S_1,S_2)) & = & \varrho(S_1,(S_1,S_2-\rho S_1)) \nonumber \\
 & \stackrel{(c)}{=} &  \varrho(S_1,S_1)
 \stackrel{(d)}{=}   \varrho(S_1,S_1)+\varrho(S_1,S_2-\rho S_1).
\label{eq:interfere1}
\end{IEEEeqnarray}
Here,
\begin{itemize}
\item[$(c)$] follows since conditioned on $S_1$, the random variable $S_2-\rho S_1$ is independent of $S_1$
and therefore
\begin{IEEEeqnarray*}{l}
{\sf{E}}\left[\varphi(S_1)|S_1,S_2-\rho S_1\right]=
{\sf{E}}\left[\varphi(S_1)|S_1\right],
\end{IEEEeqnarray*}
in which case $(c)$ follows by the fact that if $X\Markov Y\Markov Z$ then $ \varrho(X,(Y,Z))= \varrho(X,Y)$	
(see \cite[Proof of inequality (7)]{Dembo}).

\item [$(d)$] follows since $\varrho(S_1,S_2-\rho S_1)=0$ due to the fact that $S_2-\rho S_1$ is uncorrelated with $S_1$.
\end{itemize}

\medskip

Consider the maximiziation of  $\varrho(\textbf{X}_1,\textbf{X}_2)$ subject to
the conditional rate-distortion constraint following from \eqref{eq:macconf10} and \eqref{eq:macconf31},
\begin{IEEEeqnarray}{l}
R_{S_2|S_1}(D_2)\leq \frac{1}{2}\log_2\left(1+\frac{\Var{X_{2}|U}}{N}\right).
\label{eq:macconf310}
\end{IEEEeqnarray}
\begin{itemize}
\item Recall that the upper bound on the r.h.s. of \eqref{eq:macconf310} is attained by  jointly Gaussian $(X_1,U,X_2)$.

\item Conditioned on $S_1$ the energy-distortion tradeoff for attaining $R_{S_2|S_1}(D_2)$ is achieved by uncoded
transmission of $S_2-\rho S_1$ by Encoder~2. Moreover, by \eqref{eq:interfere1}
any linear function of $S_2-\rho S_1$ which Encoder~2 transmits does not interfere
with the correlation that is built via the transmission of $S_1$ by both encoders.

\item For a jointly Gaussian $(X_1,U,X_2)$ such that $X_1\Markov U\Markov X_2$ we have $\Var{X_2|U}\leq \Var{X_2|X_1}$.

\item  We use a perturbation argument to argue that uncoded transmission of $S_1$ at both encoders maximizes $\varrho(\textbf{X}_1,\textbf{X}_2)$.
This is true since by \cite[Theorem~1]{Dembo} if $Z$ is independent of the pair $(X,Y)$ then $\varrho(X,Y+\lambda Z), \lambda\in \Reals$
is continuous at $\lambda=0$.
Suppose that Encoder~2 acquires via the conference channel the sequence $\tilde{\textbf{S}}_1$
where $\tilde{S}_{1,k}=S_{1,k}+\lambda\tilde{Z}_k, \lambda\in \Reals$ and $\{\tilde{Z}_k\}$ consists of IID zero mean variance $\tilde{N}$
Gaussians that are independent of the source sequence. By Lemmas~3 and 4 uncoded transmission of $\{{S}_{1,k}\}$ by Encoder~1 and
$\{\tilde{S}_{1,k}\}$ by Encoder~2 maximizes  $\varrho(\textbf{X}_1,\textbf{X}_2)$.
Since the rate-distortion region is a continuous function of $\lambda^2\tilde{N}$,
the limit of the rate-distortion region attained by the above strategy as $\lambda\to 0$ converges to the solution to the constrained maximum of \eqref{eq:upboundrho1}.
\end{itemize}

\medskip

\begin{figure}
\centering\setlength{\unitlength}{0.25mm}
\begin{picture}(500,180)(10,10)
 \linethickness{0.1mm}


\put(290,80){\circle{30}} \put(290,80){\makebox(0,0){\small $+$}}
\put(290,125){\vector(0,-1){25}} \put(290,135){\makebox(0,0){\small
$Z_k$}} \put(325,90){\makebox(0,0){\small $Y_k$}}
\put(310,80){\vector(1,0){35}}

\put(100,155){\framebox(70,40){}} \put(135,175){\makebox(0,0){\small
$\varphi_1^{(n)}(\cdot)$}}

\put(100,15){\framebox(70,40){}}
\put(135,35){\makebox(0,0){\small $\varphi_{2,2}^{(n)}(\cdot)$}}

\put(215,170){\vector(3,-4){60}}
\put(215,100){\vector(3,-1){60}}

\put(215,40){\vector(2,1){60}}

\put(100,75){\framebox(70,40){}}
\put(135,95){\makebox(0,0){\small $\varphi_{2,1}^{(n)}(\cdot)$}}

\put(175,170){\line(1,0){40}} \put(195,180){\makebox(0,0){\small $X_{1,k}$}}
\put(175,40){\line(1,0){40}}
\put(195,50){\makebox(0,0){\small $\tilde{\tilde{X}}_{2,k}$}}

\put(175,100){\line(1,0){40}}
\put(195,110){\makebox(0,0){\small ${\tilde{X}}_{2,k}$}}

\put(350,30){\framebox(60,100){}}
\put(380,110){\makebox(0,0){\small $\phi_1^{(n)}(\cdot)$}}
\put(380,50){\makebox(0,0){\small $\phi_2^{(n)}(\cdot)$}}


\put(80,100){\makebox(0,0){\small $S_{1,k}$}}
\put(65,90){\vector(1,0){30}}
\put(80,180){\makebox(0,0){\small $S_{1,k}$}}
\put(65,170){\vector(1,0){30}}
\put(55,50){\makebox(0,0){\small $S_{2,k}-\rho S_{1,k}$}}
\put(25,40){\vector(1,0){70}}

\put(415,110){\vector(1,0){35}} \put(430,125){\makebox(0,0){\small
$\hat{S}_{1,k}$}}

\put(415,50){\vector(1,0){35}} \put(430,65){\makebox(0,0){\small
$\hat{S}_{2,k}$}}
\end{picture}

\caption{Decomposition of Encoder~2 for evaluation of the maximum correlation with
$C_{12}=\infty$}\label{fig:outb1}
\end{figure}
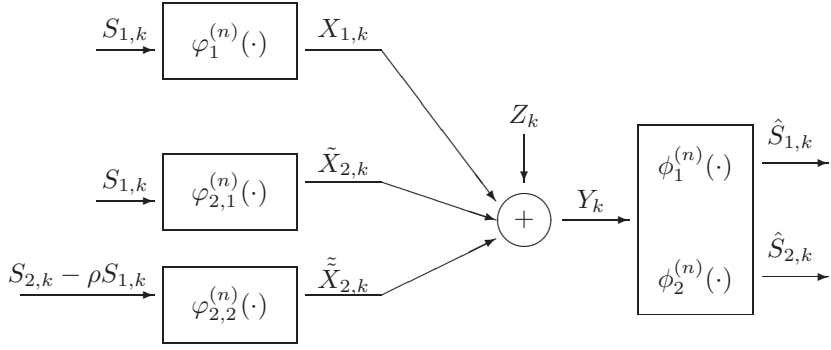

Consequently, for the solution to the constrained maximum of \eqref{eq:upboundrho1} we may assume that
Encoder~2 is split into two separate sub-encoders, with respective inputs $(S_1,S_2-\rho S_1)$,
respective outputs $(\tilde{X}_{2,k},\tilde{\tilde{X}}_{2,k})$ which are linear functions of the inputs and aggregate normalized power constraint
\begin{IEEEeqnarray*}{l}
\frac {1}{n}{\sf{E}} \left[ \sum_{k=1}^n (\tilde{X}_{2,k}+\tilde{\tilde{X}}_{2,k})^2\right] \leq 1.
\end{IEEEeqnarray*}
This decomposition is shown in Fig.~\ref{fig:outb1}.

\medskip

Thus, for $0\le \beta \le 1$, consider the linear mappings
\begin{IEEEeqnarray*}{rCl}
\varphi_1({S}_1) & = & \frac{1}{\sigma}S_1 \nonumber \\
\varphi_2({S}_1,{S}_2) & = &
\frac{\sqrt{\bar{\beta}}}{\sigma}(S_2-\rho S_1)+\frac{\sqrt{\beta(1- \rho^2)+\rho^2}}{\sigma}S_1
\nonumber \\
& = &
\frac{\sqrt{\bar{\beta}}}{\sigma}S_2+\frac{1}{\sigma}\left[\sqrt{\rho^2\bar{\beta}+\beta}-\sqrt{\rho^2\bar{\beta}}\right]S_1.
\end{IEEEeqnarray*}
It can be verified that, as required by Lemma~\ref{th:witsen1},
\begin{IEEEeqnarray*}{rCl}
{\sf{E}}\left[\varphi_1({S}_1)\right] & = & {\sf{E}}\left[\varphi_2({S}_1,S_2)\right]=0 \nonumber \\
{\sf{E}}\left[(\varphi_1({S}_1))^2\right]  & = & {\sf{E}}\left[(\varphi_2({S}_1,S_2))^2\right]=1,
\end{IEEEeqnarray*}
while
\begin{IEEEeqnarray}{l}
{\sf{E}}\left[\varphi_1({S}_1)\varphi_2({S}_2,{S}_1)\right]=
\sqrt{\rho^2\bar{\beta}+\beta}. \label{eq:mapmaxcor1}
\end{IEEEeqnarray}
Furthermore, for this set of linear mappings the random variable $U$
which satisfies \\
 $\varphi_1({S}_1)\Markov U\Markov
\varphi_2({S}_1,S_2)$, and is jointly Gaussian with
$(\varphi_1({S}_1),\varphi_2({S}_1,S_2))$, is $U=S_1$ in which case
\begin{IEEEeqnarray}{l}
\Var{X_2|U} =
\Var{\varphi_2({S}_1,S_2)|U}=\frac{\bar{\beta}}{\sigma^2}\Var{S_2|S_1}=\bar{\beta}(1-\rho^2).
\label{eq:mapmaxcor2}
\end{IEEEeqnarray}
Thus, the set of laws over which $\varrho(\textbf{X}_1,\textbf{X}_2)$ is maximized are those for which $U=X_1$
hence they simultaneously maximize the r.h.s. of \eqref{eq:macconf31} as well.

The combination of \eqref{eq:macconf1}, \eqref{eq:macconf3},
\eqref{eq:upboundrho1} and \eqref{eq:mapmaxcor1} establishes the
upper bound \eqref{eq:necessaryup1} in Theorem~\ref{th:necessary1}. \\
The combination of \eqref{eq:macconf10}, \eqref{eq:macconf31}, and
\eqref{eq:mapmaxcor2} establishes the upper bound
\eqref{eq:necessaryup10} in Theorem~\ref{th:necessary1}.

\medskip

\section{Proof of Theorem~\ref{th:vector-quantizer}}
\label{Proof of VQ scheme}

\subsection{Coding scheme}

Fix some $\epsilon>0$ and a rate tuple $(R_1,R_2,R_\textnormal{c})$.

\medskip

{\bf Code Construction:} Three codebooks ${\cal C}_1, {\cal C}_2$
and ${\cal C}_c$ are generated independently. Codebook ${\cal C}_i,
i\in {1,2}$, consists of $2^{nR_i}$ codewords
$\{\textbf{U}_i(1),\textbf{U}_i(2),\ldots,\textbf{U}_i(2^{nR_i})\}$.
The codewords are drawn independently uniformly over the surface of
the centered $\Reals^n$-sphere ${\cal S}_i$ of radius
$r_i=\sqrt{n\sigma^2(1-2^{-2R_i})}$. Codebook ${\cal C}_c$, consists
of $2^{nR_\textnormal{c}}$ codewords \\
$\{\textbf{V}(1),\textbf{V}(2),\ldots,\textbf{V}(2^{nR_\textnormal{c}})\}$. The
codewords are drawn independently uniformly over the surface of the
centered $\Reals^n$-sphere ${\cal S}_c$ of radius
$r_c=\sqrt{n\sigma^22^{-2R_1}(1-2^{-2R_\textnormal{c}})}$.

Partition randomly the codebook ${\cal C}_c$ into
${{\left( 1-{{\rho }^{2}}{{2}^{-2{{R}_{1}}}}\left( 1-{{2}^{-2{{R}_\textnormal{c}}}} \right) \right)}^{\frac{n}{2}}}{{2}^{n\left( {{R}_\textnormal{c}}+\delta(\epsilon)  \right)}}$ bins,
each of size
\begin{equation}
M_b\triangleq{{\left( 1-{{\rho }^{2}}{{2}^{-2{{R}_{1}}}}\left( 1-{{2}^{-2{{R}_\textnormal{c}}}} \right) \right)}^{-\frac{n}{2}}}{{2}^{-n\delta(\epsilon) }},
\label{eq:codebookbinsize}
\end{equation}
and for any codeword $\textbf{v}(k)$ let ${b}(k)$
denote the index of the bin containing $\textbf{v}(k)$.

\medskip

For every $\textbf{w},\textbf{v}\in \Reals^n$ where neither
$\textbf{w}$ nor $\textbf{v}$ are the zero-sequence, denote the
angle between $\textbf{w}$ and $\textbf{v}$ by
$\sphericalangle(\textbf{w},\textbf{v})$. i.e.,
\begin{equation*}
\cos \sphericalangle(\textbf{w},\textbf{v}) \triangleq \frac{\langle
\textbf{w},\textbf{v}\rangle}{\|\textbf{w}\|\|\textbf{v}\|} .
\end{equation*}

{\bf Encoding:} Given the source sequences $(\textbf{s}_1,\textbf{s}_2)$,
let ${\cal F}(\textbf{s}_i,{\cal C}_i)$ be
the set defined by
\begin{equation}\label{eq:encoding}
{\cal F}(\textbf{s}_i,{\cal C}_i) \triangleq \left\{
\textbf{u}_i\in{\cal C}_i \colon
\left|\cos\sphericalangle(\textbf{s}_i,\textbf{u}_i)-\sqrt{1-2^{-2R_i}}\right|
\right. \left. \leq \sqrt{1-2^{-2R_i}}\epsilon\right\} .
\end{equation}

Encoder~1 vector-quantizes $\textbf{s}_1$ in two steps as follows:
\begin{enumerate}
\item If ${\cal F}(\textbf{s}_1,{\cal C}_1)\neq\emptyset$
 it forms the vector $\textbf{u}_1^{*}$ by choosing it as the
 codeword $\textbf{u}_1(j^{*})\in {\cal F}(\textbf{s}_1,{\cal C}_1)$
 where $j^{*}$ minimizes $|\cos\sphericalangle(\textbf{s}_1,
 \textbf{u}_1(j))-\sqrt{1-2^{-2R_1}}|$, while if
 ${\cal F}(\textbf{s}_1,{\cal C}_1)=\emptyset$
 then $\textbf{u}_1^{*}$ is the all-zero sequence.

\item Let
\begin{equation}
\textbf{Z}_{Q_1} \triangleq \textbf{S}_1-\textbf{U}_1^{*} .
\end{equation}
Let ${\cal F}(\textbf{z}_{\textnormal{Q}_1},{\cal C}_c)$ be the set defined by
\begin{equation}\label{eq:encoding_2}
{\cal F}(\textbf{z}_{\textnormal{Q}_1},{\cal C}_c)\triangleq \left\{
\textbf{v}\in{\cal C}_c \colon \left|
\cos\sphericalangle(\textbf{z}_{\textnormal{Q}_1},\textbf{v})-\sqrt{1-2^{-2R_\textnormal{c}}}\right|
\right.  \left. \leq \sqrt{1-2^{-2R_\textnormal{c}}}\epsilon\right\} .
\end{equation}
If ${\cal F}(\textbf{z}_{\textnormal{Q}_1},{\cal C}_c)\neq\emptyset$
 it forms the vector $\textbf{v}^{*}$ by choosing it as the
 codeword $\textbf{v}(k^{*})\in {\cal F}(\textbf{z}_{\textnormal{Q}_1},{\cal C}_c)$
 where $k^{*}$ minimizes $|\cos\sphericalangle(\textbf{z}_{\textnormal{Q}_1},
 \textbf{v}(k))-\sqrt{1-2^{-2R_\textnormal{c}}}|$, while if \\
 ${\cal F}(\textbf{z}_{\textnormal{Q}_1},{\cal C}_c)=\emptyset$
 then $\textbf{v}^{*}$ is the all-zero sequence.
\end{enumerate}
The channel input $\textbf{X}_1$ is now given by \eqref{eq:inputX1}.

Since the codebooks ${\cal C}_1$ and ${\cal C}_c$ are drawn over the
centered $\Reals^n$-spheres of radii
$r_1=\sqrt{\sigma^2(1-2^{-2R_1})}$ and
$r_c=\sqrt{\sigma^22^{-2R_1}(1-2^{-2R_\textnormal{c}})}$, respectively, and (as
shown in Lemma~\ref{u1,v} ahead) the codewords $\textbf{U}_1^{*}$ and
$\textbf{V}^{*}$ are uncorrelated, the channel input $\textbf{X}_1$
satisfies the average-power constraint.

Encoder~1 informs Encoder~2 on $\textbf{v}(k^{*})$ by sending Encoder~2
the bin-index ${b}(k^{*})$ over the unidirectional conference channel.

\medskip

Encoder~2 vector-quantizes $\textbf{s}_2$ as follows:

\medskip

If ${\cal F}(\textbf{s}_2,{\cal C}_2)\neq\emptyset$ it forms the
vector $\textbf{u}_2^{*}$ by choosing it as the codeword
$\textbf{u}_2(j^{*})\in {\cal F}(\textbf{s}_2,{\cal C}_2)$ where
$j^{*}$ minimizes $|\cos\sphericalangle(\textbf{s}_2,\textbf{u}_2(j))-\sqrt{1-2^{-2R_2}}|$, while if ${\cal
F}(\textbf{s}_2,{\cal C}_2)=\emptyset$ then $\textbf{u}_2^{*}$ is
the all-zero sequence.

\medskip

Encoder~2 acquires the codeword $\textbf{v}(k^{*})$ by choosing among the
codewords within bin ${b}(k^{*})$ the codeword $\textbf{v}(k^{*})$
such that
\begin{equation*}
|{\rho}_{\textbf{v},\textbf{s}_2}-\cos\sphericalangle(\textbf{v}(k^{*}),\textbf{s}_2)| \leq
5\epsilon ,
\end{equation*}
where ${\rho}_{\textbf{v},\textbf{s}_2}\triangleq \rho \sqrt{2^{-2R_1}(1-2^{-2R_\textnormal{c}})}$.

The channel input $\textbf{X}_2$ is now given by \eqref{eq:inputX2}.
Since the codebooks ${\cal C}_2$ and ${\cal C}_c$ are drawn over the
centered $\Reals^n$-spheres of radii
$r_1=\sqrt{\sigma^2(1-2^{-2R_2})}$ and
$r_c=\sqrt{\sigma^22^{-2R_1}(1-2^{-2R_\textnormal{c}})}$, respectively, and (as
shown in Lemma~\ref{u2,v} ahead) the codewords $\textbf{U}_2^{*}$ and
$\textbf{V}^{*}$ are correlated, the channel input $\textbf{X}_2$
satisfies the average-power constraint.

\medskip

{\bf Reconstruction:} The receiver's estimate
$(\hat{\textbf{S}}_1,\hat{\textbf{S}}_2)$ of the source pair
$(\textbf{S}_1,\textbf{S}_2)$ is obtained via the channel output
$\textbf{Y}$ in two steps. First, the receiver makes a guess
$(\hat{\textbf{U}}_1,\hat{\textbf{V}},\hat{\textbf{U}}_2)$ of the
tuple $(\textbf{U}_1^{*},\textbf{V}^{*},\textbf{U}_2^{*})$ by
choosing among all ``jointly typical'' tuples
$(\textbf{u}_1,\textbf{v},\textbf{u}_2)\in {\cal C}_1\times{\cal
C}_c\times{\cal C}_2$ the tuple whose linear combination
$a_{1,1}\textbf{U}_1+a_{2,1}\textbf{U}_2+(a_{1,2}+a_{2,2})\textbf{V}$
has the smallest distance to the received sequence $\textbf{Y}$.
More formally, let $\bar{\cal F}({\cal C}_1,{\cal C}_c,{\cal C}_2)$
be the set of triplets $(\textbf{u}_1,\textbf{v},\textbf{u}_2)\in
{\cal C}_1\times{\cal C}_c\times{\cal C}_2$ such that
\begin{IEEEeqnarray}{rCl}
|\tilde{\rho}-\cos\sphericalangle(\textbf{u}_1,\textbf{u}_2)|
& \le & 7\epsilon \nonumber \\
|\bar{\rho}-\cos\sphericalangle(\textbf{v},\textbf{u}_2)|
& \le & 7 \epsilon \nonumber \\
|\cos\sphericalangle(\textbf{v},\textbf{u}_1)|
& \le & 3 \epsilon ,
\label{eq:anglconstraints}
\end{IEEEeqnarray}
where $(\tilde{\rho},\bar{\rho})$ are defined in
\eqref{eq:constants},
and for any tuple $(\textbf{u}_1,\textbf{v},\textbf{u}_2)$ define
\begin{IEEEeqnarray*}{rCl}
\textbf{X}_{\textbf{u}_1,\textbf{v},\textbf{u}_2}
& \triangleq & a_{1,1}\textbf{u}_1+a_{2,1}\textbf{u}_2+(a_{1,2}+a_{2,2})\textbf{v}\\
& = & a_{1,1}\textbf{u}_1+a_{2,1}\textbf{u}_2+\alpha\textbf{v},
\end{IEEEeqnarray*}
where $\alpha\triangleq a_{1,2}+a_{2,2}$.
Then the receiver forms its estimate by choosing
\begin{equation}
(\hat{\textbf{U}}_1,\hat{\textbf{V}},\hat{\textbf{U}}_2) = \arg
\min_{(\textbf{u}_1,\textbf{v},\textbf{u}_2)\in \bar{\cal F}({\cal
C}_1,{\cal C}_c,{\cal C}_2) }
\|\textbf{Y}-\textbf{X}_{\textbf{u}_1,\textbf{v},\textbf{u}_2}\|^2.
\label{eq:mindistdec1}
\end{equation}
If the channel output $\textbf{Y}$ and the codebooks are such that
there doesn't exist a member in $\bar{\cal F}({\cal C}_1,{\cal
C}_c,{\cal C}_2)$ that minimizes the r.h.s. in
\eqref{eq:mindistdec1}, then
$(\hat{\textbf{U}}_1,\hat{\textbf{V}},\hat{\textbf{U}}_2)$ are
chosen to be the all-zero sequences.

\medskip

In the second step, the receiver forms its estimates
$(\hat{\textbf{S}}_1,\hat{\textbf{S}}_2)$ of the source pair
$(\textbf{S}_1,\textbf{S}_2)$ according to
\eqref{eq:MMSEestimation1}.

\medskip

\subsection{Expected distortion}

Similarly to \cite{Stephan}, to analyze the expected distortion we
first show that, when the rate constraints
\eqref{eq:rateconstraints1} are satisfied, the asymptotic normalized
distortion of the proposed scheme remains the same as that of a
genie-aided scheme in which the genie provides the decoder with the
triplet $(\textbf{U}_1^{*},\textbf{V}^{*},\textbf{U}_2^{*})$. The
genie-aided decoder forms its estimate
$(\hat{\textbf{S}}_1{\hspace{-.4em}}^\textnormal{G},\hat{\textbf{S}}_2{\hspace{-.4em}}^\textnormal{G})$ based on
$(\textbf{U}_1^{*},\textbf{V}^{*},\textbf{U}_2^{*})$ according to
\eqref{eq:MMSEestimation1} and ignores its guess
$(\hat{\textbf{U}}_1,\hat{\textbf{V}},\hat{\textbf{U}}_2)$ produced
in the first decoding step.
Hence, $(\hat{\textbf{S}}_1{\hspace{-.4em}}^\textnormal{G},\hat{\textbf{S}}_2{\hspace{-.4em}}^\textnormal{G})$ is defined by
\begin{IEEEeqnarray}{rCl}\label{eq:Genie_estimation1}
\hat{\textbf{S}}_1{\hspace{-.4em}}^\textnormal{G}
& = & \gamma_{1,1}{\textbf{U}}^{*}_1+\gamma_{1,2}{\textbf{U}}^{*}_2+\gamma_{1,3}{\textbf{V}}^{*} \nonumber \\
\hat{\textbf{S}}_2{\hspace{-.4em}}^\textnormal{G}
& = & \gamma_{2,1}{\textbf{U}}^{*}_1+\gamma_{2,2}{\textbf{U}}^{*}_2+\gamma_{2,3}{\textbf{V}}^{*} ,
\end{IEEEeqnarray}
with $\gamma_{1,1},\gamma_{1,2},\gamma_{1,3},\gamma_{2,1},\gamma_{2,2},\gamma_{2,3}$
as in \eqref{eq:MMSE_coefficients}.



\medskip

\begin{proposition}\label{corollary_rates}
If $(R_1,R_2,R_\textnormal{c})$ satisfy
\begin{IEEEeqnarray}{rCl}
  {{R}_{1}} & < & \frac{1}{2}\log \left( \frac{{{\bar{\beta }}_{1}}{{P}_{1}}\left( 1-{{{\tilde{\rho }}}^{2}} \right)+N-{{\bar{\rho }}^{2}}\left( {{\bar{\beta }}_{1}}{{P}_{1}}+N \right)}{N\left( 1-{{{\tilde{\rho }}}^{2}} \right)-{{\bar{\rho }}^{2}}N} \right) \nonumber \\
  {{R}_{2}} & < & \frac{1}{2}\log \left( \frac{{{\bar{\beta }}_{2}}{{P}_{2}}\left( 1-{{{\tilde{\rho }}}^{2}}-{{\bar{\rho }}^{2}} \right)+N}{N\left( 1-{{{\tilde{\rho }}}^{2}}-{{\bar{\rho }}^{2}} \right)+{{\lambda }_{2}}} \right) \nonumber \\
  {{R}_\textnormal{c}}& < &\frac{1}{2}\log \left( \frac{{{\eta }^{2}}\left( 1-{{{\tilde{\rho }}}^{2}}-{{\bar{\rho }}^{2}} \right)+N\left( 1-{{{\tilde{\rho }}}^{2}} \right)}{N\left( 1-{{{\tilde{\rho }}}^{2}}-{{\bar{\rho }}^{2}} \right)+{{\lambda }_{c}}} \right) \nonumber \\
  {{R}_{1}}+{{R}_{2}}& < &\frac{1}{2}\log \left( \frac{{{\lambda }_{12}}-\bar{{{\beta }_{2}}}{{P}_{2}}{{\bar{\rho }}^{2}}+N}{\left( 1-\bar{{{\beta }_{2}}}{{P}_{2}}{{\bar{\rho }}^{2}}{{\lambda }_{12}}^{-1} \right)N\left( 1-{{{\tilde{\rho }}}^{2}}\  \right)} \right) \nonumber \\
  {{R}_{1}}+{{R}_\textnormal{c}}& < &\frac{1}{2}\log \left( \frac{\left( {{\lambda }_{1c}}+N \right)\left( \bar{{{\beta }_{1}}}{{P}_{1}}+{{\eta }^{2}} \right)}{{{\lambda }_{1c}}N} \right) \nonumber \\
  {{R}_{2}}+{{R}_\textnormal{c}}& < &\frac{1}{2}\log \left( \frac{{{\lambda }_{2c}}-\bar{{{\beta }_{2}}}{{P}_{2}}{{{\tilde{\rho }}}^{2}}+N}{\left( 1-\bar{{{\beta }_{2}}}{{P}_{2}}{{{\tilde{\rho }}}^{2}}{{\lambda }_{2c}}^{-1} \right)N\left( 1-{{{\bar{\rho }}}^{2}}\  \right)} \right) \nonumber \\
  {{R}_{1}}+{{R}_{2}}+{{R}_\textnormal{c}}& < &\frac{1}{2}\log \left( \frac{{{\lambda }_{12}}+2\eta \bar{\rho }\sqrt{\bar{{{\beta }_{2}}}{{P}_{2}}}+{{\eta }^{2}}+N}{N\left( 1-{{{\tilde{\rho }}}^{2}}\  \right)\left( 1-{{\bar{\rho }}^{2}}\  \right)} \right)\nonumber \\
  C_{12} & > & R_\textnormal{c}+\frac{1}{2}\log \left( 1-{{\rho
}^{2}}{{2}^{-2{{R}_{1}}}}\left( 1-{{2}^{-2{{R}_\textnormal{c}}}} \right)\right), \label{eq:rateconstraints}
\end{IEEEeqnarray}
then
\begin{equation*}
\varlimsup_{n\to\infty} \ \frac{1}{n}{\sf{E}}\left[\|\mathbf{S}_{\nu}-\hat{\mathbf{S}}_{\nu}\|^2\right] \leq
\varlimsup_{n\to\infty} \ \frac{1}{n}{\sf{E}}\left[\|\mathbf{S}_{\nu}-\hat{\mathbf{S}}_{\nu}{\hspace{-.4em}}^\textnormal{G}\|^2\right],
\ \nu=1,2 \ .
\end{equation*}
\end{proposition}

\medskip

\begin{IEEEproof}
Follows from Proposition~\ref{proposition_rates} (which appears in the Appendix) by first letting $n\rightarrow \infty$ and then
$\epsilon\rightarrow 0$ and $\delta\rightarrow 0$.
\end{IEEEproof}

\medskip

By Proposition~\ref{corollary_rates}, to analyze the distortion achievable by our scheme
it suffices to analyze the genie-aided scheme. 

\medskip

\begin{proposition}\label{proposition_genieaideddist12}
The distortion pair $(D_1,D_2)$ of the genie-aided scheme satisfies
\begin{IEEEeqnarray*}{rCl}
D_1 & \leq &
{{\sigma }^{2}}{{2}^{-2({{R}_{1}}+{{R}_\textnormal{c}})}}\frac{1-{{\rho }^{2}}\left( 1-{{2}^{-2{{R}_{2}}}} \right)}{1-{{\rho }^{2}}\left( 1-{{2}^{-2{{R}_{2}}}} \right)\left( 1-{{2}^{-2({{R}_{1}}+{{R}_\textnormal{c}})}} \right)}+\xi'(\delta,\epsilon) \nonumber \\
D_2 & \leq &
{{\sigma }^{2}}2^{-2{R}_{2}}\frac{1-{{\rho }^{2}}\left( 1-{{2}^{-2({{R}_{1}}+{{R}_\textnormal{c}})}} \right)}{1-{{\rho }^{2}}\left( 1-{{2}^{-2{{R}_{2}}}} \right)\left( 1-{{2}^{-2({{R}_{1}}+{{R}_\textnormal{c}})}} \right)}+\xi'(\delta,\epsilon),
\end{IEEEeqnarray*}
where $\underset{\delta ,\epsilon \to 0}{\mathop{\lim }}\,\xi '\left( \delta ,\epsilon  \right)=0$.
\end{proposition}

\begin{IEEEproof}
See Appendix.
\end{IEEEproof}

\medskip


\vskip.2truein

\section{Proof of Proposition~\ref{Gaussian RD with conferencing}}
\label{Proof of Gaussian RD with conferencing}

Let $(S_1,S_2)$ be a pair of zero-mean jointly Gaussian random variables
with covariance matrix as per \eqref{eq:covariancemat1}.
Let ${\cal P}(D_1,D_2)$ be the set of  triples  $(U,V,W)$
jointly Gaussian  with $(S_1,S_2)$ such that
\begin{enumerate}
\item $U \Markov (S_2,W)\Markov (S_1,W)\Markov V$ and $W \Markov S_1\Markov S_2$
are Markov chains,
\item $\sigma^{2}_{S_1|U,V,W}\leq D_1$ , $\sigma^{2}_{S_2|U,V,W}\leq D_2$.
\end{enumerate}

\medskip

This set can be defined as follows. Independently of $(S_1,S_2)$ draw a triplet  of independent
random variables $N^{( w)}\sim N( 0,\sigma _{N^{(w)}}^{2} )$,  $N^{( u)}\sim N( 0,\sigma _{N^{(u)}}^{2} )$
and $N^{( v)}\sim N( 0,\sigma _{N^{(v)}}^{2} )$, and define
\begin{IEEEeqnarray*}{rCl}
W & = & S_1+N^{( w )}  \nonumber \\
U & = & a_{1u}W+a_{2u}S_2+N^{( u )}  \nonumber \\
V & = & a_{1v}W+a_{2v}S_1+N^{( v )}.
\end{IEEEeqnarray*}
Then,
\begin{IEEEeqnarray*}{rCl,rCl}
{\cal R}_{\textnormal{G}}(D_1,D_2) = \bigcup_{(U,V,W)\in {\cal P}(D_1,D_2)}
\Biggl\{(R_0,R_1,R_2)\colon
&&& R_0 & \ge & \frac{1}{2}\log \Biggl( \frac{\sigma _{W|S_2}^{2}}{\sigma _{W|S_1}^{2}} \Biggr)  \nonumber \\
&&& R_1 & \ge & \frac{1}{2}\log \Biggl( \frac{\sigma _{\left. V \right|U,W}^{2}}{\sigma _{\left. V \right|S_1,W}^{2}} \Biggr)  \nonumber \\
&&& R_2 & \ge & \frac{1}{2}\log \Biggl( \frac{\sigma _{\left. U \right|V,W}^{2}}{\sigma _{\left. U \right|S_2,W}^{2}} \Biggr)  \nonumber \\
&&& R_1+R_2 & \ge & \frac{1}{2}\log \Biggl( \frac{\left| {\sf K}_{SS} \right|}{\left| {\sf K}_{SS|UVW} \right|} \Biggr)
  \Biggr\},
\end{IEEEeqnarray*}
where ${\sf K}_{SS|UVW}$ is the covariance matrix of $(S_1,S_2)$ conditioned on $(U,V,W)$.

Defining
$\sigma^{2}_{u} \triangleq \frac{\sigma^{2}_{N^(u)}}{{a^{2}_{2u}}},
\sigma^{2}_{v} \triangleq \frac{\sigma^{2}_{N^(v)}}{{a^{2}_{2v}}}$
and
$\sigma^{2}_{w} \triangleq {\sigma^{2}_{N^(w)}}$
the result follows since
\begin{IEEEeqnarray*}{rCl}
\frac{\sigma _{W|S_2}^{2}}{\sigma _{W|S_1}^{2}}  & = &
1+\frac{\sigma^2(1-\rho^2)}{\sigma^2_{w}} \nonumber \\
\frac{\sigma _{\left. V \right|U,W}^{2}}{\sigma _{\left. V \right|S_1,W}^{2}} & = &
1+\frac{{{\sigma }^{2}}}
{\sigma^{2}_{v}}\frac{\left( 1-\rho^{2} \right)
+\frac{\sigma^{2}_{u}}{{{\sigma}^{2}}}}
{1+ \frac{{{\sigma }^{2}}\left( 1-{{\rho }^{2}} \right)}{\sigma^{2}_{w}}
+\sigma^{2}_{u}\left( \frac{1}{\sigma^{2}_{w}}+\frac{1}{\sigma^{2}} \right)} \nonumber \\
 \frac{\sigma _{\left. U \right|V,W}^{2}}{\sigma _{\left. U \right|S_2,W}^{2}}  & = &
1+\frac{{{\sigma }^{2}}}{\sigma^2_{u}}
\frac{ 1+{{\sigma }^{2}}\left( \frac{1}{\sigma^2_{w}}+\frac{1}{\sigma^2_{v}} \right)\left( 1-{{\rho }^{2}} \right) }
{1+{{\sigma }^{2}}\left( \frac{1}{\sigma^2_{w}}+\frac{1}{\sigma^2_{v}} \right)} ,
\end{IEEEeqnarray*}
and
\begin{IEEEeqnarray*}{rCl}
\sigma^2_{S_1 |U,V,W} & = &
 {{\sigma }^{2}}\frac{1+\frac{{{\sigma }^{2}}}{\sigma^2_{u}}\left( 1-{{\rho }^{2}} \right)}{\Delta} \nonumber \\
\sigma^2_{S_2 |U,V,W} & = &
{{\sigma }^{2}}\frac{1+{{\sigma }^{2}}\left( 1-{{\rho }^{2}} \right)\left( \frac{1}{\sigma^2_{v}}+\frac{1}{\sigma^2_{w}} \right)}{\Delta}
\end{IEEEeqnarray*}
where
\begin{IEEEeqnarray*}{rCl}
 \Delta
 & \triangleq & 1+\frac{{{\sigma }^{2}}}{\sigma^2_{u}}+{{\sigma }^{2}}\left( 1+\frac{{{\sigma }^{2}}}{\sigma^2_{u}}\left( 1-{{\rho }^{2}} \right) \right)\left( \frac{1}{\sigma^2_{v}}+\frac{1}{\sigma^2_{w}} \right) \nonumber \\
 & = &  \frac{\left| {\sf K}_{SS} \right|}{\left| {\sf K}_{SS|UVW} \right|} .
\end{IEEEeqnarray*}


\medskip

\section{Proof of Theorem~\ref{th:highsnr1}}
\label{Proof of highsnr1}
The high-${\sf SNR}$ asymptotics for the multiple-access problem, when $C_{12}=\infty$,
can be obtained from the necessary condition for the achievability
of a distortion pair $(D_1,D_2)$ in Theorem~\ref{th:necessary1}, and from the sufficient conditions
for the achievability of a distortion pair $(D_1,D_2)$ derived from
Corollary~\ref{cor:VQ with unlimited capacity}.

First we recall the rate-distortion function of a bivariate
Gaussian.

\medskip

\begin{theorem}\label{th:bivgaurd1}\cite{Xiao},\cite[Theorem~III.1]{Stephan}
The rate-distortion function $R_{S_1,S_2}(D_1,D_2)$ is given by
\begin{IEEEeqnarray*}{l}
R_{S_1,S_2}(D_1,D_2)=\left\{ \begin{array}{ll}
\frac{1}{2}\log_2^{+}\left(\frac{\sigma^2}{D_{\textnormal{min}}}\right) &
\mbox{if} \ (D_1,D_2)\in{\mathscr D}_1 \\
\frac{1}{2}\log_2^{+}\left(\frac{\sigma^4(1-\rho^2)}{D_1D_2}\right)
& \mbox{if} \ (D_1,D_2)\in{\mathscr D}_2 \\
\frac{1}{2}\log_2^{+}\left(\frac{\sigma^4(1-\rho^2)}{D_1D_2-(\rho\sigma^2-\varrho(D_1,D_2))^2}\right)
& \mbox{if} \ (D_1,D_2)\in{\mathscr D}_3
\end{array} \right.
\end{IEEEeqnarray*}
where
\begin{IEEEeqnarray*}{l}
\varrho(D_1,D_2)\triangleq \sqrt{(\sigma^2-D_1)(\sigma^2-D_2)},
\end{IEEEeqnarray*}
$\log_2^+{}(x)\triangleq\max\{0,\log_2(x)\}$, and
$D_{\textnormal{min}}\triangleq\min\{D_1,D_2\}$, and the regions ${\mathscr D}_i,
i=1,2,3$ are defined by
\begin{IEEEeqnarray*}{rCl}
{\mathscr D}_1 & \triangleq & \Biggl\{ (D_1,D_2)\colon \left(0\leq
D_1\leq\upsilon, D_2\geq\upsilon+\rho^2D_1\right) \ \mbox{or}
\nonumber \\
& & \quad \left(\upsilon<D_1\leq\sigma^2, \upsilon+\rho^2D_1\leq
D_2\leq\frac{D_1-\upsilon}{\rho^2}\right) \Biggr\} \nonumber \\
{\mathscr D}_2 & \triangleq & \Biggl\{ (D_1,D_2)\colon 0\leq
D_1\leq\upsilon, 0\leq
D_2<(\upsilon-D_1)\frac{\sigma^2}{\sigma^2-D_1} \Biggr\} \nonumber \\
{\mathscr D}_3 & \triangleq & \Biggl\{ (D_1,D_2)\colon \left(0\leq
D_1\leq\upsilon, (\upsilon-D_1)\frac{\sigma^2}{\sigma^2-D_1}\leq
D_2<\upsilon+\rho^2D_1\right) \ \mbox{or}
\nonumber \\
& & \quad \left(\upsilon<D_1\leq\sigma^2,
\frac{D_1-\upsilon}{\rho^2}< D_2<\upsilon+\rho^2D_1\right) \Biggr\},
\end{IEEEeqnarray*}
with $\upsilon\triangleq \sigma^2(1-\rho^2)$.
\end{theorem}

\medskip

By Corollary~\ref{cor:VQ with unlimited capacity}, when the conferencing capacity is unlimited, it
follows that any normalized distortion pair $(d_1,d_2)$ satisfying
$d_1\leq 1,d_2\leq 1$ and
\begin{IEEEeqnarray}{rCl}
d_1 & \geq & \frac{N}{\delta_1^2} \label{eq:ref11} \\
d_2 & \geq & \frac{N}{\bar{\beta}P_2} \label{eq:ref12} \\
d_1d_2 & = &
\frac{N(1-\breve{\rho}^2)}{P_1+P_2+2\sqrt{(\breve{\rho}^2\bar{\beta}+\beta)P_1P_2}} \ ,
\qquad 0\leq\beta<1
\label{eq:ref13}
\end{IEEEeqnarray}
where $\breve{\rho}=\rho\sqrt{(1-d_1)(1-d_2)}$, is achievable.\\
Next, if
\begin{IEEEeqnarray}{l}
\lim_{N\to 0}\frac{N}{d_1P_1}=0 \  \quad \mbox{and}
\quad \lim_{N\to 0}\frac{N}{d_2P_2}=0,
\label{eq:ref14}
\end{IEEEeqnarray}
then \eqref{eq:ref11} and \eqref{eq:ref12} are satisfied for sufficiently
small $N$ and some $0\leq \beta<1$. Thus, for $N$ sufficiently small, any pair $(d_1,d_2)$ satisfying
\eqref{eq:ref13} and \eqref{eq:ref14} is achievable provided that $\beta$ satisfies the
constraint imposed by  \eqref{eq:ref12}.
However, if $(d_1,d_2)$ satisfies \eqref{eq:ref13} then
the following pair of inequalities holds
\begin{IEEEeqnarray}{l}
d_2 \leq \frac{N}{d_1P_1} \ , \quad \mbox{and} \quad d_1 \leq
\frac{N}{d_2P_2}. \label{eq:ref15}
\end{IEEEeqnarray}
Combining \eqref{eq:ref15} with the expression of $\breve{\rho}$
yields that if in addition to \eqref{eq:ref13} the pair $(d_1,d_2)$
satisfies \eqref{eq:ref14}, then $\breve{\rho}\to\rho$ as $N\to 0$.
In conclusion, the sufficient condition yields that, if a pair $(d_1,d_2)$ satisfies \eqref{eq:ref13} and
\eqref{eq:ref14}, then
\begin{IEEEeqnarray}{l}
\lim_{N\to 0}
\frac{P_1+P_2+2\sqrt{({\rho}^2\bar{\beta}+\beta)}\sqrt{P_1P_2}}{N}d_1d_2\leq
(1-{\rho}^2). \label{eq:ref16}
\end{IEEEeqnarray}

\medskip

Now, let
$(D_1^{*},D_2^{*})$
be a distortion pair of an optimal scheme. Then, by the upper bound
\eqref{eq:necessaryup1} in Theorem~\ref{th:necessary1} we have that
for some $0\leq \beta< 1$
\begin{IEEEeqnarray}{l}
R_{S_1,S_2}(D^{*}_1,D^{*}_2)\leq
\frac{1}{2}\log_2\left(1+\frac{P_1+P_2+2\sqrt{({\rho}^2\bar{\beta}+\beta)}\sqrt{P_1P_2}}{N}\right).
\label{eq:stepp1}
\end{IEEEeqnarray}
If $(D^{*}_1,D^{*}_2)$ satisfies
\begin{IEEEeqnarray}{l}
\lim_{N\to 0}\frac{N}{D^{*}_1P_1}=0 \ , \quad \mbox{and}
\quad \lim_{N\to 0}\frac{N}{D^{*}_2P_2}=0 ,
\label{eq:stepp2}
\end{IEEEeqnarray}
then for $N$ small enough
\begin{IEEEeqnarray}{l}
R_{S_1,S_2}(D^{*}_1,D^{*}_2)=\frac{1}{2}\log_2^{+}\left(\frac{\sigma^4(1-\rho^2)}{D^{*}_1D^{*}_2}\right),
\label{eq:stepp3}
\end{IEEEeqnarray}
by Theorem~\ref{th:bivgaurd1} and the fact that
$(D^{*}_1,D^{*}_2)\in{\mathscr D}_2$. The combination of
\eqref{eq:stepp1} and \eqref{eq:stepp3} implies that if
$(D^{*}_1,D^{*}_2)$ satisfies \eqref{eq:stepp2}, then
\begin{IEEEeqnarray}{l}
\lim_{N\to 0}
\frac{P_1+P_2+2\sqrt{({\rho}^2\bar{\beta}+\beta)}\sqrt{P_1P_2}}{N}d_1d_2\geq
(1-{\rho}^2). \label{eq:stepp4}
\end{IEEEeqnarray}

\medskip

\begin{remark}
To check consistency, note that for every
$(D^{*}_1,D^{*}_2)\in{\mathscr D}_2$ the rate-distortion function
$R_{S_1,S_2}(D^{*}_1,D^{*}_2)$ satisfies
\begin{IEEEeqnarray*}{rCl}
R_{S_1,S_2}(D^{*}_1,D^{*}_2) & = &
\frac{1}{2}\log_2^{+}\left(\frac{\sigma^4(1-\rho^2)}{D^{*}_1D^{*}_2}\right)
\nonumber \\
 & = &
 \frac{1}{2}\log_2^{+}\left(\frac{\sigma^2}{D^{*}_1}\right)
 +\frac{1}{2}\log_2^{+}\left(\frac{\sigma^2(1-\rho^2)}{D^{*}_2}\right)
\nonumber \\
 & = & R_{S_1}(D^{*}_1)+R_{S_2|S_1}(D^{*}_2).
\end{IEEEeqnarray*}
Consequently, as $N\to 0$, by the upper bound
\eqref{eq:necessaryup10} in Theorem~\ref{th:necessary1}
\begin{IEEEeqnarray}{l}
\frac{1}{2}\log_2^{+}\left(\frac{\sigma^2(1-\rho^2)}{D^{*}_2}\right)
\leq
\frac{1}{2}\log_2\left(\frac{\bar{\beta}P_2(1-\rho^2)}{N}\right)
\label{eq:consistency11}
\end{IEEEeqnarray}
i.e. $D^{*}_2\geq\sigma^2\frac{N}{\bar{\beta}P_2}$ as assumed in
\eqref{eq:ref12}.
\end{remark}

\medskip

The combination of \eqref{eq:ref16} with \eqref{eq:stepp4} and \eqref{eq:consistency11} implies that
the high-${\sf SNR}$ asymptotics for the Gaussian MAC with unlimited
unidirectional conferencing capacity satisfies, for some $0\leq
\beta\leq 1-\frac{N}{d^{*}_2P_2}$,
\begin{IEEEeqnarray*}{l}
\lim_{N\to 0}
\frac{P_1+P_2+2\sqrt{({\rho}^2\bar{\beta}+\beta)}\sqrt{P_1P_2}}{N}
d^{*}_1d^{*}_2=(1-\rho^2),
\end{IEEEeqnarray*}
provided that $d^{*}_1\leq 1,d^{*}_2\leq 1$, and that $(d^{*}_1,d^{*}_2)$ satisfy
\eqref{eq:highsnrvq11condition00}.
It remains to optimize the correlation $\varrho(\beta)$ over $\beta$ subject to the constraint \eqref{eq:ref12},
\begin{equation*}
\varrho^{*}=\sup_{\bar{\beta}\geq \frac{N}{d^{*}_2P_2}}\sqrt{\rho^2\bar{\beta}+\beta}
=\sqrt{\rho^2\frac{N}{d^{*}_2P_2}+(1-\frac{N}{d^{*}_2P_2})}
=\sqrt{1-\frac{N(1-\rho^2)}{d^{*}_2P_2}},
\end{equation*}
and clearly $\varrho^{*}_\textnormal{VQ}=\varrho^{*}$ --i.e. the maximal correlation attained by the VQ scheme equals $\varrho^{*}$
since it is the same function of $\beta$ and it is defined over the same domain.

This concludes the proof of Theorem~\ref{th:highsnr1}. \hfill \qed

\medskip

\section{Proof of Corollary~\ref{th:highsnr2}}
\label{Proof of highsnr2}

By Proposition~\ref{proposition_SC separation1} and \eqref{eq:unlimited conf capacity},
when the conferencing capacity is unlimited, it
follows that any normalized distortion pair $(d_1,d_2)$ satisfying
$d_1\leq 1,d_2\leq 1$ and
\begin{IEEEeqnarray}{rCl}
d_1 & \geq & \frac{N(1-\rho^2)}{P_1+P_2+2\sqrt{\beta P_1P_2}} \label{eq:refsep11} \\
d_2 & \geq & \frac{N(1-\rho^2)}{\bar{\beta}P_2} \label{eq:refsep12} \\
d_1d_2 & = &
\frac{N(1-\rho^2)\gamma(d_1,d_2)}{2(P_1+P_2+2\sqrt{\beta P_1P_2})}
\qquad 0\leq\beta<1
\label{eq:refsep13}
\end{IEEEeqnarray}
where $\gamma(d_1,d_2)=1+\sqrt{1+\frac{4\rho^2d_1d_2}{(1-\rho^2)^2}}$ is achievable.\\
Next, if
\begin{IEEEeqnarray}{l}
\lim_{N\to 0}\frac{N}{d_1P_1}=0 \ , \quad \mbox{and}
\quad \lim_{N\to 0}\frac{N}{d_2P_2}=0,
\label{eq:refsep14}
\end{IEEEeqnarray}
then \eqref{eq:refsep11} and \eqref{eq:refsep12} are satisfied for sufficiently
small $N$ and some $0\leq \beta<1$. Thus, for $N$ sufficiently small, any pair satisfying
\eqref{eq:refsep13} and \eqref{eq:refsep14} is achievable provided that $\beta$ satisfies the
constraint imposed by  \eqref{eq:refsep12}. \\
Since $\gamma(d_1,d_2)\geq 2$, a distortion pair $(d_1,d_2)$ is achievable by source-channel separation
scheme~1 if, and only if,
\begin{IEEEeqnarray*}{l}
\lim_{N\to 0}
\frac{P_1+P_2+2\sqrt{\beta P_1P_2}}{N}
d_1d_2 \geq (1-\rho^2).
\end{IEEEeqnarray*}
It remains to optimize the correlation $\varrho_{\textnormal{sep1}}(\beta)$ over $\beta$ subject to the constraint \eqref{eq:refsep12},
\begin{equation*}
\varrho^{*}_\textnormal{sep1}=\sup_{\bar{\beta}\geq \frac{N(1-\rho^2)}{d_2P_2}}\sqrt{\beta}
=\sqrt{1-\frac{N(1-\rho^2)}{d_2P_2}}.
\end{equation*}
This concludes the proof of Corollary~\ref{th:highsnr2}. \hfill \qed

\medskip

\section{Proof of Corollary~\ref{th:highsnr31}}
\label{Proof of highsnr2 fixed}

By Proposition~\ref{proposition_SC separation1} and \cite{Michelle}, for a fixed conferencing capacity $C_{12}=C$, it
follows that any normalized distortion pair $(d_1,d_2)$ satisfying
$d_1\leq 1,d_2\leq 1$ and
\begin{IEEEeqnarray}{rCl}
d_1 & \geq & \frac{N}{\bar{\beta}_1P_1}(1-\rho^2)2^{-2C} \label{eq:refsep11C12} \\
d_2 & \geq & \frac{N}{\bar{\beta_2}P_2}(1-\rho^2) \label{eq:refsep12C12} \\
d_1d_2 & = &
\frac{N(1-{\rho}^2)}
{P_1+P_2+2\sqrt{P_1P_2}\sqrt{\beta_1\beta_2}} \ ,
\quad 0\leq\beta_1,\beta_2<1
\label{eq:refsep13C12}
\end{IEEEeqnarray}
is achievable.\\
Next, if  \eqref{eq:highsnrvq11condition} holds
then \eqref{eq:refsep11C12} and \eqref{eq:refsep12C12} are satisfied for sufficiently
small $N$ and some $0\leq \beta_1,\beta_2<1$. Thus, for $N$ sufficiently small, any pair satisfying
\eqref{eq:highsnrvq11condition} and \eqref{eq:refsep13C12}, is achievable provided that $(\beta_1,\beta_2)$ satisfy the
constraints imposed by  \eqref{eq:refsep11C12} and \eqref{eq:refsep12C12}.

It remains to optimize the correlation $\varrho_{\textnormal{sep1}}(\beta_1,\beta_2)$ over $(\beta_1,\beta_2)$
subject to the constraints  \eqref{eq:refsep11C12} and \eqref{eq:refsep12C12}.
\begin{IEEEeqnarray}{rCl}
\varrho^{*}_\textnormal{sep 1} & = & \sup_{\bar{\beta}_1\geq \frac{N(1-\rho^2)}{d_1P_1}2^{-2C} \ , \ \bar{\beta}_2\geq \frac{N(1-\rho^2)}{d_2P_2}}\sqrt{\beta_1\beta_2}
\nonumber \\
& = & \sqrt{1-\frac{N(1-\rho^2)}{d_1P_1}2^{-2C}}\sqrt{1-\frac{N(1-\rho^2)}{d_2P_2}}.
\end{IEEEeqnarray}
This concludes the proof of Corollary~\ref{th:highsnr31}. \hfill \qed

\medskip

\section{Proof of Corollary~\ref{th:highsnr22}}
\label{Proof of highsnr1 fixed}

By Theorem~\ref{th:vector-quantizer}, for a fixed conferencing capacity $C_{12}=C$, it
follows that any normalized distortion pair $(d_1,d_2)$ satisfying
$d_1\leq 1,d_2\leq 1$ and
\begin{IEEEeqnarray}{rCl}
d_1 & \geq & \frac{N}{\bar{\beta}_1P_1}2^{-2C} \label{eq:ref11C12} \\
d_2 & \geq & \frac{N}{\bar{\beta_2}P_2} \label{eq:ref12C12} \\
d_1d_2 & = &
\frac{N(1-{\rho}^2)(1-\check{\rho}^2)}
{P_1+P_2+2\sqrt{P_1P_2}\left(\rho\sqrt{\bar{\beta}_1\bar{\beta}_2}+\sqrt{\beta_1\beta_2}\right)} \ ,
\quad 0\leq\beta_1,\beta_2<1
\label{eq:ref13C12}
\end{IEEEeqnarray}
is achievable.\\
Next, if  \eqref{eq:highsnrvq11condition} holds
then \eqref{eq:ref11C12} and \eqref{eq:ref12C12} are satisfied for sufficiently
small $N$ and some $0\leq \beta_1,\beta_2<1$. Thus, for $N$ sufficiently small, any pair satisfying
\eqref{eq:highsnrvq11condition} and \eqref{eq:ref13C12}, is achievable provided that $(\beta_1,\beta_2)$ satisfy the
constraints imposed by  \eqref{eq:ref11C12} and \eqref{eq:ref12C12}.

It remains to optimize the correlation $\varrho_{\textnormal{VQ}}(\beta_1,\beta_2)$ over $(\beta_1,\beta_2)$
subject to the constraints  \eqref{eq:ref11C12} and \eqref{eq:ref12C12}. Instead we compute a lower bound
on $\varrho_{\textnormal{VQ}}^{*}$ by evaluating
$\varrho_\textnormal{VQ}(\bar{\beta}_1= \frac{N}{d_1P_1}2^{-2C},\bar{\beta}_2=\frac{N}{d_2P_2})$,
\begin{IEEEeqnarray}{rCl}
\varrho^{*}_\textnormal{VQ}  & \triangleq & \sup_{\beta_1,\beta_2}
(\rho\sqrt{\bar{\beta}_1\bar{\beta}_2}+\sqrt{\beta_1\beta_2})
\nonumber \\
& \geq &  \rho \sqrt{\frac{N}{d_1P_1}2^{-2C}}\sqrt{\frac{N}{d_2P_2}}+
  \sqrt{1- \frac{N}{d_1P_1}2^{-2C}   } \sqrt{1- \frac{N}{d_2P_2}   }
\end{IEEEeqnarray}
This concludes the proof of Corollary~\ref{th:highsnr22}. \hfill \qed

\medskip

\medskip

\begin{appendix}

\medskip

The first step in the calculation of the expected distortion of the vector-quantizer scheme is  showing
that under certain rate constraints the normalized asymptotic distortion of the genie-aided scheme
is the same as for the originally proposed scheme.

\medskip

\begin{proposition}\label{proposition_rates}
For every $\delta >0 \mbox{ and } 0< \epsilon <0.3 $ there exists an
$n'(\delta, \epsilon) \mbox{ such that for all } n>n'(\delta, \epsilon),$
\begin{equation*}
\frac{1}{n}{\sf{E}}\left[\|\mathbf{S}_{\nu}-\hat{\mathbf{S}}_{\nu}\|^2\right]
\leq
\frac{1}{n}{\sf{E}}\left[\|\mathbf{S}_{\nu}-\hat{\mathbf{S}}_{\nu}{\hspace{-.4em}}^\textnormal{G}\|^2\right]
+2\sigma^2\left(\epsilon+\left(126\sqrt{1+\epsilon}+226\right)\delta \right) \ \ , \ \
\nu=1,2
\end{equation*}
whenever $(R_1,R_2,R_\textnormal{c})$ is in the rate region ${\cal R}(\epsilon)$ given by
\begin{IEEEeqnarray*}{rCl,rCl}
{\cal R}(\epsilon)=\Biggl\{
&&& R_1 & \le & \frac{1}{2}\log \left( \frac{{{\bar{\beta }}_{1}}{{P}_{1}}\left( 1-{{{\tilde{\rho }}}^{2}} \right)+N-{{\bar{\rho }}^{2}}\left( {{\bar{\beta }}_{1}}{{P}_{1}}+N \right)}{N\left( 1-{{{\tilde{\rho }}}^{2}} \right)-{{\bar{\rho }}^{2}}N}-{{\kappa}_{1}}\epsilon  \right) \\
&&& R_2 &\le & \frac{1}{2}\log \left( \frac{{{\bar{\beta }}_{2}}{{P}_{2}}\left( 1-{{{\tilde{\rho }}}^{2}}-{{\bar{\rho }}^{2}} \right)+N}{N\left( 1-{{{\tilde{\rho }}}^{2}}-{{\bar{\rho }}^{2}} \right)+{{\lambda }_{2}}}-{{\kappa}_{2}}\epsilon  \right) \\
&&& R_\textnormal{c} & \le & \frac{1}{2}\log \left( \frac{{{\eta }^{2}}\left( 1-{{{\tilde{\rho }}}^{2}}-{{\bar{\rho }}^{2}} \right)+N\left( 1-{{{\tilde{\rho }}}^{2}} \right)}{N\left( 1-{{{\tilde{\rho }}}^{2}}-{{\bar{\rho }}^{2}} \right)+{{\lambda }_{c}}}-{{\kappa}_{3}}\epsilon  \right) \\
&&& R_1+R_2 & \le & \frac{1}{2}\log \left( \frac{{{\lambda }_{12}}-\bar{{{\beta }_{2}}}{{P}_{2}}{{\bar{\rho }}^{2}}+N}{\left( 1-\bar{{{\beta }_{2}}}{{P}_{2}}{{\bar{\rho }}^{2}}{{\lambda }_{12}}^{-1} \right)N\left( 1-{{{\tilde{\rho }}}^{2}}\  \right)}-{{\kappa}_{4}}\epsilon  \right) \\
&&& R_1+R_\textnormal{c} &\le & \frac{1}{2}\log \left( \frac{\left( {{\lambda }_{1c}}+N \right)\left( \bar{{{\beta }_{1}}}{{P}_{1}}+{{\eta }^{2}} \right)}{{{\lambda }_{1c}}N}-{{\kappa}_{5}}\epsilon  \right) \\
&&& R_2+R_\textnormal{c} &\le & \frac{1}{2}\log \left( \frac{{{\lambda }_{2c}}-\bar{{{\beta }_{2}}}{{P}_{2}}{{{\tilde{\rho }}}^{2}}+N}{\left( 1-\bar{{{\beta }_{2}}}{{P}_{2}}{{{\tilde{\rho }}}^{2}}{{\lambda }_{2c}}^{-1} \right)N\left( 1-{{{\bar{\rho }}}^{2}}\  \right)}-{{\kappa}_{6}}\epsilon  \right) \\
&&& R_1+R_2+R_\textnormal{c} & \le & \frac{1}{2}\log \left( \frac{{{\lambda }_{12}}+2\eta \bar{\rho }\sqrt{\bar{{{\beta }_{2}}}{{P}_{2}}}+{{\eta }^{2}}+N}{N\left( 1-{{{\tilde{\rho }}}^{2}}\  \right)\left( 1-{{\bar{\rho }}^{2}}\  \right)}-{{\kappa}_{7}}\epsilon  \right)\\
&&& C_{12} & \ge & \ R_\textnormal{c}+\frac{1}{2}\log \left( 1-{{\rho
}^{2}}{{2}^{-2{{R}_{1}}}}\left( 1-{{2}^{-2{{R}_\textnormal{c}}}} \right)\right) \Biggr\}
\end{IEEEeqnarray*}
where $\kappa_1, \kappa_2, \kappa_3, \kappa_4, \kappa_5, \kappa_6 \mbox{ and } \kappa_7 $ depend only on
$P_1, P_2, \tilde{\rho},\bar{\rho},\beta_1,\beta_2, \mbox{ and } N.$
\end{proposition}

\medskip

{\bf  Proof of Proposition~\ref{proposition_rates}}
\label{Proof of Proposition rates}
\medskip

We show that for any $(R_1,R_2,R_\textnormal{c}) \in {\cal R}(\epsilon)$
and sufficiently large $n$, the
probability of a decoding error, and consequently
$\Pr\left[(\hat{\textbf{S}}_1,\hat{\textbf{S}}_2)\neq
(\hat{\textbf{S}}_1{\hspace{-.4em}}^\textnormal{G},\hat{\textbf{S}}_2{\hspace{-.4em}}^\textnormal{G})\right]$ is
arbitrarily small. To this end, we consider the event consisting of
all tuples $(\textbf{s}_1,\textbf{s}_2,{\cal C}_1,{\cal C}_2,{\cal
C}_c,\textbf{z})$ for which there exists a triplet
$(\tilde{\textbf{u}}_1,\tilde{\textbf{v}},\tilde{\textbf{u}}_2)\neq
(\textbf{u}_1^{*},\textbf{v}^{*},\textbf{u}_2^{*})$ in ${\cal
C}_1\times{\cal C}_c\times{\cal C}_2$ that satisfies conditions
\eqref{eq:anglconstraints} of the reconstructor, and for which the
Euclidean distance between
$\textbf{X}_{\tilde{\textbf{u}}_1,\tilde{\textbf{v}},\tilde{\textbf{u}}_2}$
and $\textbf{y}$ is smaller or equal to the Euclidean distance
between
$\textbf{X}_{\textbf{u}_1^{*},\textbf{v}^{*},\textbf{u}_2^{*}}$ and
$\textbf{y}$.

This event is split into seven sub-events:
\begin{equation*}
{{{\cal E}}_{{\hat{\textbf{U}}}}}={{{\cal E}}_{{{{\hat{\textbf{U}}}}_{1}}}}\cup {{{\cal E}}_{{{{\hat{\textbf{U}}}}_{2}}}}\cup {{{\cal E}}_{{\hat{\textbf{V}}}}}\cup {{{\cal E}}_{\left( {{{\hat{\textbf{U}}}}_{1}},{{{\hat{\textbf{U}}}}_{2}} \right)}}\cup {{{\cal E}}_{\left( {{{\hat{\textbf{U}}}}_{1}},\hat{\textbf{V}} \right)}}\cup {{{\cal E}}_{\left( {{{\hat{\textbf{U}}}}_{2}},\hat{\textbf{V}} \right)}}\cup {{{\cal E}}_{\left( {{{\hat{\textbf{U}}}}_{1}},{{{\hat{\textbf{U}}}}_{2}},\hat{\textbf{V}} \right)}}
\end{equation*}
where
\begin{IEEEeqnarray}{rCl}
{\cal E}_{\hat{\textbf{U}}_1}
& = & \Bigl\{(\textbf{s}_1,\textbf{s}_2,{\cal C}_1,{\cal C}_2,{\cal
C}_c,\textbf{z}) \colon \exists \> \tilde{\textbf{u}}_1 \in {\cal
C}_1\setminus \{\textbf{u}_1^{*}\} \ \mbox{s.t.} \
|\tilde{\rho}-\cos\sphericalangle(\tilde{\textbf{u}}_1,\textbf{u}^{*}_2)|\leq
7\epsilon,  \nonumber \\
&& \quad \mbox{and} \ |\cos\sphericalangle(\tilde{\textbf{u}}_1,\textbf{v}^{*})|\leq
3\epsilon, \ \mbox{and} \
\|\textbf{y}-\textbf{X}_{\tilde{\textbf{u}}_1,\textbf{v}^{*},\textbf{u}^{*}_2}\|^2
\leq
\|\textbf{y}-\textbf{X}_{\textbf{u}_1^{*},\textbf{v}^{*},\textbf{u}_2^{*}}\|^2
\Bigr\} \label{eq:eps_u1} \\
{\cal E}_{\hat{\textbf{U}}_2}
& = & \Bigl\{(\textbf{s}_1,\textbf{s}_2,{\cal C}_1,{\cal C}_2,{\cal
C}_c,\textbf{z}) \colon \exists \> \tilde{\textbf{u}}_2 \in {\cal
C}_2\setminus \{\textbf{u}_2^{*}\} \ \mbox{s.t.} \
|\tilde{\rho}-\cos\sphericalangle(\textbf{u}^{*}_1,\tilde{\textbf{u}}_2)|\leq
7\epsilon,  \nonumber \\
&& \quad \mbox{and} \ |\bar{\rho}-\cos\sphericalangle(\tilde{\textbf{u}}_2,\textbf{v}^{*})|\leq
7\epsilon, \ \mbox{and} \
\|\textbf{y}-\textbf{X}_{\textbf{u}^{*}_1,\textbf{v}^{*},\tilde{\textbf{u}}_2}\|^2
\leq
\|\textbf{y}-\textbf{X}_{\textbf{u}_1^{*},\textbf{v}^{*},\textbf{u}_2^{*}}\|^2
\Bigr\} \label{eq:eps_u2} \\
{\cal E}_{\hat{\textbf{V}}}
& = & \Bigl\{(\textbf{s}_1,\textbf{s}_2,{\cal C}_1,{\cal C}_2,{\cal
C}_c,\textbf{z}) \colon \exists \> \tilde{\textbf{v}} \in {\cal
C}_c\setminus \{\textbf{v}^{*}\} \ \mbox{s.t.} \
|\cos\sphericalangle(\textbf{u}^{*}_1,\tilde{\textbf{v}})|\leq
3\epsilon,  \nonumber \\
&& \quad \mbox{and} \ |\bar{\rho}-\cos\sphericalangle(\textbf{u}^{*}_2,\tilde{\textbf{v}})|\leq
7\epsilon, \ \mbox{and} \
\|\textbf{y}-\textbf{X}_{\textbf{u}^{*}_1,\tilde{\textbf{v}},\textbf{u}^{*}_2}\|^2
\leq
\|\textbf{y}-\textbf{X}_{\textbf{u}_1^{*},\textbf{v}^{*},\textbf{u}_2^{*}}\|^2
\Bigr\} \label{eq:eps_v} \\
{\cal E}_{\left( \hat{\textbf{U}}_1,\hat{\textbf{U}}_2 \right)}
& = & \Bigl\{(\textbf{s}_1,\textbf{s}_2,{\cal C}_1,{\cal C}_2,{\cal
C}_c,\textbf{z}) \colon \exists \> \tilde{\textbf{u}}_1 \in {\cal
C}_1\setminus \{\textbf{u}_1^{*}\} \ \mbox{and} \
\exists \> \tilde{\textbf{u}}_2 \in {\cal C}_2\setminus \{\textbf{u}_2^{*}\} \
\mbox{s.t.} \  \nonumber \\
&& \quad |\tilde{\rho}-\cos\sphericalangle(\tilde{\textbf{u}}_1,\tilde{\textbf{u}}_2)|\leq
7\epsilon,
\mbox{and} \ |\cos\sphericalangle(\tilde{\textbf{u}}_1,\textbf{v}^{*})|\leq
3\epsilon,  \ \nonumber \\
&& \quad  \mbox{and} \ |\bar{\rho}-\cos\sphericalangle(\tilde{\textbf{u}}_2,\textbf{v}^{*})|\leq
7\epsilon,
\mbox{and} \ |\textbf{y}-\textbf{X}_{\tilde{\textbf{u}}_1,\textbf{v}^{*},\tilde{\textbf{u}}_2}\|^2
\leq
\|\textbf{y}-\textbf{X}_{\textbf{u}_1^{*},\textbf{v}^{*},\textbf{u}_2^{*}}\|^2
\Bigr\} \label{eq:eps_u1,u2} \\
{\cal E}_{\left( \hat{\textbf{U}}_1,\hat{\textbf{V}}\right)}
& = & \Bigl\{(\textbf{s}_1,\textbf{s}_2,{\cal C}_1,{\cal C}_2,{\cal
C}_c,\textbf{z}) \colon \exists \> \tilde{\textbf{u}}_1 \in {\cal
C}_1\setminus \{\textbf{u}_1^{*}\} \ \mbox{and} \
\exists \> \tilde{\textbf{v}} \in {\cal C}_c\setminus \{\textbf{v}^{*}\} \
\mbox{s.t.} \ \nonumber \\
&& \quad |\tilde{\rho}-\cos\sphericalangle(\tilde{\textbf{u}}_1,\textbf{u}^{*}_2)|\leq
7\epsilon,
\mbox{and} \ |\cos\sphericalangle(\tilde{\textbf{u}}_1,\tilde{\textbf{v}})|\leq
3\epsilon,  \ \nonumber \\
&& \quad  \mbox{and} \ |\bar{\rho}-\cos\sphericalangle(\textbf{u}^{*}_2,\tilde{\textbf{v}})|\leq
7\epsilon,
\mbox{and} \ |\textbf{y}-\textbf{X}_{\tilde{\textbf{u}}_1,\tilde{\textbf{v}},\textbf{u}_2^{*}}\|^2
\leq
\|\textbf{y}-\textbf{X}_{\textbf{u}_1^{*},\textbf{v}^{*},\textbf{u}_2^{*}}\|^2
\Bigr\} \label{eq:eps_u1,v} \\
{\cal E}_{\left( \hat{\textbf{U}}_2,\hat{\textbf{V}}\right)}
& = & \Bigl\{(\textbf{s}_1,\textbf{s}_2,{\cal C}_1,{\cal C}_2,{\cal
C}_c,\textbf{z}) \colon \exists \> \tilde{\textbf{u}}_2 \in {\cal
C}_2\setminus \{\textbf{u}_2^{*}\} \ \mbox{and} \
\exists \> \tilde{\textbf{v}} \in {\cal C}_c\setminus \{\textbf{v}^{*}\} \
\mbox{s.t.} \ \nonumber \\
&& \quad |\tilde{\rho}-\cos\sphericalangle(\textbf{u}^{*}_1,\tilde{\textbf{u}}_2)|\leq
7\epsilon,
\mbox{and} \ |\cos\sphericalangle(\textbf{u}^{*}_1,\tilde{\textbf{v}})|\leq
3\epsilon,  \ \nonumber \\
&& \quad  \mbox{and} \ |\bar{\rho}-\cos\sphericalangle(\tilde{\textbf{u}}_2,\tilde{\textbf{v}})|\leq
7\epsilon,
\mbox{and} \ |\textbf{y}-\textbf{X}_{\textbf{u}^{*}_1,\tilde{\textbf{v}},\tilde{\textbf{u}}_2}\|^2
\leq
\|\textbf{y}-\textbf{X}_{\textbf{u}_1^{*},\textbf{v}^{*},\textbf{u}_2^{*}}\|^2
\Bigr\} \label{eq:eps_u2,v} \\
{\cal E}_{\left(\hat{\textbf{U}}_1,\hat{\textbf{U}}_2,\hat{\textbf{V}}\right)}
& = & \Bigl\{(\textbf{s}_1,\textbf{s}_2,{\cal C}_1,{\cal C}_2,{\cal
C}_c,\textbf{z}) \colon \exists \> \tilde{\textbf{u}}_1 \in {\cal
C}_1\setminus \{\textbf{u}_1^{*}\} \> \mbox{and} \>
\exists \> \tilde{\textbf{u}}_2 \in {\cal
C}_2\setminus \{\textbf{u}_2^{*}\} \> \mbox{and} \>
\exists \> \tilde{\textbf{v}} \in {\cal C}_c\setminus \{\textbf{v}^{*}\} \nonumber \\
&& \quad \  \mbox{s.t.} \
|\tilde{\rho}-\cos\sphericalangle(\tilde{\textbf{u}}_1,\tilde{\textbf{u}}_2)|\leq
7\epsilon,
\mbox{and} \ |\cos\sphericalangle(\tilde{\textbf{u}}_1,\tilde{\textbf{v}})|\leq
3\epsilon,  \ \nonumber \\
&& \quad  \mbox{and} \ |\bar{\rho}-\cos\sphericalangle(\tilde{\textbf{u}}_2,\tilde{\textbf{v}})|\leq
7\epsilon,
\mbox{and} \ |\textbf{y}-\textbf{X}_{\tilde{\textbf{u}}_1,\tilde{\textbf{v}},\tilde{\textbf{u}}_2}\|^2
\leq
\|\textbf{y}-\textbf{X}_{\textbf{u}_1^{*},\textbf{v}^{*},\textbf{u}_2^{*}}\|^2
\Bigr\}, \label{eq:eps_u1,u2,v}
\end{IEEEeqnarray}
where $\textbf{y}\triangleq
{{a}_{1,1}}\textbf{u}_{1}^{*}+{{a}_{2,1}}\textbf{u}_{2}^{*}+\alpha{{\textbf{v}}^{*}}+\textbf{z}.$

Note that a decoding error occurs only if $(\textbf{s}_1,\textbf{s}_2,{\cal C}_1,{\cal C}_2,{\cal
C}_c,\textbf{z})\in {{{\cal E}}_{{\hat{\textbf{U}}}}} $.

The main result of this section can now be stated as follows:


\begin{lemma}\label{eps_u}
For every $\delta >0 $ and $0.3>\epsilon>0$, there exists an
$n'_{4}(\delta ,\epsilon )$ such that for all $n>{{n}'_{4}}(\delta ,\epsilon )$
\begin{equation*}
\Pr\left[ {{\cal E}}_{{\hat{\mathbf{U}}}}\right]<21\delta , \quad \mbox{whenever} \left( {{R}_{1}},{{R}_{2}},{{R}_\textnormal{c}} \right)\in {\cal R}\left( \epsilon  \right).
\end{equation*}
\end{lemma}

\medskip

To prove Lemma~\ref{eps_u}, we introduce three auxiliary error events. The
first auxiliary event ${{{\cal E}}_{\textbf{S}}}$ corresponds to an atypical source output. More precisely,
\begin{IEEEeqnarray*}{rCl}
{{{\cal E}}_{\textbf{S}}}
&=&\Bigl\{\left( {\textbf{s}_{1}},{\textbf{s}_{2}} \right)\in {{\mathbb{R}}^{n}}\times {{\mathbb{R}}^{n}}\colon
\left| \frac{1}{n}{{\left\| {\textbf{s}_{1}} \right\|}^{2}}-{{\sigma }^{2}} \right|>\epsilon {{\sigma }^{2}} \\
&& \quad \mbox{or}\ \ \left| \frac{1}{n}{{\left\| {\textbf{s}_{2}} \right\|}^{2}}-{{\sigma }^{2}} \right|>\epsilon {{\sigma }^{2}}\ \
\mbox{or}\ \ \left| \cos \sphericalangle \left( {\textbf{s}_{1}},{\textbf{s}_{2}} \right)-\rho  \right|>\epsilon \rho \Bigr\}.
\end{IEEEeqnarray*}
The second auxiliary event is denoted by  ${{{\cal E}}_{\textbf{Z}}}$ and corresponds to an atypical behavior of the additive noise:
\begin{IEEEeqnarray*}{rCl}
{{{\cal E}}_{\textbf{Z}}}
&= &\biggl\{(\textbf{s}_1,\textbf{s}_2,{\cal C}_1,{\cal C}_2,{\cal
C}_c,\textbf{z}):\ \ \left| \frac{1}{n}{{\left\| \textbf{z} \right\|}^{2}}-N \right|>\epsilon N \\
&& \quad \mbox{or}\ \ \frac{1}{n}\left| \left\langle {{a}_{1,1}}\textbf{u}_{1}^{*}\left( {\textbf{s}_{1}},{{\cal C}_{1}} \right),\textbf{z} \right\rangle  \right|>\sqrt{\bar{\beta}_{1}{{P}_{1}}N}\epsilon \ \ \mbox{or}\ \ \frac{1}{n}\left| \left\langle {{a}_{2,1}}\textbf{u}_{2}^{*}\left( {\textbf{s}_{2}},{{\cal C}_{2}} \right),\textbf{z} \right\rangle  \right|>\sqrt{\bar{\beta }_{2}{{P}_{2}}N}\epsilon \  \\
&& \quad \mbox{or}\ \ \frac{1}{n}\left| \left\langle \alpha{{\textbf{v}}^{*}}\left( {{\textbf{s}}_{1}},{{\cal C}_{1}},{{\cal C}_{c}} \right),\textbf{z} \right\rangle  \right|>\frac{1}{n}\left\| \alpha{{\textbf{v}}^{*}}\left( {{\textbf{s}}_{1}},{{\cal C}_{1}},{{\cal C}_{c}} \right) \right\|\sqrt{nN}\epsilon \biggr\}.
\end{IEEEeqnarray*}
Finally, the third auxiliary event is denoted by  ${{{\cal E}}_{\textbf{X}}}$ and corresponds to irregularities at the encoders.
That is, the event that one of the codebooks contains no codeword satisfying Condition \eqref{eq:encoding} or condition \eqref{eq:encoding_2} of the vector-quantizer,
or that the quantized sequences $\textbf{u}_{1}^{*}$ and $\textbf{u}_{2}^{*}$ and
$\textbf{v}^{*}$ have an atypical angle to each other,
or that Encoder~2 recovers a codeword $\tilde{\textbf{v}}\neq \textbf{v}^{*}.$
More formally,
${\cal E}_{\textbf{X}} = {{\cal E}_{\textbf{X}_{1}}}\cup {{\cal E}_{\textbf{X}_{2}}}\cup {{\cal E}_{\textbf{X}_{v}}}
\cup {{\cal E}_{({{\textbf{X}_{1}}},{{\textbf{X}_{2}}})}}\cup {{\cal E}_{({{\textbf{X}_{1}}},{{\textbf{X}_{v}}})}}
\cup {{\cal E}_{({{\textbf{X}_{2}}},{{\textbf{X}_{v}}})}}
\cup {{\cal E}_{\textbf{X}_\textnormal{WZ}}}$
where
\begin{IEEEeqnarray}{rCl}
{{\cal E}_{\textbf{X}_{1}}} & = & \Bigl\{ (\textbf{s}_1,\textbf{s}_2,{\cal C}_1,{\cal C}_2,{\cal
C}_c) \colon \nexists \> {{\textbf{u}}_{1}}\in {{\cal C}_{1}}\
\mbox{s.t.}\ \nonumber \\
&& \quad \left| \sqrt{1-{{2}^{-2{{R}_{1}}}}}-\cos \sphericalangle \left( {\textbf{s}_{1}},{{\textbf{u}}_{1}} \right) \right|\le \epsilon \sqrt{1-{{2}^{-2{{R}_{1}}}}} \Bigr\} \label{eq:eps_x1_def} \\
{{\cal E}_{\textbf{X}_{2}}}&=&\Bigl\{ (\textbf{s}_1,\textbf{s}_2,{\cal C}_1,{\cal C}_2,{\cal C}_c)  \colon \nexists \> {{\textbf{u}}_{2}}\in {{\cal C}_{2}}\
\mbox{s.t.}\ \nonumber \\
&& \quad \left| \sqrt{1-{{2}^{-2{{R}_{2}}}}}-\cos \sphericalangle \left( {\textbf{s}_{2}},{{\textbf{u}}_{2}} \right) \right|\le \epsilon \sqrt{1-{{2}^{-2{{R}_{2}}}}} \Bigr\} \label{eq:eps_x2_def} \\
{{\cal E}_{\textbf{X}_{v}}}&=&\Bigl\{ (\textbf{s}_1,\textbf{s}_2,{\cal C}_1,{\cal C}_2,{\cal C}_c)  \colon \nexists \> \textbf{v}\in {{\cal C}_{c}}\
\mbox{s.t.}\ \nonumber \\
&& \quad \left| \sqrt{1-{{2}^{-2{{R}_\textnormal{c}}}}}-\cos \sphericalangle \left( {{\textbf{z}}_{{{Q}_{1}}}},\textbf{v} \right) \right|\le \epsilon \sqrt{1-{{2}^{-2{{R}_\textnormal{c}}}}} \Bigr\} \label{eq:eps_xv_def} \\
{{\cal E}_{({{\textbf{X}_{1}}},{{\textbf{X}_{2}}})}}&=&\Bigl\{ (\textbf{s}_1,\textbf{s}_2,{\cal C}_1,{\cal C}_2,{\cal
C}_c)  \colon  \left| \tilde{\rho }-\cos \sphericalangle \left( \textbf{u}_{1}^{*},\textbf{u}_{2}^{*} \right) \right|>7\epsilon  \Bigr\} \label{eq:eps_x1,x2_def} \\
{{\cal E}_{({{\textbf{X}_{1}}},{{\textbf{X}_{v}}})}}&=&\Bigl\{ (\textbf{s}_1,\textbf{s}_2,{\cal C}_1,{\cal C}_2,{\cal
C}_c)  \colon  \left| \cos \sphericalangle \left( \textbf{u}_{1}^{*},{{\textbf{v}}^{*}} \right) \right|>3\epsilon  \Bigr\} \label{eq:eps_x1,xv_def} \\
{{\cal E}_{({{\textbf{X}_{2}}},{{\textbf{X}_{v}}})}}&=&\Bigl\{ (\textbf{s}_1,\textbf{s}_2,{\cal C}_1,{\cal C}_2,{\cal
C}_c)  \colon \left| \bar{\rho }-\cos \sphericalangle \left( {{\textbf{v}}^{*}},\textbf{u}_{2}^{*} \right) \right|>7\epsilon  \Bigr\} \label{eq:eps_x2,xv_def}\\
{{\cal E}_{\textbf{X}_\textnormal{WZ}}}&=&\Bigl\{ (\textbf{s}_1,\textbf{s}_2,{\cal C}_1,{\cal C}_2,{\cal C}_c)  \colon
\exists \> \tilde{\textbf{v}} \in {\cal C}_c\setminus \{\textbf{v}^{*}\}
\ \mbox{s.t.}  \left|\rho_{{\textbf{v}},\textbf{s}_2}-\cos\sphericalangle(\tilde{\textbf{v}},\textbf{s}_2)\right|\leq
5\epsilon \Bigr\} \label{eq:eps_xWZ_def}.
\end{IEEEeqnarray}

\medskip

To prove Lemma~\ref{eps_u}, we now start with the decomposition
\begin{IEEEeqnarray}{rCl}\label{eq:eps_u_composition}
\Pr\left[ {{\cal E}_{{\hat{\textbf{U}}}}} \right]
& = & \Pr\left[ {{\cal E}_{{\hat{\textbf{U}}}}}\cap {\cal E}_{\textbf{S}}^{c}\cap {\cal E}_{\textbf{X}}^{c}\cap {\cal E}_{\textbf{Z}}^{c} \right]
+\Pr\left[ \left. {{\cal E }_{{\hat{\textbf{U}}}}} \right|{{\cal E}_{\textbf{S}}}\cup {{\cal E}_{\textbf{X}}}\cup {{\cal E}_{\textbf{Z}}} \right]\Pr\left[ {{\cal E}_{\textbf{S}}}\cup {{\cal E}_{\textbf{X}}}\cup {{\cal E}_{\textbf{Z}}} \right] \nonumber \\
& \le & \Pr\left[ {{\cal E}_{{\hat{\textbf{U}}}}}\cap {\cal E} _{\textbf{S}}^{c}\cap {\cal E} _{\textbf{X}}^{c}\cap
\ {\cal E}_{\textbf{Z}}^{c} \right]+\Pr\left[ {{\cal E}_{\textbf{S}}} \right]+\Pr\left[ {{\cal E}_{\textbf{X}}} \right]+\Pr\left[ {{\cal E}_{\textbf{Z}}} \right] \nonumber \\
& \le & \Pr\left[ {{\cal E}_{{{{\hat{\textbf{U}}}}_{1}}}}\cap {\cal E}_{\textbf{S}}^{c}\cap {\cal E}_{\textbf{X}}^{c}\cap {\cal E}_{\textbf{Z}}^{c} \right]+\Pr\left[ {{\cal E}_{{{{\hat{\textbf{U}}}}_{2}}}}\cap {\cal E}_{\textbf{S}}^{c}\cap
{\cal E}_{\textbf{X}}^{c}\cap {\cal E}_{\textbf{Z}}^{c} \right]+\Pr\left[ {{\cal E}_{{\hat{\textbf{V}}}}}\cap
{\cal E}_{\textbf{S}}^{c}\cap {\cal E}_{\textbf{X}}^{c}\cap {\cal E}_{\textbf{Z}}^{c} \right] \nonumber \\
&& +\Pr \left[ {{\cal E}_{\left( {{{\hat{\textbf{U}}}}_{1}},{{{\hat{\textbf{U}}}}_{2}} \right)}}\cap {\cal E}_{\textbf{S}}^{c}\cap {\cal E}_{\textbf{X}}^{c}\cap
{\cal E}_{\textbf{Z}}^{c} \right]
+\Pr\left[ {{\cal E}_{\left( {{{\hat{\textbf{U}}}}_{1}},\hat{\textbf{V}} \right)}}\cap {\cal E}_{\textbf{S}}^{c}\cap
{\cal E}_{\textbf{X}}^{c}\cap {\cal E}_{\textbf{Z}}^{c} \right]\nonumber \\
&& +\Pr\left[ {{\cal E}_{\left( {{{\hat{\textbf{U}}}}_{2}},\hat{\textbf{V}} \right)}}\cap {\cal E}_{\textbf{S}}^{c}\cap {\cal E}_{\textbf{X}}^{c}\cap {\cal E}_{\textbf{Z}}^{c} \right]
+\Pr \left[ {{\cal E }_{\left( {{{\hat{\textbf{U}}}}_{1}},{{{\hat{\textbf{U}}}}_{2}},\hat{\textbf{V}} \right)}}\cap {\cal E} _{\textbf{S}}^{c}\cap {\cal E}_{\textbf{X}}^{c}\cap
{\cal E}_{\textbf{Z}}^{c} \right]+\Pr\left[ {{\cal E}_{\textbf{S}}} \right]\nonumber \\
&& +\Pr \left[ {{\cal E}_{\textbf{X}}} \right]+\Pr\left[ {{\cal E}_{\textbf{Z}}} \right] ,
\end{IEEEeqnarray}
where we have used the shorthand notation $\Pr\left[{{\cal E}_{\nu}} \right]$
for the probability \\
$\Pr\left[(\textbf{s}_1,\textbf{s}_2,{\cal C}_1,{\cal C}_2,{\cal
C}_c,\textbf{z})\in{{\cal E}_{\nu}} \right]$, and where ${{\cal E}_{\nu}^{c}}$
denotes the complement of ${{\cal E}_{\nu}}$.
Lemma~\ref{eps_u} now follows from upper-bounding the probability terms on
the r.h.s. of \eqref{eq:eps_u_composition}.

\medskip

\begin{lemma}\label{eps_s}
For every $\delta > 0$ and $\epsilon >0$ there exists an $n'\left( \delta ,\epsilon  \right)\in \mathbb{N}$ such that for all
$n>n'\left( \delta ,\epsilon  \right)$
\begin{equation*}
\Pr\left[ {{\cal E}_{\mathbf{S}}} \right]<\delta.
\end{equation*}
\end{lemma}
\medskip

\begin{IEEEproof}
The proof follows by the weak law of large numbers.
\end{IEEEproof}

\medskip

\begin{lemma}\label{eps_z}
For every $\delta > 0$ and $\epsilon >0$ there exists an $n'\left( \delta ,\epsilon  \right)\in \mathbb{N}$ such that for all
$n>n'\left( \delta ,\epsilon  \right)$
\begin{equation*}
\Pr\left[ {{\cal E}_{\mathbf{Z}}} \right]<\delta.
\end{equation*}
\end{lemma}

\medskip

\begin{IEEEproof}
The proof follows by the weak law of large numbers, and since for every $\epsilon >0$
\begin{equation*}
\underset{\begin{smallmatrix}
 \textbf{u}\in {{\mathbb{R}}^{n}}: \\
 \left\| \textbf{u} \right\|=\sqrt{n{{\sigma }^{2}}\left( 1-{{2}^{-2{{R}_{i}}}} \right)}
\end{smallmatrix}}{\mathop{\sup }}\,\Pr \left[ \frac{1}{n}\left\langle {{a}_{i,1}}\textbf{u},\textbf{z} \right\rangle >\sqrt{{{P}_{i}}N}\epsilon  \right]\to 0 \qquad \mbox{ as } n\to \infty ,
\end{equation*}
where $i\in \left\{ 1,2 \right\}$. The same argument holds for $\textbf{v}$.
\end{IEEEproof}

\medskip

\begin{lemma}\label{eps_x}
For every $\delta > 0$ and $1>\epsilon >0$ there exists an $n'\left( \delta ,\epsilon  \right)\in \mathbb{N}$ such that for all
$n>n'\left( \delta ,\epsilon  \right)$
\begin{equation*}
\Pr\left[ {{\cal E}_{\mathbf{X}}} \right]<12\delta.
\end{equation*}
\end{lemma}

\begin{IEEEproof}
This result follows from rate-distortion theory. The detailed proof for our setting is given in
Section~\ref{proof of eps_x} of the Appendix.
\end{IEEEproof}

\medskip

\begin{lemma}\label{rates}
For every $\delta > 0$ and $\epsilon >0$ there exists an $n_{4}^{''}\left( \delta ,\epsilon  \right)\in \mathbb{N}$ such that for all $n>n_{4}^{''}\left( \delta ,\epsilon  \right)$
\begin{IEEEeqnarray}{rCl}
  & \Pr & \left[ {{\cal E}_{{{{\hat{\mathbf{U}}}}_{1}}}}\cap {\cal E}_{\mathbf{S}}^{c}\cap {\cal E}_{\mathbf{X}}^{c}\cap
   {\cal E}_{\mathbf{Z}}^{c} \right]\le \delta ,\ \ \nonumber \\
  && \mbox{if} \ {{R}_{1}}<\frac{1}{2}\log \left( \frac{{{\bar{\beta }}_{1}}{{P}_{1}}\left( 1-{{{\tilde{\rho }}}^{2}} \right)+N-{{\bar{\rho }}^{2}}\left( {{\bar{\beta }}_{1}}{{P}_{1}}+N \right)}{N\left( 1-{{{\tilde{\rho }}}^{2}} \right)-{{\bar{\rho }}^{2}}N}-{{\kappa}_{1}}\epsilon  \right) \label{eq:R1} \\
 & \Pr &\left[ {{\cal E}_{{{{\hat{\mathbf{U}}}}_{2}}}}\cap {\cal E}_{\mathbf{S}}^{c}\cap {\cal E}_{\mathbf{X}}^{c}\cap
 {\cal E}_{\mathbf{Z}}^{c} \right]\le \delta ,\ \ \nonumber \\
 && \mbox{if } \ {{R}_{2}}<\frac{1}{2}\log \left( \frac{{{\bar{\beta }}_{2}}{{P}_{2}}\left( 1-{{{\tilde{\rho }}}^{2}}-{{\bar{\rho }}^{2}} \right)+N}{N\left( 1-{{{\tilde{\rho }}}^{2}}-{{\bar{\rho }}^{2}} \right)+{{\lambda }_{2}}}-{{\kappa}_{2}}\epsilon  \right) \label{eq:R2} \\
 & \Pr & \left[ {{\cal E}_{{\hat{\mathbf{V}}}}}\cap {\cal E}_{\mathbf{S}}^{c}\cap {\cal E}_{\mathbf{X}}^{c}\cap {\cal E}_{\mathbf{Z}}^{c} \right]\le \delta ,\ \nonumber \\
  && \mbox{if } \ {{R}_\textnormal{c}}<\frac{1}{2}\log \left( \frac{{{\eta }^{2}}\left( 1-{{{\tilde{\rho }}}^{2}}-{{\bar{\rho }}^{2}} \right)+N\left( 1-{{{\tilde{\rho }}}^{2}} \right)}{N\left( 1-{{{\tilde{\rho }}}^{2}}-{{\bar{\rho }}^{2}} \right)+{{\lambda }_{c}}}-{{\kappa}_{3}}\epsilon  \right) \label{eq:Rc} \\
 & \Pr & \left[ {{\cal E}_{\left( {{{\hat{\mathbf{U}}}}_{1}},{{{\hat{\mathbf{U}}}}_{2}} \right)}}\cap
 {\cal E}_{\mathbf{S}}^{c}\cap {\cal E}_{\mathbf{X}}^{c}\cap {\cal E}_{\mathbf{Z}}^{c} \right]\le \delta ,\ \ \nonumber \\
 && \mbox{if } \ {{R}_{1}}+{{R}_{2}}<\frac{1}{2}\log \left( \frac{{{\lambda }_{12}}-\bar{\beta }_{2}{{P}_{2}}{{\bar{\rho }}^{2}}+N}{\left( 1-\bar{{{\beta }_{2}}}{{P}_{2}}{{\bar{\rho }}^{2}}{{\lambda }_{12}}^{-1} \right)N\left( 1-{{{\tilde{\rho }}}^{2}}\  \right)}-{{\kappa}_{4}}\epsilon  \right) \label{eq:R1+R2} \\
 & \Pr & \left[ {{\cal E}_{\left( {{{\hat{\mathbf{U}}}}_{1}},\hat{\mathbf{V}} \right)}}\cap {\cal E}_{\mathbf{S}}^{c}\cap
 {\cal E}_{\mathbf{X}}^{c}\cap {\cal E}_{\mathbf{Z}}^{c} \right]\le \delta ,\ \ \nonumber \\
 && \mbox{if } \ {{R}_{1}}+{{R}_\textnormal{c}}<\frac{1}{2}\log \left( \frac{\left( {{\lambda }_{1c}}+N \right)\left( \bar{\beta }_{1}{{P}_{1}}+{{\eta }^{2}} \right)}{{{\lambda }_{1c}}N}-{{\kappa}_{5}}\epsilon  \right) \label{eq:R1+Rc} \\
 & \Pr & \left[ {{\cal E}_{\left( {{{\hat{\mathbf{U}}}}_{2}},\hat{\mathbf{V}} \right)}}\cap {\cal E}_{\mathbf{S}}^{c}\cap
 {\cal E}_{\mathbf{X}}^{c}\cap {\cal E}_{\mathbf{Z}}^{c} \right]\le \delta ,\ \ \nonumber \\
 && \mbox{if } \ {{R}_{2}}+{{R}_\textnormal{c}}<\frac{1}{2}\log \left( \frac{{{\lambda }_{2c}}-\bar{\beta }_{2}{{P}_{2}}{{{\tilde{\rho }}}^{2}}+N}{\left( 1-\bar{{{\beta }_{2}}}{{P}_{2}}{{{\tilde{\rho }}}^{2}}{{\lambda }_{2c}}^{-1} \right)N\left( 1-{{{\bar{\rho }}}^{2}}\  \right)}-{{\kappa}_{6}}\epsilon  \right) \label{eq:R2+Rc} \\
 & \Pr & \left[ {{\cal E}_{\left( {{{\hat{\mathbf{U}}}}_{1}},{{{\hat{\mathbf{U}}}}_{2}},\hat{\mathbf{V}} \right)}}\cap
 {\cal E}_{\mathbf{S}}^{c}\cap {\cal E}_{\mathbf{X}}^{c}\cap {\cal E}_{\mathbf{Z}}^{c} \right]\le \delta ,\ \ \nonumber \\
 && \mbox{if } \  {{R}_{1}}+{{R}_{2}}+{{R}_\textnormal{c}}<\frac{1}{2}\log \left( \frac{{{\lambda }_{12}}+2\eta \bar{\rho }\sqrt{\bar{\beta }_{2}{{P}_{2}}}+{{\eta }^{2}}+N}{N\left( 1-{{{\tilde{\rho }}}^{2}}\  \right)\left( 1-{{\bar{\rho }}^{2}}\  \right)}-{{\kappa}_{7}}\epsilon  \right), \label{eq:R1+R2+Rc}
\end{IEEEeqnarray}
where $\kappa_{1},\kappa_{2},\kappa_{3},\kappa_{4},\kappa_{5},\kappa_{6}$ and $\kappa_{7}$ are positive constants determined by $P_{1}$, $P_{2}$, and $N$.
\end{lemma}
\medskip

The proof of this lemma appears in subsections A--G of the Appendix.
\medskip

\textbf{Concluding the proof of Proposition~\ref{proposition_rates}}

\medskip

We start with five lemmas. The first lemma upper bounds the impact
of atypical source outputs on the expected distortion.

\begin{lemma}\label{Genie_1}
For every $\epsilon > 0 $
\begin{equation*}
\frac{1}{n}{\sf{E}}\left[\|\mathbf{S}_1\|^{2}\mid {\cal E}_\mathbf{S}\right]\Pr [{\cal E}_\mathbf{S}]
\leq \sigma^{2}(\epsilon+\Pr [{\cal E}_\mathbf{S}]).
\end{equation*}
\end{lemma}

\medskip

\begin{IEEEproof}
\begin{IEEEeqnarray*}{+rCl+x*}
\frac{1}{n}{\sf{E}}\left[\|\textbf{S}_1\|^{2}\mid {\cal E}_\textbf{S}\right]\Pr [{\cal E}_\textbf{S}]
& = &\frac{1}{n}{\sf{E}}\left[\|\textbf{S}_1\|^{2}\right]-\frac{1}{n}{\sf{E}}\left[\|\textbf{S}_1\|^{2}\mid {\cal E}^{c}_\textbf{S}\right]\Pr [{\cal E}^{c}_\textbf{S}]\\
&\le & \sigma^{2}-\sigma^{2}(1-\epsilon)\Pr [{\cal E}^{c}_\textbf{S}]\\
& = & \sigma^{2}-\sigma^{2}(1-\epsilon)(1-\Pr [{\cal E}_\textbf{S}])\\
& = & \sigma^{2}\epsilon+\sigma^{2}(1-\epsilon)\Pr [{\cal E}_\textbf{S}]\\
&\le & \sigma^{2}(\epsilon+\Pr [{\cal E}_\textbf{S}]).&
\end{IEEEeqnarray*}
\end{IEEEproof}

\medskip

The second lemma considers the properties of the estimator coefficients.
\begin{lemma}\label{MMSE_coefficients_bounds}
The gain coefficients in \eqref{eq:MMSE_coefficients} satisfy
$ \gamma_{1,1},\gamma_{1,3},\gamma_{2,2}\leq 1$  and  $\gamma_{1,2},\gamma_{2,1},\gamma_{2,3}\leq \rho $ .
\end{lemma}

\medskip

\begin{IEEEproof}
The first claim is obvious, so we will first show that $\gamma_{1,2}\leq \rho $.
Note that
\begin{IEEEeqnarray*}{rCl}
\gamma_{1,2}&=&\frac{\rho 2^{-2(R_1+R_\textnormal{c})}}{1-\rho^2(1-2^{-2R_2})(1-2^{-2(R_1+R_\textnormal{c})})}\\
            &=&\frac{\rho}{2^{2(R_1+R_\textnormal{c})}-\rho^2(1-2^{-2R_2})(2^{2(R_1+R_\textnormal{c})}-1)}\\
            &=&\frac{\rho}{2^{2(R_1+R_\textnormal{c})}(1-\rho^2(1-2^{-2R_2}))+\rho^2(1-2^{-2R_2})}.
\end{IEEEeqnarray*}
Now, consider the function
\begin{equation*}
f(\alpha,\beta)=\frac{1}{\beta(1-\alpha)+\alpha},
\end{equation*}
where $0<\alpha<1$ and $\beta\geq 1$.
Note that $f(\alpha,1)=1$, and that
$\frac{\partial f(\alpha,\beta)}{\partial\beta}=-\frac{(1-\alpha)}{(\beta(1-\alpha)+\alpha)^2}<0$.
Thus $f(\alpha,\beta)$ is continuous and monotonically decreasing in $\beta$ for $\beta\geq1$.\\
On the other hand, note that $f(0,\beta)=\frac{1}{\beta}<1$ assuming $\beta>1$, and $f(1,\beta)=1$ and
$\frac{\partial f(\alpha,\beta)}{\partial\alpha}=-\frac{(1-\beta)}{(\beta(1-\alpha)+\alpha)^2}>0$.
Thus, $f(\alpha,\beta)$ is continuous and monotonically increasing in $\alpha$ for $0<\alpha<1$,
and therefore $0 < f(\alpha,\beta) \leq 1$ for $0<\alpha<1$ and $\beta >1$.

The proof that  $\gamma_{2,1},\gamma_{2,3}\leq \rho $ follows in a similar way.
\end{IEEEproof}

\bigskip

The third lemma gives upper bounds on norms related to the reconstructions
$\hat{\textbf{s}}_{1}$ and $\hat{\textbf{s}}_1{\hspace{-.4em}}^{\textnormal{G}}$.

\begin{lemma}\label{Genie_2}
Let the reconstructions $\hat{\mathbf{s}}_{1}$ and $\hat{\mathbf{s}}_1{\hspace{-.4em}}^{\textnormal{G}}$ be as defined in
\eqref{eq:MMSEestimation1} and \eqref{eq:Genie_estimation1}. Then,
\begin{equation*}
\left\|\hat{\mathbf{s}}_{1}\right\|^{2}\leq  9n\sigma^{2},
\quad \|\hat{\mathbf{s}}_1{\hspace{-.4em}}^{\textnormal{G}}\|^{2}\leq  9n\sigma^{2},
\quad \|\hat{\mathbf{s}}_1{\hspace{-.4em}}^{\textnormal{G}}-\hat{\mathbf{s}}_{1}\|^2\leq  36n\sigma^{2}.
\end{equation*}
\end{lemma}

\medskip

\begin{IEEEproof}
We start by upper-bounding the squared norm of $\hat{\textbf{s}}_{1}$
\begin{IEEEeqnarray*}{rCl}
\left\|\hat{\textbf{s}}_{1}\right\|^{2}
&=&\left\|\gamma_{1,1}\hat{\textbf{u}}_{1}+\gamma_{1,2}\hat{\textbf{u}}_{2}+\gamma_{1,3}\hat{\textbf{v}}\right\|^{2}\\
&=&\gamma_{1,1}^{2}\|\hat{\textbf{u}}_{1}\|^{2}
+2\gamma_{1,1}\gamma_{1,2}\langle \hat{\textbf{u}}_{1},\hat{\textbf{u}}_{2} \rangle
+\gamma_{1,2}^{2}\|\hat{\textbf{u}}_{2}\|^{2}
+2\gamma_{1,1}\gamma_{1,3}\langle \hat{\textbf{u}}_{1},\hat{\textbf{v}} \rangle
+2\gamma_{1,2}\gamma_{1,3}\langle \hat{\textbf{u}}_{2},\hat{\textbf{v}} \rangle \\
&& +\gamma_{1,3}^{2}\|\hat{\textbf{v}}\|^{2} \\
& \le & \gamma_{1,1}^{2}\|\hat{\textbf{u}}_{1}\|^{2}
+2\gamma_{1,1}\gamma_{1,2} \|\hat{\textbf{u}}_{1}\|\|\hat{\textbf{u}}_{2}\|
+\gamma_{1,2}^{2}\|\hat{\textbf{u}}_{2}\|^{2}
+2\gamma_{1,1}\gamma_{1,3}\| \hat{\textbf{u}}_{1}\|\|\hat{\textbf{v}}\|\\
&&  +2\gamma_{1,2}\gamma_{1,3}\| \hat{\textbf{u}}_{2}\|\|\hat{\textbf{v}}\|
+\gamma_{1,3}^{2}\|\hat{\textbf{v}}\|^{2} \\
&=&\left(\gamma_{1,1}\|\hat{\textbf{u}}_{1}\|+\gamma_{1,2}\|\hat{\textbf{u}}_{2}\|+\gamma_{1,3}\|\hat{\textbf{v}}\|\right)^{2}\\
& \overset{(a)}{\le}& n\sigma^{2}(2+\rho)^2 \\
& \le & 9n\sigma^{2} ,
\end{IEEEeqnarray*}
where in (a) we have used Lemma~\ref{MMSE_coefficients_bounds}, i.e. that
$\gamma_{1,1},\gamma_{1,3}\leq 1 $ and $\gamma_{1,2}\leq \rho$, and that
$\|\hat{\textbf{u}}_{i}\|\leq \sqrt{n\sigma^{2}},i \in \{1,2\}$ and $\|\hat{\textbf{v}}\|\leq \sqrt{n\sigma^{2}}$.
The upper bound on the squared norm of $\hat{\textbf{s}}_1{\hspace{-.4em}}^{\textnormal{G}}$ is obtained similarly. Its proof is therefore omitted.
The upper bound on the squared norm of the difference between $\hat{\textbf{s}}_{1}$ and $\hat{\textbf{s}}_1{\hspace{-.4em}}^{\textnormal{G}}$
now follows easily:
\begin{IEEEeqnarray*}{+rCl+x*}
\left\|\hat{\textbf{s}}_1{\hspace{-.4em}}^{\textnormal{G}}-\hat{\textbf{s}}_{1}\right\|^{2}
& \le & \|\hat{\textbf{s}}_1{\hspace{-.4em}}^{\textnormal{G}}\|^{2}+2\|\hat{\textbf{s}}_1{\hspace{-.4em}}^{\textnormal{G}}\|\|\hat{\textbf{s}}_{1}\|
+\|\hat{\textbf{s}}_{1}\|^{2}\\
& = & (\|\hat{\textbf{s}}_1{\hspace{-.4em}}^{\textnormal{G}}\|+\|\hat{\textbf{s}}_{1}\|)^{2} \\
& \le & 36n\sigma^{2}. &
\end{IEEEeqnarray*}
\end{IEEEproof}

\medskip

The next two lemmas are used directly in the upcoming proof of
Proposition~\ref{proposition_rates}. They rely on Lemma~\ref{Genie_1} and Lemma~\ref{Genie_2}.

\begin{lemma}\label{Genie_3}
\begin{equation*}
\frac{1}{n}{\sf{E}}\left[ \langle { \mathbf{S}_{1},\hat{\mathbf{S}}_1{\hspace{-.4em}}^\textnormal{G}-\hat{\mathbf{S}}_{1} }\rangle \right]
\leq \sigma^2 \left(\epsilon +37\Pr\left[{\cal E}_{\mathbf{S}}\right]+6\sqrt{1+\epsilon}\Pr\left[{\cal E}_{\hat{\mathbf{U}}}\right]\right).
\end{equation*}
\end{lemma}

\medskip

\begin{IEEEproof}
\begin{IEEEeqnarray}{rCl}
\frac{1}{n}{\sf{E}}\left[ \langle { \textbf{S}_{1},\hat{\textbf{S}}_1{\hspace{-.4em}}^\textnormal{G}-\hat{\textbf{S}}_{1} }\rangle \right]
&=&\frac{1}{n}{\sf{E}}\left[ \langle { \textbf{S}_{1},\hat{\textbf{S}}_1{\hspace{-.4em}}^\textnormal{G}-\hat{\textbf{S}}_{1} }\rangle \mid {\cal E}_{\textbf{S}} \right]\Pr \left[{\cal E}_{\textbf{S}}\right] \nonumber \\
&& +\frac{1}{n}{\sf{E}}\left[ \langle { \textbf{S}_{1},\hat{\textbf{S}}_1{\hspace{-.4em}}^\textnormal{G}-\hat{\textbf{S}}_{1} }\rangle
\mid {\cal E}^{c}_{\textbf{S}} \cap {\cal E}_{\hat{\textbf{U}}} \right]\Pr \left[{\cal E}^{c}_{\textbf{S}} \cap {\cal E}_{\hat{\textbf{U}}}\right] \nonumber \\
&& +\frac{1}{n}\underbrace{{\sf{E}}\left[ \langle { \textbf{S}_{1},\hat{\textbf{S}}_1{\hspace{-.4em}}^\textnormal{G}-\hat{\textbf{S}}_{1} }\rangle
\mid {\cal E}^{c}_{\textbf{S}} \cap {\cal E}^{c}_{\hat{\textbf{U}}} \right]}_{=0}\Pr \left[{\cal E}^{c}_{\textbf{S}} \cap {\cal E}^{c}_{\hat{\textbf{U}}}\right] \nonumber \\
&\overset{(a)}{\le} & \frac{1}{n}{\sf{E}}\left[\|\textbf{S}_{1}\|^2+\|\hat{\textbf{S}}_1{\hspace{-.4em}}^\textnormal{G}-\hat{\textbf{S}}_{1}\|^2 \mid {\cal E}_{\textbf{S}} \right]\Pr \left[{\cal E}_{\textbf{S}}\right]  \nonumber \\
&& +\frac{1}{n}{\sf{E}}\left[\|\textbf{S}_{1}\|\|\hat{\textbf{S}}_1{\hspace{-.4em}}^\textnormal{G}-\hat{\textbf{S}}_{1}\|
\mid {\cal E}^{c}_{\textbf{S}} \cap {\cal E}_{\hat{\textbf{U}}} \right]\Pr \left[{\cal E}_{\hat{\textbf{U}}}\right] \nonumber \\
&\overset{(b)}{\le} & \frac{1}{n}{\sf{E}}\left[\|\textbf{S}_{1}\|^2 \mid {\cal E}_{\textbf{S}} \right]\Pr \left[{\cal E}_{\textbf{S}}\right]+36\sigma^2\Pr \left[{\cal E}_{\textbf{S}}\right]  \nonumber \\
&& +\sqrt{\sigma^2(1+\epsilon)}\sqrt{36\sigma^2}\Pr \left[{\cal E}_{\hat{\textbf{U}}}\right] \nonumber \\
&\overset{(c)}{\le} & \sigma^2(\epsilon+\Pr \left[{\cal E}_{\textbf{S}}\right])+36\sigma^2\Pr \left[{\cal E}_{\textbf{S}}\right]  \nonumber \\
&& +6\sigma^2\sqrt{1+\epsilon}\Pr \left[{\cal E}_{\hat{\textbf{U}}}\right] \nonumber \\
&=& \sigma^2(\epsilon+37\Pr \left[{\cal E}_{\textbf{S}}\right])+6\sqrt{1+\epsilon}
\Pr \left[{\cal E}_{\hat{\textbf{U}}}\right].
\end{IEEEeqnarray}
In the first equality the third expectation term equals zero because by
${\cal E}^c_{\hat{\textbf{U}}}$ we have  $\|\hat{\textbf{s}}_1{\hspace{-.4em}}^{\textnormal{G}}-\textbf{s}_{1}\|=0$
and by ${\cal E}^{c}_{\textbf{s}}$ the norm $\|\textbf{s}_1\|$ is bounded.
In (a) we have used two inequalities: in the first term the inner product is upper
bounded by using \eqref{eq:inner_product_bound}.
The second term is upper bounded by the Cauchy-Schwarz inequality and by
$\Pr \left[{\cal E}^{c}_{\textbf{S}} \cap {\cal E}_{\hat{\textbf{U}}}\right]\leq
\Pr \left[ {\cal E}_{\hat{\textbf{U}}}\right]$.
In (b) we have used Lemma~\ref{Genie_2} and in (c) we have used Lemma~\ref{Genie_1}.
\end{IEEEproof}

\medskip

\begin{lemma}\label{Genie_4}
\begin{equation*}
\frac{1}{n}{\sf{E}}\left[ \|\hat{\mathbf{S}}_{1}\|^{2}-\|\hat{\mathbf{S}}_1{\hspace{-.4em}}^\textnormal{G}\|^{2}  \right]
\leq 18\sigma^2 \Pr\left[{\cal E}_{\hat{\mathbf{U}}}\right].
\end{equation*}
\end{lemma}

\medskip

\begin{IEEEproof}
\begin{IEEEeqnarray*}{rCl}
\frac{1}{n}{\sf{E}}\left[ \|\hat{\textbf{S}}_{1}\|^{2}-\|\hat{\textbf{S}}_1{\hspace{-.4em}}^\textnormal{G}\|^{2}  \right]
&=& \frac{1}{n}{\sf{E}}\left[ \|\hat{\textbf{S}}_{1}\|^{2}-\|\hat{\textbf{S}}_1{\hspace{-.4em}}^\textnormal{G}\|^{2}
\middle | {\cal E}_{\hat{\textbf{U}}} \right]
\Pr\left[{\cal E}_{\hat{\textbf{U}}} \right]\\
&& +\frac{1}{n}{\sf{E}}\left[ \|\hat{\textbf{S}}_{1}\|^{2}-\|\hat{\textbf{S}}_1{\hspace{-.4em}}^\textnormal{G}\|^{2}
\middle | {\cal E}^{c}_{\hat{\textbf{U}}} \right]
\Pr\left[{\cal E}^{c}_{\hat{\textbf{U}}} \right]\\
& \overset{(a)}{\le} & \frac{1}{n}{\sf{E}}\left[ \|\hat{\textbf{S}}_{1}\|^{2}+\|\hat{\textbf{S}}_1{\hspace{-.4em}}^\textnormal{G}\|^{2} \middle | {\cal E}_{\hat{\textbf{U}}} \right] \Pr\left[{\cal E}_{\hat{\textbf{U}}} \right]\\
& \overset{(b)}{\le} & 18\sigma^2
\Pr\left[{\cal E}_{\hat{\textbf{U}}} \right] ,
\end{IEEEeqnarray*}
where (a) follows since conditioned on ${\cal E}^{c}_{\hat{\textbf{U}}}$ we have
$\hat{\textbf{s}}_{1}=\hat{\textbf{s}}_1{\hspace{-.4em}}^{\textnormal{G}}$ and therefore
$\|\hat{\textbf{s}}_{1}\|^{2}-\|\hat{\textbf{s}}_1{\hspace{-.4em}}^{\textnormal{G}}\|=0$,
and where (b) follows by Lemma~\ref{Genie_2}.
\end{IEEEproof}

\medskip

{\it {Proof of Proposition~\ref{proposition_rates}}}.
We show that the asymptotic normalized distortion resulting from the proposed vector-quantizer scheme,
is the same as the asymptotic normalized distortion resulting from the genie-aided
version of this scheme.
\begin{IEEEeqnarray}{rCl}
\IEEEeqnarraymulticol{3}{l}{
\frac{1}{n}{\sf{E}}\left[ \|\textbf{S}_{1}-\hat{\textbf{S}}_{1}\|^{2}  \right]
-\frac{1}{n}{\sf{E}}\left[ \|\textbf{S}_{1}-\hat{\textbf{S}}_1{\hspace{-.4em}}^\textnormal{G}\|^{2}  \right]} \nonumber \\ \qquad
& = & \frac{1}{n}\Bigl(
{\sf{E}}\left[ \|\textbf{S}_{1}\|^{2}  \right]
-2{\sf{E}}\left[ \langle \textbf{S}_{1},\hat{\textbf{S}}_{1} \rangle  \right]
+{\sf{E}}\left[ \|\hat{\textbf{S}}_{1}\|^{2}  \right] \nonumber\\
&& -{\sf{E}}\left[ \|\textbf{S}_{1}\|^{2}  \right]
+2{\sf{E}}\left[ \langle \textbf{S}_{1},\hat{\textbf{S}}_1{\hspace{-.4em}}^\textnormal{G} \rangle  \right]
-{\sf{E}}\left[ \|\hat{\textbf{S}}_1{\hspace{-.4em}}^\textnormal{G}\|^{2}  \right]\Bigr) \nonumber\\
&=& 2\frac{1}{n}{\sf{E}}\left[ \langle \textbf{S}_{1},\hat{\textbf{S}}_1{\hspace{-.4em}}^\textnormal{G}-\hat{\textbf{S}}_{1} \rangle  \right]
+\frac{1}{n}{\sf{E}}\left[ \|\hat{\textbf{S}}_{1}\|^{2}-\|\hat{\textbf{S}}_1{\hspace{-.4em}}^\textnormal{G}\|^{2}  \right] \nonumber\\
& \overset{(a)}{\le} &  2\sigma^2\left(\epsilon+37\Pr \left[{\cal E}_{\textbf{S}}\right]+6\sqrt{1+\epsilon}
\Pr\left[{\cal E}_{\hat{\textbf{U}}}\right]\right)+18\sigma^2\Pr\left[{\cal E}_{\hat{\textbf{U}}}\right]\nonumber\\
& =& 2\sigma^2\left(\epsilon+37\Pr \left[{\cal E}_{\textbf{S}}\right]
+\left(6\sqrt{1+\epsilon}+9\right)
\Pr\left[{\cal E}_{\hat{\textbf{U}}}\right]\right),
\end{IEEEeqnarray}\label{eq:concluding_proposition1}
where in step (a) we have used Lemma~\ref{Genie_3} and Lemma~\ref{Genie_4}.
Combining \eqref{eq:concluding_proposition1} with Lemma~\ref{eps_u} and Lemma~\ref{eps_s}, gives that for every
$\delta>0$ and $0.3>\epsilon>0$, there exists an $n'(\delta,\epsilon)>0$ such that for all
$(R_1,R_2,R_\textnormal{c}) \in {\cal R}(\epsilon)$ and $n>n'(\delta,\epsilon)$
\begin{flalign*}
&&
\frac{1}{n}{\sf{E}}\left[ \|\textbf{S}_{1}-\hat{\textbf{S}}_{1}\|^{2}  \right]
-\frac{1}{n}{\sf{E}}\left[ \|\textbf{S}_{1}-\hat{\textbf{S}}_1{\hspace{-.4em}}^\textnormal{G}\|^{2}  \right]
<2\sigma^2\left(\epsilon+\left(126\sqrt{1+\epsilon}+226\right)\delta \right).
&&
\qed
\end{flalign*}
\subsection{Proof of rate constraint~\eqref{eq:R1}}

Define
\begin{equation}\label{eq:w_for_R1}
\textbf{w}(\textbf{s}_1,\textbf{s}_2,{\cal C}_1,{\cal C}_2,{\cal C}_c,\textbf{z})={{\varsigma }_{1}}\left( \textbf{y}-\left( \alpha{{\textbf{v}}^{*}}+{{a}_{2,1}}\textbf{u}_{2}^{*} \right) \right)+{{\varsigma }_{2}}{{a}_{2,1}}\textbf{u}_{2}^{*}+{{\varsigma }_{3}}\alpha{{\textbf{v}}^{*}},
\end{equation}
where
\begin{IEEEeqnarray}{rCl}\label{eq:w_coeff_for_R1}
{{\varsigma }_{1}}
&=&\frac{{{\sigma }^{2}}{a}_{1,1}^{2}\left( 1-{{2}^{-2{{R}_{1}}}} \right)\left( 1-{{\rho }^{2}}\left( 1-{{2}^{-2{{R}_{2}}}} \right)\left( 1-{{2}^{-2\left( {{R}_{1}}+{{R}_\textnormal{c}} \right)}} \right) \right)}{{{\sigma }^{2}}{a}_{1,1}^{2}\left( 1-{{2}^{-2{{R}_{1}}}} \right)\left( 1-{{\rho }^{2}}\left( 1-{{2}^{-2{{R}_{2}}}} \right)\left( 1-{{2}^{-2\left( {{R}_{1}}+{{R}_\textnormal{c}} \right)}} \right) \right)+N\left( 1-{{\bar{\rho }}^{2}} \right)} \nonumber \\
{{\varsigma }_{2}}
&=&\frac{{{a}_{1,1}}\rho \left( 1-{{2}^{-2{{R}_{1}}}} \right)N}{{{a}_{2,1}}\left( {{\sigma }^{2}}{a}_{1,1}^{2}\left( 1-{{2}^{-2{{R}_{1}}}} \right)\left( 1-{{\rho }^{2}}\left( 1-{{2}^{-2{{R}_{2}}}} \right)\left( 1-{{2}^{-2\left( {{R}_{1}}+{{R}_\textnormal{c}} \right)}} \right) \right)+N\left( 1-{{\bar{\rho }}^{2}} \right) \right)} \nonumber \\
{{\varsigma }_{3}}
&=&\frac{-{{a}_{1,1}}{{{\tilde{\rho }}}^{2}}N}{\alpha\left( {{\sigma }^{2}}{a}_{1,1}^{2}\left( 1-{{2}^{-2{{R}_{1}}}} \right)\left( 1-{{\rho }^{2}}\left( 1-{{2}^{-2{{R}_{2}}}} \right)\left( 1-{{2}^{-2\left( {{R}_{1}}+{{R}_\textnormal{c}} \right)}} \right) \right)+N\left( 1-{{\bar{\rho }}^{2}} \right) \right)}.\nonumber \\
\end{IEEEeqnarray}
In the remainder we shall use the shorthand notation \textbf{w} instead of
$\textbf{w}(\textbf{s}_1,\textbf{s}_2,{\cal C}_1,{\cal C}_2,{\cal C}_c,\textbf{z})$.
We now start with a lemma that will be used to prove \eqref{eq:R1}.

\medskip

\begin{lemma}\label{eps_h_u1}
Let ${{\varphi }_{j}}\in \left[ 0,\pi  \right]$  be the angle between $\mathbf{w}$ and $\mathbf{u}_{1}(j)$, and let the set ${\cal E}'_{{{{\hat{\mathbf{U}}}}_{1}}}$
be defined as
\begin{IEEEeqnarray}{rCl} \label{eq:eps_h_u1}
{\cal E}'_{{{{\hat{\mathbf{U}}}}_{1}}}
& \triangleq &
\Biggl\{ (\mathbf{s}_1,\mathbf{s}_2,{\cal C}_1,{\cal C}_2,{\cal C}_c,\mathbf{z})\colon \exists \> {{\mathbf{u}}_{1}}\left( j \right)\in {{\cal C}_{1}}\backslash \left\{ \mathbf{u}_{1}^{*} \right\}\ \mbox{s.t.} \ \nonumber \\
&& \qquad \quad \cos \left( {{\varphi }_{j}} \right)\ge \sqrt{\frac{{{\bar{\beta }}_{1}}{{P}_{1}}\left( 1-{{{\tilde{\rho }}}^{2}} \right)+N{{{\tilde{\rho }}}^{2}}-{{\bar{\rho }}^{2}}{{\bar{\beta }}_{1}}{{P}_{1}}}{{{\bar{\beta }}_{1}}{{P}_{1}}\left( 1-{{{\tilde{\rho }}}^{2}} \right)+N-{{\bar{\rho }}^{2}}\left( {{\bar{\beta }}_{1}}{{P}_{1}}+N \right)}-{{\kappa}^{''}}\epsilon }\  \Biggr\},
\end{IEEEeqnarray}
where ${{\kappa}^{''}}$ is a positive constant determined by $P_{1}, P_{2}, N,
{\varsigma}_{1}, {\varsigma}_{2} \mbox{ and } {\varsigma}_{3}$.
Then,
\begin{equation*}
{{{\cal E}}_{{{{\hat{\mathbf{U}}}}_{1}}}}\cap {\cal E}_{\mathbf{S}}^{c}\cap {\cal E}_{\mathbf{X}}^{c}\cap {\cal E}_{\mathbf{Z}}^{c}\subseteq {\cal E}'_{{{{\hat{\mathbf{U}}}}_{1}}}\cap {\cal E}_{\mathbf{S}}^{c}\cap {\cal E}_{\mathbf{X}}^{c}\cap {\cal E}_{\mathbf{Z}}^{c},
\end{equation*}
and, in particular
\begin{equation*}
\Pr\left[ {{{\cal E}}_{{{{\hat{\mathbf{U}}}}_{1}}}}\cap {\cal E}_{\mathbf{S}}^{c}\cap {\cal E}_{\mathbf{X}}^{c}\cap {\cal E}_{\mathbf{Z}}^{c} \right]\le \Pr\left[ {\cal E}'_{{{{\hat{\mathbf{U}}}}_{1}}}\cap {\cal E}_{\mathbf{S}}^{c}\cap {\cal E}_{\mathbf{X}}^{c}\cap {\cal E}_{\mathbf{Z}}^{c} \right].
\end{equation*}
\end{lemma}

\medskip

\begin{IEEEproof}
We first recall that for the event ${{\cal E}_{{{{\hat{\textbf{U}}}}_{1}}}}$
to occur, there must exist a codeword ${{\textbf{u}}_{1}}\left( j \right)\in {{{\cal C}}_{1}}\backslash \left\{ \textbf{u}_{1}^{*} \right\}$ that satisfies
the following three conditions
\begin{IEEEeqnarray}{rCl}
\left| \tilde{\rho }-\cos \sphericalangle \left( {{\textbf{u}}_{1}}\left( j \right),\textbf{u}_{2}^{*} \right) \right| & \le & 7\epsilon \\
\left| \cos \sphericalangle \left( {{\textbf{v}}^{*}},{{\textbf{u}}_{1}}\left( j \right) \right) \right| & \le & 3\epsilon \\
\|\textbf{y}-\textbf{X}_{\textbf{u}_{1}\left(j\right),\textbf{v}^{*},\textbf{u}^{*}_2}\|^2
 & \leq &
\|\textbf{y}-\textbf{X}_{\textbf{u}_1^{*},\textbf{v}^{*},\textbf{u}_2^{*}}\|^2 .
\end{IEEEeqnarray}

The proof is now based on a sequence of statements related to these three conditions:

\medskip

A) For every $(\textbf{s}_1,\textbf{s}_2,{\cal C}_1,{\cal C}_2,{\cal C}_c,\textbf{z})\in {\cal E}_{\textbf{X}}^{c}$ and every $\textbf{u}\in {\cal S}_{1}$, where ${\cal S}_{1}$ is
the surface area of the codeword sphere of ${\cal C}_{1}$ defined in the code construction,
\begin{IEEEeqnarray}{rCl}\label{eq:statement_A_R1}
&& \left| \tilde{\rho }-\cos \sphericalangle \left( \textbf{u},\textbf{u}_{2}^{*} \right) \right|\le 7\epsilon
\nonumber \\
&&\text{  }\implies \text{   }\left| n\tilde{\rho }\sqrt{\bar{{\beta}}_{1}\bar{{\beta}}_{2}{{P}_{1}}{{P}_{2}}}-\left\langle {{a}_{1,1}}\textbf{u},{{a}_{2,1}}\textbf{u}_{2}^{*} \right\rangle  \right|
\le 7n\sqrt{\bar{{\beta}}_{1}\bar{{\beta}}_{2}{{P}_{1}}{{P}_{2}}}\epsilon .
\end{IEEEeqnarray}

Statement A) follows by rewriting $\cos \sphericalangle \left( \textbf{u},\textbf{u}_{2}^{*} \right)$ as
${\left\langle \textbf{u},\textbf{u}_{2}^{*} \right\rangle }/({\left\| \textbf{u} \right\|\left\| \textbf{u}_{2}^{*} \right\|})$, and then multiplying the inequality on the l.h.s. of
\eqref{eq:statement_A_R1}  by $\left\| {{a}_{1,1}}\textbf{u} \right\|\cdot \left\| {{a}_{2,1}}\textbf{u}_{2}^{*} \right\|$
and recalling that $\left\| {{a}_{1,1}}\textbf{u} \right\|=\sqrt{n\bar{\beta}_{1}{{P}_{1}}}\ $ and that $\left\| {{a}_{2,1}}\textbf{u}_{2}^{*} \right\|=\sqrt{n\bar{\beta}_{2}{{P}_{2}}} $.

\medskip

A1) For every $(\textbf{s}_1,\textbf{s}_2,{\cal C}_1,{\cal C}_2,{\cal C}_c,\textbf{z})\in {\cal E}_{\textbf{X}}^{c}$
\begin{IEEEeqnarray}{rCl} \label{eq:statement_A1_R1}
&& \left| \cos \sphericalangle \left( \textbf{v}^{*},\textbf{u} \right) \right|\le 3\epsilon  \nonumber \\
&& \text{       }\implies \text{      }\left| \left\langle \alpha\textbf{v}^{*},{{a}_{1,1}}\textbf{u} \right\rangle  \right|\le 3\left\| \alpha\textbf{v}^{*} \right\|\sqrt{n\bar{{{\beta }_{1}}}{{P}_{1}}}\epsilon .
\end{IEEEeqnarray}

Statement A1) follows by rewriting $\cos \sphericalangle \left( \textbf{v}^{*},\textbf{u} \right)$ as ${\left\langle \textbf{v}^{*},\textbf{u} \right\rangle }/({\left\| \textbf{v}^{*} \right\|\left\| \textbf{u} \right\|})$, and then multiplying the inequality on the l.h.s. of \eqref{eq:statement_A1_R1} by
$\left\| \alpha\textbf{v}^{*} \right\|\cdot \left\| {{a}_{1,1}}\textbf{u} \right\|$.

\medskip

A2) For every $(\textbf{s}_1,\textbf{s}_2,{\cal C}_1,{\cal C}_2,{\cal C}_c,\textbf{z})\in {\cal E}_{\textbf{X}}^{c}$
\begin{IEEEeqnarray}{rCl} \label{eq:statement_A2_R1}
&& \left| \bar{\rho }-\cos \sphericalangle \left( \textbf{v}^{*},\textbf{u}_{2}^{*} \right) \right|\le 7\epsilon  \nonumber \\
&& \text{       }\implies \text{      }\left| \left\| \alpha\textbf{v}^{*} \right\|\sqrt{n\bar{{{\beta }_{2}}}{{P}_{2}}}\bar{\rho }-\left\langle \alpha\textbf{v}^{*},{{a}_{2,1}}\textbf{u}_{2}^{*} \right\rangle  \right|\le 7\epsilon \left\| \alpha\textbf{v}^{*} \right\|\sqrt{n\bar{{{\beta }_{2}}}{{P}_{2}}}.
\end{IEEEeqnarray}

Statement A2) follows by rewriting $\cos \sphericalangle \left( \textbf{v}^{*},\textbf{u}_{2}^{*} \right)$ as ${\left\langle \textbf{v}^{*},\textbf{u}_{2}^{*} \right\rangle }/({\left\| \textbf{v}^{*} \right\|\left\| \textbf{u}_{2}^{*} \right\|})$, and then multiplying the inequality on the l.h.s. of \eqref{eq:statement_A2_R1} by
$\left\| \alpha\textbf{v}^{*} \right\|\cdot \left\| {{a}_{2,1}}\textbf{u}_{2}^{*} \right\|$.

\medskip

B) For every $(\textbf{s}_1,\textbf{s}_2,{\cal C}_1,{\cal C}_2,{\cal C}_c,\textbf{z})\in {\cal E}_{\textbf{X}}^{c}\cap {\cal E}_{\textbf{Z}}^{c}$
and every $\textbf{u}\in {\cal S}_{1}$
\begin{IEEEeqnarray}{rCl} \label{eq:statement_B_R1}
&& |\textbf{y}-\textbf{X}_{\textbf{\textbf{u}},\textbf{v}^{*},\textbf{\textbf{u}}^{*}_2}\|^2
\leq
\|\textbf{y}-\textbf{X}_{\textbf{\textbf{u}}_1^{*},\textbf{v}^{*},\textbf{\textbf{u}}_2^{*}}\|^2 \nonumber \\
&& \text{                                 }\implies \left\langle \textbf{y}-\left( {{a}_{2,1}}\textbf{u}_{2}^{*}+\alpha{{\textbf{v}}^{*}} \right),{{a}_{1,1}}\textbf{u} \right\rangle \ge n\left( \bar{{{\beta }_{1}}}{{P}_{1}}-\sqrt{\bar{{{\beta }_{1}}}{{P}_{1}}N}\epsilon  \right).
\end{IEEEeqnarray}

Statement B) follows from rewriting the inequality on the l.h.s. of \eqref{eq:statement_B_R1}
as
\begin{equation*}
{{\left\| \textbf{y}-\left( {{a}_{2,1}}\textbf{u}_{2}^{*}+\alpha{{\textbf{v}}^{*}} \right)-{{a}_{1,1}}\textbf{u} \right\|}^{2}}\le {{\left\| \textbf{y}-\left( {{a}_{2,1}}\textbf{u}_{2}^{*}+\alpha{{\textbf{v}}^{*}} \right)-{{a}_{1,1}}\textbf{u}_{1}^{*} \right\|}^{2}}
\end{equation*}
or equivalently as
\begin{IEEEeqnarray}{rCl} \label{eq:statement_B_proof_R1}
\left\langle \textbf{y}-\left( {{a}_{2,1}}\textbf{u}_{2}^{*}+\alpha{{\textbf{v}}^{*}} \right),{{a}_{1,1}}\textbf{u} \right\rangle
& \ge & \left\langle \textbf{y}-\left( {{a}_{2,1}}\textbf{u}_{2}^{*}+\alpha{{\textbf{v}}^{*}} \right),{{a}_{1,1}}\textbf{u}_{1}^{*} \right\rangle  \nonumber \\
& = & \left\langle {{a}_{1,1}}\textbf{u}_{1}^{*}+\textbf{z},{{a}_{1,1}}\textbf{u}_{1}^{*} \right\rangle  \nonumber \\
& = & {{\left\| {{a}_{1,1}}\textbf{u}_{1}^{*} \right\|}^{2}}+\left\langle \textbf{z},{{a}_{1,1}}\textbf{u}_{1}^{*} \right\rangle.
\end{IEEEeqnarray}
It now follows from the equivalence of the first inequality in \eqref{eq:statement_B_R1} with
\eqref{eq:statement_B_proof_R1} that for \\
$(\textbf{s}_1,\textbf{s}_2,{\cal C}_1,{\cal C}_2,{\cal C}_c,\textbf{z})\in {\cal E}_{\textbf{Z}}^{c}$, the first inequality in \eqref{eq:statement_B_R1} can only hold if
\begin{equation*}
\left\langle \textbf{y}-\left( {{a}_{2,1}}\textbf{u}_{2}^{*}+\alpha{{\textbf{v}}^{*}} \right),{{a}_{1,1}}\textbf{u} \right\rangle \ge n\left( \bar{{{\beta }_{1}}}{{P}_{1}}-\sqrt{\bar{{{\beta }_{1}}}{{P}_{1}}N}\epsilon  \right),
\end{equation*}
thus establising B).

\medskip

C) For every $(\textbf{s}_1,\textbf{s}_2,{\cal C}_1,{\cal C}_2,{\cal C}_c,\textbf{z})\in {\cal E}_{\textbf{X}}^{c}\cap {\cal E}_{\textbf{Z}}^{c}$
and every $\textbf{u}\in {\cal S}_{1}$,
\begin{IEEEeqnarray*}{rCl}
&& \left| \tilde{\rho }-\cos \sphericalangle \left( \textbf{u},\textbf{u}_{2}^{*} \right) \right|\le 7\epsilon
\quad \mbox{  and  } \quad
|\textbf{y}-\textbf{X}_{\textbf{u},\textbf{v}^{*},\textbf{u}^{*}_2}\|^2
\leq
\|\textbf{y}-\textbf{X}_{\textbf{u}_1^{*},\textbf{v}^{*},\textbf{u}_2^{*}}\|^2 \nonumber \\
&& \qquad \qquad \implies  \\
&& {{\left\| {{a}_{1,1}}\textbf{u}-\textbf{w} \right\|}^{2}}\le
n\bar{{{\beta }_{1}}}{{P}_{1}}-2 \Biggl( n{{\varsigma }_{1}}\left( \bar{{{\beta }_{1}}}{{P}_{1}}-\sqrt{\bar{{{\beta }_{1}}}{{P}_{1}}N}\epsilon \right)
+n{{\varsigma }_{2}}\left( \sqrt{\bar{{\beta}}_{1}\bar{{\beta}}_{2}{{P}_{1}}{{P}_{2}}}\left( \tilde{\rho }-7\epsilon  \right) \right)  \\
&& \hspace{2.6cm}  -{{\varsigma }_{3}}\left(\left\| \alpha\textbf{v}^{*} \right\|\sqrt{n\bar{{{\beta }_{1}}}{{P}_{1}}}3\epsilon  \right) \Biggr)+{{\left\| \textbf{w} \right\|}^{2}}.
\end{IEEEeqnarray*}
Statement C) is obtained as follows:
\begin{IEEEeqnarray*}{l}
 {{\left\| {{a}_{1,1}}\textbf{u}-\textbf{w} \right\|}^{2}}
 = {{\left\| {{a}_{1,1}}\textbf{u} \right\|}^{2}}-2\left\langle {{a}_{1,1}}\textbf{u},\textbf{w} \right\rangle +{{\left\| \textbf{w} \right\|}^{2}} \\
\quad = {{\left\| {{a}_{1,1}}\textbf{u} \right\|}^{2}}-2\left\langle {{a}_{1,1}}\textbf{u},{{\varsigma }_{1}}\left( {{a}_{1,1}}\textbf{u}_{1}^{*}+\textbf{z} \right)+{{\varsigma }_{2}}{{a}_{2,1}}\textbf{u}_{2}^{*}+{{\varsigma }_{3}}\alpha{{\textbf{v}}^{*}} \right\rangle +{{\left\| \textbf{w} \right\|}^{2}} \\
\quad =  n\bar{{{\beta }_{1}}}{{P}_{1}}-2\left[ {{\varsigma }_{1}}\left\langle {{a}_{1,1}}\textbf{u},{{a}_{1,1}}\textbf{u}_{1}^{*}+\textbf{z} \right\rangle +{{\varsigma }_{2}}\left\langle {{a}_{1,1}}\textbf{u},{{a}_{2,1}}\textbf{u}_{2}^{*} \right\rangle
+{{\varsigma }_{3}}\left\langle {{a}_{1,1}}\textbf{u},\alpha{{\textbf{v}}^{*}} \right\rangle  \right]+{{\left\| \textbf{w} \right\|}^{2}} \\
\quad \text{ } \overset{(a)}{\mathop{\le }} \,n\bar{{{\beta }_{1}}}{{P}_{1}}-2\Biggl( n{{\varsigma }_{1}}\left( \bar{{{\beta }_{1}}}{{P}_{1}}-\sqrt{\bar{{{\beta }_{1}}}{{P}_{1}}N}\epsilon  \right)
+n{{\varsigma }_{2}}\left( \sqrt{\bar{{\beta}}_{1}\bar{{\beta}}_{2}{{P}_{1}}{{P}_{2}}}\left( \tilde{\rho }-7\epsilon  \right) \right) \\
\hspace{1.5cm} -{{\varsigma }_{3}}\left(\left\| \alpha\textbf{v}^{*} \right\|\sqrt{n\bar{{{\beta }_{1}}}{{P}_{1}}}3\epsilon \right) \Biggr)+{{\left\| \textbf{w} \right\|}^{2}},
\end{IEEEeqnarray*}
where in (a) we have used Statement A), Statement A1) and Statement B).

\medskip

D) For every $(\textbf{s}_1,\textbf{s}_2,{\cal C}_1,{\cal C}_2,{\cal C}_c,\textbf{z})\in {\cal E}_{\textbf{X}}^{c}\cap {\cal E}_{\textbf{Z}}^{c}$
\begin{IEEEeqnarray*}{rCl}
  {{\left\| \textbf{w} \right\|}^{2}}
  &\le & \,  n\Bigl( {{\varsigma }_{1}}^{2}\left( \bar{{{\beta }_{1}}}{{P}_{1}}+N \right)+2{{\varsigma }_{1}}{{\varsigma }_{2}} \sqrt{\bar{{\beta}}_{1}\bar{{\beta}}_{2}{{P}_{1}}{{P}_{2}}}\tilde{\rho } +{{\varsigma }_{2}}^{2} \bar{{{\beta }_{2}}}{{P}_{2}} +2{{\varsigma }_{2}}{{\varsigma }_{3}}\left\| \alpha\textbf{v} \right\|\sqrt{n\bar{{{\beta }_{2}}}{{P}_{2}}}\bar{\rho }\\
  && +\frac{1}{n}{{\varsigma }_{3}}^{2}{{\alpha}^{2}}{{\left\| {{\textbf{v}}^{*}} \right\|}^{2}}+\kappa\epsilon  \Bigr),
\end{IEEEeqnarray*}
where $\kappa$ depends on ${{P}_{1}},{{P}_{2}},N,{{\varsigma }_{1}},{{\varsigma }_{2}}$ and ${{\varsigma }_{3}}$ only.

\medskip

Statement D) is obtained as follows:
\begin{IEEEeqnarray*}{l}
{{\left\| \textbf{w} \right\|}^{2}}
\ ={{\left\| {{\varsigma }_{1}}\left( {{a}_{1,1}}\textbf{u}_{1}^{*}+\textbf{z} \right)+{{\varsigma }_{2}}{{a}_{2,1}}\textbf{u}_{2}^{*}+{{\varsigma }_{3}}\alpha{{\textbf{v}}^{*}} \right\|}^{2}}\\
\ ={{\varsigma }_{1}}^{2}{{\left\| {{a}_{1,1}}\textbf{u}_{1}^{*}+\textbf{z} \right\|}^{2}}+2{{\varsigma }_{1}}{{\varsigma }_{2}}\left\langle {{a}_{1,1}}\textbf{u}_{1}^{*}+\textbf{z},{{a}_{2,1}}\textbf{u}_{2}^{*} \right\rangle +{{\varsigma }_{2}}^{2}{{\left\| {{a}_{2,1}}\textbf{u}_{2}^{*} \right\|}^{2}} \\
\hspace{1.5cm} +2{{\varsigma }_{1}}{{\varsigma }_{3}}\left\langle {{a}_{1,1}}\textbf{u}_{1}^{*}+\textbf{z},\alpha{{\textbf{v}}^{*}} \right\rangle 
 +2{{\varsigma }_{2}}{{\varsigma }_{3}}\left\langle {{a}_{2,1}}\textbf{u}_{2}^{*},\alpha{{\textbf{v}}^{*}} \right\rangle +{{\varsigma }_{3}}^{2}{{\left\| \alpha{{\textbf{v}}^{*}} \right\|}^{2}} \\
\ =  {{\varsigma }_{1}}^{2}\left( {{\left\| {{a}_{1,1}}\textbf{u}_{1}^{*} \right\|}^{2}}+2\left\langle {{a}_{1,1}}\textbf{u}_{1}^{*},\textbf{z} \right\rangle +{{\left\| \textbf{z} \right\|}^{2}} \right)+2{{\varsigma }_{1}}{{\varsigma }_{2}}\left( \left\langle {{a}_{1,1}}\textbf{u}_{1}^{*},{{a}_{2,1}}\textbf{u}_{2}^{*} \right\rangle +\left\langle \textbf{z},{{a}_{2,1}}\textbf{u}_{2}^{*} \right\rangle  \right)\\
\quad +{{\varsigma }_{2}}^{2}\left( n\bar{{{\beta }_{2}}}{{P}_{2}} \right) +2{{\varsigma }_{1}}{{\varsigma }_{3}}\left\langle \textbf{z},\alpha{{\textbf{v}}^{*}} \right\rangle +2{{\varsigma }_{2}}{{\varsigma }_{3}}\left\langle {{a}_{2,1}}\textbf{u}_{2}^{*},\alpha{{\textbf{v}}^{*}} \right\rangle +{{\varsigma }_{3}}^{2}{{\alpha}^{2}}{{\left\| {{\textbf{v}}^{*}} \right\|}^{2}} \\
\ \text{ } \overset{(a)}{\mathop{\le }}  n\Bigl( {{\varsigma }_{1}}^{2}\left( \bar{{{\beta }_{1}}}{{P}_{1}}+N \right)+2{{\varsigma }_{1}}{{\varsigma }_{2}} \sqrt{\bar{{\beta}}_{1}\bar{{\beta}}_{2}{{P}_{1}}{{P}_{2}}}\tilde{\rho } +{{\varsigma }_{2}}^{2}\left( \bar{{{\beta }_{2}}}{{P}_{2}} \right)+2{{\varsigma }_{2}}{{\varsigma }_{3}}\left\| \alpha\textbf{v} \right\|\sqrt{n\bar{{{\beta }_{2}}}{{P}_{2}}}\bar{\rho }  \\
\hspace{1.5cm}  +\frac{1}{n}{{\varsigma }_{3}}^{2}{\alpha^{2}}{{\left\| {{\textbf{v}}^{*}} \right\|}^{2}}+\kappa\epsilon  \Bigr),
\end{IEEEeqnarray*}
where in (a) we have used that $(\textbf{s}_1,\textbf{s}_2,{\cal C}_1,{\cal C}_2,{\cal C}_c,\textbf{z})\in
{\cal E}^{c}_{\textbf{X}}$, and statements A) and A2).

\medskip

E) For every$(\textbf{s}_1,\textbf{s}_2,{\cal C}_1,{\cal C}_2,{\cal C}_c,\textbf{z})\in {\cal E}_{\textbf{X}}^{c}\cap {\cal E}_{\textbf{Z}}^{c}$ and an arbitrary $\textbf{u}\in {\cal S}_{1}$,
\begin{IEEEeqnarray*}{rCl}
&& \left| \tilde{\rho }-\cos \sphericalangle \left( \textbf{u},\textbf{u}_{2}^{*} \right) \right|\le 7\epsilon
\text{  and  }\left| \bar{\rho }-\cos \sphericalangle \left( \textbf{v}^{*},\textbf{u}_{2}^{*} \right) \right|\le 7\epsilon \text{  and} \\
&& {{\left\| \textbf{y}-\left( {{a}_{1,1}}\textbf{u}+{{a}_{2,1}}\textbf{u}_{2}^{*}+\left( {{a}_{1,2}}+{{a}_{2,2}} \right)
{{\textbf{v}}^{*}} \right) \right\|}^{2}}\le {{\left\| \textbf{y}-\left( {{a}_{1,1}}\textbf{u}_{1}^{*}+{{a}_{2,1}}\textbf{u}_{2}^{*}+\alpha{{\textbf{v}}^{*}} \right) \right\|}^{2}} \\
&& \implies \text{        }{{\left\| {{a}_{1,1}}\textbf{u}-\textbf{w} \right\|}^{2}}\le \Upsilon (\epsilon ),
\end{IEEEeqnarray*}
where
\begin{equation*}
\Upsilon (\epsilon )=n\frac{\bar{{{\beta }_{1}}}{{P}_{1}}N\left( 1-{{{\tilde{\rho }}}^{2}}-{{\bar{\rho }}^{2}} \right)}{{{\bar{\beta }}_{1}}{{P}_{1}}\left( 1-{{{\tilde{\rho }}}^{2}} \right)+N-{{\bar{\rho }}^{2}}\left( {{\bar{\beta }}_{1}}{{P}_{1}}+N \right)}+n{{\kappa}'}\epsilon ,
\end{equation*}
and where ${{\kappa}'}$ only depends on ${{P}_{1}},{{N}_{1}},{{N}_{2}},{{\varsigma }_{1}},{{\varsigma }_{2}}$  and $ {{\varsigma }_{3}}$.

\medskip

Statement E) follows from combining Statement C) with Statement
D) and the explicit values of ${{\varsigma }_{1}},{{\varsigma }_{2}}$  and $ {{\varsigma }_{3}}$ given in \eqref{eq:w_coeff_for_R1}.

\medskip

F) For every $\textbf{u}\in {\cal S}_{1}$, denote by $\varphi \in \left[ 0,\pi  \right]$ the angle between \textbf{u} and \textbf{w}, and let
\begin{IEEEeqnarray*}{l}
{\cal B}(\textbf{s}_1,\textbf{s}_2,\textbf{u}_1^{*},\textbf{v}^{*},\textbf{u}_2^{*},\textbf{z})\triangleq
\Biggl\{\textbf{u}\in S_{1}^{\left( n \right)} \colon
\cos \left( \varphi  \right)
 \ge  \sqrt{1-\frac{\Upsilon (\epsilon )}{n{{\bar{\beta }}_{1}}{{P}_{1}}}} \\
\hspace{5.5cm} =  \sqrt{\frac{{{\bar{\beta }}_{1}}{{P}_{1}}\left( 1-{{{\tilde{\rho }}}^{2}} \right)+N{{{\tilde{\rho }}}^{2}}-{{\bar{\rho }}^{2}}{{\bar{\beta }}_{1}}{{P}_{1}}}{{{\bar{\beta }}_{1}}{{P}_{1}}\left( 1-{{{\tilde{\rho }}}^{2}} \right)+N-{{\bar{\rho }}^{2}}\left( {{\bar{\beta }}_{1}}{{P}_{1}}+N \right)}-{{\kappa}^{''}}\epsilon } \Biggr\},
\end{IEEEeqnarray*}
where ${{\kappa}^{''}}$ only depends on ${{P}_{1}},{{N}_{1}},{{N}_{2}},{{\varsigma }_{1}},{{\varsigma }_{2}}$  and $ {{\varsigma }_{3}}$, and where we
assume $\epsilon $ sufficiently small such that
\begin{equation*}
\frac{{{\bar{\beta }}_{1}}{{P}_{1}}\left( 1-{{{\tilde{\rho }}}^{2}} \right)+N{{{\tilde{\rho }}}^{2}}-{{\bar{\rho }}^{2}}{{\bar{\beta }}_{1}}{{P}_{1}}}{{{\bar{\beta }}_{1}}{{P}_{1}}\left( 1-{{{\tilde{\rho }}}^{2}} \right)+N-{{\bar{\rho }}^{2}}\left( {{\bar{\beta }}_{1}}{{P}_{1}}+N \right)}-{{\kappa}^{''}}\epsilon >0.
\end{equation*}
Then, for every $(\textbf{s}_1,\textbf{s}_2,{\cal C}_1,{\cal C}_2,{\cal C}_c,\textbf{z})\in {\cal E}_{\textbf{X}}^{c}\cap {\cal E}_{\textbf{Z}}^{c}$,
\begin{IEEEeqnarray}{rCl}\label{eq:statement_F_R1}
 && \left| \tilde{\rho }-\cos \sphericalangle \left( \textbf{u},\textbf{u}_{2}^{*} \right) \right|\le 7\epsilon \quad \mbox{  and  }\quad \left| \bar{\rho }-\cos \sphericalangle \left( \textbf{v}^{*},\textbf{u}_{2}^{*} \right) \right|\le 7\epsilon \quad \mbox{  and  } \nonumber \\
 && {{\left\| \textbf{y}-\left( {{a}_{1,1}}\textbf{u}+{{a}_{2,1}}\textbf{u}_{2}^{*}+\alpha{{\textbf{v}}^{*}} \right) \right\|}^{2}}\le {{\left\| \textbf{y}-\left( {{a}_{1,1}}\textbf{u}_{1}^{*}+{{a}_{2,1}}\textbf{u}_{2}^{*}+\alpha{{\textbf{v}}^{*}} \right) \right\|}^{2}} \nonumber \\
 && \qquad \implies \text{ }\textbf{u}\in {\cal B}(\textbf{s}_1,\textbf{s}_2,\textbf{u}_1^{*},\textbf{v}^{*},\textbf{u}_2^{*},\textbf{z}).
\end{IEEEeqnarray}

\medskip

Statement F) follows from Statement E) by noting that if $\mathbf{w}\ne 0$ and
$1-\frac{\Upsilon (\epsilon )}{n\bar{{{\beta }_{1}}}{{P}_{1}}}>0$, then
\begin{IEEEeqnarray}{rCl}\label{eq:statement_F_proof_R1}
\left.
\begin{array}{l}
{{\left\| {{a}_{1,1}}\textbf{u} \right\|}^{2}}=n\bar{{{\beta }_{1}}}{{P}_{1}}\\
 {{\left\| {{a}_{1,1}}\textbf{u}-\textbf{w} \right\|}^{2}}\le \Upsilon (\epsilon ) \\
\end{array}
\right\}
\implies   \cos \sphericalangle \left( \textbf{u},\textbf{w} \right)\ge \sqrt{1-\frac{\Upsilon (\epsilon )}{n\bar{{{\beta }_{1}}}{{P}_{1}}}} .
\end{IEEEeqnarray}
To see this, first note that every ${{a}_{1,1}}\textbf{u}$, where $\textbf{u}\in {\cal S}_{1}$, satisfying the
condition on the l.h.s. of \eqref{eq:statement_F_R1} lies within a sphere of radius
$\sqrt{\Upsilon (\epsilon )}$ centered at $\mathbf{w}$. In addition, for every $\textbf{u}\in {\cal S}_{1}$
we have that ${{a}_{1,1}}\textbf{u}$ also lies on the
centered ${{\mathbb{R}}^{n}}$-sphere of radius $\sqrt{n\bar{{{\beta }_{1}}}{{P}_{1}}}$.
Hence, every $\textbf{u}\in {\cal S}_{1}$ satisfying
the condition on the l.h.s. of \eqref{eq:statement_F_R1} lies in the intersection of these two
regions, which is a polar cap on the centered sphere of radius $\sqrt{n\bar{{{\beta }_{1}}}{{P}_{1}}}$.
The area of this polar cap is outer bounded as follows. Let $\mathbf{r}$ be an arbitrary point on
the boundary of this polar cap. The half-angle of the polar cap would be
maximized if $\mathbf{w}$ and $\mathbf{r}-\mathbf{w}$ would lie perpendicular to each other.
Hence, every $\textbf{u}\in {\cal S}_{1}^{\left( n \right)}$
satisfying the upper conditions of \eqref{eq:statement_F_R1}, also satisfies
\begin{IEEEeqnarray*}{rCl}
\cos \varphi
& \ge & \sqrt{1-\frac{\Upsilon (\epsilon )}{n{{\bar{\beta }}_{1}}{{P}_{1}}}} \\
& = & \sqrt{\frac{{{\bar{\beta }}_{1}}{{P}_{1}}\left( 1-{{{\tilde{\rho }}}^{2}} \right)+N{{{\tilde{\rho }}}^{2}}-{{\bar{\rho }}^{2}}{{\bar{\beta }}_{1}}{{P}_{1}}}{{{\bar{\beta }}_{1}}{{P}_{1}}\left( 1-{{{\tilde{\rho }}}^{2}} \right)+N-{{\bar{\rho }}^{2}}\left( {{\bar{\beta }}_{1}}{{P}_{1}}+N \right)}-{{\kappa}^{''}}\epsilon } \ ,
\end{IEEEeqnarray*}
where we assume $\epsilon $ sufficiently small such that $1-\frac{\Upsilon (\epsilon )}{n\bar{{{\beta }_{1}}}{{P}_{1}}}>0$ and where ${{\kappa}^{''}}=\frac{{{\kappa}'}}{n{{\bar{\beta }}_{1}}{{P}_{1}}}$.

\medskip

The proof of Lemma~\ref{eps_h_u1} is now concluded by noticing that the set
${\cal E}'_{{\hat{\textbf{U}}}_{1}}$, defined in \eqref{eq:eps_h_u1},
is the set of tuples
$(\textbf{s}_1,\textbf{s}_2,{\cal C}_1,{\cal C}_2,{\cal C}_c,\textbf{z})$
for which there exists a ${{\textbf{u}}_{1}}\left( j \right)\in {{{\cal C}}_{1}}\backslash \left\{ \textbf{u}_{1}^{*} \right\}$ such that
${{\textbf{u}}_{1}}\left( j \right)\in
{\cal B}(\textbf{s}_1,\textbf{s}_2,\textbf{u}_1^{*},\textbf{v}^{*},\textbf{u}_2^{*},\textbf{z})$.
Thus, by Statement F) and by the definition of ${{{\cal E}}_{{{{\hat{\textbf{U}}}}_{1}}}}$
in \eqref{eq:eps_u1} it follows that
\begin{equation*}
{{{\cal E}}_{{{{\hat{\textbf{U}}}}_{1}}}}\cap {\cal E}_{\textbf{S}}^{c}\cap {\cal E}_{\textbf{X}}^{c}\cap {\cal E}_{\textbf{Z}}^{c}\subseteq {\cal E}'_{{{{\hat{\textbf{U}}}}_{1}}}\cap {\cal E}_{\textbf{S}}^{c}\cap {\cal E}_{\textbf{X}}^{c}\cap {\cal E}_{\textbf{Z}}^{c},
\end{equation*}
and therefore
\begin{flalign*}
&&
\Pr\left[ {{{\cal E}}_{{{{\hat{\textbf{U}}}}_{1}}}}\cap {\cal E}_{\textbf{S}}^{c}\cap {\cal E}_{\textbf{X}}^{c}\cap {\cal E}_{\textbf{Z}}^{c} \right]\le \Pr\left[ {\cal E}'_{{{{\hat{\textbf{U}}}}_{1}}}\cap {\cal E}_{\textbf{S}}^{c}\cap {\cal E}_{\textbf{X}}^{c}\cap {\cal E}_{\textbf{Z}}^{c} \right].
&&
\end{flalign*}
\end{IEEEproof}

We now state one more lemma that will be used for the proof of \eqref{eq:R1}.
\begin{lemma} \label{Rate proof}
For every $\Delta \in (0,1]$, let the set ${\cal G}$ be given by
\begin{equation*}
{\cal G} =\left\{ \left( {\mathbf{s}_{1}},{\mathbf{s}_{2}},{{{\cal C}}_{1}},{{{\cal C}}_{2}},{{{\cal C}}_{c}},\mathbf{z} \right):\ \exists \> {{\mathbf{u}}_{1}}(j)\in {{{\cal C}}_{1}}\backslash \left\{ \mathbf{u}_{1}^{*} \right\}\ s.t.\  \cos \sphericalangle \left( \mathbf{w},{{\mathbf{u}}_{1}}(j) \right)\ge \Delta  \right\},
\end{equation*}
where $\mathbf{w}$ is defined in \eqref{eq:w_for_R1}.
Then,
\begin{equation*}
{{R}_{1}}<-\frac{1}{2}\log \left( 1-{{\Delta }^{2}} \right)
\quad \implies \quad
\left( \underset{n\to \infty }{\mathop{\lim }}\,\ {\Pr}\left[ {\cal G} |{\cal E}_{{\mathbf{X}}_{1}}^{c} \right]=0,\epsilon >0 \right),
\end{equation*}
\end{lemma}
where ${\cal E}_{\mathbf{X}_1}$ is defined in \eqref{eq:eps_x1_def}.

\medskip

\begin{IEEEproof}
The proof follows from upper-bounding in every point on ${\cal S}_1$ the
density of every ${{\textbf{u}}_{1}}\left( j \right)\in {{{\cal C}}_{1}}\backslash \left\{ \textbf{u}_{1}^{*} \right\}$
and then using a standard argument from sphere-packing.
See \cite[Appendix D-E2]{Stephan}.
\end{IEEEproof}

\medskip

Next,
\begin{IEEEeqnarray}{rCl} \label{eq:R1_proof}
\Pr\left[ {{{\cal E}}_{{{{\hat{\textbf{U}}}}_{1}}}}\cap {\cal E}_{\textbf{S}}^{c}\cap {\cal E}_{\textbf{X}}^{c}\cap {\cal E}_{\textbf{Z}}^{c} \right]
& \overset{(a)}{\mathop{\le }}&\,\Pr\left[ {{{\cal E}}'}_{{{{\hat{\textbf{U}}}}_{1}}}\cap {\cal E}_{\textbf{S}}^{c}\cap {\cal E}_{\textbf{X}}^{c}\cap {\cal E}_{\textbf{Z}}^{c} \right] \nonumber\\
& \overset{(b)}{\mathop{\le }}&\,\Pr\left[ {{{\cal E}}'}_{{{{\hat{\textbf{U}}}}_{1}}}\middle|{\cal E}_{{\textbf{X}}_{1}}^{c} \right]
\end{IEEEeqnarray}
where (a) follows by Lemma~\ref{eps_h_u1} and (b) follows because ${\cal E}_{\textbf{X}}^{c}\subseteq {\cal E}_{{\textbf{X}}_{1}}^{c}$.
The proof of \eqref{eq:R1} is now completed by combining  \eqref{eq:R1_proof} with Lemma~\ref{Rate proof}.
This gives that for every $\delta >0$ and every $\epsilon >0$ there exists some
$n'(\delta ,\epsilon )$
such that for all $n>n'(\delta ,\epsilon )$, we have
\begin{equation*}
\ \Pr\left[ {{{\cal E}}_{{{{\hat{\textbf{U}}}}_{1}}}}\cap {\cal E}_{\textbf{S}}^{c}\cap {\cal E}_{\textbf{X}}^{c}\cap {\cal E}_{\textbf{Z}}^{c} \right]\le
\Pr\left[ {{{\cal E}}'}_{{{{\hat{\textbf{U}}}}_{1}}}|{\cal E}_{{\textbf{X}}_{1}}^{c} \right]<\delta ,
\end{equation*}
whenever
\begin{equation*}
{{R}_{1}}<\frac{1}{2}\log \left( \frac{{{\bar{\beta }}_{1}}{{P}_{1}}\left( 1-{{{\tilde{\rho }}}^{2}}-{{\bar{\rho }}^{2}} \right)+N\left( 1-{{\bar{\rho }}^{2}} \right)}{N\left( 1-{{{\tilde{\rho }}}^{2}} \right)-{{\bar{\rho }}^{2}}N}-{{\kappa}_{1}}\epsilon  \right),
\end{equation*}
where ${{\kappa}_{1}}$ is a positive constant determined by ${{P}_{1}},{{P}_{2}},N,{{\varsigma }_{1}},{{\varsigma }_{2}}\text{ and }{{\varsigma }_{3}}$. \hfill\qed

\vskip.2truein

\subsection{Proof of rate constraint \eqref{eq:R2}}
Define
\begin{equation}\label{eq:w_for_R2}
 \textbf{w}(\textbf{s}_1,\textbf{s}_2,{\cal C}_1,{\cal C}_2,{\cal C}_c,\textbf{z})={{\varsigma }_{1}}\left( \textbf{y}-\left( {{a}_{1,1}}\textbf{u}_{1}^{*}+\alpha{{\textbf{v}}^{*}} \right) \right)+{{\varsigma }_{2}}{{a}_{1,1}}\textbf{u}_{1}^{*}+{{\varsigma }_{3}}\alpha{{\textbf{v}}^{*}},
\end{equation}
where
\begin{IEEEeqnarray}{rCl}\label{eq:w_coeff_for_R2}
{{\varsigma }_{1}}&=&\frac{{{\sigma }^{2}}{a}_{2,1}^{2}\left( 1-{{2}^{-2{{R}_{2}}}} \right)\left( 1-{{\bar{\rho }}^{2}}-{{{\tilde{\rho }}}^{2}} \right)}{{{\sigma }^{2}}{a}_{2,1}^{2}\left( 1-{{2}^{-2{{R}_{2}}}} \right)\left( 1-{{\bar{\rho }}^{2}}-{{{\tilde{\rho }}}^{2}} \right)+N} \nonumber \\
{{\varsigma }_{2}}&=&\frac{{{a}_{2,1}}\left( 1-{{2}^{-2{{R}_{2}}}} \right)\rho \left( 1-{{\bar{\rho }}^{2}} \right)N}{{{a}_{1,1}}\left( {{\sigma }^{2}}{a}_{2,1}^{2}\left( 1-{{2}^{-2{{R}_{2}}}} \right)\left( 1-{{\bar{\rho }}^{2}}-{{{\tilde{\rho }}}^{2}} \right)+N \right)} \nonumber \\
{{\varsigma }_{3}}&=&\frac{{{a}_{2,1}}\left( 1-{{2}^{-2{{R}_{2}}}} \right)\rho \left( 1-{{{\tilde{\rho }}}^{2}} \right)N}{\alpha\left( {{\sigma }^{2}}{a}_{2,1}^{2}\left( 1-{{2}^{-2{{R}_{2}}}} \right)\left( 1-{{\bar{\rho }}^{2}}-{{{\tilde{\rho }}}^{2}} \right)+N \right)}.
\end{IEEEeqnarray}

We now start with a lemma that will be used to prove \eqref{eq:R2}.
\begin{lemma}\label{eps_h_u2}
Let ${{\varphi }_{j}}\in \left[ 0,\pi  \right]$  be the angle between $\mathbf{w}$ and $\mathbf{u}_{2}(j)$, and let the set ${\cal E}'_{{{{\hat{\mathbf{U}}}}_{2}}}$
be defined as
\begin{IEEEeqnarray}{rCl}\label{eq:eps_h_u2}
&&{\cal E}'_{{{{\hat{\mathbf{U}}}}_{2}}}
\triangleq
\Biggl\{(\mathbf{s}_1,\mathbf{s}_2,{\cal C}_1,{\cal C}_2,
{\cal C}_c,\mathbf{z})\colon \exists \> {{\mathbf{u}}_{2}}\left( j \right)\in {{\cal C}_{2}}\backslash \left\{ \mathbf{u}_{2}^{*} \right\}\ \mbox{s.t.}\
\nonumber \\
&&\cos \left( {{\varphi }_{j}} \right)\ge
\sqrt{\frac{{{\bar{\beta }}_{2}}{{P}_{2}}\left( 1-{{{\tilde{\rho }}}^{2}} \right)+N{{{\tilde{\rho }}}^{2}}-{{\bar{\rho }}^{2}}\left( {{\bar{\beta }}_{2}}{{P}_{2}}-N \right)}{{{\bar{\beta }}_{2}}{{P}_{2}}\left[ 1-{{{\tilde{\rho }}}^{2}} \right]+N-{{\bar{\rho }}^{2}}{{\bar{\beta }}_{2}}{{P}_{2}}}-\frac{{{N}^{2}}{{{\tilde{\rho }}}^{2}}{{\bar{\rho }}^{2}}\left( 2+{{{\tilde{\rho }}}^{2}} \right)}{{{\left( {{\bar{\beta }}_{2}}{{P}_{2}}\left[ 1-{{\bar{\rho }}^{2}}-{{{\tilde{\rho }}}^{2}} \right]+N \right)}^{2}}}-{{\kappa}^{''}}\epsilon }
\Biggr\},
\nonumber \\*
\end{IEEEeqnarray}
where ${{\kappa}^{''}}$ is a positive constant determined by $P_{1}, P_{2},N ,
{\varsigma}_{1}, {\varsigma}_{2} \mbox{ and } {\varsigma }_{3}$.
Then,
\begin{equation*}
{{\cal E}_{{{{\hat{\mathbf{U}}}}_{2}}}}\cap {\cal E}_{\mathbf{S}}^{c}\cap {\cal E}_{\mathbf{X}}^{c}\cap {\cal E}_{\mathbf{Z}}^{c}\subseteq
{\cal E}'_{{{{\hat{\mathbf{U}}}}_{2}}}\cap {\cal E}_{\mathbf{S}}^{c}\cap {\cal E}_{\mathbf{X}}^{c}\cap {\cal E}_{\mathbf{Z}}^{c},
\end{equation*}
and, in particular
\begin{equation*}
\Pr\left[ {{\cal E}_{{{{\hat{\mathbf{U}}}}_{2}}}}\cap {\cal E}_{\mathbf{S}}^{c}\cap {\cal E}_{\mathbf{X}}^{c}\cap
{\cal E}_{\mathbf{Z}}^{c} \right]\le \Pr\left[ {{\cal E}'}_{{{{\hat{\mathbf{U}}}}_{2}}}\cap {\cal E}_{\mathbf{S}}^{c}\cap
{\cal E}_{\mathbf{X}}^{c}\cap {\cal E}_{\mathbf{Z}}^{c}  \right].
\end{equation*}
\end{lemma}

\medskip

\begin{IEEEproof}
We first recall that for the event ${{\cal E}_{{{{\hat{\textbf{U}}}}_{2}}}}$
to occur, there must exist a codeword ${{\textbf{u}}_{2}}\left( j \right)\in {{{\cal C}}_{2}}\backslash \left\{ \textbf{u}_{2}^{*} \right\}$ that satisfies
the following three conditions
\begin{IEEEeqnarray}{rCl}
\left| \tilde{\rho }-\cos \sphericalangle \left( \textbf{u}_{1}^{*},{{\textbf{u}}_{2}}\left( j \right) \right) \right| & \le & 7\epsilon \\
\left| \bar{\rho}-\cos \sphericalangle \left( {{\textbf{v}}^{*}},{{\textbf{u}}_{2}}\left( j \right) \right) \right| & \le & 7\epsilon \\
\|\textbf{y}-\textbf{X}_{\textbf{u}_{1}^{*},\textbf{\textbf{v}}^{*},\textbf{u}_{2}\left(j\right)}\|^2
& \leq &
\|\textbf{y}-\textbf{X}_{\textbf{u}_1^{*},\textbf{\textbf{v}}^{*},\textbf{u}_2^{*}}\|^2.
\end{IEEEeqnarray}

The proof is now based on a sequence of statements related to these three conditions.

\medskip

A) For every $(\textbf{s}_1,\textbf{s}_2,{\cal C}_1,{\cal C}_2,{\cal C}_c,\textbf{z})\in {\cal E}_{\textbf{X}}^{c}$ and every $\textbf{u}\in {\cal S}_{2}$, where ${\cal S}_{2}$ is
the surface area of the codeword sphere of ${\cal C}_{2}$ defined in the code
construction,
\begin{IEEEeqnarray}{rCl}\label{eq:statement_A_R2}
&& \left| \tilde{\rho }-\cos \sphericalangle \left( \textbf{u}_{1}^{*},\textbf{u} \right) \right|\le 7\epsilon  \nonumber\\
&& \text{           }\implies \left| n\tilde{\rho }\sqrt{\bar{{\beta}}_{1}\bar{{\beta}}_{2}{{P}_{1}}{{P}_{2}}}-\left\langle {{a}_{1,1}}\textbf{u}_{1}^{*},{{a}_{2,1}}\textbf{u} \right\rangle  \right|\le 7n\sqrt{\bar{{\beta}}_{1}\bar{{\beta}}_{2}{{P}_{1}}{{P}_{2}}}\epsilon .
\end{IEEEeqnarray}

Statement A) follows by rewriting $\cos \sphericalangle \left( \textbf{u}_{1}^{*},\textbf{u} \right)$ as ${\left\langle \textbf{u}_{1}^{*},\textbf{u} \right\rangle }/({\left\| \textbf{u}_{1}^{*} \right\|\left\| \textbf{u} \right\|})$, and then multiplying the inequality on the l.h.s.
of \eqref{eq:statement_A_R2} by
$\left\| {{a}_{1,1}}\textbf{u}_{1}^{*} \right\|\cdot \left\| {{a}_{2,1}}\textbf{u} \right\|$
and recalling that $\left\| {{a}_{1,1}}\textbf{u}_{1}^{*} \right\|=\sqrt{n\bar{\beta }_{1}{{P}_{1}}}\ $ and that $\ \left\| {{a}_{2,1}}\textbf{u} \right\|=\sqrt{n\bar{\beta }_{2}{{P}_{2}}}$.

\medskip

A1) For every $(\textbf{s}_1,\textbf{s}_2,{\cal C}_1,{\cal C}_2,{\cal C}_c,\textbf{z})\in {\cal E}_{\textbf{X}}^{c}$,
\begin{IEEEeqnarray}{rCl} \label{eq:statement_A1_R2}
&& \left| \cos \sphericalangle \left( \textbf{v}^{*},\textbf{u}_{1}^{*} \right) \right|\le 3\epsilon  \nonumber \\
&& \text{       }
\implies
\text{      }\left| \left\langle \alpha\textbf{v}^{*},{{a}_{1,1}}\textbf{u}_{1}^{*} \right\rangle  \right|\le 3\left\| \alpha\textbf{v}^{*} \right\|\sqrt{n\bar{{{\beta }_{1}}}{{P}_{1}}}\epsilon .
\end{IEEEeqnarray}

Statement A1) follows by rewriting $\cos \sphericalangle \left( \textbf{v}^{*},\textbf{u}_{1}^{*} \right)$ as ${\left\langle \textbf{v}^{*},\textbf{u}_{1}^{*} \right\rangle }/({\left\| \textbf{v}^{*} \right\|\left\| \textbf{u}_{1}^{*} \right\|})$, and then multiplying the inequality on the l.h.s. of \eqref{eq:statement_A1_R2} by
$\left\| \alpha\textbf{v}^{*} \right\|\cdot \left\| {{a}_{1,1}}\textbf{u}_{1}^{*} \right\|$.

\medskip

A2) For every $(\textbf{s}_1,\textbf{s}_2,{\cal C}_1,{\cal C}_2,{\cal C}_c,\textbf{z})\in {\cal E}_{\textbf{X}}^{c}$ and every $\textbf{u}\in {\cal S}_{2}$,
\begin{IEEEeqnarray}{rCl}\label{eq:statement_A2_R2}
&& \left| \bar{\rho }-\cos \sphericalangle \left( {{\textbf{v}}^{*}},\textbf{u} \right) \right|\le 7\epsilon  \nonumber\\
&& \text{  }\implies \left| \left\| \alpha{{\textbf{v}}^{*}} \right\|\sqrt{n\bar{{{\beta }_{2}}}{{P}_{2}}}\bar{\rho }-\left\langle \alpha{{\textbf{v}}^{*}},{{a}_{2,1}}\textbf{u} \right\rangle  \right|\le 7\epsilon \left\| \alpha{{\textbf{v}}^{*}} \right\|\sqrt{n\bar{{{\beta }_{2}}}{{P}_{2}}}.
\end{IEEEeqnarray}

Statement A2) follows by rewriting $\cos \sphericalangle \left( \textbf{v}^{*},\textbf{u}_{2} \right)$ as ${\left\langle \textbf{v}^{*},\textbf{u}_{2} \right\rangle }/({\left\| \textbf{v}^{*} \right\|\left\| \textbf{u}_{2}\right\|})$,
and then multiplying the inequality on the l.h.s.
of \eqref{eq:statement_A2_R2} by
$\left\| \alpha{{\textbf{v}}^{*}} \right\|\cdot \left\| {{a}_{2,1}}\textbf{u} \right\|$.

\medskip

B) For every $(\textbf{s}_1,\textbf{s}_2,{\cal C}_1,{\cal C}_2,{\cal C}_c,\textbf{z})\in {\cal E}_{\textbf{X}}^{c}\cap
{\cal E}_{\textbf{Z}}^{c}$
and every $\textbf{u}\in {\cal S}_{2}$
\begin{IEEEeqnarray}{rCl}\label{eq:statement_B_R2}
&& |\textbf{y}-\textbf{X}_{\textbf{u}^{*}_1,\textbf{v}^{*},\textbf{u}}\|^2
\leq
\|\textbf{y}-\textbf{X}_{\textbf{u}_1^{*},\textbf{v}^{*},\textbf{u}_2^{*}}\|^2 \nonumber \\
&& \qquad \implies \left\langle {{a}_{2,1}}\textbf{u}_{2}^{*}+\textbf{z},{{a}_{2,1}}\textbf{u} \right\rangle \ge n\left( \bar{{{\beta }_{2}}}{{P}_{2}}-\sqrt{\bar{{{\beta }_{2}}}{{P}_{2}}N}\epsilon  \right).
\end{IEEEeqnarray}

Statement B) follows from rewriting the inequality on the l.h.s. of \eqref{eq:statement_B_R2} as
\begin{equation*}
 {{\left\| \left( \textbf{y}-{{a}_{1,1}}\textbf{u}_{1}^{*}-\alpha{{\textbf{v}}^{*}} \right)
 - {{a}_{2,1}}\textbf{u}  \right\|}^{2}}\le {{\left\| \left( \textbf{y}-{{a}_{1,1}}\textbf{u}_{1}^{*}-\alpha{{\textbf{v}}^{*}} \right)
 - {{a}_{2,1}}\textbf{u}_{2}^{*}  \right\|}^{2}},
 \end{equation*}
 or equivalently as
\begin{IEEEeqnarray}{rCl}\label{eq:statement_B_R2_proof}
  \left\langle \textbf{y}-{{a}_{1,1}}\textbf{u}_{1}^{*}-\alpha{{\textbf{v}}^{*}},{{a}_{2,1}}\textbf{u} \right\rangle
  & \ge & \left\langle \textbf{y}-{{a}_{1,1}}\textbf{u}_{1}^{*}-\alpha{{\textbf{v}}^{*}},{{a}_{2,1}}\textbf{u}_{2}^{*} \right\rangle  \nonumber \\
 &=& \left\langle {{a}_{2,1}}\textbf{u}_{2}^{*}+\textbf{z},{{a}_{2,1}}\textbf{u} \right\rangle  \nonumber \\
 &=&{{\left\| {{a}_{2,1}}\textbf{u}_{2}^{*} \right\|}^{2}}+\left\langle \textbf{z},{{a}_{2,1}}\textbf{u}_{2}^{*} \right\rangle  \nonumber \\
 &\geq &n\left( \bar{{{\beta }_{2}}}{{P}_{2}}-\sqrt{\bar{{{\beta }_{2}}}{{P}_{2}}N}\epsilon  \right),
\end{IEEEeqnarray}
thus establishing B).

\medskip

C) For every $(\textbf{s}_1,\textbf{s}_2,{\cal C}_1,{\cal C}_2,{\cal C}_c,\textbf{z})\in {\cal E}_{\textbf{X}}^{c}\cap
{\cal E}_{\textbf{Z}}^{c}$
and every $\textbf{u}\in {\cal S}_{2}$,
\begin{IEEEeqnarray*}{rCl}
&& \left| \tilde{\rho }-\cos \sphericalangle \left(\textbf{u}^{*}_1,\textbf{u} \right) \right|\le 7\epsilon
\mbox{   and  }
\left| \bar{\rho }-\cos \sphericalangle \left(\textbf{u}_2,\textbf{v}^{*} \right) \right|\le 7\epsilon
\mbox{   and  }
|\textbf{y}-\textbf{X}_{\textbf{u}^{*}_1,\textbf{v}^{*},\textbf{u}}\|^2
\leq
\|\textbf{y}-\textbf{X}_{\textbf{u}_1^{*},\textbf{v}^{*},\textbf{u}_2^{*}}\|^2  \\
&& \qquad \qquad \implies  \\
&& {{\left\| {{a}_{2,1}}\textbf{u}-\textbf{w} \right\|}^{2}}
\le n\bar{{{\beta }_{2}}}{{P}_{2}}-2\Biggl( n{{\varsigma }_{1}}\left( \bar{{{\beta }_{2}}}{{P}_{2}}-\sqrt{\bar{{{\beta }_{2}}}{{P}_{2}}N}\epsilon  \right)+n{{\varsigma }_{2}}\left( \sqrt{\bar{{\beta}}_{1}\bar{{\beta}}_{2}{{P}_{1}}{{P}_{2}}}\left( \tilde{\rho }-7\epsilon  \right) \right) \\
&& \hspace{2.6cm} +n{{\varsigma }_{3}}\left( \frac{1}{n}\left\| \alpha{{\textbf{v}}^{*}} \right\|\sqrt{n\bar{{{\beta }_{2}}}{{P}_{2}}}\left( \bar{\rho }-7\epsilon  \right) \right) \Biggr)
+{{\left\| \textbf{w} \right\|}^{2}}.
\end{IEEEeqnarray*}

Statement C) is obtained as follows:
\begin{IEEEeqnarray*}{l}
 {{\left\| {{a}_{2,1}}\textbf{u}-\textbf{w} \right\|}^{2}}
 \quad = {{\left\| {{a}_{2,1}}\textbf{u} \right\|}^{2}}-2\left\langle {{a}_{2,1}}\textbf{u},\textbf{w} \right\rangle +{{\left\| \textbf{w} \right\|}^{2}} \\
 \quad =  {{\left\| {{a}_{2,1}}\textbf{u} \right\|}^{2}}-2\left\langle {{a}_{2,1}}\textbf{u},{{\varsigma }_{1}}\left( {{a}_{2,1}}\textbf{u}_{2}^{*}+\textbf{z} \right)+{{\varsigma }_{2}}{{a}_{1,1}}\textbf{u}_{1}^{*}+{{\varsigma }_{3}}\alpha{{\textbf{v}}^{*}} \right\rangle +{{\left\| \textbf{w} \right\|}^{2}} \\
 \quad =  n\bar{{{\beta }_{2}}}{{P}_{2}}-2\left[ {{\varsigma }_{1}}\left\langle {{a}_{2,1}}\textbf{u},{{a}_{2,1}}\textbf{u}_{2}^{*}+\textbf{z} \right\rangle +{{\varsigma }_{2}}\left\langle {{a}_{2,1}}\textbf{u},{{a}_{1,1}}\textbf{u}_{1}^{*} \right\rangle +{{\varsigma }_{3}}\left\langle {{a}_{2,1}}\textbf{u},\alpha{{\textbf{v}}^{*}} \right\rangle  \right]+{{\left\| \textbf{w} \right\|}^{2}} \\
 \quad \overset{(a)}{\mathop{\le }}  \,n\bar{{{\beta }_{2}}}{{P}_{2}}-2n\Biggl({{\varsigma }_{1}}\left( \bar{{{\beta }_{2}}}{{P}_{2}}-\sqrt{\bar{{{\beta }_{2}}}{{P}_{2}}N}\epsilon  \right)+{{\varsigma }_{2}}\left( \sqrt{\bar{{\beta}}_{1}\bar{{\beta}}_{2}{{P}_{1}}{{P}_{2}}}\left( \tilde{\rho }-7\epsilon  \right) \right)\\
 \hspace{1.5cm} +{{\varsigma }_{3}}\left( \frac{1}{n}\left\| \alpha{{\textbf{v}}^{*}} \right\|\sqrt{n\bar{{{\beta }_{2}}}{{P}_{2}}}\left( \bar{\rho }-7\epsilon  \right) \right) \Biggr)
 +{{\left\| \textbf{w} \right\|}^{2}},
\end{IEEEeqnarray*}

where in (a) we have used Statement A), Statement A2) and Statement B).

\medskip

D) For every $(\textbf{s}_1,\textbf{s}_2,{\cal C}_1,{\cal C}_2,{\cal C}_c,\textbf{z})\in {\cal E}_{\textbf{X}}^{c}\cap{\cal E}_{\textbf{Z}}^{c}$
\begin{IEEEeqnarray*}{rCl}
 {{\left\| \textbf{w} \right\|}^{2}}
 &{\mathop{\le }}&\,n\Bigl( {{\varsigma }_{1}}^{2}\left( \bar{{{\beta }_{2}}}{{P}_{2}}+N \right)+2{{\varsigma }_{1}}{{\varsigma }_{2}} \sqrt{\bar{{\beta}}_{1}\bar{{\beta}}_{2}{{P}_{1}}{{P}_{2}}}\tilde{\rho }
 +{{\varsigma }_{2}}^{2}\left( \bar{{{\beta }_{1}}}{{P}_{1}} \right)\\
 && +\frac{1}{n}2{{\varsigma }_{1}}{{\varsigma }_{3}}\left\| \alpha{{\textbf{v}}^{*}} \right\|\sqrt{n\bar{{{\beta }_{2}}}{{P}_{2}}}\bar{\rho }+\frac{1}{n}{{\varsigma }_{3}}^{2}{{\alpha}^{2}}{{\left\| {{\textbf{v}}^{*}} \right\|}^{2}}+\kappa\epsilon  \Bigr),
\end{IEEEeqnarray*}
where $\kappa$ depends on ${{P}_{1}},{{P}_{2}},N,{{\varsigma }_{1}},{{\varsigma }_{2}}$ and ${{\varsigma }_{3}}$ only.

\medskip

Statement D) is obtained as follows
\begin{IEEEeqnarray*}{rCl}
{{\left\| \textbf{w} \right\|}^{2}}
& = & {{\left\| {{\varsigma }_{1}}\left( {{a}_{2,1}}\textbf{u}_{2}^{*}+\textbf{z} \right)+{{\varsigma }_{2}}{{a}_{1,1}}\textbf{u}_{1}^{*}+{{\varsigma }_{3}}\alpha{{\textbf{v}}^{*}} \right\|}^{2}}\\
& = & {{\varsigma }_{1}}^{2}{{\left\| {{a}_{2,1}}\textbf{u}_{2}^{*}+\textbf{z} \right\|}^{2}}+2{{\varsigma }_{1}}{{\varsigma }_{2}}\left\langle {{a}_{2,1}}\textbf{u}_{2}^{*}+\textbf{z},{{a}_{1,1}}\textbf{u}_{1}^{*} \right\rangle +{{\varsigma }_{2}}^{2}{{\left\| {{a}_{1,1}}\textbf{u}_{1}^{*} \right\|}^{2}}+2{{\varsigma }_{1}}{{\varsigma }_{3}}\left\langle {{a}_{2,1}}\textbf{u}_{2}^{*}+\textbf{z},\alpha{{\textbf{v}}^{*}} \right\rangle \\
&& +2{{\varsigma }_{2}}{{\varsigma }_{3}}\left\langle {{a}_{1,1}}\textbf{u}_{1}^{*},\alpha{{\textbf{v}}^{*}} \right\rangle +{{\varsigma }_{3}}^{2}{{\left\| \alpha{{\textbf{v}}^{*}} \right\|}^{2}} \\
& = & {{\varsigma }_{1}}^{2}\left( {{\left\| {{a}_{2,1}}\textbf{u}_{2}^{*} \right\|}^{2}}+2\left\langle {{a}_{2,1}}\textbf{u}_{2}^{*},\textbf{z} \right\rangle +{{\left\| \textbf{z} \right\|}^{2}} \right)+2{{\varsigma }_{1}}{{\varsigma }_{2}}\left( \left\langle {{a}_{2,1}}\textbf{u}_{2}^{*},{{a}_{1,1}}\textbf{u}_{1}^{*} \right\rangle +\left\langle \textbf{z},{{a}_{1,1}}\textbf{u}_{1}^{*} \right\rangle  \right)\\
&& +{{\varsigma }_{2}}^{2}\left( n\bar{{{\beta }_{1}}}{{P}_{1}} \right) +2{{\varsigma }_{1}}{{\varsigma }_{3}}\left[ \left\langle {{a}_{2,1}}\textbf{u}_{2}^{*},\alpha{{\textbf{v}}^{*}} \right\rangle +\left\langle \textbf{z},\alpha{{\textbf{v}}^{*}} \right\rangle  \right]+2{{\varsigma }_{2}}{{\varsigma }_{3}}\left\langle {{a}_{1,1}}\textbf{u}_{1}^{*},\alpha{{\textbf{v}}^{*}} \right\rangle +{{\varsigma }_{3}}^{2}{{\alpha}^{2}}{{\left\| {{\textbf{v}}^{*}} \right\|}^{2}} \\
& \text{ } \overset{(a)}{\mathop{\le }}&\,
n\Bigl( {{\varsigma }_{1}}^{2}\left( \bar{{{\beta }_{2}}}{{P}_{2}}+N \right)
+2{{\varsigma }_{1}}{{\varsigma }_{2}} \sqrt{\bar{{\beta}}_{1}\bar{{\beta}}_{2}{{P}_{1}}{{P}_{2}}}\tilde{\rho } +{{\varsigma }_{2}}^{2}\left( \bar{{{\beta }_{1}}}{{P}_{1}} \right)+\frac{1}{n}2{{\varsigma }_{1}}{{\varsigma }_{3}}\left\| \alpha{{\textbf{v}}^{*}} \right\|\sqrt{n\bar{{{\beta }_{2}}}{{P}_{2}}}\bar{\rho }\\
&& +\frac{1}{n}{{\varsigma }_{3}}^{2}{{\alpha}^{2}}{{\left\| {{\textbf{v}}^{*}} \right\|}^{2}}+k\epsilon  \Bigr),
\end{IEEEeqnarray*}
where in (a) we have used that $(\textbf{s}_1,\textbf{s}_2,{\cal C}_1,{\cal C}_2,{\cal C}_c,\textbf{z})\in {\cal E}_{\textbf{X}}^{c}$, and statements A),A1) and A2).

\medskip

E) For every$(\textbf{s}_1,\textbf{s}_2,{\cal C}_1,{\cal C}_2,{\cal C}_c,\textbf{z})\in {\cal E}_{\textbf{X}}^{c}\cap{\cal E}_{\textbf{Z}}^{c}$ and an arbitrary $\textbf{u}\in {\cal S}_{2}$,
\begin{IEEEeqnarray*}{rCl}
&& \left| \tilde{\rho }-\cos \sphericalangle \left( \textbf{u}_1^{*},\textbf{u} \right) \right|\le 7\epsilon
\text{  and  }\left| \bar{\rho }-\cos \sphericalangle \left( \textbf{v}^{*},\textbf{u} \right) \right|\le 7\epsilon \text{  and  }
|\textbf{y}-\textbf{X}_{\textbf{u}^{*}_1,\textbf{v}^{*},\textbf{u}}\|^2
\leq  \|\textbf{y}-\textbf{X}_{\textbf{u}_1^{*},\textbf{v}^{*},\textbf{u}_2^{*}}\|^2 \\
&& \implies \text{        }{{\left\| {{a}_{2,1}}\textbf{u}-\textbf{w} \right\|}^{2}}\le \Upsilon (\epsilon ),
\end{IEEEeqnarray*}
where
\begin{equation*}
\Upsilon (\epsilon )=n\left( \frac{\bar{{{\beta }_{2}}}{{P}_{2}}N\left( 1-{{\bar{\rho }}^{2}}-{{{\tilde{\rho }}}^{2}} \right)}{{{\bar{\beta }}_{2}}{{P}_{2}}\left( 1-{{\bar{\rho }}^{2}}-{{{\tilde{\rho }}}^{2}} \right)+N}+\frac{\bar{{{\beta }_{2}}}{{P}_{2}}{{N}^{2}}{{{\tilde{\rho }}}^{2}}{{\bar{\rho }}^{2}}\left( 2+{{{\tilde{\rho }}}^{2}} \right)}{{{\left( {{\bar{\beta }}_{2}}{{P}_{2}}\left( 1-{{\bar{\rho }}^{2}}-{{{\tilde{\rho }}}^{2}} \right)+N \right)}^{2}}} \right)+n{{\kappa}'}\epsilon ,
\end{equation*}
and where ${{\kappa}'}$ only depends on ${{P}_{2}},{{N}_{1}},{{N}_{2}},{{\varsigma }_{1}},{{\varsigma }_{2}}$  and $ {{\varsigma }_{3}}$.

\medskip

Statement E) follows from combining Statement C) with Statement
D) and the explicit values of ${{\varsigma }_{1}},{{\varsigma }_{2}}$  and $ {{\varsigma }_{3}}$ given in
\eqref{eq:w_coeff_for_R2}.

\medskip

F) For every $\textbf{u}\in {\cal S}_{2}$, denote by $\varphi \in \left[ 0,\pi  \right]$ the angle between \textbf{u} and \textbf{w}, and let
\begin{IEEEeqnarray*}{rCl}
&&  {\cal B}(\textbf{s}_1,\textbf{s}_2,\textbf{u}_1^{*},\textbf{v}^{*},\textbf{u}_2^{*},\textbf{z})
\triangleq \Biggl\{ \textbf{u}\in S_{2}^{\left( n \right)}: \\
&& \cos \left( \varphi  \right)\ge
\sqrt{\frac{{{\bar{\beta }}_{2}}{{P}_{2}}\left[ 1-{{{\tilde{\rho }}}^{2}} \right]+N{{{\tilde{\rho }}}^{2}}-{{\bar{\rho }}^{2}}\left( {{\bar{\beta }}_{2}}{{P}_{2}}-N \right)}{{{\bar{\beta }}_{2}}{{P}_{2}}\left( 1-{{{\tilde{\rho }}}^{2}} \right)+N-{{\bar{\rho }}^{2}}{{\bar{\beta }}_{2}}{{P}_{2}}}-\frac{{{N}^{2}}{{{\tilde{\rho }}}^{2}}{{\bar{\rho }}^{2}}\left( 2+{{{\tilde{\rho }}}^{2}} \right)}{{{\left( {{\bar{\beta }}_{2}}{{P}_{2}}\left[ 1-{{\bar{\rho }}^{2}}-{{{\tilde{\rho }}}^{2}} \right]+N \right)}^{2}}}-{{\kappa}^{''}}\epsilon } \Biggr\},
\end{IEEEeqnarray*}
where ${{\kappa}^{''}}$ only depends on ${{P}_{1}},{{N}_{1}},{{N}_{2}},{{\varsigma }_{1}},{{\varsigma }_{2}}$  and $ {{\varsigma }_{3}}$. \\
Then, for every $(\textbf{s}_1,\textbf{s}_2,{\cal C}_1,{\cal C}_2,{\cal C}_c,\textbf{z})\in {\cal E}_{\textbf{X}}^{c}\cap{\cal E}_{\textbf{Z}}^{c}$,
\begin{IEEEeqnarray}{rCl}\label{eq:statement_F_R2}
&& \left| \tilde{\rho }-\cos \sphericalangle \left( \textbf{u}_{1}^{*},\textbf{u} \right) \right|\le 7\epsilon \quad \mbox{  and  } \quad \left| \bar{\rho }-\cos \sphericalangle \left( \textbf{v}^{*},\textbf{u} \right) \right|\le 7\epsilon \quad \mbox{  and  } \nonumber \\
&& {{\left\| \textbf{y}-\left( {{a}_{1,1}}\textbf{u}_1^{*}+{{a}_{2,1}}\textbf{u}+\alpha{{\textbf{v}}^{*}} \right) \right\|}^{2}}\le {{\left\| \textbf{y}-\left( {{a}_{1,1}}\textbf{u}_{1}^{*}+{{a}_{2,1}}\textbf{u}_{2}^{*}+\alpha{{\textbf{v}}^{*}} \right) \right\|}^{2}} \nonumber \\
&& \qquad \implies \text{    }\textbf{u}\in {\cal B}(\textbf{s}_1,\textbf{s}_2,\textbf{u}_1^{*},\textbf{v}^{*},\textbf{u}_2^{*},\textbf{z}).
\end{IEEEeqnarray}

\medskip

Statement F) follows from Statement E) by noting that if $\mathbf{w}\ne 0$ and
$1-\frac{\Upsilon (\epsilon )}{n\bar{{{\beta }_{2}}}{{P}_{2}}}>0$, then
\begin{IEEEeqnarray*}{rCl}
\left.
\begin{array}{l}
{{\left\| {{a}_{2,1}}\textbf{u} \right\|}^{2}}=n\bar{{{\beta }_{2}}}{{P}_{2}}\\
 {{\left\| {{a}_{2,1}}\textbf{u}-\textbf{w} \right\|}^{2}}\le \Upsilon (\epsilon ) \\
\end{array}
\right\}
\implies \cos \sphericalangle \left( \textbf{u},\textbf{w} \right)\ge \sqrt{1-\frac{\Upsilon (\epsilon )}{n\bar{{{\beta }_{2}}}{{P}_{2}}}} ,
\end{IEEEeqnarray*}
which follows by the same argument as \eqref{eq:statement_F_proof_R1}.

\medskip

The proof of Lemma~\ref{eps_h_u2} is now concluded by noticing that the set
${\cal E}'_{{{{\hat{\textbf{U}}}}_{2}}}$ , defined in \eqref{eq:eps_h_u2},
is the set of tuples
$(\textbf{s}_1,\textbf{s}_2,{\cal C}_1,{\cal C}_2,{\cal C}_c,\textbf{z})$
for which there exists a ${{\textbf{u}}_{2}}\left( j \right)\in {{{\cal C}}_{2}}\backslash \left\{ \textbf{u}_{2}^{*} \right\}$ such that
${{\textbf{u}}_{2}}\left( j \right)\in
{\cal B}(\textbf{s}_1,\textbf{s}_2,\textbf{u}_1^{*},\textbf{v}^{*},\textbf{u}_2^{*},\textbf{z})$.
Thus, by Statement F) and by the definition of ${{\cal E}_{{{{\hat{\textbf{U}}}}_{2}}}}$
in \eqref{eq:eps_u2} it follows that
\begin{equation*}
{{\cal E}_{{{{\hat{\textbf{U}}}}_{2}}}}\cap {\cal E}_{\textbf{S}}^{c}\cap {\cal E}_{\textbf{X}}^{c}\cap {\cal E}_{\textbf{Z}}^{c}
\subseteq {{\cal E} }'_{{{{\hat{\textbf{U}}}}_{2}}}\cap {\cal E}_{\textbf{S}}^{c}\cap {\cal E}_{\textbf{X}}^{c}\cap {\cal E}_{\textbf{Z}}^{c},
\end{equation*}
and therefore
\begin{flalign*}
&&
\Pr\left[ {{{\cal E}}_{{{{\hat{\textbf{U}}}}_{2}}}}\cap {\cal E}_{\textbf{S}}^{c}\cap {\cal E}_{\textbf{X}}^{c}\cap {\cal E}_{\textbf{Z}}^{c} \right]
\le \Pr\left[ {\cal E}'_{{{{\hat{\textbf{U}}}}_{2}}}\cap {\cal E}_{\textbf{S}}^{c}\cap {\cal E}_{\textbf{X}}^{c}\cap {\cal E}_{\textbf{Z}}^{c} \right].
&&
\end{flalign*}
\end{IEEEproof}

Next,
\begin{IEEEeqnarray}{rCl}\label{eq:R2_proof}
 \ \Pr\left[ {{\cal E}_{{{{\hat{\textbf{U}}}}_{2}}}}\cap {\cal E}_{\textbf{S}}^{c}\cap {\cal E}_{\textbf{X}}^{c}\cap {\cal E}_{\textbf{Z}}^{c} \right]
 & \overset{(a)}{\mathop{\le }}& \,\Pr\left[
 {\cal E}'_{{{{\hat{\textbf{U}}}}_{2}}}\cap {\cal E}_{\textbf{S}}^{c}\cap {\cal E}_{\textbf{X}}^{c}\cap {\cal E}_{\textbf{Z}}^{c} \right] \nonumber \\
 &  \overset{(b)}{\mathop{\le }}&\,\Pr\left[ {\cal E}'_{{{{\hat{\textbf{U}}}}_{2}}}|{\cal E}_{{{\textbf{X}}_{1}}}^{c} \right],
\end{IEEEeqnarray}
where (a) follows by Lemma~\ref{eps_h_u2} and (b) follows because ${\cal E}_{\textbf{X}}^{c}\subseteq {\cal E}_{{{\textbf{X}}_{2}}}^{c}$.
The proof of \eqref{eq:R2} is now completed by combining \eqref{eq:R2_proof} with Lemma~\ref{Rate proof}.
This gives that for every $\delta >0$ and every $\epsilon >0$ there exists some
$n'(\delta ,\epsilon )$
such that for all $n>n'(\delta ,\epsilon )$,
we have
\begin{equation*}
\Pr\left[ {{\cal E}_{{{{\hat{\textbf{U}}}}_{2}}}}\cap {\cal E}_{\textbf{S}}^{c}\cap {\cal E}_{\textbf{X}}^{c}\cap{\cal E}_{\textbf{Z}}^{c} \right]\le \Pr\left[ {\cal E}'_{{{{\hat{\textbf{U}}}}_{2}}}|{\cal E}_{{{\textbf{X}}_{2}}}^{c} \right]<\delta,
\end{equation*}
whenever
\begin{equation*}
{{R}_{2}}<\frac{1}{2}\log \left( \frac{{{\bar{\beta }}_{2}}{{P}_{2}}\left( 1-{{{\tilde{\rho }}}^{2}}-{{\bar{\rho }}^{2}} \right)+N}{N\left( 1-{{{\tilde{\rho }}}^{2}}-{{\bar{\rho }}^{2}} \right)+{{\lambda }_{2}}}-\kappa_{2}\epsilon \right),
\end{equation*}
where ${{\kappa}_{2}}$ is a positive constant determined by ${{P}_{1}},{{P}_{2}},N,{{\varsigma }_{1}},{{\varsigma }_{2}}\text{ and }{{\varsigma }_{3}}$. \hfill\qed

\vskip.2truein

\subsection{Proof of rate constraint \eqref{eq:Rc}}
Define
\begin{equation}\label{eq:w_for_Rc}
 \textbf{w}(\textbf{s}_1,\textbf{s}_2,{\cal C}_1,{\cal C}_2,{\cal C}_c,\textbf{z})={{\varsigma }_{1}}\left( \textbf{y}-\left( {{a}_{1,1}}\textbf{u}_{1}^{*}+{{a}_{2,1}}\textbf{u}_{2}^{*} \right) \right)+{{\varsigma }_{2}}{{a}_{1,1}}\textbf{u}_{1}^{*}+{{\varsigma }_{3}}{{a}_{2,1}}\textbf{u}_{2}^{*},
\end{equation}
where
\begin{IEEEeqnarray}{rCl}\label{eq:w_coeff_for_Rc}
{{\varsigma }_{1}}
&=&\frac{{{\sigma }^{2}}{{\alpha}^{2}}\left( 1-{{2}^{-2{{R}_\textnormal{c}}}} \right){{2}^{-2{{R}_{1}}}}\left( 1-{{{\tilde{\rho }}}^{2}}-{{\bar{\rho }}^{2}} \right)}{{{\sigma }^{2}}{{\alpha}^{2}}\left( 1-{{2}^{-2{{R}_\textnormal{c}}}} \right){{2}^{-2{{R}_{1}}}}\left( 1-{{{\tilde{\rho }}}^{2}}-{{\bar{\rho }}^{2}} \right)+N\left( 1-{{{\tilde{\rho }}}^{2}} \right)} \nonumber \\
{{\varsigma }_{2}}
&=&-\frac{\alpha{{\bar{\rho }}^{2}}N}{{{a}_{1,1}}\left( {{\sigma }^{2}}{{\alpha}^{2}}\left( 1-{{2}^{-2{{R}_\textnormal{c}}}} \right){{2}^{-2{{R}_{1}}}}\left( 1-{{{\tilde{\rho }}}^{2}}-{{\bar{\rho }}^{2}} \right)+N\left( 1-{{{\tilde{\rho }}}^{2}} \right) \right)} \nonumber \\
{{\varsigma }_{3}}
&=&\frac{\alpha\rho \left( 1-{{2}^{-2{{R}_\textnormal{c}}}} \right){{2}^{-2{{R}_{1}}}}N}{{{a}_{2,1}}\left( {{\sigma }^{2}}{{\alpha}^{2}}\left( 1-{{2}^{-2{{R}_\textnormal{c}}}} \right){{2}^{-2{{R}_{1}}}}\left( 1-{{{\tilde{\rho }}}^{2}}-{{\bar{\rho }}^{2}} \right)+N\left( 1-{{{\tilde{\rho }}}^{2}} \right) \right)}.
\end{IEEEeqnarray}

We now start with a lemma that will be used to prove \eqref{eq:Rc}.
\begin{lemma}\label{eps_h_v}
Let ${{\varphi }_{j}}\in \left[ 0,\pi  \right]$  be the angle between $\mathbf{w}$ and $\mathbf{v}(j)$,
and let the set ${\cal E}'_{{{{\hat{\mathbf{V}}}}}}$
be defined as
\begin{IEEEeqnarray}{rCl}\label{eq:eps_h_v}
{\cal E}'_{{{{\hat{\mathbf{V}}}}}}
\triangleq
&&\Biggl\{(\mathbf{s}_1,\mathbf{s}_2,{\cal C}_1,{\cal C}_2,{\cal C}_c,\mathbf{z})\colon \exists \> {{\mathbf{v}}}\left( j \right)\in {{\cal C}_{c}}\backslash \left\{ \mathbf{v}^{*} \right\}\
\mbox{s.t.}\ \ \nonumber\\
&&  \qquad \quad \cos \left( {{\varphi }_{j}} \right)\ge \sqrt{1-\frac{\Upsilon (\epsilon )}{n{{\alpha}^{2}}{{\sigma }^{2}}{{2}^{-2{{R}_{1}}}}\left( 1-{{2}^{-2{{R}_\textnormal{c}}}} \right)}}\  \Biggr\},
\end{IEEEeqnarray}
where $\Upsilon(\epsilon)$ is defined in \eqref{eq:upsilon_Rc}.
Then,
\begin{equation*}
{{\cal E}_{{\hat{\mathbf{V}}}}}\cap {\cal E}_{\mathbf{S}}^{c}\cap {\cal E}_{\mathbf{X}}^{c}\cap {\cal E}_{\mathbf{Z}}^{c}\subseteq
{{\cal E}'}_{{\hat{\mathbf{V}}}}\cap {\cal E}_{\mathbf{S}}^{c}\cap {\cal E}_{\mathbf{X}}^{c}\cap {\cal E}_{\mathbf{Z}}^{c},
\end{equation*}
and, in particular
\begin{equation*}
\Pr\left[ {{\cal E}_{{\hat{\mathbf{V}}}}}\cap {\cal E}_{\mathbf{S}}^{c}\cap {\cal E}_{\mathbf{X}}^{c}\cap {\cal E}_{\mathbf{Z}}^{c} \right]\le \Pr\left[ {{\cal E}'}_{{\hat{\mathbf{V}}}}\cap {\cal E}_{\mathbf{S}}^{c}\cap {\cal E}_{\mathbf{X}}^{c}\cap{\cal E}_{\mathbf{Z}}^{c} \right].
\end{equation*}
\end{lemma}

\medskip

\begin{IEEEproof}
We first recall that for the event ${{\cal E}_{{{{\hat{\textbf{V}}}}}}}$
to occur, there must exist a codeword ${{\textbf{v}}}\left(j \right)\in {{\cal C}_{c}}\backslash \left\{ \textbf{v}^{*} \right\}$ that satisfies
the following three conditions
\begin{IEEEeqnarray}{rCl}
\left| \bar{\rho }-\cos \sphericalangle \left(\textbf{v}\left(j\right),\textbf{u}_{2}^{*} \right) \right| & \le &  7\epsilon \\
\left| \cos \sphericalangle \left(\textbf{v}\left(j\right),\textbf{u}_{1}^{*} \right) \right| & \le & 3\epsilon \\
\|\textbf{y}-\textbf{X}_{\textbf{u}_{1}^{*},\textbf{v}\left(j\right),\textbf{u}_{2}^{*}}\|^2
&  \leq &
\|\textbf{y}-\textbf{X}_{\textbf{u}_1^{*},\textbf{v}^{*},\textbf{u}_2^{*}}\|^2 .
\end{IEEEeqnarray}

The proof is now based on a sequence of statements related to these three conditions.

\medskip

A) For every $(\textbf{s}_1,\textbf{s}_2,{\cal C}_1,{\cal C}_2,{\cal C}_c,\textbf{z})\in {\cal E}_{\textbf{X}}^{c}$ and every $\textbf{v}\in {\cal S}_{c}$, where $S_{c}$ is
the surface area of the codeword sphere of ${\cal C}_{c}$ defined in the code construction,
\begin{IEEEeqnarray}{rCl}\label{eq:statement_A_Rc}
&& \left| \bar{\rho }-\cos \sphericalangle \left( \textbf{v},\textbf{u}_{2}^{*} \right) \right|\le 7\epsilon \nonumber \\
&& \implies \left| \left\| \alpha{{\textbf{v}}^{*}} \right\|\sqrt{n\bar{{{\beta }_{2}}}{{P}_{2}}}\bar{\rho }-\left\langle \alpha\textbf{v},{{a}_{2,1}}\textbf{u}_{2}^{*} \right\rangle  \right|\le \left\| \alpha{{\textbf{v}}^{*}} \right\|\sqrt{n\bar{{{\beta }_{2}}}{{P}_{2}}}7\epsilon .
\end{IEEEeqnarray}

Statement A) follows by rewriting $\cos \sphericalangle \left( \textbf{v},\textbf{u}_{2}^{*} \right)$ as ${\left\langle \textbf{v},\textbf{u}_{2}^{*} \right\rangle }/({\left\| \textbf{v} \right\|\left\| \textbf{u}_{2}^{*} \right\|})$, and then multiplying the inequality on the l.h.s.
of \eqref{eq:statement_A_Rc} by
$\left\| \alpha{{\textbf{v}}^{*}} \right\|\cdot \left\| {{a}_{2,1}}\textbf{u}_{2}^{*} \right\|$.

\medskip

A1) For every $(\textbf{s}_1,\textbf{s}_2,{\cal C}_1,{\cal C}_2,{\cal C}_c,\textbf{z})\in {\cal E}_{\textbf{X}}^{c}$,
\begin{IEEEeqnarray}{rCl} \label{eq:statement_A1_Rc}
&& \left| \cos \sphericalangle \left( \textbf{v},\textbf{u}_{1}^{*} \right) \right|\le 3\epsilon  \nonumber \\
&& \text{       }\implies \text{      }\left| \left\langle \alpha\textbf{v},{{a}_{1,1}}\textbf{u}_{1}^{*} \right\rangle  \right|\le 3\left\| \alpha\textbf{v} \right\|\sqrt{n\bar{{{\beta }_{1}}}{{P}_{1}}}\epsilon .
\end{IEEEeqnarray}

Statement A1) follows by rewriting $\cos \sphericalangle \left( \textbf{v},\textbf{u}_{1}^{*} \right)$ as ${\left\langle \textbf{v},\textbf{u}_{1}^{*} \right\rangle }/({\left\| \textbf{v}^{*} \right\|\left\| \textbf{u}_{1}^{*} \right\|})$, and then multiplying the inequality on the l.h.s. of \eqref{eq:statement_A1_Rc} by
$\left\| \alpha\textbf{v} \right\|\cdot \left\| {{a}_{1,1}}\textbf{u}_{1}^{*} \right\|$
and recalling that $\left\| {{a}_{1,1}}\textbf{u}_{1}^{*} \right\|=\sqrt{n\bar{{{\beta }_{1}}}{{P}_{1}}}$.

\medskip

B) For every $(\textbf{s}_1,\textbf{s}_2,{\cal C}_1,{\cal C}_2,{\cal C}_c,\textbf{z})\in {\cal E}_{\textbf{X}}^{c}\cap
{\cal E}_{\textbf{Z}}^{c}$
and every $\textbf{v}\in {\cal S}_{c}$
\begin{IEEEeqnarray}{rCl}\label{eq:statement_B_Rc}
&& |\textbf{y}-\textbf{X}_{\textbf{u}^{*}_1,\textbf{\textbf{v}},\textbf{u}^{*}_2}\|^2
 \leq
\|\textbf{y}-\textbf{X}_{\textbf{u}_1^{*},\textbf{\textbf{v}}^{*},\textbf{u}_2^{*}}\|^2 \nonumber \\
&& \text{  }\implies \left\langle \textbf{y}-\left( {{a}_{1,1}}\textbf{u}_{1}^{*}-{{a}_{2,1}}\textbf{u}_{2}^{*} \right),\alpha\textbf{v} \right\rangle \ge {{\left\| \alpha{{\textbf{v}}^{*}} \right\|}^{2}}-\left\| \alpha{{\textbf{v}}^{*}} \right\|\sqrt{nN}\epsilon .
\end{IEEEeqnarray}

Statement B) follows from rewriting the inequality on the l.h.s. of \eqref{eq:statement_B_Rc} as
\begin{equation*}
 {{\left\| \left( \textbf{y}-{{a}_{1,1}}\textbf{u}_{1}^{*}-{{a}_{2,1}}\textbf{u}_{2}^{*} \right)-\alpha\textbf{v} \right\|}^{2}}\le {{\left\| \left( \textbf{y}-{{a}_{1,1}}\textbf{u}_{1}^{*}-{{a}_{2,1}}\textbf{u}_{2}^{*} \right)-\alpha{{\textbf{v}}^{*}} \right\|}^{2}},
 \end{equation*}
 or equivalently as
\begin{IEEEeqnarray}{rCl}\label{eq:statement_B_proof_Rc}
\left\langle \textbf{y}-{{a}_{1,1}}\textbf{u}_{1}^{*}-{{a}_{2,1}}\textbf{u}_{2}^{*},\alpha\textbf{v} \right\rangle
& \ge & \left\langle \textbf{y}-{{a}_{1,1}}\textbf{u}_{1}^{*}-{{a}_{2,1}}\textbf{u}_{2}^{*},\alpha{{\textbf{v}}^{*}} \right\rangle  \nonumber \\
& = & \left\langle \alpha
{{\textbf{v}}^{*}}+\textbf{z},\alpha{{\textbf{v}}^{*}} \right\rangle  \nonumber  \\
& = & {{\left\| \alpha{{\textbf{v}}^{*}} \right\|}^{2}}+\left\langle \textbf{z},\alpha{{\textbf{v}}^{*}} \right\rangle  \nonumber  \\
& \ge &  {{\left\| \alpha{{\textbf{v}}^{*}} \right\|}^{2}}-\left\| \alpha{{\textbf{v}}^{*}} \right\|\sqrt{nN}\epsilon ,
\end{IEEEeqnarray}
thus establising B).

\medskip

C) For every $(\textbf{s}_1,\textbf{s}_2,{\cal C}_1,{\cal C}_2,{\cal C}_c,\textbf{z})\in {\cal E}_{\textbf{X}}^{c}\cap{\cal E}_{\textbf{Z}}^{c}$
and every $\textbf{v}\in {\cal S}_{c}$,
\begin{IEEEeqnarray*}{rCl}
&& \left| \bar{\rho }-\cos \sphericalangle \left( \textbf{v},\textbf{u}_{2}^{*} \right) \right|\le 7\epsilon \text{  and  }\left| \cos \sphericalangle \left( \textbf{v},\textbf{u}_{1}^{*} \right) \right|\le 3\epsilon \text{  and  }
|\textbf{y}-\textbf{X}_{\textbf{u}^{*}_1,\textbf{v},\textbf{u}^{*}_2}\|^2
\leq
\|\textbf{y}-\textbf{X}_{\textbf{u}_1^{*},\textbf{v}^{*},\textbf{u}_2^{*}}\|^2 \nonumber \\
&& \implies  \\
&& {{\left\| \alpha\textbf{v}-\textbf{w} \right\|}^{2}}
\le
{{\left\| \alpha\textbf{v} \right\|}^{2}}-2n\Biggl( {{\varsigma }_{1}}\left( \frac{1}{n}{{\left\| \alpha{{\textbf{v}}^{*}} \right\|}^{2}}-\frac{1}{n}\left\| \alpha{{\textbf{v}}^{*}} \right\|\sqrt{nN}\epsilon  \right)-{{\varsigma }_{2}}\left\| \alpha\textbf{v} \right\|\sqrt{n\bar{{{\beta }_{1}}}{{P}_{1}}}3\epsilon\\
&& \hspace{2.2cm} +{{\varsigma }_{3}}\left( \frac{1}{n}\left\| \alpha{{\textbf{v}}^{*}} \right\|\sqrt{n\bar{{{\beta }_{2}}}{{P}_{2}}}\left( \bar{\rho }-7\epsilon  \right) \right) \Biggr)
+{{\left\| \textbf{w} \right\|}^{2}}.
\end{IEEEeqnarray*}

Statement C) is obtained as follows:
\begin{IEEEeqnarray*}{rCl}
 {{\left\| \alpha\textbf{v}-\textbf{w} \right\|}^{2}}
 & = &{{\left\| \alpha\textbf{v} \right\|}^{2}}-2\left\langle \alpha\textbf{v},\textbf{w} \right\rangle +{{\left\| \textbf{w} \right\|}^{2}} \\
 & = & {{\left\| \alpha\textbf{v} \right\|}^{2}}-2\left\langle \alpha\textbf{v},{{\varsigma }_{1}}\left( \alpha{{\textbf{v}}^{*}}+\textbf{z} \right)+{{\varsigma }_{2}}{{a}_{1,1}}\textbf{u}_{1}^{*}+{{\varsigma }_{3}}{{a}_{2,1}}\textbf{u}_{2}^{*} \right\rangle +{{\left\| \textbf{w} \right\|}^{2}} \\
 & = & {{\left\| \alpha\textbf{v} \right\|}^{2}}-2\Bigl( {{\varsigma }_{1}}\left\langle \alpha\textbf{v},\alpha{{\textbf{v}}^{*}}+\textbf{z} \right\rangle +{{\varsigma }_{2}}\left\langle \alpha\textbf{v},{{a}_{1,1}}\textbf{u}_{1}^{*} \right\rangle \\
 && +{{\varsigma }_{3}}\left\langle \alpha\textbf{v},{{a}_{2,1}}\textbf{u}_{2}^{*} \right\rangle  \Bigr)+{{\left\| \textbf{w} \right\|}^{2}} \\
 & \text{ } \overset{(a)}{\mathop{\le }} & \,{{\left\| \alpha\textbf{v} \right\|}^{2}}-2\Biggl( n{{\varsigma }_{1}}\left( \frac{1}{n}{{\left\| \alpha{{\textbf{v}}^{*}} \right\|}^{2}}-\frac{1}{n}\left\| \alpha{{\textbf{v}}^{*}} \right\|\sqrt{nN}\epsilon  \right)-{{\varsigma }_{2}}\left\| \alpha\textbf{v} \right\|\sqrt{n\bar{{{\beta }_{1}}}{{P}_{1}}}3\epsilon\\
 && +n{{\varsigma }_{3}}\left( \frac{1}{n}\left\| \alpha{{\textbf{v}}^{*}} \right\|\sqrt{n\bar{{{\beta }_{2}}}{{P}_{2}}}\left( \bar{\rho }-7\epsilon  \right) \right) \Biggr)+{{\left\| \textbf{w} \right\|}^{2}},
\end{IEEEeqnarray*}
where in (a) we have used Statement A), A1) and Statement B).

\medskip

D) For every $(\textbf{s}_1,\textbf{s}_2,{\cal C}_1,{\cal C}_2,{\cal C}_c,\textbf{z})\in {\cal E}_{\textbf{X}}^{c}\cap{\cal E}_{\textbf{Z}}^{c}$
\begin{IEEEeqnarray*}{rCl}
{{\left\| \textbf{w} \right\|}^{2}}
& \le & n\Bigl( \frac{1}{n}{{\varsigma }_{1}}^{2}
{{\left\| \alpha{{\textbf{v}}^{*}} \right\|}^{2}}
+\frac{2}{n}{{\varsigma }_{1}}{{\varsigma }_{3}} \left\| \alpha{{\textbf{v}}^{*}} \right\|\sqrt{n\bar{{{\beta }_{2}}}{{P}_{2}}}\bar{\rho }
+{{\varsigma }_{1}}^{2}N+{{\varsigma }_{2}}^{2} \bar{{{\beta }_{1}}}{{P}_{1}} \\
&& +2{{\varsigma }_{2}}{{\varsigma }_{3}}\sqrt{\bar{{\beta}}_{1}\bar{{\beta}}_{2}{{P}_{1}}{{P}_{2}}}\tilde{\rho }+{{\varsigma }_{3}}^{2} \bar{{{\beta }_{2}}}{{P}_{2}} +\kappa\epsilon  \Bigr),
\end{IEEEeqnarray*}
where $\kappa$ depends on ${{P}_{1}},{{P}_{2}},N,{{\varsigma }_{1}},{{\varsigma }_{2}}$ and ${{\varsigma }_{3}}$ only.

Statement D) is obtained as follows
\begin{IEEEeqnarray*}{l}
{{\left\| \textbf{w} \right\|}^{2}}
 =  {{\left\| {{\varsigma }_{1}}\left( \alpha{{\textbf{v}}^{*}}+\textbf{z} \right)+{{\varsigma }_{2}}{{a}_{1,1}}\textbf{u}_{1}^{*}+{{\varsigma }_{3}}{{a}_{2,1}}\textbf{u}_{2}^{*} \right\|}^{2}}\\
\ =  {{\varsigma }_{1}}^{2}{{\left\| \alpha  {{\textbf{v}}^{*}}+\textbf{z} \right\|}^{2}}+2{{\varsigma }_{1}}{{\varsigma }_{2}}\left\langle \alpha{{\textbf{v}}^{*}}+\textbf{z},{{a}_{1,1}}\textbf{u}_{1}^{*} \right\rangle +{{\varsigma }_{2}}^{2}{{\left\| {{a}_{1,1}}\textbf{u}_{1}^{*} \right\|}^{2}} \\
\hspace{1.0cm} +2{{\varsigma }_{1}}{{\varsigma }_{3}}\left\langle \alpha{{\textbf{v}}^{*}}+\textbf{z},{{a}_{2,1}}\textbf{u}_{2}^{*} \right\rangle +2{{\varsigma }_{2}}{{\varsigma }_{3}}\left\langle {{a}_{1,1}}\textbf{u}_{1}^{*},{{a}_{2,1}}\textbf{u}_{2}^{*} \right\rangle +{{\varsigma }_{3}}^{2}{{\left\| {{a}_{2,1}}\textbf{u}_{2}^{*} \right\|}^{2}} \\
\ =  {{\varsigma }_{1}}^{2}\left( {{\left\| \left[ {{a}_{1,2}}+{{a}_{2,2}} \right]
{{\textbf{v}}^{*}} \right\|}^{2}}+2\left\langle \alpha{{\textbf{v}}^{*}},\textbf{z} \right\rangle +{{\left\| \textbf{z} \right\|}^{2}} \right)+2{{\varsigma }_{1}}{{\varsigma }_{2}}\left( \left\langle \alpha{{\textbf{v}}^{*}},{{a}_{1,1}}\textbf{u}_{1}^{*} \right\rangle +\left\langle \textbf{z},{{a}_{1,1}}\textbf{u}_{1}^{*} \right\rangle  \right)\\
\qquad +{{\varsigma }_{2}}^{2}\left( n\bar{{{\beta }_{1}}}{{P}_{1}} \right) +2{{\varsigma }_{1}}{{\varsigma }_{3}}\left[ \left\langle \alpha{{\textbf{v}}^{*}},{{a}_{2,1}}\textbf{u}_{2}^{*} \right\rangle +\left\langle \textbf{z},{{a}_{2,1}}\textbf{u}_{2}^{*} \right\rangle  \right]+2{{\varsigma }_{2}}{{\varsigma }_{3}}\left\langle {{a}_{1,1}}\textbf{u}_{1}^{*},{{a}_{2,1}}\textbf{u}_{2}^{*} \right\rangle +{{\varsigma }_{3}}^{2}\left( n\bar{{{\beta }_{2}}}{{P}_{2}} \right) \\
\ \text{ } \overset{(a)}{\mathop{\le }} \,
n\Bigl( \frac{1}{n}{{\varsigma }_{1}}^{2} {{\left\| \alpha{{\textbf{v}}^{*}} \right\|}^{2}} +\frac{2}{n}{{\varsigma }_{1}}{{\varsigma }_{3}} \left\| \alpha{{\textbf{v}}^{*}} \right\|\sqrt{n\bar{{{\beta }_{2}}}{{P}_{2}}}\bar{\rho } +{{\varsigma }_{1}}^{2}N+{{\varsigma }_{2}}^{2} \bar{{{\beta }_{1}}}{{P}_{1}} \\
\hspace{1.0cm} +2{{\varsigma }_{2}}{{\varsigma }_{3}}\sqrt{\bar{{\beta}}_{1}\bar{{\beta}}_{2}{{P}_{1}}{{P}_{2}}}\tilde{\rho }+{{\varsigma }_{3}}^{2} \bar{{{\beta }_{2}}}{{P}_{2}} +\kappa \epsilon  \Bigr),
\end{IEEEeqnarray*}

where in (a) we have used that $(\textbf{s}_1,\textbf{s}_2,{\cal C}_1,{\cal C}_2,{\cal C}_c,\textbf{z})\in {\cal E}_{\textbf{X}}^{c}$, and Statement A) and Statement A1).

\medskip

E) For every$(\textbf{s}_1,\textbf{s}_2,{\cal C}_1,{\cal C}_2,{\cal C}_c,\textbf{z})\in {\cal E}_{\textbf{X}}^{c}\cap{\cal E}_{\textbf{Z}}^{c}$ and an arbitrary $\textbf{v}\in {\cal S}_{c}$,
\begin{IEEEeqnarray*}{rCl}
&& \left| \bar{\rho }-\cos \sphericalangle \left( \textbf{v},\textbf{u}_{2}^{*} \right) \right|\le 7\epsilon \text{  and  }\left| \cos \sphericalangle \left( \textbf{v},\textbf{u}_{1}^{*} \right) \right|\le 7\epsilon \text{  and} \\
&& {{\left\| \textbf{y}-\left( {{a}_{1,1}}\textbf{u}_{1}^{*}+{{a}_{2,1}}\textbf{u}_{2}^{*}+\alpha\textbf{v} \right) \right\|}^{2}}\le {{\left\| \textbf{y}-\left( {{a}_{1,1}}\textbf{u}_{1}^{*}+{{a}_{2,1}}\textbf{u}_{2}^{*}+\alpha{{\textbf{v}}^{*}} \right) \right\|}^{2}} \\
&& \text{        }\implies \text{  }{{\left\| \alpha\textbf{v}-\textbf{w} \right\|}^{2}}\le \Upsilon (\epsilon ),
\end{IEEEeqnarray*}
where
\begin{IEEEeqnarray}{rCl}\label{eq:upsilon_Rc}
  \Upsilon (\epsilon )& = & n\frac{{{\sigma }^{2}}{{\alpha}^{2}}\left( 1-{{2}^{-2{{R}_\textnormal{c}}}} \right){{2}^{-2{{R}_{1}}}}\left\{ 1-{{{\tilde{\rho }}}^{2}}-{{\bar{\rho }}^{2}} \right\}N}{\left[ {{\sigma }^{2}}{{\alpha}^{2}}\left( 1-{{2}^{-2{{R}_\textnormal{c}}}} \right){{2}^{-2{{R}_{1}}}}\left[ 1-{{{\tilde{\rho }}}^{2}}-{{\bar{\rho }}^{2}} \right]+N\left[ 1-{{{\tilde{\rho }}}^{2}} \right] \right]} \nonumber \\
 && +n{{\alpha}^{2}}{{N}^{2}}{{\bar{\rho }}^{2}}\frac{\left( {{\bar{\rho }}^{2}}\bar{{{\beta }_{1}}}{{P}_{1}}-{{\sigma }^{2}}{{{\tilde{\rho }}}^{2}}\left( 1-{{2}^{-2{{R}_\textnormal{c}}}} \right){{2}^{-2{{R}_{1}}}} \right)}{{{\left[ {{\sigma }^{2}}{{\alpha}^{2}}\left( 1-{{2}^{-2{{R}_\textnormal{c}}}} \right){{2}^{-2{{R}_{1}}}}\left[ 1-{{{\tilde{\rho }}}^{2}}-{{\bar{\rho }}^{2}} \right]+N\left[ 1-{{{\tilde{\rho }}}^{2}} \right] \right]}^{2}}}+n{{\kappa}'}\epsilon,
\end{IEEEeqnarray}
and where ${{\kappa}'}$ depends only on ${{P}_{2}},{{N}_{1}},{{N}_{2}},{{\varsigma }_{1}},{{\varsigma }_{2}}$  and $ {{\varsigma }_{3}}$.

Statement E) follows from combining Statement C) with Statement
D) and the explicit values of ${{\varsigma }_{1}},{{\varsigma }_{2}}$  and $ {{\varsigma }_{3}}$ given in
 \eqref{eq:w_coeff_for_Rc}.

\medskip

F) For every $\textbf{v}\in {\cal S}_{c}$, denote by $\varphi \in \left[ 0,\pi  \right]$ the angle between \textbf{v} and \textbf{w}, and let
\begin{equation*}
{\cal B}(\textbf{s}_1,\textbf{s}_2,\textbf{u}_1^{*},\textbf{v}^{*},\textbf{u}_2^{*},\textbf{z})\triangleq \left\{ \textbf{v}\in S_{c}^{\left( n \right)}:\ \ \cos \left( \varphi  \right)\ge \sqrt{1-\frac{\Upsilon (\epsilon )}{n{{\alpha}^{2}}{{\sigma }^{2}}{{2}^{-2{{R}_{1}}}}\left( 1-{{2}^{-2{{R}_\textnormal{c}}}} \right)}} \right\},
\end{equation*}
where $\epsilon$ is sufficiently large such that the term inside the square is non-negative. \\
Then, for every $(\textbf{s}_1,\textbf{s}_2,{\cal C}_1,{\cal C}_2,{\cal C}_c,\textbf{z})\in {\cal E}_{\textbf{X}}^{c}\cap
{\cal E}_{\textbf{Z}}^{c}$,
\begin{IEEEeqnarray}{rCl}\label{eq:statement_F_Rc}
&& \left| \bar{\rho }-\cos \sphericalangle \left( \textbf{v},\textbf{u}_{2}^{*} \right) \right|\le 7\epsilon
\quad \mbox{  and  }\quad \left| \cos \sphericalangle \left( \textbf{v},\textbf{u}_{1}^{*} \right) \right|\le 7\epsilon \quad \mbox{  and} \nonumber \\
&& {{\left\| \textbf{y}-\left( {{a}_{1,1}}\textbf{u}_{1}^{*}+{{a}_{2,1}}\textbf{u}_{2}^{*}+\alpha\textbf{v} \right) \right\|}^{2}}\le {{\left\| \textbf{y}-\left( {{a}_{1,1}}\textbf{u}_{1}^{*}+{{a}_{2,1}}\textbf{u}_{2}^{*}+\left( {{a}_{1,2}}+{{a}_{2,2}} \right)
{{\textbf{v}}^{*}} \right) \right\|}^{2}} \nonumber \\
&& \qquad \implies \textbf{v}\in {\cal B}(\textbf{s}_1,\textbf{s}_2,\textbf{u}_1^{*},\textbf{v}^{*},\textbf{u}_2^{*},\textbf{z}).
\end{IEEEeqnarray}

Statement F) follows from Statement E) by noting that if $\mathbf{w}\ne 0$ and
$$1-\frac{\Upsilon (\epsilon )}{n{{\alpha}^{2}}{{\sigma }^{2}}{{2}^{-2{{R}_{1}}}}\left( 1-{{2}^{-2{{R}_\textnormal{c}}}} \right)}>0,$$
then
\begin{IEEEeqnarray*}{l}
\left.
\begin{array}{l}
{{\left\| \alpha\textbf{v} \right\|}^{2}}=n{{\alpha}^{2}}{{\sigma }^{2}}{{2}^{-2{{R}_{1}}}}\left( 1-{{2}^{-2{{R}_\textnormal{c}}}} \right)\\
 {{\left\| \alpha\textbf{v}-\textbf{w} \right\|}^{2}}\le \Upsilon (\epsilon ) \\
\end{array}
\right\} \nonumber \\
 \implies \cos \sphericalangle \left( \textbf{u},\textbf{w} \right)\ge \sqrt{1-\frac{\Upsilon (\epsilon )}{n{{\alpha}^{2}}{{\sigma }^{2}}{{2}^{-2{{R}_{1}}}}\left( 1-{{2}^{-2{{R}_\textnormal{c}}}} \right)}} ,
\end{IEEEeqnarray*}
which follows by the same argument as \eqref{eq:statement_F_proof_R1}.

\medskip

The proof of Lemma~\ref{eps_h_v} is now concluded by noticing that the set
${\cal E}'_{{{{\hat{\textbf{V}}}}}}$ , defined in \eqref{eq:eps_h_v},
is the set of tuples
$(\textbf{s}_1,\textbf{s}_2,{\cal C}_1,{\cal C}_2,{\cal C}_c,\textbf{z})$
for which there exists a $\textbf{v}\left( j \right)\in {{{\cal C}}_{c}}\backslash \left\{ {{\textbf{v}}^{*}} \right\}$
such that
$\textbf{v}\left( j \right)\in
{\cal B}(\textbf{s}_1,\textbf{s}_2,\textbf{u}_1^{*},\textbf{v}^{*},\textbf{u}_2^{*},\textbf{z})$.
Thus, by Statement F) and by the definition of ${{\cal E}_{{{{\hat{\textbf{V}}}}}}}$
in \eqref{eq:eps_v} it follows that
\begin{equation*}
{{\cal E}_{{\hat{\textbf{V}}}}}\cap {\cal E}_{\textbf{S}}^{c}\cap {\cal E}_{\textbf{X}}^{c}\cap {\cal E}_{\textbf{Z}}^{c}\subseteq
{{\cal E}'}_{{\hat{\textbf{V}}}}\cap {\cal E}_{\textbf{S}}^{c}\cap {\cal E}_{\textbf{X}}^{c}\cap {\cal E}_{\textbf{Z}}^{c},
\end{equation*}
and therefore
\begin{flalign*}
&&
\Pr\left[ {{\cal E}_{{\hat{\textbf{V}}}}}\cap {\cal E}_{\textbf{S}}^{c}\cap {\cal E}_{\textbf{X}}^{c}\cap {\cal E}_{\textbf{Z}}^{c} \right]
 \le \Pr\left[ {{\cal E}'}_{{\hat{\textbf{V}}}}\cap {\cal E}_{\textbf{S}}^{c}\cap {\cal E}_{\textbf{X}}^{c}\cap{\cal E}_{\textbf{V}}^{c} \right].
&&
\end{flalign*}
\end{IEEEproof}

Next,
\begin{IEEEeqnarray}{rCl}\label{eq:Rc_proof}
   \ \Pr\left[ {{\cal E}_{{\hat{\textbf{V}}}}}\cap {\cal E}_{\textbf{S}}^{c}\cap {\cal E}_{\textbf{X}}^{c}\cap
  {\cal E}_{\textbf{Z}}^{c} \right]
&\overset{(a)}{\mathop{\le }}&\,\Pr\left[ {{\cal E}'}_{{\hat{\textbf{V}}}}\cap
  {\cal E}_{\textbf{S}}^{c}\cap {\cal E}_{\textbf{X}}^{c}\cap {\cal E}_{\textbf{Z}}^{c} \right] \nonumber \\
 & \overset{(b)}{\mathop{\le }}&\,\Pr\left[ {{\cal E}'}_{{\hat{\textbf{V}}}}|{\cal E}_{{{\textbf{X}}_{v}}}^{c} \right],
\end{IEEEeqnarray}
where (a) follows by Lemma~\ref{eps_h_v} and (b) follows because ${\cal E}_{\textbf{X}}^{c}\subseteq {\cal E}_{{{\textbf{X}}_{v}}}^{c}$.
The proof of \eqref{eq:Rc} is now completed by combining \eqref{eq:Rc_proof} with Lemma~\ref{Rate proof}.
This gives that for every $\delta >0$ and every $\epsilon >0$ there exists some
$n'(\delta ,\epsilon )$
such that for all $n>n'(\delta ,\epsilon )$,
we have
\begin{equation*}
\ \Pr\left[ {{\cal E}_{{\hat{\textbf{V}}}}}\cap {\cal E}_{\textbf{S}}^{c}\cap {\cal E}_{\textbf{X}}^{c}\cap {\cal E}_{\textbf{Z}}^{c} \right]\le \Pr\left[ {{\cal E}'}_{{\hat{\textbf{V}}}}|{\cal E}_{{{\textbf{X}}_{V}}}^{c} \right]<\delta ,
\end{equation*}
whenever
\begin{equation*}
{{R}_\textnormal{c}}<\frac{1}{2}\log \left( \frac{{{\eta }^{2}}\left( 1-{{{\tilde{\rho }}}^{2}}-{{\bar{\rho }}^{2}} \right)+N\left( 1-{{{\tilde{\rho }}}^{2}} \right)}{N\left( 1-{{{\tilde{\rho }}}^{2}}-{{\bar{\rho }}^{2}} \right)+{{\lambda }_{c}}}-{{\kappa}_{3}}\epsilon  \right),
\end{equation*}
where ${{\kappa}_{3}}$ is a positive constant determined by ${{P}_{1}},{{P}_{2}},N,{{\varsigma }_{1}},{{\varsigma }_{2}}\text{ and }{{\varsigma }_{3}}$. \hfill\qed

\vskip.2truein

\subsection{Proof of rate constraint \eqref{eq:R1+R2}}

Define
\begin{equation*}
\textbf{w}(\textbf{s}_1,\textbf{s}_2,{\cal C}_1,{\cal C}_2,{\cal C}_c,\textbf{z})=
{{\varsigma }_{1}}\left( \textbf{y}- \alpha{{\textbf{v}}^{*}} \right)
+{{\varsigma }_{2}}\alpha{{\textbf{v}}^{*}},
\end{equation*}
where
\begin{IEEEeqnarray}{l}
{{\varsigma }_{1}} = \frac{{a}_{1,1}^{2}\left( 1-{{2}^{-2{{R}_{1}}}} \right)+2{{a}_{1,1}}{{a}_{2,1}}\rho \left( 1-{{2}^{-2{{R}_{1}}}} \right)\left( 1-{{2}^{-2{{R}_{2}}}} \right)+{a}_{2,1}^{2}\left( 1-{{2}^{-2{{R}_{2}}}} \right)\left( 1-{{\bar{\rho }}^{2}} \right)}{{a}_{1,1}^{2}\left( 1-{{2}^{-2{{R}_{1}}}} \right)+2{{a}_{1,1}}{{a}_{2,1}}\rho \left( 1-{{2}^{-2{{R}_{1}}}} \right)\left( 1-{{2}^{-2{{R}_{2}}}} \right)+{a}_{2,1}^{2}\left( 1-{{2}^{-2{{R}_{2}}}} \right)\left( 1-{{\bar{\rho }}^{2}} \right)+\frac{N}{{{\sigma }^{2}}}} \nonumber \\
{{\varsigma }_{2}} =  \nonumber \\
\frac{\frac{N}{{{\sigma }^{2}}}{{a}_{2,1}}\rho \left( 1-{{2}^{-2{{R}_{2}}}} \right)}{\alpha\left( {a}_{1,1}^{2}\left( 1-{{2}^{-2{{R}_{1}}}} \right)+2{{a}_{1,1}}{{a}_{2,1}}\rho \left( 1-{{2}^{-2{{R}_{1}}}} \right)\left( 1-{{2}^{-2{{R}_{2}}}} \right)+{a}_{2,1}^{2}\left( 1-{{2}^{-2{{R}_{2}}}} \right)\left( 1-{{\bar{\rho }}^{2}} \right)+\frac{N}{{{\sigma }^{2}}} \right)}.
\nonumber \\
\label{eq:w_for_R1+R2}
\end{IEEEeqnarray}

In the remainder we shall use the shorthand notation \textbf{w} instead of
$\textbf{w}(\textbf{s}_1,\textbf{s}_2,{\cal C}_1,{\cal C}_2,{\cal C}_c,\textbf{z})$.
We now start with a lemma that will be used to prove \eqref{eq:R1+R2}.

\medskip

\begin{lemma}\label{eps_h_u1,u2}
Let ${{\varphi }_{j,l}}\in \left[ 0,\pi  \right]$  be the angle between $\mathbf{w}$ and ${a}_{1,1}\mathbf{u}_{1}(j)+{a}_{2,1}\mathbf{u}_{2}(l)$, and let the set
${\cal E}'_{({\hat{\mathbf{U}}}_{1},{\hat{\mathbf{U}}}_{2})}$
be defined as
\begin{IEEEeqnarray}{l}
{\cal E}'_{({\hat{\mathbf{U}}}_{1},{\hat{\mathbf{U}}}_{2})}
\triangleq
\biggl\{ (\mathbf{s}_1,\mathbf{s}_2,{\cal C}_1,{\cal C}_2,{\cal C}_c,\mathbf{z})\colon \exists \> {{\mathbf{u}}_{1}}\left( j \right)\in {{\cal C}_{1}}\backslash \left\{ \mathbf{u}_{1}^{*} \right\}
\mbox{ and } \exists \> {{\mathbf{u}}_{2}}\left( j \right)\in {{{\cal C}}_{2}}\backslash \left\{ \mathbf{u}_{2}^{*} \right\}  \nonumber \\
\hspace{3.5cm} \mbox{s.t.} \  \cos \left( {{\varphi }_{j,l}} \right)\ge \sqrt{1-\tilde{\Upsilon}-\kappa^{''}\epsilon} \biggr\}
\label{eq:eps_h_u1,u2}
\end{IEEEeqnarray}
where
\begin{IEEEeqnarray*}{l}
\tilde{\Upsilon}\triangleq
\frac{\left( \bar{{{\beta }_{1}}}{{P}_{1}}+2\sqrt{\bar{{{\beta }_{1}}}\bar{{{\beta }_{2}}}{{P}_{1}}{{P}_{2}}}\tilde{\rho }+\bar{{{\beta }_{2}}}{{P}_{2}}\left( 1-{{\bar{\rho }}^{2}} \right) \right)N}{\left( \bar{{{\beta }_{1}}}{{P}_{1}}+2\sqrt{\bar{{{\beta }_{1}}}{{P}_{1}}\bar{{{\beta }_{2}}}{{P}_{2}}}\tilde{\rho }+\bar{{{\beta }_{2}}}{{P}_{2}}\left[ 1-{{\bar{\rho }}^{2}} \right]+N \right)\left( \bar{{{\beta }_{1}}}{{P}_{1}}+2\tilde{\rho }\sqrt{\bar{{{\beta }_{1}}}{{P}_{1}}\bar{{{\beta }_{2}}}{{P}_{2}}}+\bar{{{\beta }_{2}}}{{P}_{2}} \right)} ,
\end{IEEEeqnarray*}
and ${{\kappa}^{''}}$ is a positive constant determined by $P_{1}$, $P_{2}$, $N$,
${{\varsigma }_{1}}$ and ${{\varsigma }_{2}}$.
Then,
\begin{equation*}
{{{\cal E}}}_{({\hat{\mathbf{U}}}_{1},{\hat{\mathbf{U}}}_{2})}\cap {\cal E}_{\mathbf{S}}^{c}\cap {\cal E}_{\mathbf{X}}^{c}\cap {\cal E}_{\mathbf{Z}}^{c}\subseteq {\cal E}'_{({\hat{\mathbf{U}}}_{1},{\hat{\mathbf{U}}}_{2})}\cap {\cal E}_{\mathbf{S}}^{c}\cap {\cal E}_{\mathbf{X}}^{c}\cap {\cal E}_{\mathbf{Z}}^{c},
\end{equation*}
and, in particular
\begin{equation*}
\Pr\left[ {{{\cal E}}}_{({\hat{\mathbf{U}}}_{1},{\hat{\mathbf{U}}}_{2})}\cap {\cal E}_{\mathbf{S}}^{c}\cap {\cal E}_{\mathbf{X}}^{c}\cap {\cal E}_{\mathbf{Z}}^{c} \right]\le \Pr\left[ {\cal E}'_{({\hat{\mathbf{U}}}_{1},{\hat{\mathbf{U}}}_{2})}\cap {\cal E}_{\mathbf{S}}^{c}\cap {\cal E}_{\mathbf{X}}^{c}\cap {\cal E}_{\mathbf{Z}}^{c} \right].
\end{equation*}
\end{lemma}

\medskip

\begin{IEEEproof}
We first recall that, for the event ${{{\cal E}}}_{({\hat{\textbf{U}}}_{1},{\hat{\textbf{U}}}_{2})}$
to occur, there must exist codewords ${{\textbf{u}}_{1}}\left( j \right)\in {{{\cal C}}_{1}}\backslash \left\{ \textbf{u}_{1}^{*} \right\}  \mbox{ and } {{\textbf{u}}_{2}}\left( l \right)\in {{{\cal C}}_{2}}\backslash \left\{ \textbf{u}_{2}^{*} \right\}$
that satisfy the following four conditions
\begin{IEEEeqnarray}{rCl}
\left| \tilde{\rho }-\cos \sphericalangle \left( {{\textbf{u}}_{1}}\left( j \right),\textbf{u}_{2}\left( l \right) \right) \right| &\le & 7\epsilon \\
\left| \bar{\rho }-\cos \sphericalangle \left( {{\textbf{v}^{*}}},\textbf{u}_{2}\left( l \right) \right) \right| & \le & 7\epsilon \\
\left| \cos \sphericalangle \left( {{\textbf{v}}^{*}},{{\textbf{u}}_{1}}\left( j \right) \right) \right| & \le &  3\epsilon  \\
\|\textbf{y}-\textbf{X}_{\textbf{u}_{1}\left(j\right),\textbf{v}^{*},\textbf{u}_2\left( l \right)}\|^2
&  \leq &
\|\textbf{y}-\textbf{X}_{\textbf{u}_1^{*},\textbf{v}^{*},\textbf{u}_2^{*}}\|^2.
\end{IEEEeqnarray}

The proof is now based on a sequence of statements related to these conditions:

\medskip

A) For every $(\textbf{s}_1,\textbf{s}_2,{\cal C}_1,{\cal C}_2,{\cal C}_c,\textbf{z})\in {\cal E}_{\textbf{X}}^{c}$ and every $\textbf{u}_1\in {\cal S}_{1}\mbox{ and } \textbf{u}_2\in {\cal S}_{2} $,
\begin{IEEEeqnarray}{l}
\left| \tilde{\rho }-\cos \sphericalangle \left( \textbf{u}_{1},\textbf{u}_{2} \right) \right|\le 7\epsilon \text{  }
\implies \text{   }\left| n\tilde{\rho }\sqrt{\bar{{\beta}}_{1}\bar{{\beta}}_{2}{{P}_{1}}{{P}_{2}}}-\left\langle {{a}_{1,1}}\textbf{u}_{1},{{a}_{2,1}}\textbf{u}_{2} \right\rangle  \right|\le 7n\sqrt{\bar{{\beta}}_{1}\bar{{\beta}}_{2}{{P}_{1}}{{P}_{2}}}\epsilon .\nonumber \\
\label{eq:statement_A_R1+R2}
\end{IEEEeqnarray}

Statement A) follows by rewriting $\cos \sphericalangle \left( \textbf{u}_{1},\textbf{u}_{2} \right)$ as ${\left\langle \textbf{u}_{1},\textbf{u}_{2} \right\rangle }/({\left\| \textbf{u}_1 \right\|\left\| \textbf{u}_{2} \right\|})$, and then multiplying the inequality on the l.h.s. of
\eqref{eq:statement_A_R1+R2}  by $\left\| {{a}_{1,1}}\textbf{u}_1 \right\|\cdot \left\| {{a}_{2,1}}\textbf{u}_{2} \right\|$
and recalling that $\left\| {{a}_{1,1}}\textbf{u}_1 \right\|=\sqrt{n\bar{{{\beta }_{1}}}{{P}_{1}}}\ $ and that $\left\| {{a}_{2,1}}\textbf{u}_{2} \right\|=\sqrt{n\bar{{{\beta }_{2}}}{{P}_{2}}} $.

\medskip

A1) For every $(\textbf{s}_1,\textbf{s}_2,{\cal C}_1,{\cal C}_2,{\cal C}_c,\textbf{z})\in {\cal E}_{\textbf{X}}^{c}$,
\begin{IEEEeqnarray}{rCl} \label{eq:statement_A1_R1+R2}
&& \left| \cos \sphericalangle \left( \textbf{v}^{*},\textbf{u}_{1} \right) \right|\le 3\epsilon  \nonumber \\
&& \text{       }\implies \text{      }\left| \left\langle \alpha\textbf{v}^{*},{{a}_{1,1}}\textbf{u}_{1} \right\rangle  \right|\le 3\left\| \alpha\textbf{v}^{*} \right\|\sqrt{n\bar{{{\beta }_{1}}}{{P}_{1}}}\epsilon .
\end{IEEEeqnarray}

Statement A1) follows by rewriting $\cos \sphericalangle \left( \textbf{v}^{*},\textbf{u}_{1}^{*} \right)$ as ${\left\langle \textbf{v}^{*},\textbf{u}_{1} \right\rangle }/({\left\| \textbf{v}^{*} \right\|\left\| \textbf{u}_{1} \right\|})$, and then multiplying the inequality on the l.h.s. of \eqref{eq:statement_A1_R1+R2} by
$\left\| \alpha\textbf{v}^{*} \right\|\cdot \left\| {{a}_{1,1}}\textbf{u}_{1} \right\|$.

\medskip

A2) For every $(\textbf{s}_1,\textbf{s}_2,{\cal C}_1,{\cal C}_2,{\cal C}_c,\textbf{z})\in {\cal E}_{\textbf{X}}^{c}$ and every $\textbf{u}_2\in {\cal S}_{2} $,
\begin{IEEEeqnarray}{rCl} \label{eq:statement_A2_R1+R2}
&& \left| \bar{\rho }-\cos \sphericalangle \left( \textbf{v}^{*},\textbf{u}_{2} \right) \right|\le 7\epsilon  \nonumber \\
&& \text{       }\implies \text{      }\left| \left\| \alpha\textbf{v}^{*} \right\|\sqrt{n\bar{{{\beta }_{2}}}{{P}_{2}}}\bar{\rho }-\left\langle \alpha\textbf{v}^{*},{{a}_{2,1}}\textbf{u}_{2} \right\rangle  \right|\le 7\epsilon \left\| \alpha\textbf{v}^{*} \right\|\sqrt{n\bar{{{\beta }_{2}}}{{P}_{2}}}.
\end{IEEEeqnarray}

Statement A2) follows by rewriting $\cos \sphericalangle \left( \textbf{v}^{*},\textbf{u}_{2} \right)$ as ${\left\langle \textbf{v}^{*},\textbf{u}_{2} \right\rangle }/({\left\| \textbf{v}^{*} \right\|\left\| \textbf{u}_{2} \right\|})$, and then multiplying the inequality on the l.h.s. of \eqref{eq:statement_A2_R1+R2} by
$\left\| \alpha\textbf{v}^{*} \right\|\cdot \left\| {{a}_{2,1}}\textbf{u}_{2} \right\|$.

\medskip

B) For every $(\textbf{s}_1,\textbf{s}_2,{\cal C}_1,{\cal C}_2,{\cal C}_c,\textbf{z})\in {\cal E}_{\textbf{X}}^{c}\cap {\cal E}_{\textbf{Z}}^{c}$
and every $\textbf{u}_1\in {\cal S}_{1}\mbox{ and } \textbf{u}_2\in {\cal S}_{2}$,
\begin{IEEEeqnarray}{rCl} \label{eq:statement_B_R1+R2}
&& |\textbf{y}-\textbf{X}_{\textbf{u}_1,\textbf{v}^{*},\textbf{u}_2}\|^2
\leq
\|\textbf{y}-\textbf{X}_{\textbf{\textbf{u}}_1^{*},\textbf{v}^{*},\textbf{\textbf{u}}_2^{*}}\|^2 \nonumber \\
&& \text{                                 }\implies \left\langle \textbf{y}- \alpha{{\textbf{v}}^{*}} ,{{a}_{1,1}}\textbf{u}_1+{{a}_{2,1}}\textbf{u}_2 \right\rangle
\ge n\left( \bar{{{\beta }_{1}}}{{P}_{1}}
+2\sqrt{\bar{{\beta}}_{1}\bar{{\beta}}_{2}{P}_{1}{P}_{2}}(\widetilde{\rho}-7\epsilon)
+\bar{{{\beta }_{2}}}{{P}_{2}}-\kappa \epsilon  \right).
\nonumber\\*
\end{IEEEeqnarray}

Statement B) follows from rewriting the inequality on the l.h.s. of \eqref{eq:statement_B_R1+R2}
as
\begin{equation*}
{{\left\|\left( \textbf{y}- \alpha{{\textbf{v}}^{*}} \right)-({{a}_{1,1}}\textbf{u}_1+
{{a}_{2,1}}\textbf{u}_{2}) \right\|}^{2}}\le
 {{\left\|\left( \textbf{y}- \alpha{{\textbf{v}}^{*}} \right)-({{a}_{1,1}}\textbf{u}_{1}^{*}+{{a}_{2,1}}\textbf{u}_{2}^{*}) \right\|}^{2}},
 \end{equation*}
or equivalently as
\begin{IEEEeqnarray}{rCl} \label{eq:statement_B_proof_R1+R2}
\left\langle \left(\textbf{y}- \alpha{{\textbf{v}}^{*}} \right),
{{a}_{1,1}}\textbf{u}_1+{{a}_{2,1}}\textbf{u}_{2} \right\rangle
& \ge &
\left\langle \left( \textbf{y}- \alpha{{\textbf{v}}^{*}} \right),{{a}_{1,1}}\textbf{u}_{1}^{*}+{{a}_{2,1}}\textbf{u}_{2}^{*} \right\rangle  \nonumber \\
& = & \left\langle {{a}_{1,1}}\textbf{u}_{1}^{*}+{{a}_{2,1}}\textbf{u}_{2}^{*}+\textbf{z},
{{a}_{1,1}}\textbf{u}_{1}^{*}+{{a}_{2,1}}\textbf{u}_{2}^{*} \right\rangle  \nonumber \\
& = & {{\left\| {{a}_{1,1}}\textbf{u}_{1}^{*}+{{a}_{2,1}}\textbf{u}_{2}^{*} \right\|}^{2}}
+\left\langle \textbf{z},{{a}_{1,1}}\textbf{u}_{1}^{*}+{{a}_{2,1}}\textbf{u}_{2}^{*} \right\rangle .
\end{IEEEeqnarray}
It now follows from the equivalence of the first inequality in \eqref{eq:statement_B_R1+R2} with
\eqref{eq:statement_B_proof_R1+R2} that for \\
$(\textbf{s}_1,\textbf{s}_2,{\cal C}_1,{\cal C}_2,{\cal C}_c,\textbf{z})\in {\cal E}_{\textbf{Z}}^{c}$, the first inequality in \eqref{eq:statement_B_R1+R2} can only hold if
\begin{equation*}
\left\langle \textbf{y}- \alpha{{\textbf{v}}^{*}},
{{a}_{1,1}}\textbf{u}_1+{{a}_{2,1}}\textbf{u}_{2} \right\rangle \ge
n\left( \bar{{{\beta }_{1}}}{{P}_{1}}
 +2\sqrt{\bar{{\beta}}_{1}\bar{{\beta}}_{2}{P}_{1}{P}_{2}}(\widetilde{\rho}-7\epsilon)
 +\bar{{{\beta }_{2}}}{{P}_{2}}-\kappa \epsilon  \right),
 \end{equation*}
thus establishing B).

\medskip

C) For every $(\textbf{s}_1,\textbf{s}_2,{\cal C}_1,{\cal C}_2,{\cal C}_c,\textbf{z})\in {\cal E}_{\textbf{X}}^{c}\cap {\cal E}_{\textbf{Z}}^{c}$
and every $\textbf{u}_1\in {\cal S}_{1}\mbox{ and } \textbf{u}_2\in {\cal S}_{2} $,
\begin{IEEEeqnarray*}{rCl}
&& \left| \tilde{\rho }-\cos \sphericalangle \left( \textbf{u}_1,\textbf{u}_{2} \right) \right|\le 7\epsilon
\ \mbox{  and  } \
\left| \bar{\rho }-\cos \sphericalangle \left( \textbf{u}_2,\textbf{v}^{*} \right) \right|\le 7\epsilon
\  \mbox{  and  } \
\left| \cos \sphericalangle \left( \textbf{u}_1,\textbf{v}^{*} \right) \right|\le 7\epsilon \\
&& \mbox{  and  } \
|\textbf{y}-\textbf{X}_{\textbf{u}_1,\textbf{v}^{*},\textbf{u}_2}\|^2
\leq
\|\textbf{y}-\textbf{X}_{\textbf{u}_1^{*},\textbf{v}^{*},\textbf{u}_2^{*}}\|^2 \nonumber \\
&& \qquad \qquad \implies  \\
&& {{\left\| {{a}_{1,1}}\textbf{u}_1+{{a}_{2,1}}\textbf{u}_2-\textbf{w} \right\|}^{2}}
\le
n\Biggl( \left( \bar{{\beta }}_{1}{{P}_{1}}+2\sqrt{\bar{{\beta}}_{1}\bar{{\beta }}_{2}{{P}_{1}}{{P}_{2}}}\tilde{\rho }+\bar{{\beta}}_{2}{{P}_{2}} \right)\left( 1-2{{\varsigma }_{1}} \right)\\
&& \hspace{4.2cm}  -2 {{\varsigma }_{2}} \frac{1}{n}\left\| \alpha{{v}^{*}} \right\|\sqrt{n\bar{{\beta }}_{2}{{P}_{2}}}\bar{\rho }  \Biggr)+{{\left\| \textbf{w} \right\|}^{2}}+n{{k}'}\epsilon .
\end{IEEEeqnarray*}

Statement C) is obtained as follows:
\begin{IEEEeqnarray*}{rCl}
\IEEEeqnarraymulticol{3}{l}{
{{\left\| {{a}_{1,1}}\textbf{u}_1+{{a}_{2,1}}\textbf{u}_2-\textbf{w} \right\|}^{2}}
={{\left\| {{a}_{1,1}}\textbf{u}_1+{{a}_{2,1}}\textbf{u}_2 \right\|}^{2}}
-2\left\langle {{a}_{1,1}}\textbf{u}_1+{{a}_{2,1}}\textbf{u}_2,\textbf{w} \right\rangle +{{\left\| \textbf{w} \right\|}^{2}} }\\ \quad
&=&{{\left\| {{a}_{1,1}}\textbf{u}_1+{{a}_{2,1}}\textbf{u}_2 \right\|}^{2}}
-2\left\langle {{a}_{1,1}}\textbf{u}_1+{{a}_{2,1}}\textbf{u}_2,
{{\varsigma }_{1}}\left( {{a}_{1,1}}\textbf{u}_{1}^{*}+{{a}_{2,1}}\textbf{u}_{2}^{*}+\textbf{z} \right)
+{{\varsigma }_{2}}\alpha{{\textbf{v}}^{*}} \right\rangle +{{\left\| \textbf{w} \right\|}^{2}} \\
& \text{ } \overset{(a)}{\mathop{\le }}&\,n
\Biggl( \left( \bar{{\beta}}_{1}{{P}_{1}}+2\sqrt{\bar{{\beta }}_{1}\bar{{\beta}}_{2}{{P}_{1}}{{P}_{2}}}\tilde{\rho }+\bar{{\beta}}_{2}{{P}_{2}} \right)\left( 1-2{{\varsigma }_{1}} \right)-2 {{\varsigma }_{2}} \frac{1}{n}\left\| \alpha{{v}^{*}} \right\|\sqrt{n\bar{{\beta}}_{2}{{P}_{2}}}\bar{\rho } \Biggr)+{{\left\| \textbf{w} \right\|}^{2}} \nonumber \\
&& +n{{\kappa}'}\epsilon ,
\end{IEEEeqnarray*}
where in (a) we have used Statement A), A1), A2) and Statement B).

\medskip

D) For every $(\textbf{s}_1,\textbf{s}_2,{\cal C}_1,{\cal C}_2,{\cal C}_c,\textbf{z})\in {\cal E}_{\textbf{X}}^{c}\cap {\cal E}_{\textbf{Z}}^{c}$
\begin{IEEEeqnarray*}{rCl}
{{\left\| \textbf{w} \right\|}^{2}}
&\le &\, n\Biggl( {{\varsigma }_{1}}^{2}\left( \bar{{{\beta }_{1}}}{{P}_{1}}+2\sqrt{\bar{{\beta}}_{1}\bar{{\beta}}_{2}{{P}_{1}}{{P}_{2}}}\tilde{\rho}
+\bar{{{\beta }_{2}}}{{P}_{2}}+N \right)
+\frac{1}{n}2{{\varsigma }_{1}}{{\varsigma }_{2}}{\alpha}{\left\| {{\textbf{v}}^{*}} \right\|}\sqrt{n\bar{{{\beta }_{2}}}{{P}_{2}}}\bar{\rho }\\
&\quad& +\frac{1}{n}{{\varsigma }_{2}}^{2}{{\alpha}^{2}}{{\left\| {{\textbf{v}}^{*}} \right\|}^{2}}+k\epsilon  \Biggr),
\end{IEEEeqnarray*}
where $k$ depends on ${{P}_{1}},{{P}_{2}},N,{{\varsigma }_{1}}$ and ${{\varsigma }_{2}}$ only.

Statement D) is obtained as follows:
\begin{IEEEeqnarray*}{rCl}
{{\left\| \textbf{w} \right\|}^{2}}
&=&{{\left\| {{\varsigma }_{1}}\left( {{a}_{1,1}}\textbf{u}_{1}^{*}+{{a}_{2,1}}\textbf{u}_{2}^{*}+\textbf{z} \right)
+{{\varsigma }_{2}}\alpha{{\textbf{v}}^{*}} \right\|}^{2}}\\
&=&{{\varsigma }_{1}}^{2}{{\left\| {{a}_{1,1}}\textbf{u}_{1}^{*}+{{a}_{2,1}}\textbf{u}_{2}^{*}+\textbf{z} \right\|}^{2}}
+2{{\varsigma }_{1}}{{\varsigma }_{2}}\left\langle
{{a}_{1,1}}\textbf{u}_{1}^{*}+{{a}_{2,1}}\textbf{u}_{2}^{*}+\textbf{z},\alpha{{\textbf{v}}^{*}} \right\rangle
+{{\varsigma }_{2}}^{2}{{\left\| \alpha{{\textbf{v}}^{*}} \right\|}^{2}} \\
&=&{{\varsigma }_{1}}^{2}\left( {{\left\| {{a}_{1,1}}\textbf{u}_{1}^{*}+{{a}_{2,1}}\textbf{u}_{2}^{*} \right\|}^{2}}
+2\left\langle {{a}_{1,1}}\textbf{u}_{1}^{*}+{{a}_{2,1}}\textbf{u}_{2}^{*},\textbf{z} \right\rangle +{{\left\| \textbf{z} \right\|}^{2}} \right)\\
&& +2{{\varsigma }_{1}}{{\varsigma }_{2}}\left\langle {{a}_{1,1}}\textbf{u}_{1}^{*}+{{a}_{2,1}}\textbf{u}_{2}^{*}
+\textbf{z},\alpha{{\textbf{v}}^{*}} \right\rangle+{{\varsigma }_{2}}^{2}{{\alpha}^{2}}{{\left\| {{\textbf{v}}^{*}} \right\|}^{2}} \\
&\text{ } \overset{(a)}{\mathop{\le }}&\,n
\Biggl( {{\varsigma }_{1}}^{2}\left( \bar{{{\beta }_{1}}}{{P}_{1}}+2\sqrt{\bar{{\beta}}_{1}\bar{{\beta}}_{2}{{P}_{1}}{{P}_{2}}}\tilde{\rho}
+\bar{{{\beta }_{2}}}{{P}_{2}}+N \right)
+\frac{1}{n}2{{\varsigma }_{1}}{{\varsigma }_{2}}{\alpha}{\left\| {{\textbf{v}}^{*}} \right\|}\sqrt{n\bar{{{\beta }_{2}}}{{P}_{2}}}\bar{\rho }\\
&& +\frac{1}{n}{{\varsigma }_{2}}^{2}{{\alpha}^{2}}{{\left\| {{\textbf{v}}^{*}} \right\|}^{2}}+\kappa\epsilon  \Biggr),
\end{IEEEeqnarray*}
where in (a) we have used that $(\textbf{s}_1,\textbf{s}_2,{\cal C}_1,{\cal C}_2,{\cal C}_c,\textbf{z})\in {\cal E}_{\textbf{X}}^{c}$, and statements A), A1) and A2).

\medskip

E) For every$(\textbf{s}_1,\textbf{s}_2,{\cal C}_1,{\cal C}_2,{\cal C}_c,\textbf{z})\in {\cal E}_{\textbf{X}}^{c}\cap {\cal E}_{\textbf{Z}}^{c}$ and an arbitrary $\textbf{u}_1\in {\cal S}_{1}\mbox{ and } \textbf{u}_2\in {\cal S}_{2} $,
\begin{IEEEeqnarray*}{rCl}
&& \left| \tilde{\rho }-\cos \sphericalangle \left( \textbf{u}_1,\textbf{u}_{2} \right) \right|\le 7\epsilon \text{  and  }\left| \bar{\rho }-\cos \sphericalangle \left( \textbf{v},\textbf{u}_{2} \right) \right|\le 7\epsilon
\text{  and }
|\textbf{y}-\textbf{X}_{\textbf{u}_1,\textbf{v}^{*},\textbf{u}_2}\|^2
\leq
\|\textbf{y}-\textbf{X}_{\textbf{u}_1^{*},\textbf{v}^{*},\textbf{u}_2^{*}}\|^2 \\
&& \implies \text{        }{{\left\| {{a}_{1,1}}\textbf{u}_1+{{a}_{2,1}}\textbf{u}_2-\textbf{w} \right\|}^{2}}\le \Upsilon (\epsilon ),
\end{IEEEeqnarray*}
where
\begin{equation*}
\Upsilon (\epsilon )=
n\frac{\left( {{\bar{\beta }}_{1}}{{P}_{1}}+2\sqrt{{{\bar{\beta }}_{1}}{{\bar{\beta }}_{2}}{{P}_{1}}{{P}_{2}}}\tilde{\rho }+{{\bar{\beta }}_{2}}{{P}_{2}}\left( 1-{{\bar{\rho }}^{2}} \right) \right)N}{{{\bar{\beta }}_{1}}{{P}_{1}}+2\sqrt{{{\bar{\beta }}_{1}}{{\bar{\beta }}_{2}}{{P}_{1}}{{P}_{2}}}\tilde{\rho }+{{\bar{\beta }}_{2}}{{P}_{2}}\left( 1-{{\bar{\rho }}^{2}} \right)+N}+n{{k}'}\epsilon,
\end{equation*}
and where ${{k}'}$ only depends on ${{P}_{1}},{{N}_{1}},{{N}_{2}},{{\varsigma }_{1}}$  and $ {{\varsigma }_{2}}$.

Statement E) follows from combining Statement C) with Statement
D) and the explicit values of ${{\varsigma }_{1}}$  and $ {{\varsigma }_{2}}$ given in \eqref{eq:w_for_R1+R2}.

\medskip

F) For every $\textbf{u}_1\in {\cal S}_{1},\textbf{u}_2\in {\cal S}_{2}$, denote by $\varphi \in \left[ 0,\pi  \right]$ the angle between ${{a}_{1,1}}\textbf{u}_1+{{a}_{2,1}}\textbf{u}_{2}$ and \textbf{w}, and let
\begin{IEEEeqnarray*}{rCl}
{\cal B}(\textbf{s}_1,\textbf{s}_2,\textbf{u}_1^{*},\textbf{v}^{*},\textbf{u}_2^{*},\textbf{z})\triangleq
&& \Biggl\{ \textbf{u}_1\in S_{1}^{\left( n \right)},\textbf{u}_2\in S_{2}^{\left( n \right)}: \nonumber \\
&& \quad \cos \left( \varphi  \right)\ge \sqrt{1-\frac{\Upsilon (\epsilon )}{n\left( \bar{{{\beta }_{1}}}{{P}_{1}}+2\tilde{\rho }\sqrt{\bar{{{\beta }_{1}}}{{P}_{1}}\bar{{{\beta }_{2}}}{{P}_{2}}}+\bar{{{\beta }_{2}}}{{P}_{2}} \right)}}\Biggr\},
\end{IEEEeqnarray*}
where $\epsilon$ is sufficiently small such that the term inside the square is non-negative.
Then, for every $(\textbf{s}_1,\textbf{s}_2,{\cal C}_1,{\cal C}_2,{\cal C}_c,\textbf{z})\in {\cal E}_{\textbf{X}}^{c}\cap {\cal E}_{\textbf{Z}}^{c}$,
\begin{IEEEeqnarray}{rCl}\label{eq:statement_F_R1+R2}
&& \left| \tilde{\rho }-\cos \sphericalangle \left( \textbf{u}_1,\textbf{u}_{2}\right) \right|\le 7\epsilon \
\quad \mbox{  and  }\quad \left| \bar{\rho }-\cos \sphericalangle \left( \textbf{v}^{*},\textbf{u}_{2} \right) \right|\le 7\epsilon
\quad \mbox{  and  }\quad \nonumber \\
&& \left| \cos \sphericalangle \left( \textbf{v}^{*},\textbf{u}_{1} \right) \right|\le 3\epsilon
\quad \mbox{  and  }\quad
|\textbf{y}-\textbf{X}_{\textbf{u}_1,\textbf{v}^{*},\textbf{u}_2}\|^2
\leq
\|\textbf{y}-\textbf{X}_{\textbf{u}_1^{*},\textbf{v}^{*},\textbf{u}_2^{*}}\|^2 \nonumber \\
&& \qquad \implies \text{    }{a_{1,1}}\textbf{u}_{1}+{a_{2,1}}\textbf{u}_{2}\in
{\cal B}(\textbf{s}_1,\textbf{s}_2,\textbf{u}_1^{*},\textbf{v}^{*},\textbf{u}_2^{*},\textbf{z}).
\end{IEEEeqnarray}

Statement F) follows from Statement E) by noting that if $\mathbf{w}\ne 0$ and
$$1-\frac{\Upsilon (\epsilon )}{n\left( \bar{{{\beta }_{1}}}{{P}_{1}}+2\tilde{\rho }\sqrt{\bar{{{\beta }_{1}}}{{P}_{1}}\bar{{{\beta }_{2}}}{{P}_{2}}}+\bar{{{\beta }_{2}}}{{P}_{2}} \right)}>0,$$
then
\begin{IEEEeqnarray}{l}
   {{\left\| {{a}_{1,1}}\textbf{u}_1+{{a}_{2,1}}\textbf{u}_2 \right\|}^{2}}=n\left( \bar{{{\beta }_{1}}}{{P}_{1}}+2\tilde{\rho }\sqrt{\bar{{{\beta }_{1}}}{{P}_{1}}\bar{{{\beta }_{2}}}{{P}_{2}}}+\bar{{{\beta }_{2}}}{{P}_{2}} \right)  \nonumber \\
  \mbox{ and  }{{\left\| {{a}_{1,1}}\textbf{u}_1+{{a}_{2,1}}\textbf{u}_2-\textbf{w} \right\|}^{2}}\le \Upsilon (\epsilon ) \nonumber \\
  \quad \implies \cos \sphericalangle \left( {{a}_{1,1}}\textbf{u}_1+{{a}_{2,1}}\textbf{u}_2,\textbf{w} \right)\ge \sqrt{1-\frac{\Upsilon (\epsilon )}{n\left( \bar{{{\beta }_{1}}}{{P}_{1}}+2\tilde{\rho }\sqrt{\bar{{{\beta }_{1}}}{{P}_{1}}\bar{{{\beta }_{2}}}{{P}_{2}}}+\bar{{{\beta }_{2}}}{{P}_{2}} \right)}}.
\label{eq:statement_F_proof_R1+R2}
\end{IEEEeqnarray}
To see this, first note that every ${{a}_{1,1}}\textbf{u}_1+{{a}_{2,1}}\textbf{u}_2$, where $\textbf{u}_1\in {\cal S}_{1},\textbf{u}_2\in {\cal S}_{2}$, satisfying the
condition on the l.h.s. of \eqref{eq:statement_F_R1+R2} lies within a sphere of radius
$\sqrt{\Upsilon (\epsilon )}$ centered at $\mathbf{w}$. In addition, for every $\textbf{u}_1\in {\cal S}_{1},\textbf{u}_2\in {\cal S}_{2}$ we have that ${{a}_{1,1}}\textbf{u}_1+{{a}_{2,1}}\textbf{u}_2$ also lies on the
centered ${{\mathbb{R}}^{n}}$-sphere of radius
$n\left( \bar{{{\beta }_{1}}}{{P}_{1}}+2\tilde{\rho }\sqrt{\bar{{{\beta }_{1}}}{{P}_{1}}\bar{{{\beta }_{2}}}{{P}_{2}}}+\bar{{{\beta }_{2}}}{{P}_{2}} \right)$.
Hence, every $\textbf{u}_1\in {\cal S}_{1},\textbf{u}_2\in {\cal S}_{2}$ satisfying
the condition on the l.h.s. of \eqref{eq:statement_F_R1+R2} lies in the intersection of these two
regions, which is a polar cap on the centered sphere of radius $n\left( \bar{{{\beta }_{1}}}{{P}_{1}}+2\tilde{\rho }\sqrt{\bar{{{\beta }_{1}}}{{P}_{1}}\bar{{{\beta }_{2}}}{{P}_{2}}}+\bar{{{\beta }_{2}}}{{P}_{2}} \right)$.
Hence, every $\textbf{u}_1\in {\cal S}_{1}^{\left( n \right)},\textbf{u}_2\in {\cal S}_{2}^{\left( n \right)}$
satisfying the upper conditions of \eqref{eq:statement_F_R1+R2}, also satisfies
\begin{IEEEeqnarray*}{rCl}
 \cos \varphi & \ge & \sqrt{1-\frac{\Upsilon (\epsilon )}{n\left( \bar{{{\beta }_{1}}}{{P}_{1}}+2\tilde{\rho }\sqrt{\bar{{{\beta }_{1}}}{{P}_{1}}\bar{{{\beta }_{2}}}{{P}_{2}}}+\bar{{{\beta }_{2}}}{{P}_{2}} \right)}} \nonumber \\
& = & \sqrt{1-\tilde{\Upsilon}-\kappa^{''}\epsilon},
\end{IEEEeqnarray*}
where
\begin{IEEEeqnarray*}{rCl}
\tilde{\Upsilon} & \triangleq &
 \frac{\left( \bar{{{\beta }_{1}}}{{P}_{1}}+2\sqrt{\bar{{{\beta }_{1}}}\bar{{{\beta }_{2}}}{{P}_{1}}{{P}_{2}}}\tilde{\rho }+\bar{{{\beta }_{2}}}{{P}_{2}}\left( 1-{{\bar{\rho }}^{2}} \right) \right)N}{\left( \bar{{{\beta }_{1}}}{{P}_{1}}+2\sqrt{\bar{{{\beta }_{1}}}{{P}_{1}}\bar{{{\beta }_{2}}}{{P}_{2}}}\tilde{\rho }+\bar{{{\beta }_{2}}}{{P}_{2}}\left( 1-{{\bar{\rho }}^{2}} \right)+N \right)\left( \bar{{{\beta }_{1}}}{{P}_{1}}+2\tilde{\rho }\sqrt{\bar{{{\beta }_{1}}}{{P}_{1}}\bar{{{\beta }_{2}}}{{P}_{2}}}+\bar{{{\beta }_{2}}}{{P}_{2}} \right)}  \nonumber \\
{{\kappa}^{''}} & \triangleq & \frac{{{k}'}}{n\left( \bar{{{\beta }_{1}}}{{P}_{1}}+2\tilde{\rho }\sqrt{\bar{{{\beta }_{1}}}{{P}_{1}}\bar{{{\beta }_{2}}}{{P}_{2}}}+\bar{{{\beta }_{2}}}{{P}_{2}} \right)}.
\end{IEEEeqnarray*}

\medskip

The proof of Lemma~\ref{eps_h_u1,u2} is now concluded by noticing that the set
${{\cal E }'}_{({\hat{\textbf{U}}}_{1},{\hat{\textbf{U}}}_{2})}$, defined in \eqref{eq:eps_h_u1,u2},
is the set of tuples
$(\textbf{s}_1,\textbf{s}_2,{\cal C}_1,{\cal C}_2,{\cal C}_c,\textbf{z})$
for which there exists a ${{\textbf{u}}_{1}}\left( j \right)\in {{{\cal C}}_{1}}\backslash \left\{ \textbf{u}_{1}^{*} \right\}$
and $ {{\textbf{u}}_{2}}\left( l \right)\in {{{\cal C}}_{2}}\backslash \left\{ \textbf{u}_{2}^{*} \right\}$
such that
${{a}_{1,1}}{\textbf{u}}_{1}\left( j \right)+{{a}_{2,1}}{\textbf{u}}_{2}\left( l \right)\in
{\cal B}(\textbf{s}_1,\textbf{s}_2,\textbf{u}_1^{*},\textbf{v}^{*},\textbf{u}_2^{*},\textbf{z})$.
Thus, by Statement F) and by the definition of ${{\cal E}}_{({\hat{\textbf{U}}}_{1},{\hat{\textbf{U}}}_{2})}$
in \eqref{eq:eps_u1,u2} it follows that
\begin{equation*}
{{\cal E}}_{({\hat{\textbf{U}}}_{1},{\hat{\textbf{U}}}_{2})}\cap {\cal E}_{\textbf{S}}^{c}\cap {\cal E}_{\textbf{X}}^{c}\cap {\cal E}_{\textbf{Z}}^{c}\subseteq {\cal E}'_{({\hat{\textbf{U}}}_{1},{\hat{\textbf{U}}}_{2})}\cap {\cal E}_{\textbf{S}}^{c}\cap {\cal E}_{\textbf{X}}^{c}\cap {\cal E}_{\textbf{Z}}^{c},
\end{equation*}
and therefore
\begin{flalign*}
&&
\Pr\left[ {{{\cal E}}_{({\hat{\textbf{U}}}_{1},{\hat{\textbf{U}}}_{2})}}\cap {\cal E}_{\textbf{S}}^{c}\cap {\cal E}_{\textbf{X}}^{c}\cap {\cal E}_{\textbf{Z}}^{c} \right]\le \Pr\left[ {\cal E}'_{({\hat{\textbf{U}}}_{1},{\hat{\textbf{U}}}_{2})}\cap {\cal E}_{\textbf{S}}^{c}\cap {\cal E}_{\textbf{X}}^{c}\cap {\cal E}_{\textbf{Z}}^{c} \right].
&&
\end{flalign*}
\end{IEEEproof}

\medskip

We now state one more lemma that will be used for the proof of \eqref{eq:R1+R2}.
\begin{lemma} \label{two-Rates proof}
For every $\Theta \in (0,1]$ and $\Delta \in (0,1]$, let the set $\cal G$ be given by
\begin{IEEEeqnarray*}{rCl}
{\cal G} =\Bigl\{ \left( {\mathbf{s}_{1}},{\mathbf{s}_{2}},{{{\cal C}}_{1}},{{{\cal C}}_{2}},{{{\cal C}}_{c}},\mathbf{z} \right):\ \exists \> {{\mathbf{u}}_{1}}(j)\in {{{\cal C}}_{1}}\backslash \left\{ \mathbf{u}_{1}^{*} \right\}
\mbox{  and  } \exists \> {{\mathbf{u}}_{2}}(l)\in {{{\cal C}}_{2}}\backslash \left\{ \mathbf{u}_{2}^{*} \right\}\ \mbox{s.t.} \ \\
 \cos \sphericalangle \left( \mathbf{w},{{{{a}_{1,1}}\mathbf{u}}_{1}}(j)+{{{{a}_{2,1}}\mathbf{u}}_{2}}(l) \right)\ge \Delta  \mbox{ and }\cos \sphericalangle \left( {\mathbf{u}_{1}}(j),{\mathbf{u}_{2}}(l) \right)\ge \Theta \  \Bigr\}.
\end{IEEEeqnarray*}
Then,
\begin{equation*}
{{R}_{1}}+{{R}_{2}}<-\frac{1}{2}\log \left( \left( 1-{{\Theta }^{2}} \right)\left( 1-{{\Delta }^{2}} \right) \right)\implies \left( \underset{n\to \infty }{\mathop{\lim }}\,\ {\Pr}\left[ {\cal G} |{\cal E}_{{\mathbf{X}}_{1}}^{c}\cap {\cal E}_{{\mathbf{X}}_{2}}^{c} \right]=0,\epsilon >0 \right).
\end{equation*}
\end{lemma}

\medskip

\begin{IEEEproof}
The proof follows from upper-bounding in every point on ${\cal S}_1,{\cal S}_2$ the
density of every ${{\textbf{u}}_{1}}\left( j \right)\in {{{\cal C}}_{1}}\backslash \left\{ \textbf{u}_{1}^{*} \right\},{{\textbf{u}}_{2}}\left( l \right)\in {{{\cal C}}_{2}}\backslash \left\{ \textbf{u}_{2}^{*} \right\}$
and then using a standard argument from sphere-packing.
\end{IEEEproof}

Next,
\begin{IEEEeqnarray}{rCl} \label{eq:R1+R2_proof}
\Pr\left[ {{\cal E}}_{({{\hat{\textbf{U}}}}_{1},{{\hat{\textbf{U}}}}_{2})}\cap {\cal E}_{\textbf{S}}^{c}\cap {\cal E}_{\textbf{X}}^{c}\cap {\cal E}_{\textbf{Z}}^{c} \right]
& \overset{(a)}{\mathop{\le }}&\,
\Pr\left[ {{{\cal E}}'}_{({{\hat{\textbf{U}}}}_{1},{{\hat{\textbf{U}}}}_{2})}\cap {\cal E}_{\textbf{S}}^{c}\cap {\cal E}_{\textbf{X}}^{c}\cap {\cal E}_{\textbf{Z}}^{c} \right] \nonumber\\
& \overset{(b)}{\mathop{\le }}&\,\Pr\left[ {{{\cal E}}'}_{({{\hat{\textbf{U}}}}_{1},{{\hat{\textbf{U}}}}_{2})}
\middle|
{\cal E}_{{{\textbf{X}}_{1}}}^{c}\cap {\cal E}_{{{\textbf{X}}_{2}}}^{c} \right],
\end{IEEEeqnarray}
where (a) follows by Lemma~\ref{eps_h_u1,u2} and (b) follows because ${\cal E}_{\textbf{X}}^{c}\subseteq {\cal E}_{{{\textbf{X}}_{1}}}^{c}\cap {\cal E}_{{{\textbf{X}}_{2}}}^{c}$. \\
The proof of \eqref{eq:R1+R2} is now completed by combining  \eqref{eq:R1+R2_proof} with Lemma~\ref{two-Rates proof}.
This gives that for every $\delta >0$ and every $\epsilon >0$ there exists some
$n'(\delta ,\epsilon )$
such that for all $n>n'(\delta ,\epsilon )$,
we have
\begin{equation*}
\ \Pr\left[ {{\cal E}}_{({{\hat{\textbf{U}}}}_{1},{{\hat{\textbf{U}}}}_{2})}\cap {\cal E}_{\textbf{S}}^{c}\cap {\cal E}_{\textbf{X}}^{c}\cap {\cal E}_{\textbf{Z}}^{c} \right]\le
\Pr\left[ {{{\cal E}}'}_{({{\hat{\textbf{U}}}}_{1},{{\hat{\textbf{U}}}}_{2})}|{\cal E}_{{{\textbf{X}}_{1}}}^{c}\cap {\cal E}_{{{\textbf{X}}_{2}}}^{c} \right]<\delta ,
\end{equation*}
whenever
\begin{equation*}
{{R}_{1}}+{{R}_{2}}<\frac{1}{2}\log \left( \frac{{{\lambda }_{12}}-\bar{{{\beta }_{2}}}{{P}_{2}}{{\bar{\rho }}^{2}}+N}{\left( 1-\bar{{{\beta }_{2}}}{{P}_{2}}{{\bar{\rho }}^{2}}{{\lambda }_{12}}^{-1} \right)N\left( 1-{{{\tilde{\rho }}}^{2}}\  \right)}-{{\kappa}_{4}}\epsilon  \right),
\end{equation*}
where ${{\kappa}_{4}}$ is a positive constant determined by ${{P}_{1}},{{P}_{2}},N,{{\varsigma }_{1}}\text{ and }{{\varsigma }_{2}}$. \hfill \qed

\vskip.2truein

\subsection{Proof of rate constraint \eqref{eq:R1+Rc}}

Define
\begin{equation*}
\textbf{w}(\textbf{s}_1,\textbf{s}_2,{\cal C}_1,{\cal C}_2,{\cal C}_c,\textbf{z})=
{{\varsigma }_{1}}\left( \textbf{y}- {{a}_{2,1}} {{\textbf{u}_2}^{*}}  \right)
+{{\varsigma }_{2}}{{a}_{2,1}} {{\textbf{u}_2}^{*}},
\end{equation*}
where
\begin{IEEEeqnarray}{rCl}\label{eq:w_for_R1+Rc}
{{\varsigma }_{1}}&=&\frac{{{\bar{\beta }}_{1}}{{P}_{1}}\left( 1-{{{\tilde{\rho }}}^{2}} \right)+{{\left\| \alpha{{\textbf{v}}^{*}} \right\|}^{2}}\left( 1-{{\bar{\rho }}^{2}} \right)-2\sqrt{{{\bar{\beta }}_{1}}{{P}_{1}}{{\sigma }^{2}}\left( 1-{{2}^{-2{{R}_{1}}}} \right)}\alpha{{\bar{\rho }}^{2}}}{{{\bar{\beta }}_{1}}{{P}_{1}}\left( 1-{{{\tilde{\rho }}}^{2}} \right)+{{\left\| \alpha{{\textbf{v}}^{*}} \right\|}^{2}}\left( 1-{{\bar{\rho }}^{2}} \right)-2\sqrt{{{\bar{\beta }}_{1}}{{P}_{1}}{{\sigma }^{2}}\left( 1-{{2}^{-2{{R}_{1}}}} \right)}\alpha{{\bar{\rho }}^{2}}+N} \nonumber \\
{{\varsigma }_{2}}&=&\frac{\rho N\left( {{a}_{1,1}}\left( 1-{{2}^{-2{{R}_{1}}}} \right)+\alpha\left( 1-{{2}^{-2{{R}_\textnormal{c}}}} \right){{2}^{-2{{R}_{1}}}} \right)}{{{a}_{2,1}}\left( {{\bar{\beta }}_{1}}{{P}_{1}}\left( 1-{{{\tilde{\rho }}}^{2}} \right)+{{\left\| \alpha{{\textbf{v}}^{*}} \right\|}^{2}}\left( 1-{{\bar{\rho }}^{2}} \right)-2\sqrt{{{\bar{\beta }}_{1}}{{P}_{1}}{{\sigma }^{2}}\left( 1-{{2}^{-2{{R}_{1}}}} \right)}\alpha{{\bar{\rho }}^{2}}+N \right)}. \nonumber \\
\end{IEEEeqnarray}

In the remainder we shall use the shorthand notation \textbf{w} instead of
$\textbf{w}(\textbf{s}_1,\textbf{s}_2,{\cal C}_1,{\cal C}_2,{\cal C}_c,\textbf{z})$.
We now start with a lemma that will be used to prove \eqref{eq:R1+Rc}.

\medskip

\begin{lemma}\label{eps_h_u1,v}
Let ${{\varphi }_{j,l}}\in \left[ 0,\pi  \right]$  be the angle between $\mathbf{w}$ and ${a}_{1,1}\mathbf{u}_{1}(j)+\alpha{{\mathbf{v}}}(l)$, and let the set
${\cal E}'_{({\hat{\mathbf{U}}}_{1},{\hat{\mathbf{V}}})}$
be defined as
\begin{IEEEeqnarray}{l}\label{eq:eps_h_u1,v}
{\cal E}'_{({\hat{\mathbf{U}}}_{1},{\hat{\mathbf{V}}})}
\triangleq
\Biggl\{ (\mathbf{s}_1,\mathbf{s}_2,{\cal C}_1,{\cal C}_2,{\cal C}_c,\mathbf{z})\colon \exists \> {{\mathbf{u}}_{1}}\left( j \right)\in {{{\cal C}}_{1}}\backslash \left\{ \mathbf{u}_{1}^{*} \right\}
\mbox{ and } \exists \> {{\mathbf{v}}}\left( j \right)\in {{{\cal C}}_{c}}\backslash \left\{ \mathbf{v}^{*} \right\} \mbox{s.t.} \nonumber \\
\hspace{3.5cm}  \ \cos \left( {{\varphi }_{j}} \right)\ge \sqrt{1-\frac{\Upsilon (\epsilon )}{n\left( \bar{{{\beta }_{1}}}{{P}_{1}}+\frac{1}{n}{{\left\| \alpha \mathbf{v} \right\|}^{2}} \right)}} \ \ \Biggr\},
\end{IEEEeqnarray}
where $\Upsilon(\epsilon)$ is defined in \eqref{eq:upsilon_R1+Rc}, and $\epsilon$ is sufficiently small such that the term inside the square is non-negative.
Then,
\begin{equation*}
{{{\cal E}}}_{({\hat{\mathbf{U}}}_{1},{\hat{\mathbf{V}}})}\cap {\cal E}_{\mathbf{S}}^{c}\cap {\cal E}_{\mathbf{X}}^{c}\cap {\cal E}_{\mathbf{Z}}^{c}\subseteq {\cal E}'_{({\hat{\mathbf{U}}}_{1},{\hat{\mathbf{V}}})}\cap {\cal E}_{\mathbf{S}}^{c}\cap {\cal E}_{\mathbf{X}}^{c}\cap {\cal E}_{\mathbf{Z}}^{c},
\end{equation*}
and, in particular
\begin{equation*}
\Pr\left[ {{{\cal E}}}_{({\hat{\mathbf{U}}}_{1},{\hat{\mathbf{V}}})}\cap {\cal E}_{\mathbf{S}}^{c}\cap {\cal E}_{\mathbf{X}}^{c}\cap {\cal E}_{\mathbf{Z}}^{c} \right]\le \Pr\left[ {\cal E}'_{({\hat{\mathbf{U}}}_{1},{\hat{\mathbf{V}}})}\cap {\cal E}_{\mathbf{S}}^{c}\cap {\cal E}_{\mathbf{X}}^{c}\cap {\cal E}_{\mathbf{Z}}^{c} \right].
\end{equation*}
\end{lemma}

\medskip

\begin{IEEEproof}
We first recall that for the event ${{{\cal E}}}_{({\hat{\textbf{U}}}_{1},{\hat{\textbf{V}}})}$
to occur, there must exist codewords ${{\textbf{u}}_{1}}\left( j \right)\in {{{\cal C}}_{1}}\backslash \left\{ \textbf{u}_{1}^{*} \right\}  \mbox{and } {{\textbf{v}}}\left( l \right)\in {{{\cal C}}_{c}}\backslash \left\{ \textbf{v}^{*} \right\}$
that satisfy the following four conditions
\begin{IEEEeqnarray}{rCl}
\left| \tilde{\rho }-\cos \sphericalangle \left( {{\textbf{u}}_{1}}\left( j \right),\textbf{u}_{2}^{*} \right) \right| & \le & 7\epsilon \\
\left| \bar{\rho }-\cos \sphericalangle \left( {{\textbf{v}}}\left( l \right),\textbf{u}_{2}^{*} \right) \right| & \le & 7\epsilon  \\
\left| \cos \sphericalangle \left( {{\textbf{v}}}\left( l \right),{{\textbf{u}}_{1}}\left( j \right) \right) \right| & \le & 7\epsilon \\
\|\textbf{y}-\textbf{X}_{\textbf{u}_{1}\left(j\right),\textbf{v}\left( l \right),\textbf{u}_2^{*}}\|^2
&  \leq &
\|\textbf{y}-\textbf{X}_{\textbf{u}_1^{*},\textbf{v}^{*},\textbf{u}_2^{*}}\|^2.
\end{IEEEeqnarray}


The proof is now based on a sequence of statements related to these conditions:

\medskip

A) For every $(\textbf{s}_1,\textbf{s}_2,{\cal C}_1,{\cal C}_2,{\cal C}_c,\textbf{z})\in {\cal E}_{\textbf{X}}^{c}$ and every $\textbf{u}_1\in {\cal S}_{1}\mbox{ and } \textbf{v}\in {\cal S}_{c} $,
\begin{equation}\label{eq:statement_A_R1+Rc}
\left| \cos \sphericalangle \left( \textbf{u}_{1},\textbf{v} \right) \right|\le 7\epsilon \text{  }
\implies \text{   }
\left| \left\langle {{a}_{1,1}}{\textbf{u}_{1}},\alpha\textbf{v} \right\rangle  \right|\le 7\sqrt{n\bar{{{\beta }_{1}}}{{P}_{1}}} \left\| \alpha\textbf{v} \right\|\epsilon  .
\end{equation}

Statement A) follows by rewriting $\cos \sphericalangle \left( \textbf{u}_{1},\textbf{v} \right)$ as
${\left\langle \textbf{u}_{1},\textbf{v} \right\rangle }/({\left\| \textbf{u}_1 \right\|\left\| \textbf{v} \right\|})$, and then multiplying the inequality on the l.h.s. of
\eqref{eq:statement_A_R1+Rc}  by $\left\| {{a}_{1,1}}{{\textbf{u}}_{1}} \right\|\cdot \left\| \alpha\textbf{v} \right\|
\mbox{ and recalling that } \left\| {{a}_{1,1}}\textbf{u}_1 \right\|=\sqrt{n\bar{{{\beta }_{1}}}{{P}_{1}}}\ $.

\medskip

A1) For every $(\textbf{s}_1,\textbf{s}_2,{\cal C}_1,{\cal C}_2,{\cal C}_c,\textbf{z})\in {\cal E}_{\textbf{X}}^{c}$ and every $\textbf{u}_1\in {\cal S}_{1}\mbox{ and } \textbf{v}\in {\cal S}_{c} $,
\begin{IEEEeqnarray}{rCl} \label{eq:statement_A2_R1+Rc}
&& \left| \bar{\rho }-\cos \sphericalangle \left( \textbf{v},\textbf{u}_{2}^{*} \right) \right|\le 7\epsilon  \nonumber \\
&& \text{       }\implies \text{      }\left| \left\| \alpha\textbf{v} \right\|\sqrt{n\bar{{{\beta }_{2}}}{{P}_{2}}}\bar{\rho }-\left\langle \alpha\textbf{v},{{a}_{2,1}}\textbf{u}_{2}^{*} \right\rangle  \right|\le 7\epsilon \left\| \alpha\textbf{v} \right\|\sqrt{n\bar{{{\beta }_{2}}}{{P}_{2}}}.
\end{IEEEeqnarray}

Statement A1) follows by rewriting $\cos \sphericalangle \left( \textbf{v},\textbf{u}_{2}^{*} \right)$ as ${\left\langle \textbf{v},\textbf{u}_{2}^{*} \right\rangle }/({\left\| \textbf{v} \right\|\left\| \textbf{u}_{2}^{*} \right\|})$, and then multiplying the inequality on the l.h.s. of \eqref{eq:statement_A2_R1+Rc} by
$\left\| \alpha\textbf{v} \right\|\cdot \left\| {{a}_{2,1}}\textbf{u}_{2}^{*} \right\|$
and recalling that $\left\| {{a}_{2,1}}\textbf{u}_{2}^{*} \right\|=\sqrt{n\bar{{{\beta }_{2}}}{{P}_{2}}}$.

\medskip

B) For every $(\textbf{s}_1,\textbf{s}_2,{\cal C}_1,{\cal C}_2,{\cal C}_c,\textbf{z})\in {\cal E}_{\textbf{X}}^{c}\cap {\cal E}_{\textbf{Z}}^{c}$
and every $\textbf{u}_1\in {\cal S}_{1}\mbox{ and } \textbf{v}\in {\cal S}_{c} $
\begin{IEEEeqnarray}{rCl}\label{eq:statement_B_R1+Rc}
&& |\textbf{y}-\textbf{X}_{\textbf{u}_1,\textbf{v},\textbf{u}_2^{*}}\|^2
\leq
\|\textbf{y}-\textbf{X}_{\textbf{u}_1^{*},\textbf{v}^{*},\textbf{u}_2^{*}}\|^2 \nonumber \\
&& \text{                                 }\implies
\left\langle \textbf{y}-{{a}_{2,1}}\textbf{u}_2^{*} ,{{a}_{1,1}}\textbf{u}_1+\alpha{{\textbf{v}}} \right\rangle
\ge n\left( \bar{{{\beta }_{1}}}{{P}_{1}}+\frac{1}{n}{{\left\| \alpha{{\textbf{v}}^{*}} \right\|}^{2}}-\kappa\epsilon  \right).
\end{IEEEeqnarray}

Statement B) follows from rewriting the inequality on the l.h.s. of \eqref{eq:statement_B_R1+Rc} as
\begin{equation*}
{{\left\|\left( \textbf{y}-{{a}_{2,1}}\textbf{u}_{2}^{*}   \right)-({{a}_{1,1}}\textbf{u}_1+\alpha{{\textbf{v}}})
\right\|}^{2}}\le
 {{\left\|\left( \textbf{y}-{{a}_{2,1}}\textbf{u}_{2}^{*}  \right)-({{a}_{1,1}}\textbf{u}_{1}^{*}+\alpha{{\textbf{v}}^{*}}) \right\|}^{2}},
 \end{equation*}
or equivalently as
\begin{IEEEeqnarray}{rCl} \label{eq:statement_B_proof_R1+Rc}
\left\langle \left(\textbf{y}-{{a}_{2,1}}\textbf{u}_{2}^{*}  \right),
{{a}_{1,1}}\textbf{u}_1+\alpha{{\textbf{v}}} \right\rangle
& \ge &
\left\langle \left( \textbf{y}-{{a}_{2,1}}\textbf{u}_{2}^{*}  \right),{{a}_{1,1}}\textbf{u}_{1}^{*}+\alpha{{\textbf{v}}^{*}} \right\rangle  \nonumber \\
& = & \left\langle {{a}_{1,1}}\textbf{u}_{1}^{*}+\alpha{{\textbf{v}}^{*}}+\textbf{z},
{{a}_{1,1}}\textbf{u}_{1}^{*}+\alpha{{\textbf{v}}^{*}} \right\rangle  \nonumber \\
& = & {{\left\| {{a}_{1,1}}\textbf{u}_{1}^{*}+\alpha{{\textbf{v}}^{*}} \right\|}^{2}}
+\left\langle \textbf{z},{{a}_{1,1}}\textbf{u}_{1}^{*}+\alpha{{\textbf{v}}^{*}} \right\rangle.
\end{IEEEeqnarray}
It now follows from the equivalence of the first inequality in \eqref{eq:statement_B_R1+Rc} with
\eqref{eq:statement_B_proof_R1+Rc} that for \\
$(\textbf{s}_1,\textbf{s}_2,{\cal C}_1,{\cal C}_2,{\cal C}_c,\textbf{z})\in {\cal E}_{\textbf{Z}}^{c}$, the first inequality in \eqref{eq:statement_B_R1+Rc} can only hold if
\begin{equation*}
\left\langle \textbf{y}-{{a}_{2,1}}\textbf{u}_2^{*} ,{{a}_{1,1}}\textbf{u}_1+\alpha{{\textbf{v}}} \right\rangle
 \ge n\left( \bar{{{\beta }_{1}}}{{P}_{1}}+\frac{1}{n}{{\left\| \alpha{{\textbf{v}}^{*}} \right\|}^{2}}-\kappa\epsilon  \right),
 \end{equation*}
thus establishing B).

\medskip

C) For every $(\textbf{s}_1,\textbf{s}_2,{\cal C}_1,{\cal C}_2,{\cal C}_c,\textbf{z})\in {\cal E}_{\textbf{X}}^{c}\cap {\cal E}_{\textbf{Z}}^{c}$
and every $\textbf{u}_1\in {\cal S}_{1}\mbox{ and } \textbf{v}\in {\cal S}_{c} $,
\begin{IEEEeqnarray*}{rCl}
&& \left| \tilde{\rho }-\cos \sphericalangle \left( \textbf{u}_1,\textbf{u}_{2}^{*} \right) \right|\le 7\epsilon \text{   and  }
|\textbf{y}-\textbf{X}_{\textbf{u}_1,\textbf{v},\textbf{u}_2^{*}}\|^2
\leq
\|\textbf{y}-\textbf{X}_{\textbf{u}_1^{*},\textbf{v}^{*},\textbf{u}_2^{*}}\|^2 \nonumber \\
&& \qquad \qquad \implies  \\
&& {{\left\| {{a}_{1,1}}\textbf{u}_1+\alpha{{\textbf{v}}}-\textbf{w} \right\|}^{2}}
\le
n\Biggl( \left( \bar{{{\beta }_{1}}}{{P}_{1}}+\frac{1}{n}{{\left\| \alpha \textbf{v} \right\|}^{2}} \right)\left( 1-2{{\varsigma }_{1}} \right)-2{{\varsigma }_{2}}\bigl( \sqrt{\bar{{{\beta }_{1}}}\bar{{{\beta }_{2}}}{{P}_{1}}{{P}_{2}}}\tilde{\rho }\\
&& \hspace{3.65cm} +\frac{1}{n}\left\| \alpha{\textbf{v}^{*}} \right\|\sqrt{n\bar{{{\beta }_{2}}}{{P}_{2}}}\bar{\rho } \bigr) \Biggr)+{{\left\| \textbf{w} \right\|}^{2}}+n{{\kappa}'}\epsilon .
\end{IEEEeqnarray*}
Statement C) is obtained as follows:
\begin{IEEEeqnarray*}{rCl}
{{\left\| {{a}_{1,1}}\textbf{u}_1+\alpha{{\textbf{v}}}-\textbf{w} \right\|}^{2}}
& = & {{\left\| {{a}_{1,1}}\textbf{u}_1+\alpha{{\textbf{v}}} \right\|}^{2}}
-2\left\langle {{a}_{1,1}}\textbf{u}_1+\alpha{{\textbf{v}}},\textbf{w} \right\rangle +{{\left\| \textbf{w} \right\|}^{2}} \\
& = & {{\left\| {{a}_{1,1}}\textbf{u}_1+\alpha{{\textbf{v}}} \right\|}^{2}}\\
&& -2\left\langle {{a}_{1,1}}\textbf{u}_1+\alpha{{\textbf{v}}},
{{\varsigma }_{1}}\left( {{a}_{1,1}}\textbf{u}_{1}^{*}+\alpha{{\textbf{v}^{*}}}+\textbf{z} \right)
+{{\varsigma }_{2}}{{a}_{2,1}}\textbf{u}_{2}^{*} \right\rangle +{{\left\| \textbf{w} \right\|}^{2}} \\
& \text{ } \overset{(a)}{\mathop{\le }} & \, n
\Biggl( \left( \bar{{{\beta }_{1}}}{{P}_{1}}+\frac{1}{n}{{\left\| \alpha \textbf{v} \right\|}^{2}} \right)\left( 1-2{{\varsigma }_{1}} \right)-2{{\varsigma }_{2}}\bigl( \sqrt{\bar{{{\beta }_{1}}}\bar{{{\beta }_{2}}}{{P}_{1}}{{P}_{2}}}\tilde{\rho }\\
&& +\frac{1}{n}\left\| \alpha{\textbf{v}^{*}} \right\|\sqrt{n\bar{{{\beta }_{2}}}{{P}_{2}}}\bar{\rho } \bigr) \Biggr)+{{\left\| \textbf{w} \right\|}^{2}}+n{{\kappa}'}\epsilon  ,
\end{IEEEeqnarray*}
where in (a) we have used Statement A) and Statement B).

\medskip

D) For every $(\textbf{s}_1,\textbf{s}_2,{\cal C}_1,{\cal C}_2,{\cal C}_c,\textbf{z})\in {\cal E}_{\textbf{X}}^{c}\cap {\cal E}_{\textbf{Z}}^{c}$
\begin{IEEEeqnarray*}{rCl}
{{\left\| \textbf{w} \right\|}^{2}}
& \le &\, n\Biggl( {{\varsigma }_{1}}^{2}\left( \bar{{{\beta }_{1}}}{{P}_{1}}+N+\frac{1}{n}{{\left\| \alpha{{\textbf{v}}^{*}} \right\|}^{2}} \right)+2{{\varsigma }_{1}}{{\varsigma }_{2}}\left( \sqrt{\bar{{{\beta }_{1}}}\bar{{{\beta }_{2}}}{{P}_{1}}{{P}_{2}}}\tilde{\rho }+\frac{1}{n}\left\| \alpha{{\textbf{v}}^{*}} \right\|\sqrt{n\bar{{{\beta }_{2}}}{{P}_{2}}}\bar{\rho } \right)\\
&& +{{\varsigma }_{2}}^{2}\bar{{{\beta }_{2}}}{{P}_{2}}+\kappa\epsilon  \Biggr),
\end{IEEEeqnarray*}
where $\kappa$ depends on ${{P}_{1}},{{P}_{2}},N,{{\varsigma }_{1}}$ and ${{\varsigma }_{2}}$ only.

Statement D) is obtained as follows:
\begin{IEEEeqnarray*}{rCl}
{{\left\| \textbf{w} \right\|}^{2}}
&=&{{\left\| {{\varsigma }_{1}}\left( {{a}_{1,1}}\textbf{u}_{1}^{*}+\alpha{{\textbf{v}}^{*}}+\textbf{z} \right)
+{{\varsigma }_{2}}{{a}_{2,1}}\textbf{u}_{2}^{*} \right\|}^{2}}\\
&=&{{\varsigma }_{1}}^{2}{{\left\| {{a}_{1,1}}\textbf{u}_{1}^{*}+\alpha{{\textbf{v}}^{*}}+\textbf{z} \right\|}^{2}}
+2{{\varsigma }_{1}}{{\varsigma }_{2}}\left\langle
{{a}_{1,1}}\textbf{u}_{1}^{*}+\alpha{{\textbf{v}}^{*}}+\textbf{z},{{a}_{2,1}}\textbf{u}_{2}^{*} \right\rangle
+{{\varsigma }_{2}}^{2}{{\left\| {{a}_{2,1}}\textbf{u}_{2}^{*} \right\|}^{2}} \\
&=&{{\varsigma }_{1}}^{2}\left( {{\left\| {{a}_{1,1}}\textbf{u}_{1}^{*}+\alpha{{\textbf{v}}^{*}} \right\|}^{2}}
+2\left\langle {{a}_{1,1}}\textbf{u}_{1}^{*}+\alpha{{\textbf{v}}^{*}},\textbf{z} \right\rangle +{{\left\| \textbf{z} \right\|}^{2}} \right)\\
&& +2{{\varsigma }_{1}}{{\varsigma }_{2}}\left\langle
{{a}_{1,1}}\textbf{u}_{1}^{*}+\alpha{{\textbf{v}}^{*}}+\textbf{z},{{a}_{2,1}}\textbf{u}_{2}^{*} \right\rangle +{{\varsigma }_{2}}^{2}n\bar{{{\beta }_{2}}}{{P}_{2}} \\
&\text{ } \overset{(a)}{\mathop{\le }} & \,
n\Biggl( {{\varsigma }_{1}}^{2}\left( \bar{{{\beta }_{1}}}{{P}_{1}}+N+\frac{1}{n}{{\left\| \alpha{{\textbf{v}}^{*}} \right\|}^{2}} \right)+2{{\varsigma }_{1}}{{\varsigma }_{2}}\left( \sqrt{\bar{{{\beta }_{1}}}\bar{{{\beta }_{2}}}{{P}_{1}}{{P}_{2}}}\tilde{\rho }+\frac{1}{n}\left\| \alpha{{\textbf{v}}^{*}} \right\|\sqrt{n\bar{{{\beta }_{2}}}{{P}_{2}}}\bar{\rho } \right)\\
&& +{{\varsigma }_{2}}^{2}\bar{{{\beta }_{2}}}{{P}_{2}}+k\epsilon  \Biggr),
\end{IEEEeqnarray*}
where in (a) we have used that $(\textbf{s}_1,\textbf{s}_2,{\cal C}_1,{\cal C}_2,{\cal C}_c,\textbf{z})\in {\cal E}_{\textbf{X}}^{c}$, and statements A) and A1).

\medskip

E) For every$(\textbf{s}_1,\textbf{s}_2,{\cal C}_1,{\cal C}_2,{\cal C}_c,\textbf{z})\in {\cal E}_{\textbf{X}}^{c}\cap {\cal E}_{\textbf{Z}}^{c}$ and an arbitrary $\textbf{u}_1\in {\cal S}_{1}\mbox{ and } \textbf{v}\in {\cal S}_{c} $,
\begin{IEEEeqnarray*}{rCl}
&& \left| \tilde{\rho }-\cos \sphericalangle \left( \textbf{u}_1,\textbf{u}_{2}^{*} \right) \right|\le 7\epsilon \text{  and  }\left| \bar{\rho }-\cos \sphericalangle \left( \textbf{v},\textbf{u}_{2}^{*} \right) \right|\le 7\epsilon \text{  and  }
|\textbf{y}-\textbf{X}_{\textbf{u}_1,\textbf{v},\textbf{u}_2^{*}}\|^2
\leq
\|\textbf{y}-\textbf{X}_{\textbf{u}_1^{*},\textbf{v}^{*},\textbf{u}_2^{*}}\|^2 \\
&& \implies \text{        }{{\left\| {{a}_{1,1}}\textbf{u}_1+\alpha{{\textbf{v}}}-\textbf{w} \right\|}^{2}}\le \Upsilon (\epsilon ),
\end{IEEEeqnarray*}
where
\begin{equation}\label{eq:upsilon_R1+Rc}
\Upsilon (\epsilon )=n\frac{\left( \bar{{{\beta }_{1}}}{{P}_{1}}\left( 1-{{{\tilde{\rho }}}^{2}} \right)+\frac{1}{n}{{\left\| \alpha\textbf{v} \right\|}^{2}}\left( 1-{{\bar{\rho }}^{2}} \right)-2\sqrt{\bar{{{\beta }_{1}}}{{P}_{1}}{{\sigma }^{2}}\left( 1-{{2}^{-2{{R}_{1}}}} \right)}\alpha{{\bar{\rho }}^{2}} \right)N}{{{\bar{\beta }}_{1}}{{P}_{1}}\left( 1-{{{\tilde{\rho }}}^{2}} \right)+\frac{1}{n}{{\left\| \alpha{{\textbf{v}}^{*}} \right\|}^{2}}\left( 1-{{\bar{\rho }}^{2}} \right)-2\sqrt{{{\bar{\beta }}_{1}}{{P}_{1}}{{\sigma }^{2}}\left( 1-{{2}^{-2{{R}_{1}}}} \right)}\alpha{{\bar{\rho }}^{2}}+N}+n{{\kappa}'}\epsilon ,
\end{equation}
and where ${{\kappa}'}$ only depends on ${{P}_{1}},{{N}_{1}},{{N}_{2}},{{\varsigma }_{1}}$  and $ {{\varsigma }_{2}}$.

Statement E) follows from combining Statement C) with Statement
D) and the explicit values of ${{\varsigma }_{1}}$  and $ {{\varsigma }_{2}}$ given in \eqref{eq:w_for_R1+Rc}.

\medskip

F) For every $\textbf{u}_1\in {\cal S}_{1},\textbf{v}\in {\cal S}_{c}$, denote by $\varphi \in \left[ 0,\pi  \right]$ the angle between ${{a}_{1,1}}\textbf{u}_1+\alpha{{\textbf{v}}}$ and \textbf{w}, and let
\begin{equation*}
{\cal B}(\textbf{s}_1,\textbf{s}_2,\textbf{u}_1^{*},\textbf{v}^{*},\textbf{u}_2^{*},\textbf{z})\triangleq
\left\{ \textbf{u}_1\in S_{1}^{\left( n \right)},\textbf{v}\in S_{c}^{\left( n \right)}:
\ \ \ \cos \left( \varphi  \right)\ge \sqrt{1-\frac{\Upsilon (\epsilon )}{n\left( \bar{{{\beta }_{1}}}{{P}_{1}}+\frac{1}{n}{{\left\| \alpha \textbf{v} \right\|}^{2}} \right)}}\right\},
\end{equation*}
where $\epsilon$ is sufficiently small such that the term inside the square is non-negative.
Then, for every $(\textbf{s}_1,\textbf{s}_2,{\cal C}_1,{\cal C}_2,{\cal C}_c,\textbf{z})\in {\cal E}_{\textbf{X}}^{c}\cap {\cal E}_{\textbf{Z}}^{c}$,
\begin{IEEEeqnarray}{rCl}\label{eq:statement_F_R1+Rc}
&& \left| \tilde{\rho }-\cos \sphericalangle \left( \textbf{u}_1,\textbf{u}_{2}^{*}\right) \right|\le 7\epsilon \ \quad \mbox{  and  }\quad \left| \bar{\rho }-\cos \sphericalangle \left( \textbf{v},\textbf{u}_{2}^{*} \right) \right|\le 7\epsilon \quad \mbox{  and  }\nonumber \\
&& |\textbf{y}-\textbf{X}_{\textbf{u}_1,\textbf{v},\textbf{u}_2^{*}}\|^2
\leq
\|\textbf{y}-\textbf{X}_{\textbf{u}_1^{*},\textbf{v}^{*},\textbf{u}_2^{*}}\|^2 \nonumber \\
&& \implies \text{    }{{a}_{1,1}}\textbf{u}_1+\alpha{{\textbf{v}}}\in
{\cal B}(\textbf{s}_1,\textbf{s}_2,\textbf{u}_1^{*},\textbf{v}^{*},\textbf{u}_2^{*},\textbf{z}) .
\end{IEEEeqnarray}

Statement F) follows from Statement E) by noting that if $\mathbf{w}\ne 0$ and
$$1-\frac{\Upsilon (\epsilon )}{n\left( \bar{{{\beta }_{1}}}{{P}_{1}}+\frac{1}{n}{{\left\| \alpha\textbf{v} \right\|}^{2}} \right)}>0,$$
then
\begin{IEEEeqnarray*}{rCl}
&& {{\left\| {{a}_{1,1}}\textbf{u}_1+\alpha{{\textbf{v}}} \right\|}^{2}}
={n\left( \bar{{{\beta }_{1}}}{{P}_{1}}+\frac{1}{n}{{\left\| \alpha\textbf{v} \right\|}^{2}} \right)}
\mbox{  and   }{{\left\| {{a}_{1,1}}\textbf{u}_1+\alpha{{\textbf{v}}}-\textbf{w} \right\|}^{2}}\le \Upsilon (\epsilon ) \\
&& \qquad \implies \cos \sphericalangle \left( {{a}_{1,1}}\textbf{u}_1+\alpha{{\textbf{v}}},\textbf{w} \right)\ge \sqrt{1-\frac{\Upsilon (\epsilon )}{n\left( \bar{{{\beta }_{1}}}{{P}_{1}}+\frac{1}{n}{{\left\| \alpha\textbf{v} \right\|}^{2}} \right)}},
\end{IEEEeqnarray*}
which follows by the same argument as \eqref{eq:statement_F_proof_R1+R2}.

\medskip

The proof of Lemma~\ref{eps_h_u1,v} is now concluded by noticing that the set
${{\cal E }'}_{({\hat{\textbf{U}}}_{1},{\hat{\textbf{V}}})}$, defined in \eqref{eq:eps_h_u1,v},
is the set of tuples
$(\textbf{s}_1,\textbf{s}_2,{\cal C}_1,{\cal C}_2,{\cal C}_c,\textbf{z})$
for which there exists a ${{\textbf{u}}_{1}}\left( j \right)\in {{{\cal C}}_{1}}\backslash \left\{ \textbf{u}_{1}^{*} \right\}$
 and  ${{\textbf{v}}}\left( l \right)\in {{{\cal C}}_{c}}\backslash \left\{ \textbf{v}^{*} \right\}$
such that
${{a}_{1,1}}{\textbf{u}}_{1}\left( j \right)+\alpha\textbf{v}\left( l \right)\in
{\cal B}(\textbf{s}_1,\textbf{s}_2,\textbf{u}_1^{*},\textbf{v}^{*},\textbf{u}_2^{*},\textbf{z})$.
Thus, by Statement F) and by the definition of ${{\cal E}}_{({\hat{\textbf{U}}}_{1},{\hat{\textbf{V}}})}$
in \eqref{eq:eps_u1,v} it follows that
\begin{equation*}
{{\cal E}}_{({\hat{\textbf{U}}}_{1},{\hat{\textbf{V}}})}\cap {\cal E}_{\textbf{S}}^{c}\cap {\cal E}_{\textbf{X}}^{c}\cap {\cal E}_{\textbf{Z}}^{c}\subseteq {\cal E}'_{({\hat{\textbf{U}}}_{1},{\hat{\textbf{V}}})}\cap {\cal E}_{\textbf{S}}^{c}\cap {\cal E}_{\textbf{X}}^{c}\cap {\cal E}_{\textbf{Z}}^{c},
\end{equation*}
and therefore
\begin{flalign*}
&&
\Pr\left[ {{{\cal E}}_{({\hat{\textbf{U}}}_{1},{\hat{\textbf{V}}})}}\cap {\cal E}_{\textbf{S}}^{c}\cap {\cal E}_{\textbf{X}}^{c}\cap {\cal E}_{\textbf{Z}}^{c} \right]\le \Pr\left[ {\cal E}'_{({\hat{\textbf{U}}}_{1},{\hat{\textbf{V}}})}\cap {\cal E}_{\textbf{S}}^{c}\cap {\cal E}_{\textbf{X}}^{c}\cap {\cal E}_{\textbf{Z}}^{c} \right].
&&
\end{flalign*}
\end{IEEEproof}

Next,
\begin{IEEEeqnarray}{rCl} \label{eq:R1+Rc_proof}
Pr\left[ {\cal E}_{({\hat{\textbf{U}}}_{1},{\hat{\textbf{V}}})}\cap {\cal E}_{\textbf{S}}^{c}\cap {\cal E}_{\textbf{X}}^{c}\cap {\cal E}_{\textbf{Z}}^{c} \right]
& \overset{(a)}{\mathop{\le }}&\,
\Pr\left[ {\cal E}'_{({\hat{\textbf{U}}}_{1},{\hat{\textbf{V}}})}\cap {\cal E}_{\textbf{S}}^{c}\cap {\cal E}_{\textbf{X}}^{c}\cap {\cal E}_{\textbf{Z}}^{c} \right] \nonumber\\
& \overset{(b)}{\mathop{\le }}&\,\Pr\left[ {\cal E}'_{({\hat{\textbf{U}}}_{1},{\hat{\textbf{V}}})}|
{\cal E}_{{\textbf{X}}_{1}}^{c}\cap {\cal E}_{{\textbf{X}}_{2}}^{c} \right],
\end{IEEEeqnarray}
where (a) follows by Lemma~\ref{eps_h_u1,v} and (b) follows because ${\cal E}_{\textbf{X}}^{c}\subseteq {\cal E}_{{\textbf{X}}_{1}}^{c}\cap {\cal E}_{{\textbf{X}}_{v}}^{c}$. \\
The proof of \eqref{eq:R1+Rc} is now completed by combining  \eqref{eq:R1+Rc_proof} with Lemma~\ref{two-Rates proof}.
This gives that for every $\delta >0$ and every $\epsilon >0$ there exists some
$n'(\delta ,\epsilon )$
such that for all $n>n'(\delta ,\epsilon )$,
we have
\begin{equation*}
\Pr\left[ {{\cal E}}_{({{\hat{\textbf{U}}}}_{1},{{\hat{\textbf{V}}}})}\cap {\cal E}_{\textbf{S}}^{c}\cap {\cal E}_{\textbf{X}}^{c}\cap {\cal E}_{\textbf{Z}}^{c} \right]\le
\Pr\left[ {{{\cal E}}'}_{({{\hat{\textbf{U}}}}_{1},{{\hat{\textbf{V}}}})}|{\cal E}_{{{\textbf{X}}_{1}}}^{c}\cap {\cal E}_{{{\textbf{X}}_{v}}}^{c} \right]<\delta ,
\end{equation*}
whenever
\begin{equation*}
{{R}_{1}}+{{R}_\textnormal{c}}<\frac{1}{2}\log \left( \frac{\left( {{\lambda }_{1c}}+N \right)\left( \bar{{{\beta }_{1}}}{{P}_{1}}+{{\eta }^{2}} \right)}{{{\lambda }_{1c}}N}-{{\kappa}_{5}}\epsilon  \right),
\end{equation*}
where ${{\kappa}_{5}}$ is a positive constant determined by ${{P}_{1}},{{P}_{2}},N,{{\varsigma }_{1}}\text{ and }{{\varsigma }_{2}}$. \hfill \qed

\vskip.2truein

\subsection{Proof of rate constraint \eqref{eq:R2+Rc}}

Define
\begin{equation*}
\textbf{w}(\textbf{s}_1,\textbf{s}_2,{\cal C}_1,{\cal C}_2,{\cal C}_c,\textbf{z})=
{{\varsigma }_{1}}\left( \textbf{y}- {{a}_{1,1}} {{\textbf{u}_1}^{*}}  \right)
+{{\varsigma }_{2}}{{a}_{1,1}} {{\textbf{u}_1}^{*}},
\end{equation*}
where
\begin{IEEEeqnarray}{rCl}\label{eq:w_for_R2+Rc}
{{\varsigma }_{1}}&=&\frac{\bar{{{\beta }_{2}}}{{P}_{2}}\left( 1-{{{\tilde{\rho }}}^{2}} \right)+2\left\| \alpha\textbf{v} \right\|\sqrt{n\bar{{{\beta }_{2}}}{{P}_{2}}}\bar{\rho }+{{\left\| \alpha\textbf{v} \right\|}^{2}}}{\bar{{{\beta }_{2}}}{{P}_{2}}\left( 1-{{{\tilde{\rho }}}^{2}} \right)+2\left\| \alpha\textbf{v} \right\|\sqrt{n\bar{{{\beta }_{2}}}{{P}_{2}}}\bar{\rho }+{{\left\| \alpha\textbf{v} \right\|}^{2}}+N} \nonumber \\
{{\varsigma }_{2}}&=&\frac{{{a}_{2,1}}\rho \left( 1-{{2}^{-2{{R}_{2}}}} \right)N}{{{a}_{1,1}}\left( \bar{{{\beta }_{2}}}{{P}_{2}}\left( 1-{{{\tilde{\rho }}}^{2}} \right)+2\left\| \alpha\textbf{v} \right\|\sqrt{n\bar{{{\beta }_{2}}}{{P}_{2}}}\bar{\rho }+{{\left\| \alpha\textbf{v} \right\|}^{2}}+N \right)}.
\end{IEEEeqnarray}

In the remainder we shall use the shorthand notation \textbf{w} instead of
$\textbf{w}(\textbf{s}_1,\textbf{s}_2,{\cal C}_1,{\cal C}_2,{\cal C}_c,\textbf{z})$.
We now start with a lemma that will be used to prove \eqref{eq:R2+Rc}.

\medskip

\begin{lemma}\label{eps_h_u2,v}
Let ${{\varphi }_{j,l}}\in \left[ 0,\pi  \right]$  be the angle between $\mathbf{w}$ and ${a}_{2,1}\mathbf{u}_{2}(j)+\alpha{{\mathbf{v}}^{*}}(l)$, and let the set
${\cal E}'_{({\hat{\mathbf{U}}}_{1},{\hat{\mathbf{V}}})}$
be defined as
\begin{IEEEeqnarray}{rCl}\label{eq:eps_h_u2,v}
{\cal E}'_{({\hat{\mathbf{U}}}_{2},{\hat{\mathbf{V}}})}
&&\triangleq
\Biggl\{ (\mathbf{s}_1,\mathbf{s}_2,{\cal C}_1,{\cal C}_2,{\cal C}_c,\mathbf{z})\colon \exists \> {{\mathbf{u}}_{2}}\left( j \right)\in {{{\cal C}}_{2}}\backslash \left\{ \mathbf{u}_{2}^{*} \right\}
\mbox{ and } \exists \> {{\mathbf{v}}}\left( l \right)\in {{{\cal C}}_{c}}\backslash \left\{ \mathbf{v}^{*} \right\}
\ s.t.\ \nonumber \\
&& \qquad \quad \cos \left( {{\varphi }_{j,l}} \right)\ge \sqrt{1-\frac{\Upsilon (\epsilon )}{n\left( \bar{{{\beta }_{2}}}{{P}_{2}}+\frac{2}{n}\left\| \alpha\mathbf{v} \right\|\sqrt{n\bar{{{\beta }_{2}}}{{P}_{2}}}\bar{\rho }+\frac{1}{n}{{\left\| \alpha\mathbf{v} \right\|}^{2}} \right)}} \Biggr\},
\end{IEEEeqnarray}
where $\Upsilon(\epsilon)$ is defined in \eqref{eq:upsilon_R2+Rc} and $\epsilon$ is sufficiently small such that the term inside the square is non-negative.
Then,
\begin{equation*}
{{{\cal E}}}_{({\hat{\mathbf{U}}}_{2},{\hat{\mathbf{V}}})}\cap {\cal E}_{\mathbf{S}}^{c}\cap {\cal E}_{\mathbf{X}}^{c}\cap {\cal E}_{\mathbf{Z}}^{c}\subseteq {\cal E}'_{({\hat{\mathbf{U}}}_{2},{\hat{\mathbf{V}}})}\cap {\cal E}_{\mathbf{S}}^{c}\cap {\cal E}_{\mathbf{X}}^{c}\cap {\cal E}_{\mathbf{Z}}^{c},
\end{equation*}
and, in particular
\begin{equation*}
\Pr\left[ {{{\cal E}}}_{({\hat{\mathbf{U}}}_{2},{\hat{\mathbf{V}}})}\cap {\cal E}_{\mathbf{S}}^{c}\cap {\cal E}_{\mathbf{X}}^{c}\cap {\cal E}_{\mathbf{Z}}^{c} \right]\le \Pr\left[ {\cal E}'_{({\hat{\mathbf{U}}}_{2},{\hat{\mathbf{V}}})}\cap {\cal E}_{\mathbf{S}}^{c}\cap {\cal E}_{\mathbf{X}}^{c}\cap {\cal E}_{\mathbf{Z}}^{c} \right].
\end{equation*}
\end{lemma}

\medskip

\begin{IEEEproof}
We first recall that for the event ${{{\cal E}}}_{({\hat{U}}_{2},{\hat{V}})}$
to occur, there must exist codewords ${{\textbf{u}}_{2}}\left( j \right)\in {{{\cal C}}_{2}}\backslash \left\{ \textbf{u}_{2}^{*} \right\}$  and ${{\textbf{v}}}\left( l \right)\in {{{\cal C}}_{c}}\backslash \left\{ \textbf{v}^{*} \right\}$
that satisfy the following four conditions
\begin{IEEEeqnarray}{rCl}
\left| \tilde{\rho }-\cos \sphericalangle \left( {{\textbf{u}}_{1}^{*}},\textbf{u}_{2}\left( j \right) \right) \right| & \le & 7\epsilon \\
\left| \bar{\rho }-\cos \sphericalangle \left( {{\textbf{v}}}\left( l \right),\textbf{u}_{2}\left( j \right) \right) \right| & \le & 7\epsilon  \\
\left| \cos \sphericalangle \left( {{\textbf{v}}}\left( l \right),{{\textbf{u}}_{1}^{*}} \right) \right| & \le & 3\epsilon \\
\|\textbf{y}-\textbf{X}_{\textbf{u}_{1}^{*},\textbf{v}\left( l \right),\textbf{u}_2\left(j\right)}\|^2
&  \leq &
\|\textbf{y}-\textbf{X}_{\textbf{u}_1^{*},\textbf{v}^{*},\textbf{u}_2^{*}}\|^2 .
\end{IEEEeqnarray}


The proof is now based on a sequence of statements related to these conditions:

\medskip

A) For every $(\textbf{s}_1,\textbf{s}_2,{\cal C}_1,{\cal C}_2,{\cal C}_c,\textbf{z})\in {\cal E} _{X}^{c}$ and every $\textbf{u}_2\in {\cal S}_{2}\mbox{ and } \textbf{v}\in {\cal S}_{c} $,
\begin{equation}\label{eq:statement_A_R2+Rc}
\left| \tilde{\rho }-\cos \sphericalangle \left( \textbf{u}_{1}^{*},\textbf{u}_{2} \right) \right|\le 7\epsilon
\text{  }
\implies \text{   }\left| n\tilde{\rho }\sqrt{\bar{{\beta}}_{1}\bar{{\beta}}_{2}{{P}_{1}}{{P}_{2}}}-\left\langle {{a}_{1,1}}\textbf{u}_{1}^{*},{{a}_{2,1}}\textbf{u}_{2} \right\rangle  \right|\le 7n\sqrt{\bar{{\beta}}_{1}\bar{{\beta}}_{2}{{P}_{1}}{{P}_{2}}}\epsilon.
\end{equation}

Statement A) follows by rewriting $\cos \sphericalangle \left( \textbf{u}_{1}^{*},\textbf{u}_{2} \right)$ as ${\left\langle \textbf{u}_{1}^{*},\textbf{u}_{2} \right\rangle }/({\left\| \textbf{u}_1^{*} \right\|\left\| \textbf{u}_{2} \right\|})$, and then multiplying the inequality on the l.h.s. of
\eqref{eq:statement_A_R2+Rc}  by $\left\| {{a}_{1,1}}{{\textbf{u}}_{1}^{*}} \right\|\cdot \left\| {{a}_{2,1}}{{\textbf{u}}_{2}} \right\| $.

\medskip

A1) For every $(\textbf{s}_1,\textbf{s}_2,{\cal C}_1,{\cal C}_2,{\cal C}_c,\textbf{z})\in {\cal E}_{\textbf{X}}^{c}$,
\begin{IEEEeqnarray}{rCl} \label{eq:statement_A1_R2+Rc}
&& \left| \cos \sphericalangle \left( \textbf{v},\textbf{u}_{1}^{*} \right) \right|\le 3\epsilon  \nonumber \\
&& \text{       }\implies \text{      }\left| \left\langle \alpha\textbf{v},{{a}_{1,1}}\textbf{u}_{1}^{*} \right\rangle  \right|\le 3\left\| \alpha\textbf{v} \right\|\sqrt{n\bar{{{\beta }_{1}}}{{P}_{1}}}\epsilon .
\end{IEEEeqnarray}

Statement A1) follows by rewriting $\cos \sphericalangle \left( \textbf{v},\textbf{u}_{1}^{*} \right)$ as ${\left\langle \textbf{v},\textbf{u}_{1}^{*} \right\rangle }/({\left\| \textbf{v} \right\|\left\| \textbf{u}_{1}^{*} \right\|})$, and then multiplying the inequality on the l.h.s. of \eqref{eq:statement_A1_R2+Rc} by
$\left\| \alpha\textbf{v} \right\|\cdot \left\| {{a}_{1,1}}\textbf{u}_{1}^{*} \right\|$
and recalling that $\left\| {{a}_{1,1}}\textbf{u}_{1}^{*} \right\|=\sqrt{n\bar{{{\beta }_{1}}}{{P}_{1}}}$.

\medskip

A2) For every $(\textbf{s}_1,\textbf{s}_2,{\cal C}_1,{\cal C}_2,{\cal C}_c,\textbf{z})\in {\cal E}_{\textbf{X}}^{c}$ and every $\textbf{u}_2\in {\cal S}_{2}\mbox{ and } \textbf{v}\in {\cal S}_{c} $,
\begin{IEEEeqnarray}{rCl} \label{eq:statement_A2_R2+Rc}
&& \left| \bar{\rho }-\cos \sphericalangle \left( \textbf{v},\textbf{u}_{2} \right) \right|\le 7\epsilon  \nonumber \\
&& \text{       }\implies \text{      }\left| \left\| \alpha\textbf{v} \right\|\sqrt{n\bar{{{\beta }_{2}}}{{P}_{2}}}\bar{\rho }-\left\langle \alpha\textbf{v},{{a}_{2,1}}\textbf{u}_{2} \right\rangle  \right|\le 7\epsilon \left\| \alpha\textbf{v} \right\|\sqrt{n\bar{{{\beta }_{2}}}{{P}_{2}}}.
\end{IEEEeqnarray}

Statement A2) follows by rewriting $\cos \sphericalangle \left( \textbf{v},\textbf{u}_{2} \right)$ as ${\left\langle \textbf{v},\textbf{u}_{2} \right\rangle }/({\left\| \textbf{v} \right\|\left\| \textbf{u}_{2}^{*} \right\|})$, and then multiplying the inequality on the l.h.s. of \eqref{eq:statement_A2_R2+Rc} by
$\left\| \alpha\textbf{v} \right\|\cdot \left\| {{a}_{2,1}}\textbf{u}_{2}\right\|$
and recalling that $\left\| {{a}_{2,1}}\textbf{u}_{2} \right\|=\sqrt{n\bar{{{\beta }_{2}}}{{P}_{2}}}$.

\medskip

B) For every $(\textbf{s}_1,\textbf{s}_2,{\cal C}_1,{\cal C}_2,{\cal C}_c,\textbf{z})\in {\cal E}_{\textbf{X}}^{c}\cap {\cal E}_{\textbf{Z}}^{c}$
and every $\textbf{u}_2\in {\cal S}_{2}\mbox{ and } \textbf{v}\in {\cal S}_{c} $
\begin{IEEEeqnarray}{rCl}
&& |\textbf{y}-\textbf{X}_{\textbf{u}_1^{*},\textbf{v},\textbf{u}_2}\|^2
\leq
\|\textbf{y}-\textbf{X}_{\textbf{u}_1^{*},\textbf{v}^{*},\textbf{u}_2^{*}}\|^2  \nonumber \\
&& \implies
\left\langle \textbf{y}-{{a}_{1,1}}\textbf{u}_1^{*} ,{{a}_{2,1}}\textbf{u}_2+\alpha{{\textbf{v}}} \right\rangle
\ge n\left( \bar{{{\beta }_{2}}}{{P}_{2}}+\frac{2}{n}\left\| \alpha\textbf{v} \right\|\sqrt{n\bar{{{\beta }_{2}}}{{P}_{2}}}\bar{\rho }+\frac{1}{n}{{\left\| \alpha{{\textbf{v}}^{*}} \right\|}^{2}}-\kappa\epsilon  \right).
\IEEEeqnarraynumspace
\label{eq:statement_B_R2+Rc}
\end{IEEEeqnarray}

Statement B) follows from rewriting the inequality on the l.h.s. of \eqref{eq:statement_B_R2+Rc}
as
\begin{equation*}
{{\left\|\left( \textbf{y}-{{a}_{1,1}}\textbf{u}_{1}^{*}   \right)-({{a}_{2,1}}\textbf{u}_2+\alpha{{\textbf{v}}})
\right\|}^{2}}\le
 {{\left\|\left( \textbf{y}-{{a}_{1,1}}\textbf{u}_{1}^{*}  \right)-({{a}_{2,1}}\textbf{u}_{1}^{*}+\alpha{{\textbf{v}}^{*}}) \right\|}^{2}},
 \end{equation*}
or equivalently as
\begin{IEEEeqnarray}{rCl} \label{eq:statement_B_proof_R2+Rc}
\left\langle \left(\textbf{y}-{{a}_{1,1}}\textbf{u}_{1}^{*}  \right),
{{a}_{2,1}}\textbf{u}_2+\alpha{{\textbf{v}}} \right\rangle
& \ge &
\left\langle \left( \textbf{y}-{{a}_{1,1}}\textbf{u}_{1}^{*}  \right),{{a}_{2,1}}\textbf{u}_{2}^{*}+\alpha{{\textbf{v}}^{*}} \right\rangle  \nonumber \\
& = & \left\langle {{a}_{2,1}}\textbf{u}_{2}^{*}+\alpha{{\textbf{v}}^{*}}+\textbf{z},
{{a}_{2,1}}\textbf{u}_{2}^{*}+\alpha{{\textbf{v}}^{*}} \right\rangle  \nonumber \\
& = & {{\left\| {{a}_{2,1}}\textbf{u}_{2}^{*}+\alpha{{\textbf{v}}^{*}} \right\|}^{2}}
+\left\langle \textbf{z},{{a}_{2,1}}\textbf{u}_{2}^{*}+\alpha{{\textbf{v}}^{*}} \right\rangle .
\end{IEEEeqnarray}

It now follows from the equivalence of the first inequality in \eqref{eq:statement_B_R2+Rc} with
\eqref{eq:statement_B_proof_R2+Rc} that for $(\textbf{s}_1,\textbf{s}_2,{\cal C}_1,{\cal C}_2,{\cal C}_c,\textbf{z})\in {\cal E}_{\textbf{Z}}^{c}$, the first inequality in \eqref{eq:statement_B_R2+Rc} can only hold if
\begin{equation*}
\left\langle \textbf{y}-{{a}_{1,1}}\textbf{u}_1^{*} ,{{a}_{2,1}}\textbf{u}_2+\alpha{{\textbf{v}}} \right\rangle
 \ge n\left( \bar{{{\beta }_{2}}}{{P}_{2}}+\frac{2}{n}\left\| \alpha\textbf{v} \right\|\sqrt{n\bar{{{\beta }_{2}}}{{P}_{2}}}\bar{\rho }+\frac{1}{n}{{\left\| \alpha{{\textbf{v}}^{*}} \right\|}^{2}}-\kappa\epsilon  \right),
 \end{equation*}
thus establishing B).

\medskip

C) For every $(\textbf{s}_1,\textbf{s}_2,{\cal C}_1,{\cal C}_2,{\cal C}_c,\textbf{z})\in {\cal E}_{\textbf{X}}^{c}\cap {\cal E}_{\textbf{Z}}^{c}$
and every $\textbf{u}_2\in {\cal S}_{2}\mbox{ and } \textbf{v}\in {\cal S}_{c} $,
\begin{IEEEeqnarray*}{rCl}
&& \left| \tilde{\rho }-\cos \sphericalangle \left( \textbf{u}_1^{*},\textbf{u}_{2} \right) \right|\le 7\epsilon \text{   and  }
|\textbf{y}-\textbf{X}_{\textbf{u}_1^{*},\textbf{v},\textbf{u}_2}\|^2
\leq
\|\textbf{y}-\textbf{X}_{\textbf{u}_1^{*},\textbf{v}^{*},\textbf{u}_2^{*}}\|^2 \nonumber \\
&& \qquad \qquad \implies  \\
&&{{\left\| {{a}_{2,1}}\textbf{u}_2+\alpha{{\textbf{v}}}-\textbf{w} \right\|}^{2}}
\le
n\Biggl( \left( \bar{{{\beta }_{2}}}{{P}_{2}}+\frac{2}{n}\left\| \alpha \textbf{v} \right\|\sqrt{n\bar{{{\beta }_{2}}}{{P}_{2}}}\bar{\rho }+\frac{1}{n}{{\left\| \alpha \textbf{v} \right\|}^{2}} \right)\left( 1-2{{\varsigma }_{1}} \right)\\
&& \hspace{3.65cm} -2{{\varsigma }_{2}}\sqrt{\bar{{{\beta }_{1}}}\bar{{{\beta }_{2}}}{{P}_{1}}{{P}_{2}}}\tilde{\rho } \Biggr)+{{\left\| \textbf{w} \right\|}^{2}}+n{{\kappa}'}\epsilon .
\end{IEEEeqnarray*}
Statement C) is obtained as follows:
\begin{IEEEeqnarray*}{rCl}
{{\left\| {{a}_{2,1}}\textbf{u}_2+\alpha {{\textbf{v}}}-\textbf{w} \right\|}^{2}}
& =&{{\left\| {{a}_{2,1}}\textbf{u}_2+\alpha{{\textbf{v}}} \right\|}^{2}}
-2\left\langle {{a}_{2,1}}\textbf{u}_2+\alpha{{\textbf{v}}},\textbf{w} \right\rangle +{{\left\| \textbf{w} \right\|}^{2}} \\
& =&{{\left\| {{a}_{2,1}}\textbf{u}_2+\alpha{{\textbf{v}}} \right\|}^{2}}\\
&& -2\left\langle {{a}_{2,1}}\textbf{u}_2+\alpha{{\textbf{v}}},
{{\varsigma }_{1}}\left( {{a}_{2,1}}\textbf{u}_{2}^{*}+\alpha{{\textbf{v}^{*}}}+\textbf{z} \right)
+{{\varsigma }_{2}}{{a}_{1,1}}\textbf{u}_{1}^{*} \right\rangle +{{\left\| \textbf{w} \right\|}^{2}} \\
& \text{ } \overset{(a)}{\mathop{\le }}&\, n
\Biggl( \left( \bar{{{\beta }_{2}}}{{P}_{2}}+\frac{2}{n}\left\| \alpha\textbf{v} \right\|\sqrt{n\bar{{{\beta }_{2}}}{{P}_{2}}}\bar{\rho }+\frac{1}{n}{{\left\| \alpha \textbf{v} \right\|}^{2}} \right)\left( 1-2{{\varsigma }_{1}} \right)\\
&& -2{{\varsigma }_{2}}\sqrt{\bar{{{\beta }_{1}}}\bar{{{\beta }_{2}}}{{P}_{1}}{{P}_{2}}}\tilde{\rho } \Biggr)+{{\left\| \textbf{w} \right\|}^{2}}+n{{\kappa}'}\epsilon   ,
\end{IEEEeqnarray*}
where in (a) we have used Statement A), A1) and Statement B).

\medskip

D) For every $(\textbf{s}_1,\textbf{s}_2,{\cal C}_1,{\cal C}_2,{\cal C}_c,\textbf{z})\in {\cal E}_{\textbf{X}}^{c}\cap {\cal E}_{\textbf{Z}}^{c}$
\begin{IEEEeqnarray*}{rCl}
{{\left\| \textbf{w} \right\|}^{2}}
& \le & \,n\Biggl( {{\varsigma }_{1}}^{2}\left( \bar{{{\beta }_{2}}}{{P}_{2}}+N \right)+{{\varsigma }_{1}}^{2}\left( \frac{2}{n}\left\| \alpha\textbf{v}^{*} \right\|\sqrt{n\bar{{{\beta }_{2}}}{{P}_{2}}}\bar{\rho }+\frac{1}{n}{{\left\| \alpha{{\textbf{v}}^{*}} \right\|}^{2}} \right)\\
&& +2{{\varsigma }_{1}}{{\varsigma }_{2}} \sqrt{\bar{{{\beta }_{1}}}{{P}_{1}}\bar{{{\beta }_{2}}}{{P}_{2}}}\tilde{\rho } +{{\varsigma }_{2}}^{2}\bar{{{\beta }_{1}}}{{P}_{1}}+\kappa\epsilon  \Biggr),   \end{IEEEeqnarray*}
where $k$ depends on ${{P}_{1}},{{P}_{2}},N,{{\varsigma }_{1}}$ and ${{\varsigma }_{2}}$ only.

\medskip

Statement D) is obtained as follows:
\begin{IEEEeqnarray*}{rCl}
{{\left\| \textbf{w} \right\|}^{2}}
&=&{{\left\| {{\varsigma }_{1}}\left( {{a}_{2,1}}\textbf{u}_{2}^{*}+\alpha{{\textbf{v}}^{*}}+\textbf{z} \right)
+{{\varsigma }_{2}}{{a}_{1,1}}\textbf{u}_{1}^{*} \right\|}^{2}}\\
&=&{{\varsigma }_{1}}^{2}{{\left\| {{a}_{2,1}}\textbf{u}_{2}^{*}+\alpha{{\textbf{v}}^{*}}+\textbf{z} \right\|}^{2}}
+2{{\varsigma }_{1}}{{\varsigma }_{2}}\left\langle
{{a}_{2,1}}\textbf{u}_{2}^{*}+\alpha{{\textbf{v}}^{*}}+\textbf{z},{{a}_{1,1}}\textbf{u}_{1}^{*} \right\rangle
+{{\varsigma }_{2}}^{2}{{\left\| {{a}_{1,1}}\textbf{u}_{1}^{*} \right\|}^{2}} \\
& =&{{\varsigma }_{1}}^{2}\left( {{\left\| {{a}_{2,1}}\textbf{u}_{2}^{*}+\alpha{{\textbf{v}}^{*}} \right\|}^{2}}
+2\left\langle {{a}_{2,1}}\textbf{u}_{2}^{*}+\alpha{{\textbf{v}}^{*}},\textbf{z} \right\rangle +{{\left\| \textbf{z} \right\|}^{2}} \right)\\
&& +2{{\varsigma }_{1}}{{\varsigma }_{2}}\left\langle
{{a}_{2,1}}\textbf{u}_{2}^{*}+\alpha{{\textbf{v}}^{*}}+\textbf{z},{{a}_{1,1}}\textbf{u}_{1}^{*} \right\rangle +{{\varsigma }_{2}}^{2}n\bar{{{\beta }_{1}}}{{P}_{1}} \\
&\text{ } \overset{(a)}{\mathop{\le }}&\,
n\Biggl( {{\varsigma }_{1}}^{2}\left( \bar{{{\beta }_{2}}}{{P}_{2}}+N \right)+{{\varsigma }_{1}}^{2}\left( \frac{2}{n}\left\| \alpha\text{v} \right\|\sqrt{n\bar{{{\beta }_{2}}}{{P}_{2}}}\bar{\rho }+\frac{1}{n}{{\left\| \alpha{{\text{v}}^{*}} \right\|}^{2}} \right)\\
&& +2{{\varsigma }_{1}}{{\varsigma }_{2}} \sqrt{\bar{{{\beta }_{1}}}{{P}_{1}}\bar{{{\beta }_{2}}}{{P}_{2}}}\tilde{\rho } +{{\varsigma }_{2}}^{2}\bar{{{\beta }_{1}}}{{P}_{1}}+k\epsilon  \Biggr),
\end{IEEEeqnarray*}
where in (a) we have used that $(\textbf{s}_1,\textbf{s}_2,{\cal C}_1,{\cal C}_2,{\cal C}_c,\textbf{z})\in {\cal E}_{\textbf{X}}^{c}$, and statements A) and A2).

\medskip

E) For every$(\textbf{s}_1,\textbf{s}_2,{\cal C}_1,{\cal C}_2,{\cal C}_c,\textbf{z})\in {\cal E}_{\textbf{X}}^{c}\cap {\cal E}_{\textbf{Z}}^{c}$ and an arbitrary $\textbf{u}_2\in {\cal S}_{2}\mbox{ and } \textbf{v}\in {\cal S}_{c} $,
\begin{IEEEeqnarray*}{rCl}
&& \left| \tilde{\rho }-\cos \sphericalangle \left( \textbf{u}_1^{*},\textbf{u}_{2} \right) \right|\le 7\epsilon \text{  and  }\left| \bar{\rho }-\cos \sphericalangle \left( \textbf{v},\textbf{u}_{2} \right) \right|\le 7\epsilon
\text{  and  }
|\textbf{y}-\textbf{X}_{\textbf{u}_1^{*},\textbf{v},\textbf{u}_2}\|^2
\leq
\|\textbf{y}-\textbf{X}_{\textbf{u}_1^{*},\textbf{v}^{*},\textbf{u}_2^{*}}\|^2 \\
&& \implies \text{        }{{\left\| {{a}_{2,1}}\textbf{u}_2+\alpha{{\textbf{v}}}-\textbf{w} \right\|}^{2}}\le \Upsilon (\epsilon ),
\end{IEEEeqnarray*}
where
\begin{equation}\label{eq:upsilon_R2+Rc}
\Upsilon (\epsilon )=n\frac{\left( \bar{{{\beta }_{2}}}{{P}_{2}}\left( 1-{{{\tilde{\rho }}}^{2}} \right)+\frac{2}{n}\left\| \alpha\textbf{v} \right\|\sqrt{n\bar{{{\beta }_{2}}}{{P}_{2}}}\bar{\rho }+\frac{1}{n}{{\left\| \alpha\textbf{v} \right\|}^{2}} \right)N}{\bar{{{\beta }_{2}}}{{P}_{2}}\left( 1-{{{\tilde{\rho }}}^{2}} \right)+\frac{2}{n}\left\| \alpha\textbf{v} \right\|\sqrt{n\bar{{{\beta }_{2}}}{{P}_{2}}}\bar{\rho }+\frac{1}{n}{{\left\| \alpha\textbf{v} \right\|}^{2}}+N}+n{{\kappa}'}\epsilon ,
\end{equation}
and where ${{\kappa}'}$ only depends on ${{P}_{1}},{{N}_{1}},{{N}_{2}},{{\varsigma }_{1}}$  and $ {{\varsigma }_{2}}$.

\medskip

Statement E) follows from combining Statement C) with Statement
D) and the explicit values of ${{\varsigma }_{1}}$  and $ {{\varsigma }_{2}}$ given in \eqref{eq:w_for_R2+Rc}.

\medskip

F) For every $\textbf{u}_2\in {\cal S}_{2},\textbf{v}\in {\cal S}_{c}$, denote by $\varphi \in \left[ 0,\pi  \right]$ the angle between ${{a}_{2,1}}\textbf{u}_2+\alpha{{\textbf{v}}}$ and \textbf{w}, and let
\begin{IEEEeqnarray*}{rCl}
{\cal B}(\textbf{s}_1,\textbf{s}_2,\textbf{u}_1^{*},\textbf{v}^{*},\textbf{u}_2^{*},\textbf{z}) \triangleq
&& \Biggl\{\textbf{u}_2\in S_{2}^{\left( n \right)},\textbf{v}\in S_{c}^{\left( n \right)}: \\
&& \quad \cos \left( \varphi  \right)\ge \sqrt{1-\frac{\Upsilon (\epsilon )}{n\left( \bar{{{\beta }_{2}}}{{P}_{2}}+\frac{2}{n}\left\| \alpha\textbf{v} \right\|\sqrt{n\bar{{{\beta }_{2}}}{{P}_{2}}}\bar{\rho }+\frac{1}{n}{{\left\| \alpha\textbf{v} \right\|}^{2}} \right)}}\Biggr\},
\end{IEEEeqnarray*}
where $\epsilon$ is sufficiently small such that the term in the square is non-negative. \\
Then, for every $(\textbf{s}_1,\textbf{s}_2,{\cal C}_1,{\cal C}_2,{\cal C}_c,\textbf{z})\in {\cal E}_{\textbf{X}}^{c}\cap {\cal E}_{\textbf{Z}}^{c}$,
\begin{IEEEeqnarray}{rCl}\label{eq:statement_F_R2+Rc}
&& \left| \tilde{\rho }-\cos \sphericalangle \left( \textbf{u}_1^{*},\textbf{u}_{2}\right) \right|\le 7\epsilon \ \quad \mbox{  and  }\quad \left| \bar{\rho }-\cos \sphericalangle \left( \textbf{v},\textbf{u}_{2} \right) \right|\le 7\epsilon \quad \mbox{  and  } \nonumber \\
&& |\textbf{y}-\textbf{X}_{\textbf{u}_1^{*},\textbf{v},\textbf{u}_2}\|^2
\leq
\|\textbf{y}-\textbf{X}_{\textbf{u}_1^{*},\textbf{v}^{*},\textbf{u}_2^{*}}\|^2 \nonumber \\
&& \qquad \implies \text{    }{{a}_{2,1}}\textbf{u}_2+\alpha{{\textbf{v}}}\in
{\cal B}(\textbf{s}_1,\textbf{s}_2,\textbf{u}_1^{*},\textbf{v}^{*},\textbf{u}_2^{*},\textbf{z}) .
\end{IEEEeqnarray}

\medskip

Statement F) follows from Statement E) by noting that if $\mathbf{w}\ne 0$ and
$$ 1-\frac{\Upsilon (\epsilon )}{n\left( \bar{{{\beta }_{2}}}{{P}_{2}}+\frac{2}{n}\left\| \alpha\textbf{v} \right\|
\sqrt{n\bar{{{\beta }_{2}}}{{P}_{2}}}\bar{\rho }+\frac{1}{n}{{\left\| \alpha\textbf{v} \right\|}^{2}} \right)}>0,$$ then
\begin{IEEEeqnarray*}{l}
 {{\left\| {{a}_{2,1}}\textbf{u}_2+\alpha{{\textbf{v}}} \right\|}^{2}}=
{n\left( \bar{{{\beta }_{2}}}{{P}_{2}}+\frac{2}{n}\left\| \alpha\textbf{v} \right\|\sqrt{n\bar{{{\beta }_{2}}}{{P}_{2}}}\bar{\rho }+\frac{1}{n}{{\left\| \alpha\textbf{v} \right\|}^{2}} \right)}
\nonumber \\
\hspace{0.5cm} \mbox{ and } \
{{\left\| {{a}_{2,1}}\textbf{u}_2+\alpha{{\textbf{v}}}-\textbf{w} \right\|}^{2}}\le \Upsilon (\epsilon ) \nonumber \\
  \qquad \implies \cos \sphericalangle \left( {{a}_{2,1}}\textbf{u}_2+\alpha{{\textbf{v}}},\textbf{w} \right)\ge \sqrt{1-\frac{\Upsilon (\epsilon )}{n\left( \bar{{{\beta }_{2}}}{{P}_{2}}+\frac{2}{n}\left\| \alpha\textbf{v} \right\|\sqrt{n\bar{{{\beta }_{2}}}{{P}_{2}}}\bar{\rho }+\frac{1}{n}{{\left\| \alpha\textbf{v} \right\|}^{2}} \right)}},
\end{IEEEeqnarray*}
which follows by the same argument as \eqref{eq:statement_F_proof_R1+R2}.

\medskip

The proof of Lemma~\ref{eps_h_u2,v} is now concluded by noticing that the set
${{\cal E }'}_{({\hat{U}}_{2},{\hat{V}})}$, defined in \eqref{eq:eps_h_u2,v},
is the set of tuples
$(\textbf{s}_1,\textbf{s}_2,{\cal C}_1,{\cal C}_2,{\cal C}_c,\textbf{z})$
for which there exists a ${{\textbf{u}}_{2}}\left( j \right)\in {{{\cal C}}_{2}}\backslash \left\{ \textbf{u}_{1}^{*} \right\}
\mbox{  and  } {{\textbf{v}}}\left( l \right)\in {{{\cal C}}_{c}}\backslash \left\{ \textbf{v}^{*} \right\}$
such that \\
${{a}_{2,1}}{\textbf{u}}_{2}\left( j \right)+\alpha\textbf{v}\left( l \right)\in
B(\textbf{s}_1,\textbf{s}_2,\textbf{u}_1^{*},\textbf{v}^{*},\textbf{u}_2^{*},\textbf{z})$.
Thus, by Statement F) and by the definition of ${\cal E}_{({\hat{\textbf{U}}}_{2},{\hat{\textbf{V}}})}$
in \eqref{eq:eps_u2,v} it follows that
\begin{equation*}
{\cal E}_{({\hat{\textbf{U}}}_{2},{\hat{\textbf{V}}})}\cap {\cal E}_{\textbf{S}}^{c}\cap {\cal E}_{\textbf{X}}^{c}\cap {\cal E}_{\textbf{Z}}^{c}\subseteq {\cal E}'_{({\hat{\textbf{U}}}_{2},{\hat{\textbf{V}}})}\cap {\cal E}_{\textbf{S}}^{c}\cap {\cal E}_{\textbf{X}}^{c}\cap {\cal E}_{\textbf{Z}}^{c},
\end{equation*}
and therefore
\begin{flalign*}
&&
\Pr\left[ {{\cal E}_{({\hat{\textbf{U}}}_{2},{\hat{\textbf{V}}})}}\cap {\cal E}_{\textbf{S}}^{c}\cap {\cal E}_{\textbf{X}}^{c}\cap {\cal E}_{\textbf{Z}}^{c} \right]\le \Pr\left[ {\cal E}'_{({\hat{\textbf{U}}}_{2},{\hat{\textbf{V}}})}\cap {\cal E}_{\textbf{S}}^{c}\cap {\cal E}_{\textbf{X}}^{c}\cap {\cal E}_{\textbf{Z}}^{c} \right].
&&
\end{flalign*}
\end{IEEEproof}

Next,
\begin{IEEEeqnarray}{rCl} \label{eq:R2+Rc_proof}
\Pr\left[ {\cal E}_{({\hat{\textbf{U}}}_{2},{\hat{\textbf{V}}})}\cap {\cal E}_{\textbf{S}}^{c}\cap {\cal E}_{\textbf{X}}^{c}\cap {\cal E}_{\textbf{Z}}^{c} \right]
& \overset{(a)}{\mathop{\le }}&\,
\Pr\left[ {\cal E}'_{({\hat{\textbf{U}}}_{2},{\hat{\textbf{V}}})}\cap {\cal E}_{\textbf{S}}^{c}\cap {\cal E}_{\textbf{X}}^{c}\cap {\cal E}_{\textbf{Z}}^{c} \right] \nonumber\\
& \overset{(b)}{\mathop{\le }}&\,\Pr\left[ {\cal E}'_{({\hat{\textbf{U}}}_{2},{\hat{\textbf{V}}})}|
{\cal E}_{{\textbf{X}}_{2}}^{c}\cap {\cal E}_{{\textbf{X}}_{v}}^{c} \right],
\end{IEEEeqnarray}
where (a) follows by Lemma~\ref{eps_h_u2,v} and (b) follows because ${\cal E}_{\textbf{X}}^{c}\subseteq
{\cal E}_{{\textbf{X}}_{2}}^{c}\cap {\cal E}_{{\textbf{X}}_{v}}^{c}$. \\
The proof of \eqref{eq:R2+Rc} is now completed by combining  \eqref{eq:R2+Rc_proof} with Lemma~\ref{two-Rates proof}.
This gives that for every $\delta >0$ and every $\epsilon >0$ there exists some
$n'(\delta ,\epsilon )$
such that for all $n>n'(\delta ,\epsilon )$,
we have
\begin{equation*}
\Pr\left[ {\cal E}_{({\hat{\textbf{U}}}_{2},{\hat{\textbf{V}}})}\cap {\cal E}_{\textbf{S}}^{c}\cap {\cal E}_{\textbf{X}}^{c}\cap {\cal E}_{\textbf{Z}}^{c} \right]\le
\Pr\left[ {\cal E}'_{({\hat{\textbf{U}}}_{2},{\hat{\textbf{V}}})}|{\cal E}_{{\textbf{X}}_{2}}^{c}\cap {\cal E}_{{\textbf{X}}_{v}}^{c} \right]<\delta ,
\end{equation*}
whenever
\begin{equation*}
{{R}_{2}}+{{R}_\textnormal{c}}<\frac{1}{2}\log \left( \frac{{{\lambda }_{2c}}-\bar{{{\beta }_{2}}}{{P}_{2}}{{{\tilde{\rho }}}^{2}}+N}{\left( 1-\bar{{{\beta }_{2}}}{{P}_{2}}{{{\tilde{\rho }}}^{2}}{{\lambda }_{2c}}^{-1} \right)N\left( 1-{{{\bar{\rho }}}^{2}}\  \right)}-{{\kappa}_{6}}\epsilon  \right),
\end{equation*}
where ${{\kappa}_{6}}$ is a positive constant determined by ${{P}_{1}},{{P}_{2}},N,{{\varsigma }_{1}}\text{ and }{{\varsigma }_{2}}$. \hfill \qed

\vskip.2truein

\subsection{Proof of rate constraint \eqref{eq:R1+R2+Rc}}

\begin{lemma}\label{eps_h_u1,u2,v}
For every sufficiently small $\epsilon >0$, define the set
${\cal E}'_{\left( {{{\hat{\mathbf{U}}}}_{1}},{{{\hat{\mathbf{U}}}}_{2}},\hat{\mathbf{V}} \right)}$
as
\begin{IEEEeqnarray*}{rCl}
&&{\cal E}'_{\left( {{{\hat{\mathbf{U}}}}_{1}},{{{\hat{\mathbf{U}}}}_{2}},\hat{\mathbf{V}} \right)}\triangleq
\Bigl\{(\mathbf{s}_1,\mathbf{s}_2,{\cal C}_1,{\cal C}_2,{\cal C}_c,\mathbf{z}):
\exists \> {\mathbf{u}_{1}}\left( j \right)\in {{{\cal C}}_{1}}\backslash \left\{ \mathbf{u}_{1}^{*} \right\}\ \mbox{and}
\ \exists \> {\mathbf{u}_{2}}\left( l \right)\in {{{\cal C}}_{2}}\backslash \left\{ \mathbf{u}_{2}^{*} \right\} \\
&& \hspace{0.1cm} \mbox{and } \ \exists \> \mathbf{v}(k)\in {{{\cal C}}_{c}}\backslash \left\{ {\mathbf{v}^{*}} \right\}\mbox{ s.t. }  \cos \sphericalangle \left( {\mathbf{u}_{1}}\left( j \right),{\mathbf{u}_{2}}\left( l \right) \right)\ge \tilde{\rho }-7\epsilon \ \mbox{and}\ \cos \sphericalangle \left( {\mathbf{u}_{1}}\left( j \right),\mathbf{v}(k) \right)\ge -3\epsilon \  \\
&& \hspace{0.1cm} \mbox{and }\ \cos \sphericalangle \left( \mathbf{v}(k),{\mathbf{u}_{2}}\left( l \right) \right)\ge \bar{\rho }-7\epsilon
\mbox{ and }\ \cos \sphericalangle \left( \mathbf{y},{{a}_{1,1}}{\mathbf{u}_{1}}\left( j \right)+{{a}_{2,1}}{\mathbf{u}_{2}}\left( l \right)+\alpha\mathbf{v}(k) \right)\ge \Lambda (\epsilon )
\Bigr\},
\end{IEEEeqnarray*}
where
\begin{equation*}
\Lambda (\epsilon )=\sqrt{\frac{{{\bar{\beta }}_{1}}{{P}_{1}}+2\sqrt{{{\bar{\beta }}_{1}}{{P}_{1}}{{\bar{\beta }}_{2}}{{P}_{2}}}\tilde{\rho }+{{\bar{\beta }}_{2}}{{P}_{2}}+2\frac{1}{n}\left\| \alpha{\mathbf{v}^{*}} \right\|\sqrt{n\bar{{{\beta }_{2}}}{{P}_{2}}}\bar{\rho }+\frac{1}{n}{{\left\| \alpha{\mathbf{v}^{*}} \right\|}^{2}}-{{\xi }'}\epsilon }{{{\bar{\beta }}_{1}}{{P}_{1}}+2\sqrt{{{\bar{\beta }}_{1}}{{P}_{1}}{{\bar{\beta }}_{2}}{{P}_{2}}}\tilde{\rho }+{{\bar{\beta }}_{2}}{{P}_{2}}+\frac{2}{n}\left\| \alpha{\mathbf{v}^{*}} \right\|\sqrt{n\bar{{{\beta }_{2}}}{{P}_{2}}}\bar{\rho }+\frac{1}{n}{{\left\| \alpha{\mathbf{v}^{*}} \right\|}^{2}}+N+{{\xi }_{2}}\epsilon }},
\end{equation*}
and where ${{\xi }'} \mbox{ and } {{\xi }_{2}}$  depend only on $P_1, P_2 \mbox{ and } N$. Then, for sufficiently small $\epsilon >0$
\begin{equation*}
{{\cal E}_{\left( {{{\hat{\mathbf{U}}}}_{1}},{{{\hat{\mathbf{U}}}}_{2}},\hat{\mathbf{V}} \right)}}\cap {\cal E}_{\mathbf{S}}^{c}\cap {\cal E}_{\mathbf{X}}^{c}\cap {\cal E}_{\mathbf{Z}}^{c}
\subseteq
{\cal E}'_{\left( {{{\hat{\mathbf{U}}}}_{1}},{{{\hat{\mathbf{U}}}}_{2}},\hat{\mathbf{V}} \right)}\cap
{\cal E}_{\mathbf{S}}^{c}\cap {\cal E}_{\mathbf{X}}^{c}\cap {\cal E}_{\mathbf{Z}}^{c},
\end{equation*}
and, in particular
\begin{equation*}
\Pr\left[ {{\cal E}_{\left( {{{\hat{\mathbf{U}}}}_{1}},{{{\hat{\mathbf{U}}}}_{2}},\hat{\mathbf{V}} \right)}}\cap
{\cal E}_{\mathbf{S}}^{c}\cap {\cal E}_{\mathbf{X}}^{c}\cap {\cal E}_{\mathbf{Z}}^{c} \right]
\le
\Pr \left( {\cal E}'_{\left( {{{\hat{\mathbf{U}}}}_{1}},{{{\hat{\mathbf{U}}}}_{2}},\hat{\mathbf{V}} \right)}\cap {\cal E}_{\mathbf{S}}^{c}\cap {\cal E}_{\mathbf{X}}^{c}\cap {\cal E}_{\mathbf{Z}}^{c} \right).
\end{equation*}
\end{lemma}

\medskip

\begin{IEEEproof}
We first recall that for the event ${{\cal E}_{\left( {{{\hat{U}}}_{1}},{{{\hat{U}}}_{2}},\hat{V} \right)}}$
to occur, there must exist codewords
$ {\textbf{u}_{1}}\left( j \right)\in {{{\cal C}}_{1}}\backslash \left\{ \textbf{u}_{1}^{*} \right\}$
and
${\textbf{u}_{2}}\left( l \right)\in {{{\cal C}}_{2}}\backslash \left\{ \textbf{u}_{2}^{*} \right\}$
and
$\textbf{v}(k)\in {{{\cal C}}_{c}}\backslash \left\{ {\textbf{v}^{*}} \right\}$
such that the following inequalities are simultaneously satisfied
\begin{IEEEeqnarray*}{rCl}
\left| \tilde{\rho }-\cos \sphericalangle \left( {\textbf{u}_{1}}\left( j \right),{\textbf{u}_{2}}\left( l \right) \right) \right| & \le & 7\epsilon \nonumber \\
\left| \cos \sphericalangle \left( \textbf{v}(k),{\textbf{u}_{1}}\left( j \right) \right) \right| & \le &  3\epsilon  \nonumber \\
\left| \bar{\rho }-\cos \sphericalangle \left( \textbf{v}(k),{\textbf{u}_{2}}\left( l \right) \right) \right|  & \le & 7\epsilon  \nonumber \\
|\textbf{y}-\textbf{X}_{\textbf{u}_1\left( j \right),\textbf{v}\left( k \right),\textbf{u}_2\left( l \right)}\|^2
& \leq  &
\|\textbf{y}-\textbf{X}_{\textbf{u}_1^{*},\textbf{v}^{*},\textbf{u}_2^{*}}\|^2 .
\end{IEEEeqnarray*}


The proof is now based on a sequence of statements related to these conditions.

\medskip

A) For every $(\textbf{s}_1,\textbf{s}_2,{\cal C}_1,{\cal C}_2,{\cal C}_c,\textbf{z})\in {\cal E}_{\textbf{X}}^{c}\bigcap
{\cal E}_{\textbf{Z}}^{c}$,
\begin{IEEEeqnarray}{rCl} \label{eq:statement_A_R1+R2+Rc}
&&|\textbf{y}-\textbf{X}_{\textbf{u}_1\left( j \right),\textbf{v}\left( k \right),\textbf{u}_2\left( l \right)}\|^2
 \leq
\|\textbf{y}-\textbf{X}_{\textbf{u}_1^{*},\textbf{v}^{*},\textbf{u}_2^{*}}\|^2 \nonumber \\
&& \qquad \implies  \nonumber \\
&& \left\langle \textbf{y},{{a}_{1}}{\textbf{u}_{1}}(j)+{{a}_{2}}{\textbf{u}_{2}}(l)+\alpha\textbf{v}(k) \right\rangle \nonumber \\
&& \quad \ge n\left( {{\bar{\beta }}_{1}}{{P}_{1}}+2\sqrt{{{\bar{\beta }}_{1}}{{P}_{1}}{{\bar{\beta }}_{2}}{{P}_{2}}}\tilde{\rho }+{{\bar{\beta }}_{2}}{{P}_{2}}+2\frac{1}{n}\left\| \alpha{\textbf{v}^{*}} \right\|\sqrt{n\bar{{{\beta }_{2}}}{{P}_{2}}}\bar{\rho }+\frac{1}{n}{{\left\| \alpha{\textbf{v}^{*}} \right\|}^{2}}-{{\xi }_{1}}\epsilon  \right),
\IEEEeqnarraynumspace
\end{IEEEeqnarray}
where ${{\xi }_{1}}$ depends only on ${{P}_{1}},{{P}_{2}} \mbox{  and  } N$.

Statement A) follows by rewriting the l.h.s. of \eqref{eq:statement_A_R1+R2+Rc} as
\begin{IEEEeqnarray}{rCl}
\IEEEeqnarraymulticol{3}{l}{
2\left\langle y,{{a}_{1,1}}{\textbf{u}_{1}}(j)+{{a}_{2,1}}{\textbf{u}_{2}}(l)+\alpha\textbf{v}(k) \right\rangle  }\nonumber \\ \quad
& \ge & 2\left\langle \textbf{y},{{a}_{1,1}}\textbf{u}_{1}^{*}+{{a}_{2,1}}\textbf{u}_{2}^{*}+\alpha{\textbf{v}^{*}} \right\rangle +{{\left\| {{a}_{1,1}}{\textbf{u}_{1}}\left( j \right)+{{a}_{2,1}}{\textbf{u}_{2}}(l)+\alpha\textbf{v}(k) \right\|}^{2}} \nonumber \\
&& -{{\left\| {{a}_{1,1}}\textbf{u}_{1}^{*}+{{a}_{2,1}}\textbf{u}_{2}^{*}+\alpha{\textbf{v}^{*}} \right\|}^{2}} \nonumber \nonumber \\
& = & {{\left\| {{a}_{1,1}}\textbf{u}_{1}^{*}+{{a}_{2,1}}\textbf{u}_{2}^{*}+\alpha{\textbf{v}^{*}} \right\|}^{2}}+2\left\langle \textbf{z},{{a}_{1,1}}\textbf{u}_{1}^{*}+{{a}_{2,1}}\textbf{u}_{2}^{*}+\alpha{\textbf{v}^{*}} \right\rangle \nonumber \\
&& +{{\left\| {{a}_{1,1}}{\textbf{u}_{1}}\left( j \right)+{{a}_{2,1}}{\textbf{u}_{2}}(l)+\alpha\textbf{v}(k) \right\|}^{2}} \nonumber \\
& \overset{(a)}{\mathop{\ge }}& \,2n\left( {{\bar{\beta }}_{1}}{{P}_{1}}+2\sqrt{{{\bar{\beta }}_{1}}{{P}_{1}}{{\bar{\beta }}_{2}}{{P}_{2}}}\tilde{\rho }+{{\bar{\beta }}_{2}}{{P}_{2}}+2\frac{1}{n}\left\| \alpha{\textbf{v}^{*}} \right\|\sqrt{n\bar{{{\beta }_{2}}}{{P}_{2}}}\bar{\rho }+\frac{1}{n}{{\left\| \alpha{\textbf{v}^{*}} \right\|}^{2}}-{{\xi }_{1}}\epsilon  \right),
\IEEEeqnarraynumspace
\end{IEEEeqnarray}
where in (a) we have used that $(\textbf{s}_1,\textbf{s}_2,{\cal C}_1,{\cal C}_2,{\cal C}_c,\textbf{z})\in {\cal E}_{\textbf{X}}^{c}\bigcap {\cal E}_{\textbf{Z}}^{c}$.

\medskip

B) For every $(\textbf{s}_1,\textbf{s}_2,{\cal C}_1,{\cal C}_2,{\cal C}_c,\textbf{z})\in {\cal E}_{\textbf{X}}^{c}\bigcap
{\cal E}_{\textbf{Z}}^{c}$,
\begin{equation*}
{{\left\| \textbf{y} \right\|}^{2}}\le n\left( {{\bar{\beta }}_{1}}{{P}_{1}}+2\sqrt{{{\bar{\beta }}_{1}}{{P}_{1}}{{\bar{\beta }}_{2}}{{P}_{2}}}\tilde{\rho }+{{\bar{\beta }}_{2}}{{P}_{2}}+\frac{2}{n}\left\| \alpha{\textbf{v}^{*}} \right\|\sqrt{n\bar{{{\beta }_{2}}}{{P}_{2}}}\bar{\rho }+\frac{1}{n}{{\left\| \alpha{\textbf{v}^{*}} \right\|}^{2}}+N+{{\xi }_{2}}\epsilon  \right),
\end{equation*}
where  ${{\xi }_{2}}$ depends only on ${{P}_{1}},{{P}_{2}}\mbox{  and  } N$.

\medskip

Statement B) is obtained as follows:
\begin{IEEEeqnarray*}{rCl}
{{\left\| \textbf{y} \right\|}^{2}}
& = & {{\left\| {{a}_{1,1}}\textbf{u}_{1}^{*}+{{a}_{2,1}}\textbf{u}_{2}^{*}+\alpha{\textbf{v}^{*}}+\mathbf{z} \right\|}^{2}}\\
&=& {{\left\| {{a}_{1,1}}\textbf{u}_{1}^{*} \right\|}^{2}}+2\left\langle {{a}_{1,1}}\textbf{u}_{1}^{*},{{a}_{2,1}}\textbf{u}_{2}^{*} \right\rangle +{{\left\| {{a}_{2,1}}\textbf{u}_{2}^{*} \right\|}^{2}}+2\left\langle {{a}_{2,1}}\textbf{u}_{2}^{*},\alpha{\textbf{v}^{*}} \right\rangle +{{\left\| \alpha{\textbf{v}^{*}} \right\|}^{2}}\\
&& +2\left( \left\langle {{a}_{1,1}}\textbf{u}_{1}^{*},\textbf{z} \right\rangle +\left\langle {{a}_{2,1}}\textbf{u}_{2}^{*},\textbf{z} \right\rangle +\left\langle \alpha{\textbf{v}^{*}},\textbf{z} \right\rangle  \right)+{{\left\| \textbf{z} \right\|}^{2}} \\
& \ \overset{(a)}{\mathop{\le }} & \,n{{\bar{\beta }}_{1}}{{P}_{1}}+2n\sqrt{{{\bar{\beta }}_{1}}{{P}_{1}}{{\bar{\beta }}_{2}}{{P}_{2}}}(\tilde{\rho }+7\epsilon )+{{\bar{\beta }}_{2}}{{P}_{2}}+2\left\| \alpha{\textbf{v}^{*}} \right\|\sqrt{n\bar{{{\beta }_{2}}}{{P}_{2}}}\left( \bar{\rho }+7\epsilon  \right)+{{\left\| \alpha{\textbf{v}^{*}} \right\|}^{2}}\\
&& +2n\left( \sqrt{\bar{{{\beta }_{1}}}{{P}_{1}}N}\epsilon +\sqrt{\bar{{{\beta }_{2}}}{{P}_{2}}N}\epsilon +\left\| \alpha{\textbf{v}^{*}} \right\|\sqrt{nN}\epsilon  \right)+nN\left( 1+\epsilon  \right) \\
& = & n\left( {{\bar{\beta }}_{1}}{{P}_{1}}+2\sqrt{{{\bar{\beta }}_{1}}{{P}_{1}}{{\bar{\beta }}_{2}}{{P}_{2}}}\tilde{\rho }+{{\bar{\beta }}_{2}}{{P}_{2}}+\frac{2}{n}\left\| \alpha{\textbf{v}^{*}} \right\|\sqrt{n\bar{{{\beta }_{2}}}{{P}_{2}}}\bar{\rho }+\frac{1}{n}{{\left\| \alpha{\textbf{v}^{*}} \right\|}^{2}}+N+{{\xi }_{2}}\epsilon  \right),
\end{IEEEeqnarray*}
where in (a) we have used that $(\textbf{s}_1,\textbf{s}_2,{\cal C}_1,{\cal C}_2,{\cal C}_c,\textbf{z})\in {\cal E}_{\textbf{X}}^{c}\bigcap {\cal E}_{\textbf{Z}}^{c}$.

\medskip

C) For every $(\textbf{s}_1,\textbf{s}_2,{\cal C}_1,{\cal C}_2,{\cal C}_c,\textbf{z})$,
\begin{IEEEeqnarray}{l}
\left| \tilde{\rho }-\left\langle \frac{{\textbf{u}_{1}}(j)}{\left\| {\textbf{u}_{1}}(j) \right\|},\frac{{\textbf{u}_{2}}(l)}{\left\| {\textbf{u}_{2}}(l) \right\|} \right\rangle  \right|<7\epsilon
\mbox{ and } \left| \bar{\rho }-\left\langle \frac{{\textbf{u}_{2}}(l)}{\left\| {\textbf{u}_{2}}(l) \right\|},\frac{\textbf{v}(k)}{\left\| \textbf{v}(k) \right\|} \right\rangle  \right|<7\epsilon  \nonumber \\
\mbox{ and } \left| \left\langle \frac{{\textbf{u}_{1}}(j)}{\left\| {\textbf{u}_{1}}(j) \right\|},\frac{{\textbf{v}}(k)}{\left\| {\textbf{v}}(k) \right\|} \right\rangle  \right|<3\epsilon    \nonumber \\
\hspace{0.5cm} \implies 
{{\left\| {{a}_{1,1}}{\textbf{u}_{1}}(j)+{{a}_{2,1}}{\textbf{u}_{2}}(l)+\alpha\textbf{v}(k) \right\|}^{2}} \nonumber \\
\quad \le n\left( {{\bar{\beta }}_{1}}{{P}_{1}}+2\sqrt{{{\bar{\beta }}_{1}}{{P}_{1}}\bar{{{\beta }_{2}}}{{P}_{2}}}\tilde{\rho }+\bar{{{\beta }_{2}}}{{P}_{2}}+2\frac{1}{n}\left\| \alpha{\textbf{v}^{*}} \right\|\sqrt{n\bar{{{\beta }_{2}}}{{P}_{2}}}\bar{\rho }+\frac{1}{n}{{\left\| \alpha\textbf{v}(k) \right\|}^{2}}+{{\xi }_{3}}\epsilon  \right). \IEEEeqnarraynumspace
\label{eq:statement_C_R1+R2+Rc}
\end{IEEEeqnarray}

Statement C) follows by
\begin{IEEEeqnarray*}{rCl}
\IEEEeqnarraymulticol{3}{l}{
{{\left\| {{a}_{1,1}}{{u}_{1}}(j)+{{a}_{2,1}}{{u}_{2}}(l)+\alpha\textbf{v}(k) \right\|}^{2}}} \\ \quad
& =&{{\left\| {{a}_{1,1}}{{u}_{1}}(j) \right\|}^{2}}+2\left\langle {{a}_{1,1}}{\textbf{u}_{1}}(j),{{a}_{2,1}}{\textbf{u}_{2}}(l) \right\rangle +{{\left\| {{a}_{2,1}}{\textbf{u}_{2}}(l) \right\|}^{2}}+2\left\langle {{a}_{2,1}}{\textbf{u}_{2}}(l),\alpha\textbf{v}(k) \right\rangle +{{\left\| \alpha\textbf{v}(k) \right\|}^{2}} \\
& \overset{(a)}{\mathop{\le }}&\,n{{\bar{\beta }}_{1}}{{P}_{1}}+2n\sqrt{{{\bar{\beta }}_{1}}{{P}_{1}}\bar{{{\beta }_{2}}}{{P}_{2}}}(\tilde{\rho }+7\epsilon )+n\bar{{{\beta }_{2}}}{{P}_{2}}+2\left\| \alpha{\textbf{v}^{*}} \right\|\sqrt{n\bar{{{\beta }_{2}}}{{P}_{2}}}\left( \bar{\rho }+7\epsilon  \right)+{{\left\| \alpha\textbf{v}(k) \right\|}^{2}} \\
&=&n\left( {{\bar{\beta }}_{1}}{{P}_{1}}+2\sqrt{{{\bar{\beta }}_{1}}{{P}_{1}}\bar{{{\beta }_{2}}}{{P}_{2}}}\tilde{\rho }+\bar{{{\beta }_{2}}}{{P}_{2}}+\frac{2}{n}\left\| \alpha{\textbf{v}^{*}} \right\|\sqrt{n\bar{{{\beta }_{2}}}{{P}_{2}}}\bar{\rho }+\frac{1}{n}{{\left\| \alpha\textbf{v}(k) \right\|}^{2}}+{{\xi }_{3}}\epsilon  \right),
\end{IEEEeqnarray*}
where in (a) we have used that multiplying the first inequality on the l.h.s. of
\eqref{eq:statement_C_R1+R2+Rc} by $\left\| {{a}_{1,1}}{\textbf{u}_{1}}(j) \right\|\cdot \left\| {{a}_{2,1}}{\textbf{u}_{2}}(l) \right\|$ and recalling that $\left\| {{a}_{1,1}}{\textbf{u}_{1}}(j) \right\|\le \sqrt{n\bar{{{\beta }_{1}}}{{P}_{1}}}\mbox{  and that  }\left\| {{a}_{2,1}}{\textbf{u}_{2}}(l) \right\|\le \sqrt{n\bar{{{\beta }_{2}}}{{P}_{2}}}$
gives
\begin{equation*}
\left| n\tilde{\rho }\sqrt{{{\bar{\beta }}_{1}}{{P}_{1}}\bar{{{\beta }_{2}}}{{P}_{2}}}-\left\langle {{a}_{1}}{\textbf{u}_{1}}(j),{{a}_{2}}{\textbf{u}_{2}}(l) \right\rangle  \right|<7n\sqrt{{{\bar{\beta }}_{1}}{{P}_{1}}\bar{{{\beta }_{2}}}{{P}_{2}}}\epsilon ,
\end{equation*}
and thus
\begin{equation*}
n\sqrt{{{\bar{\beta }}_{1}}{{P}_{1}}\bar{{{\beta }_{2}}}{{P}_{2}}}\left( \tilde{\rho }-7\epsilon  \right)< \left\langle {{a}_{1,1}}{\textbf{u}_{1}}(j),{{a}_{2,1}}{\textbf{u}_{2}}(l) \right\rangle < n\sqrt{{{\bar{\beta }}_{1}}{{P}_{1}}\bar{{{\beta }_{2}}}{{P}_{2}}}\left( \tilde{\rho }+7\epsilon  \right).
\end{equation*}
In a similar manner, we have used that  multiplying the second inequality on the l.h.s. of \eqref{eq:statement_C_R1+R2+Rc} by
$\left\| {{a}_{2,1}}{\textbf{u}_{2}}(l) \right\|\cdot \left\| \alpha\textbf{v}(k) \right\|$
gives
\begin{equation*}
\left| \sqrt{n\bar{{{\beta }_{2}}}{{P}_{2}}}\bar{\rho }\left\| \alpha\textbf{v}(k) \right\|-\left\langle {{a}_{2}}{\textbf{u}_{2}}(l),\alpha\textbf{v}(k) \right\rangle  \right|<7\sqrt{n\bar{{{\beta }_{2}}}{{P}_{2}}}\left\| \alpha\textbf{v}(k) \right\|\epsilon ,
\end{equation*}
and thus
\begin{equation*}
\sqrt{n\bar{{{\beta }_{2}}}{{P}_{2}}}\left\| \alpha\textbf{v}(k) \right\|\left( \bar{\rho }-7\epsilon  \right)<\left\langle {{a}_{2}}{\textbf{u}_{2}}(l),\alpha\textbf{v}(k) \right\rangle <\sqrt{n\bar{{{\beta }_{2}}}{{P}_{2}}}\left\| \alpha\textbf{v}(k) \right\|\left( \bar{\rho }+7\epsilon  \right),
\end{equation*}
thus establishing C).

\medskip

D) For every $(\textbf{s}_1,\textbf{s}_2,{\cal C}_1,{\cal C}_2,{\cal C}_c,\textbf{z})\in {\cal E}_{\textbf{X}}^{c}\bigcap
{\cal E}_{\textbf{Z}}^{c}$,
\begin{IEEEeqnarray*}{rCl}
&& \Bigl( \left| \tilde{\rho }-\cos \sphericalangle \left( {\textbf{u}_{1}}\left( j \right),{\textbf{u}_{2}}\left( l \right) \right) \right|\le 7\epsilon
\quad \mbox{  and  }\quad
\left| \cos \sphericalangle \left( \textbf{v}(k),{\textbf{u}_{1}}\left( j \right) \right) \right|\le 7\epsilon \nonumber \\
&& \quad \mbox{  and  }\quad \left| \bar{\rho }-\cos \sphericalangle \left( \textbf{v}(k),{\textbf{u}_{2}}\left( l \right) \right) \right|\le 7\epsilon
\quad \mbox{  and  } \quad
|\textbf{y}-\textbf{X}_{\textbf{u}_1\left( j \right),\textbf{v}\left( k \right),\textbf{u}_2\left( l \right)}\|^2
\leq
\|\textbf{y}-\textbf{X}_{\textbf{u}_1^{*},\textbf{v}^{*},\textbf{u}_2^{*}}\|^2 \Bigr) \nonumber \\
&&\qquad \implies \quad
\cos \sphericalangle \left( \textbf{y},{{a}_{1,1}}{\textbf{u}_{1}}(j)+{{a}_{2,1}}{\textbf{u}_{2}}(l)+\alpha\textbf{v}(k) \right)\ge \Lambda (\epsilon ).
\end{IEEEeqnarray*}

\medskip

Statement D) follows by rewriting
$\cos \sphericalangle \left( \textbf{y},{{a}_{1,1}}{\textbf{u}_{1}}(j)+{{a}_{2,1}}{\textbf{u}_{2}}(l)+\alpha\textbf{v}(k) \right)$
as
\begin{equation*}
\cos \sphericalangle \left( \textbf{y},{{a}_{1,1}}{\textbf{u}_{1}}(j)+{{a}_{2,1}}{\textbf{u}_{2}}(l)+\alpha\textbf{v}(k) \right)=\frac{\left\langle \textbf{y},{{a}_{1,1}}{\textbf{u}_{1}}(j)+{{a}_{2,1}}{\textbf{u}_{2}}(l)+\alpha\textbf{v}(k) \right\rangle }{\left\| \textbf{y} \right\|\cdot \left\| {{a}_{1,1}}{\textbf{u}_{1}}(j)+{{a}_{2,1}}{\textbf{u}_{2}}(l)+\alpha\textbf{v}(k) \right\|},
\end{equation*}
and then lower bounding $\left\langle y,{{a}_{1,1}}{\textbf{u}_{1}}(j)+{{a}_{2,1}}{\textbf{u}_{2}}(l)+\alpha\textbf{v}(k) \right\rangle $ using A),
and upper-bounding $\left\| \textbf{y} \right\|$
and
$\left\| {{a}_{1,1}}{\textbf{u}_{1}}(j)+{{a}_{2,1}}{\textbf{u}_{2}}(l)+\alpha\textbf{v}(k) \right\|$ using B) and C) respectively.\\
This  yields that for every $(\textbf{s}_1,\textbf{s}_2,{\cal C}_1,{\cal C}_2,{\cal C}_c,\textbf{z})\in {\cal E}_{\textbf{X}}^{c}\bigcap {\cal E}_{\textbf{Z}}^{c}$,
\begin{IEEEeqnarray*}{rCl}
\IEEEeqnarraymulticol{3}{l}{
\cos \sphericalangle \left( \textbf{y},{{a}_{1,1}}{\textbf{u}_{1}}(j)+{{a}_{2,1}}{\textbf{u}_{2}}(l)+\alpha\textbf{v}(k) \right)} \\ \quad
& \ge &\frac{n\left( {{\bar{\beta }}_{1}}{{P}_{1}}+2\sqrt{{{\bar{\beta }}_{1}}{{P}_{1}}{{\bar{\beta }}_{2}}{{P}_{2}}}\tilde{\rho }+{{\bar{\beta }}_{2}}{{P}_{2}}+2\frac{1}{n}\left\| \alpha{\textbf{v}^{*}} \right\|\sqrt{n\bar{{{\beta }_{2}}}{{P}_{2}}}\bar{\rho }+\frac{1}{n}{{\left\| \alpha{\textbf{v}^{*}} \right\|}^{2}}-{{\xi }_{1}}\epsilon  \right)}{\sqrt{n\left( {{\bar{\beta }}_{1}}{{P}_{1}}+2\sqrt{{{\bar{\beta }}_{1}}{{P}_{1}}{{\bar{\beta }}_{2}}{{P}_{2}}}\tilde{\rho }+{{\bar{\beta }}_{2}}{{P}_{2}}+\frac{2}{n}\left\| \alpha{\textbf{v}^{*}} \right\|\sqrt{n\bar{{{\beta }_{2}}}{{P}_{2}}}\bar{\rho }+\frac{1}{n}{{\left\| \alpha{\textbf{v}^{*}} \right\|}^{2}}+N+{{\xi }_{2}}\epsilon  \right)}} \\
&& \cdot \frac{1}{\sqrt{n\left( {{\bar{\beta }}_{1}}{{P}_{1}}+2\sqrt{{{\bar{\beta }}_{1}}{{P}_{1}}\bar{{{\beta }_{2}}}{{P}_{2}}}\tilde{\rho }+\bar{{{\beta }_{2}}}{{P}_{2}}+\frac{2}{n}\left\| \alpha{\textbf{v}^{*}} \right\|\sqrt{n\bar{{{\beta }_{2}}}{{P}_{2}}}\bar{\rho }+\frac{1}{n}{{\left\| \alpha\textbf{v}(k) \right\|}^{2}}+{{\xi }_{3}}\epsilon  \right)}} \\
& \ge & \sqrt{\frac{{{\bar{\beta }}_{1}}{{P}_{1}}+2\sqrt{{{\bar{\beta }}_{1}}{{P}_{1}}{{\bar{\beta }}_{2}}{{P}_{2}}}\tilde{\rho }+{{\bar{\beta }}_{2}}{{P}_{2}}+2\frac{1}{n}\left\| \alpha{\textbf{v}^{*}} \right\|\sqrt{n\bar{{{\beta }_{2}}}{{P}_{2}}}\bar{\rho }+\frac{1}{n}{{\left\| \alpha{\textbf{v}^{*}} \right\|}^{2}}-{{\xi }'}\epsilon }{{{\bar{\beta }}_{1}}{{P}_{1}}+2\sqrt{{{\bar{\beta }}_{1}}{{P}_{1}}{{\bar{\beta }}_{2}}{{P}_{2}}}\tilde{\rho }+{{\bar{\beta }}_{2}}{{P}_{2}}+\frac{2}{n}\left\| \alpha{\textbf{v}^{*}} \right\|\sqrt{n\bar{{{\beta }_{2}}}{{P}_{2}}}\bar{\rho }+\frac{1}{n}{{\left\| \alpha{\textbf{v}^{*}} \right\|}^{2}}+N+{{\xi }_{2}}\epsilon }} \\
&=&\Lambda (\epsilon ) .
\end{IEEEeqnarray*}

\medskip

Lemma~\ref{eps_h_u1,u2,v} now follows by D) which gives
\begin{equation*}
{{\cal E}_{\left( {{{\hat{\textbf{U}}}}_{1}},{{{\hat{\textbf{U}}}}_{2}},\hat{\textbf{V}} \right)}}\cap {\cal E}_{\textbf{S}}^{c}\cap {\cal E}_{\textbf{X}}^{c}\cap
{\cal E}_{\textbf{Z}}^{c}
\subseteq
{\cal E}'_{\left( {{{\hat{\textbf{U}}}}_{1}},{{{\hat{\textbf{U}}}}_{2}},\hat{\textbf{V}} \right)}\cap {\cal E}_{\textbf{S}}^{c}\cap {\cal E}_{\textbf{X}}^{c}\cap {\cal E}_{\textbf{Z}}^{c},
\end{equation*}
and therefore
\begin{flalign*}
&&
 \Pr\left[ {{\cal E}_{\left( {{{\hat{\textbf{U}}}}_{1}},{{{\hat{\textbf{U}}}}_{2}},\hat{\textbf{V}} \right)}}|\epsilon _{S}^{c}\cap {\cal E}_{\textbf{X}}^{c}\cap {\cal E}_{\textbf{Z}}^{c} \right]\le \Pr\left[ {\cal E}'_{\left( {{{\hat{\textbf{U}}}}_{1}},{{{\hat{\textbf{U}}}}_{2}},\hat{\textbf{V}} \right)}|{\cal E}_{\textbf{S}}^{c}\cap {\cal E}_{\textbf{X}}^{c}\cap {\cal E}_{\textbf{Z}}^{c} \right].
&&
\end{flalign*}
\end{IEEEproof}

\medskip

We now state the second lemma needed for the proof of \eqref{eq:R1+R2+Rc}.
\begin{lemma} \label{three-Rates proof}
For every $\Theta_{i} \in (0,1] , i=1,2 $ and $\Delta \in (0,1]$, let the set $\cal G$ be given by
\begin{IEEEeqnarray*}{rCl}
{\cal G}&=&\Bigl\{ \left( {\mathbf{s}_{1}},{\mathbf{s}_{2}},{{{\cal C}}_{1}},{{{\cal C}}_{2}},{{{\cal C}}_{c}},\mathbf{z} \right) \colon
\exists \> {{\mathbf{u}}_{1}}(j)\in {{{\cal C}}_{1}}\backslash \left\{ \mathbf{u}_{1}^{*} \right\}
\mbox{  and  } \exists \> {{\mathbf{u}}_{2}}(l)\in {{{\cal C}}_{2}}\backslash \left\{ \mathbf{u}_{2}^{*} \right\} \nonumber \\
&& \quad \mbox{  and  } \exists \> {{\mathbf{v}}}(k)\in {{{\cal C}}_{c}}\backslash \left\{ \mathbf{v}^{*} \right\}
\ \mbox{s.t.} \ \cos \sphericalangle \left( {\mathbf{u}_{1}}(j),{\mathbf{u}_{2}}(l) \right)\ge {{\Theta }_{1}} ,
\cos \sphericalangle \left( {\mathbf{u}_{2}}(l) ,  \mathbf{v}(l) \right)\ge {{\Theta }_{2}} , \nonumber \\
&& \quad \mbox{  and  } \cos \sphericalangle \left( \mathbf{y},{{a}_{1}}{\mathbf{u}_{1}}(j)+{{a}_{2}}{\mathbf{u}_{2}}(l)+\alpha
\mathbf{v}(k) \right)\ge \Delta \Big\}.
\end{IEEEeqnarray*}
 Then,
\begin{IEEEeqnarray*}{rCl}
{{R}_{1}}+{{R}_{2}}+{{R}_\textnormal{c}}<-\frac{1}{2}\log \left( \left( 1-{{\Theta }_{1}}^{2} \right)\left( 1-{{\Theta }_{2}}^{2} \right)\left( 1-{{\Delta }^{2}} \right) \right) \nonumber \\
\implies \left( \underset{n\to \infty }{\mathop{\lim }}\,\ {\Pr}\left[ {\cal G} \mid
{\cal E}^{c}_{\mathbf{X}_{1}}\cap {\cal E}^{c}_{\mathbf{X}_{2}}\cap {\cal E}^{c}_{\mathbf{X}_{V}} \right]=0,\epsilon >0 \right).
\end{IEEEeqnarray*}
\end{lemma}

\medskip

\begin{IEEEproof}
The proof follows from upper-bounding in every point on ${\cal S}_{i} , i\in {1,2}$ and ${\cal S}_c$,
the density of every ${{\textbf{u}}_{i}}(j)\in {{{\cal C}}_{i}}\backslash \left\{ \textbf{u}_{i}^{*} \right\}$
and every ${{\textbf{v}}}(k)\in {{{\cal C}}_{c}}\backslash \left\{ \textbf{v}^{*} \right\}$ and then using a standard argument from sphere-packing.
This follows similarly as the proof of lemma D.9 in \cite{Stephan}, using Lemma~\ref{conditional density} ahead.
\end{IEEEproof}

\medskip

Now we can turn to the proof of \eqref{eq:R1+R2+Rc}.
\begin{IEEEeqnarray}{rCl} \label{eq:R1+R2+Rc_proof}
\Pr\left[ {{\cal E}_{\left( {{{\hat{\textbf{U}}}}_{1}},{{{\hat{\textbf{U}}}}_{2}},\hat{\textbf{V}} \right)}}\cap {\cal E}_{\textbf{S}}^{c}\cap {\cal E}_{\textbf{X}}^{c}\cap {\cal E}_{\textbf{Z}}^{c} \right]
& \overset{(a)}{\mathop{\le }}&\,\Pr\left[ {\cal E}'_{\left( {{{\hat{\textbf{U}}}}_{1}},{{{\hat{\textbf{U}}}}_{2}},\hat{\textbf{V}} \right)}\cap {\cal E}_{\textbf{S}}^{c}\cap {\cal E}_{\textbf{X}}^{c}\cap {\cal E}_{\textbf{Z}}^{c} \right] \nonumber \\
& \overset{(b)}{\mathop{\le }}&\,\Pr\left[ {\cal E}'_{\left( {{{\hat{\textbf{U}}}}_{1}},{{{\hat{\textbf{U}}}}_{2}},\hat{\textbf{V}} \right)}
\middle|{\cal E}_{{\textbf{X}}_{1}}^{c}\cap {\cal E}_{{\textbf{X}}_{2}}^{c}\cap {\cal E}_{Xv}^{c} \right],
\end{IEEEeqnarray}
where (a) follows by Lemma~\ref{eps_h_u1,u2,v} and (b) follows because ${\cal E}_{\textbf{X}}^{c}\subseteq \left( {\cal E}_{{\textbf{X}}_{1}}^{c}\cap {\cal E}_{{\textbf{X}}_{2}}^{c}\cap {\cal E}_{Xv}^{c} \right)$.

\medskip

The proof of \eqref{eq:R1+R2+Rc} is now completed by combining Inequality \eqref{eq:R1+R2+Rc_proof}
with Lemma~\ref{three-Rates proof}, which gives that for every
$\delta >0$ and every $\epsilon >0$ , there exists some $n'(\delta ,\epsilon )$
such that for all $n>n'(\delta ,\epsilon )$ , we have
\begin{equation*}
\Pr\left[ {{\cal E}_{\left( {{{\hat{\textbf{U}}}}_{1}},{{{\hat{\textbf{U}}}}_{2}},\hat{\textbf{V}} \right)}}\cap {\cal E}_{\textbf{S}}^{c}\cap {\cal E}_{\textbf{X}}^{c}\cap {\cal E}_{\textbf{Z}}^{c} \right]\le \Pr\left[ {\cal E}'_{\left( {{{\hat{\textbf{U}}}}_{1}},{{{\hat{\textbf{U}}}}_{2}},\hat{\textbf{V}} \right)}|{\cal E}_{{\textbf{X}}_{1}}^{c}\cap {\cal E}_{{\textbf{X}}_{2}}^{c} \right]<\delta ,
\end{equation*}
whenever
\begin{equation*}
{{R}_{1}}+{{R}_{2}}+{{R}_\textnormal{c}}<\frac{1}{2}\log \left( \frac{{{\lambda }_{12}}+2\eta \bar{\rho }\sqrt{\bar{{{\beta }_{2}}}{{P}_{2}}}+{{\eta }^{2}}+N}{N\left( 1-{{{\tilde{\rho }}}^{2}}\  \right)\left( 1-{{\bar{\rho }}^{2}}\  \right)}-{{\kappa}_{7}}\epsilon  \right),
\end{equation*}
where $\kappa_7$ is is a positive constant determined by $P_1, P_2 \mbox{  and  } N$.

The proof of Lemma~\ref{rates} is now completed. \hfill \qed

\medskip

The proof of Lemma~\ref{eps_u} now follows straight forwardly:

\textbf{Proof of Lemma~\ref{eps_u}}:
Combining \eqref{eq:eps_u_composition} with Lemma~\ref{eps_s}, Lemma~\ref{eps_z},
Lemma~\ref{eps_x} and Lemma~\ref{rates}
yields that
for every $\delta >0 $ and $0<\epsilon<0.3$, there exists an
$n'(\delta ,\epsilon )\in \mathbb{N} $ such that for all $n>n'(\delta ,\epsilon )$
\begin{flalign*}
&&
\Pr\left[ {{\cal E}_{{\hat{\textbf{U}}}}} \right]<21\delta, \quad \mbox{ if } \left( {{R}_{1}},{{R}_{2}},{{R}_\textnormal{c}} \right)\in {\cal R}\left( \epsilon  \right).
&&
\qed
\end{flalign*}

\vskip.2truein
\subsection{Proof of Lemma~\ref{eps_x}}
\label{proof of eps_x}

The proofs in this section rely on bounds from the geometry of sphere
packing. To this end, we denote by ${{C}_{n}}\left( \varphi  \right)$ the surface area of a polar
cap of half angle $\varphi$ on an ${{\mathbb{R}}^{n}}$-sphere of unit radius.
Upper and lower bounds on the surface area ${{C}_{n}}\left( \varphi  \right)$ are given in the following lemma.

\begin{lemma}\label{Cn}
For any $\varphi \in \left[ 0,\pi /2 \right]$,
\begin{equation*}
\frac{\Gamma \left( \frac{n}{2}+1 \right){{\sin }^{(n-1)}}\varphi }{n\Gamma \left( \frac{n+1}{2} \right)\sqrt{\pi }\cos \varphi }\left( 1-\frac{1}{n}{{\tan }^{2}}\varphi  \right)\le \frac{{{C}_{n}}\left( \varphi  \right)}{{{C}_{n}}\left( \pi  \right)}\le \frac{\Gamma \left( \frac{n}{2}+1 \right){{\sin }^{(n-1)}}\varphi }{n\Gamma \left( \frac{n+1}{2} \right)\sqrt{\pi }\cos \varphi }.
\end{equation*}
\end{lemma}

\medskip

\begin{IEEEproof}
See \cite[Inequality (27)]{Shannon}.
\end{IEEEproof}

\medskip

The ratio of the two gamma functions that appears in the upper
bound and the lower bound of Lemma~\ref{Cn} has the following asymptotic
series.

\medskip

\begin{lemma}\label{gamma_ratio}
\begin{equation*}
\frac{\Gamma \left( x+\frac{1}{2} \right)}{\Gamma \left( x \right)}
=\sqrt{x}\left( 1-\frac{1}{8x}+\frac{1}{128{{x}^{2}}}+\frac{5}{1024{{x}^{3}}}-\frac{21}{32678{{x}^{4}}}+... \right),
\end{equation*}
and in particular
\begin{equation*}
\underset{x\to \infty }{\mathop{\lim }}\,\frac{\Gamma \left( x+\frac{1}{2} \right)}{\Gamma \left( x \right)}=1.
\end{equation*}
\end{lemma}

\medskip

\begin{IEEEproof}
See \cite[Appendix D-E]{Stephan}.
\end{IEEEproof}

\medskip

Before starting with the proofs of this section, we give one more
lemma. To this end, whenever the vector-quantizer of Encoder~1 does not
produce the all-zero sequence, denote by ${{\varsigma }_{1}}\left( {\textbf{s}_{1}},{{\cal C}_{1}} \right)$
the index of $\textbf{u}_{1}^{*}$ in its codebook ${{\cal C}_{1}}$.
And whenever the vector-quantizer of Encoder~1 produces the all-zero sequence, let
${{\varsigma }_{1}}\left( {\textbf{s}_{1}},{{\cal C}_{1}} \right)=0 $.
Further, let ${{\lambda }_{1}}\left( \cdot  \right)$ denote the
measure on the codeword sphere ${{\cal S}_{1}}$ induced by the uniform distribution,
and let ${{f}^{{{\lambda }_{1}}}}\left( \cdot  \right)$ denote the density on ${{\cal S}_{1}}$
with respect to ${{\lambda }_{1}}\left( \cdot  \right)$.
Similarly, for Encoder 2 define ${{\varsigma }_{2}}\left( {\textbf{s}_{2}},{{\cal C}_{2}} \right)$
and ${{f}^{{{\lambda }_{2}}}}\left( \cdot  \right)$ accordingly.

\medskip

\begin{lemma}\label{conditional density}
Conditioned on ${{\varsigma }_{1}}\left( {\mathbf{s}_{1}},{{\mathscr C}_{1}} \right) = 1$, the density of
${\mathbf{U}_{1}}\left( j \right)$ is upper bounded for every $j\in \left\{ 2,3,...,{{2}^{n{{R}_{1}}}} \right\}$
and at every point $\mathbf{u}\in {{\cal S}_{1}}$ by twice the uniform density:
\begin{equation*}
{{f}^{{{\lambda }_{1}}}}\left( \left. {\mathbf{U}_{1}}\left( j \right)=\mathbf{u} \right|{{\varsigma }_{1}}\left( {\mathbf{s}_{1}},{{\cal C}_{1}} \right)=1 \right)\le 2\cdot \frac{1}{r_{1}^{n-1}{{C}_{n}}\left( \pi  \right)},
\end{equation*}
and similarly for Encoder~2.
\end{lemma}

\medskip

\begin{IEEEproof}
See \cite[Appendix D-E]{Stephan}
\end{IEEEproof}

\medskip

\textbf{Proof of Lemma~\ref{eps_x}}:

We begin with the following decomposition
\begin{IEEEeqnarray*}{rCl}
\IEEEeqnarraymulticol{3}{l}{
\Pr\left[ {{\cal E}_\textbf{X}} \right] =\Pr\left[ {{\cal E}_\textbf{X}}\cap {{\cal E}_\textbf{S}} \right]+\Pr\left[ {{\cal E}_\textbf{X}}\cap {\cal E}_\textbf{S}^{c} \right] }\\
& \le &\,\Pr\left[ {{\cal E}_\textbf{S}} \right]+\Pr\left[ {{\cal E}_{{\textbf{X}_{1}}}}\cap {\cal E}_\textbf{S}^{c} \right]
+\Pr\left[ {{\cal E}_{{\textbf{X}_{2}}}}\cap {\cal E}_\textbf{S}^{c} \right]
+\Pr\left[ {{\cal E}_{\textbf{X}_{v}}}\cap {\cal E}_\textbf{S}^{c} \right]
+\Pr\left[ {{\cal E}_{\left( {\textbf{X}_{1}},{\textbf{X}_{2}} \right)}}\cap {\cal E}_{{\textbf{X}_{1}}}^{c}\cap {\cal E}_{{\textbf{X}_{2}}}^{c}\cap {\cal E}_\textbf{S}^{c} \right] \\
&& +\Pr\left[ {{\cal E}_{\left( {\textbf{X}_{1}},{\textbf{X}_{v}} \right)}}\cap {\cal E}_{{\textbf{X}_{1}}}^{c}
\cap {\cal E}_{\textbf{X}_{v}}^{c}\cap {\cal E}_\textbf{S}^{c} \right]
+\Pr\left[ {{\cal E}_{\left( {\textbf{X}_{2}},{\textbf{X}_{v}} \right)}}
\cap {\cal E}_{{\textbf{X}_{2}}}^{c}\cap {\cal E}_{\textbf{X}_{v}}^{c}\cap {\cal E}_\textbf{S}^{c} \right]
+\Pr\left[ {{\cal E}_{\textbf{X}_\textnormal{WZ}}}\cap {\cal E}_{{\textbf{X}_{v}}}^{c}\cap {\cal E}_\textbf{S}^{c} \right]\\
& \le & \,\Pr\left[ {{\cal E}_\textbf{S}} \right]+\Pr\left[ {{\cal E}_{{\textbf{X}_{1}}}} \right]+\Pr\left[ {{\cal E}_{{\textbf{X}_{2}}}} \right]+\Pr\left[ {{\cal E}_{\textbf{X}_{v}}} \right]
+\Pr\left[ {{\cal E}_{\left( {\textbf{X}_{1}},{\textbf{X}_{2}} \right)}}\cap {\cal E}_{{\textbf{X}_{1}}}^{c}\cap {\cal E}_{{\textbf{X}_{2}}}^{c}\cap {\cal E}_\textbf{S}^{c} \right] \\
&& +\Pr\left[ {{\cal E}_{\left( {\textbf{X}_{1}},{\textbf{X}_{v}} \right)}}\cap {\cal E}_{{\textbf{X}_{1}}}^{c}\cap {\cal E}_{\textbf{X}_{v}}^{c}\cap {\cal E}_\textbf{S}^{c} \right]
+\Pr\left[ {{\cal E}_{\left( {\textbf{X}_{2}},{\textbf{X}_{v}} \right)}}\cap {\cal E}_{{\textbf{X}_{2}}}^{c}\cap {\cal E}_{\textbf{X}_{v}}^{c}\cap {\cal E}_\textbf{S}^{c} \right]
+\Pr\left[ {{\cal E}_{\textbf{X}_\textnormal{WZ}}}\cap {\cal E}_{{\textbf{X}_{v}}}^{c}\cap {\cal E}_\textbf{S}^{c} \right].
\end{IEEEeqnarray*}

The proof of Lemma~\ref{eps_x} now follows by showing that for every $\delta>0$ and
$0.3>\epsilon>0$ there exists an $n'_{2}(\delta,\epsilon) > 0$ such that for all
$n>n'_{2}(\delta,\epsilon) > 0$
\begin{IEEEeqnarray}{rCl}
\Pr\left[ {{\cal E}_{\textbf{X}_{i}}} \right]
&\le & \delta,\quad i \in \left\{1,2,v\right\}  \label{eq:eps_x}     \\
\Pr\left[ {{\cal E}_{\left( {\textbf{X}_{1}},{\textbf{X}_{2}} \right)}}\cap {\cal E}_{{\textbf{X}_{1}}}^{c}\cap {\cal E}_{{\textbf{X}_{2}}}^{c}\cap {\cal E}_\textbf{S}^{c} \right]
&\le & 3\delta,      \label{eq:eps_x1,x2} \\
\Pr\left[ {{\cal E}_{\left( {\textbf{X}_{1}},{\textbf{X}_{v}} \right)}}\cap {\cal E}_{{\textbf{X}_{1}}}^{c}\cap
{\cal E}_{\textbf{X}_{v}}^{c}\cap {\cal E}_\textbf{S}^{c} \right]
&\le & \delta,  \label{eq:eps_x1,xv} \\
\Pr\left[ {{\cal E}_{\left( {\textbf{X}_{2}},{\textbf{X}_{v}} \right)}}\cap {\cal E}_{{\textbf{X}_{2}}}^{c}\cap {\cal E}_{\textbf{X}_{v}}^{c}\cap {\cal E}_\textbf{S}^{c} \right]
&\le & 3\delta,        \label{eq:eps_x2,xv}\\
\Pr\left[ {{\cal E}_{\textbf{X}_\textnormal{WZ}}}\cap {\cal E}_{{\textbf{X}_{v}}}^{c}\cap {\cal E}_\textbf{S}^{c} \right]
&\le & \delta .
\label{eq:eps_xWZ}
\end{IEEEeqnarray}

\medskip

\textbf{Proof of \eqref{eq:eps_x}:}
We give the proof for ${\cal E}_{{\textbf{X}_{1}}}$ . Due to the symmetry the
proof for ${\cal E}_{{\textbf{X}_{2}}} \mbox{ and } {\cal E}_{{\textbf{V}}}$ then follows by similar arguments.
Let ${\cal E}_{{\textbf{X}_{1}}}(j)$ be the event that $\textbf{U}_{1}(j)$ does not have a typical angle to
$\textbf{S}_1$, i.e.
\begin{equation*}
{{\cal E}_{{\textbf{X}_{1}}}}(j)=\left\{ \left( {\textbf{s}_{1}},{\textbf{s}_{2}},{{\cal C}_{1}},{{\cal C}_{2}},{{\cal C}_{c}} \right):\ \ \left| \cos \sphericalangle \left( {\textbf{u}_{1}}(j),{\textbf{s}_{1}} \right)-\sqrt{1-{{2}^{-2{{R}_{1}}}}} \right|>\epsilon \sqrt{1-{{2}^{-2{{R}_{1}}}}} \right\}.
\end{equation*}
Then,
\begin{IEEEeqnarray}{rCl}\label{eq:proof_eps_x}
\Pr\left[ {{\cal E}_{{\textbf{X}_{1}}}} \right]
& = & \Pr\left[ \left. {{\cal E}_{{\textbf{X}_{1}}}} \right|{\textbf{S}_{1}}={\textbf{s}_{1}} \right] \nonumber \\
& = &\Pr\left[ \left. \bigcap\limits_{j=1}^{{{2}^{n{{R}_{1}}}}}{{{\cal E}_{{\textbf{X}_{1}}}}(j)} \right|{\textbf{S}_{1}}={\textbf{s}_{1}} \right] \nonumber \\
& = & \prod\limits_{j=1}^{{{2}^{n{{R}_{1}}}}}{\Pr\left[ \left. {{\cal E}_{{\textbf{X}_{1}}}}(j) \right|{\textbf{S}_{1}}={\textbf{s}_{1}} \right]} \nonumber \\
& \overset{(a)} = & \prod\limits_{j=1}^{{{2}^{n{{R}_{1}}}}}{\Pr\left[ {{\cal E}_{{\textbf{X}_{1}}}}(j) \right]} \nonumber \\
& \overset{(b)}= & {{\left( \Pr\left[ {{\cal E}_{{\textbf{X}_{1}}}}(1) \right] \right)}^{{{2}^{n{{R}_{1}}}}}} \nonumber \\
& = & {{\left( 1-\Pr\left[{\cal E}_{{\textbf{X}_{1}}}^{c}(1) \right] \right)}^{{{2}^{n{{R}_{1}}}}}},
\end{IEEEeqnarray}
where in (a) we have used that the probability of ${{\cal E}_{{\textbf{X}_{1}}}}(j)$ does not depend
on $\textbf{S}_{1} = \textbf{s}_{1}$, and in (b) we have used that all $\textbf{U}_{1}(j)$ have the same
distribution. To upper-bound \eqref{eq:proof_eps_x} we now rewrite
${\cal E}_{{\textbf{X}_{1}}}^{c}(1)$ as
\begin{IEEEeqnarray*}{rCl}
\IEEEeqnarraymulticol{3}{l}{
{\cal E}_{{\textbf{X}_{1}}}^{c}(1)}\\\quad
&=&\left\{ \left( {\textbf{s}_{1}},{\textbf{s}_{2}},{{\cal C}_{1}},{{\cal C}_{2}},{{\cal C}_{c}} \right):\ \ \left| \cos \sphericalangle \left( {\textbf{u}_{1}}(1),{\textbf{s}_{1}} \right)-\sqrt{1-{{2}^{-2{{R}_{1}}}}} \right|\le \epsilon \sqrt{1-{{2}^{-2{{R}_{1}}}}} \right\} \\
&=&\left\{ \left( {\textbf{s}_{1}},{\textbf{s}_{2}},{{\cal C}_{1}},{{\cal C}_{2}},{{\cal C}_{c}} \right):\ \ \sqrt{1-{{2}^{-2{{R}_{1}}}}}\left( 1-\epsilon  \right)\le \cos \sphericalangle \left( {\textbf{u}_{1}}(1),{\textbf{s}_{1}} \right)\le \sqrt{1-{{2}^{-2{{R}_{1}}}}}\left( 1+\epsilon  \right) \right\} \\
&=&\left\{ \left( {\textbf{s}_{1}},{\textbf{s}_{2}},{{\cal C}_{1}},{{\cal C}_{2}},{{\cal C}_{c}} \right):\ \ \cos \,{{\theta }_{1,\max }}\le \cos \sphericalangle \left( {\textbf{u}_{1}}(1),{\textbf{s}_{1}} \right)\le \cos \,{{\theta }_{1,\min }} \right\},
\end{IEEEeqnarray*}
where we have used the notation
\begin{IEEEeqnarray*}{rCl}
 \cos \,{{\theta }_{1,\max }} & \triangleq & \sqrt{1-{{2}^{-2{{R}_{1}}}}}\left( 1-\epsilon  \right)  \nonumber \\
 \cos \,{{\theta }_{1,\min }} & \triangleq & \sqrt{1-{{2}^{-2{{R}_{1}}}}}\left( 1+\epsilon  \right).
 \end{IEEEeqnarray*}
Hence, since  $\textbf{U}_{1}(1)$ is generated independently of $\textbf{S}_{1}$ and distributed
uniformly on ${\cal S}_{1}$,
\begin{equation}\label{eq:proof_eps_x_c}
\Pr \left[ {\cal E}_{{\textbf{X}_{1}}}^{c}(1) \right]=\frac{{{C}_{n}}\left( {{\theta }_{1,\max }} \right)-{{C}_{n}}\left( {{\theta }_{1,\min }} \right)}{{{C}_{n}}\left( \pi  \right)}.
\end{equation}
Combining \eqref{eq:proof_eps_x_c} with \eqref{eq:proof_eps_x} gives, as reported in  \cite[Appendix D-E1]{Stephan},
\begin{IEEEeqnarray}{l}
\Pr\left[ {{\cal E}_{{\textbf{X}_{1}}}} \right]
 \leq \exp \biggl( -\frac{\Gamma \left( \frac{n}{2}+1 \right)}{n\Gamma \left( \frac{n+1}{2} \right)\sqrt{\pi }}\biggl[ \frac{{{2}^{n\left( {{R}_{1}}+{{\log }_{2}}\left( \sin {{\theta }_{1,\max }} \right) \right)}}}{\sin {{\theta }_{1,\max }}\cos {{\theta }_{1,\max }}}\Bigl( 1-\frac{1}{n}{{\tan }^{2}}{{\theta }_{1,\max }} \Bigr)   \nonumber \\
 \hspace{6.0cm} -\frac{{{2}^{n\left( {{R}_{1}}+{{\log }_{2}}\left( \sin {{\theta }_{1,\min }} \right) \right)}}}{\sin {{\theta }_{1,\min }}\cos {{\theta }_{1,\min }}} \biggr] \biggr).
\label{eq:eps_x1_proof}
\end{IEEEeqnarray}
It now follows from sphere-packing and covering, that for every
$\epsilon >0$ we have $\Pr\left[ {{\epsilon }_{{{x}_{1}}}} \right]{\to 0} $
as ${n\to \infty }$, as reported in  \cite[Appendix D-E1]{Stephan}.
\hfill \qed

\medskip

\textbf{Proof of inequality \eqref{eq:eps_x1,x2}:}
By the notation in \eqref{eq:ortogonalization} we have
\begin{IEEEeqnarray}{rCl} \label{eq:cos(u1,u2)}
\cos \sphericalangle \left( \textbf{u}_{1}^{*},\textbf{u}_{2}^{*} \right)
&=&\frac{\left\langle \textbf{u}_{1}^{*},\textbf{u}_{2}^{*} \right\rangle }{\left\| \textbf{u}_{1}^{*} \right\|\left\| \textbf{u}_{2}^{*} \right\|} \nonumber \\
&=&\frac{\left\langle {{\nu}_{1}}{\textbf{s}_{1}}+{\textbf{w}_{1}},{{\nu}_{2}}{\textbf{s}_{2}}+{\textbf{w}_{2}} \right\rangle }{\left\| \textbf{u}_{1}^{*} \right\|\left\| \textbf{u}_{2}^{*} \right\|} \nonumber \\
&=&\frac{{{\nu}_{1}}{{\nu}_{2}}\left\langle {\textbf{s}_{1}},{\textbf{s}_{2}} \right\rangle +{{\nu}_{1}}\left\langle {\textbf{s}_{1}},{\textbf{w}_{2}} \right\rangle +{{\nu}_{2}}\left\langle {\textbf{w}_{1}},{\textbf{s}_{2}} \right\rangle +\left\langle {\textbf{w}_{1}},{\textbf{w}_{2}} \right\rangle }{\left\| \textbf{u}_{1}^{*} \right\|\left\| \textbf{u}_{2}^{*} \right\|},
\end{IEEEeqnarray}
where we recall that ${\nu}_{1}$ is a function of $\left\| \textbf{s}_{1} \right\|$ and
$\cos \sphericalangle \left( \textbf{s}_{1},\textbf{u}_{1}^{*} \right)$ and similarly
${\nu}_{2}$ is a function of $\left\| \textbf{s}_{2} \right\|$ and
$\cos \sphericalangle \left( \textbf{s}_{2},\textbf{u}_{2}^{*} \right)$.
Now, define the four events
\begin{IEEEeqnarray*}{rCl}
{{{\cal A}}_{1}}& = & \left\{ \left( {\textbf{s}_{1}},{\textbf{s}_{2}},{{\cal C}_{1}},{{\cal C}_{2}},{{\cal C}_{c}} \right):\left| \tilde{\rho }-\frac{{{\nu}_{1}}{{\nu}_{2}} }{\left\| \textbf{u}_{1}^{*} \right\|\left\| \textbf{u}_{2}^{*} \right\|}\left\langle {\textbf{s}_{1}},{\textbf{s}_{2}} \right\rangle \right|>4\epsilon  \right\} \\
{{{\cal A}}_{2}}&=&\left\{ \left( {\textbf{s}_{1}},{\textbf{s}_{2}},{{\cal C}_{1}},{{\cal C}_{2}},{{\cal C}_{c}} \right):\left| \frac{{{\nu}_{1}}}{\left\| \textbf{u}_{1}^{*} \right\|\left\| \textbf{u}_{2}^{*} \right\|}\left\langle {\textbf{s}_{1}},{\textbf{w}_{2}} \right\rangle  \right|>\epsilon  \right\} \\
{{{\cal A}}_{3}}&=&\left\{ \left( {\textbf{s}_{1}},{\textbf{s}_{2}},{{\cal C}_{1}},{{\cal C}_{2}},{{\cal C}_{c}} \right):\left| \frac{{{\nu}_{2}}}{\left\| \textbf{u}_{1}^{*} \right\|\left\| \textbf{u}_{2}^{*} \right\|}\left\langle {\textbf{w}_{1}},{\textbf{s}_{2}} \right\rangle  \right|>\epsilon  \right\} \\
{{{\cal A}}_{4}}&=&\left\{ \left( {\textbf{s}_{1}},{\textbf{s}_{2}},{{\cal C}_{1}},{{\cal C}_{2}},{{\cal C}_{c}} \right):\left| \frac{1}{\left\| \textbf{u}_{1}^{*} \right\|\left\| \textbf{u}_{2}^{*} \right\|}\left\langle {\textbf{w}_{1}},{\textbf{w}_{2}} \right\rangle  \right|>\epsilon  \right\} .
\end{IEEEeqnarray*}
Note that by \eqref{eq:cos(u1,u2)},\\
${{\cal E}_{\left( {\textbf{X}_{1}},{\textbf{X}_{2}} \right)}}=\left\{ \left( {\textbf{s}_{1}},{\textbf{s}_{2}},{{\cal C}_{1}},{{\cal C}_{2}},{{\cal C}_{c}} \right):\left| \tilde{\rho }-\cos \sphericalangle \left( \textbf{u}_{1}^{*},\textbf{u}_{2}^{*} \right) \right|>7\epsilon  \right\}\subset \left( {{{\cal A}}_{1}}\cup {{{\cal A}}_{2}}\cup {{{\cal A}}_{3}}\cup {{{\cal A}}_{4}} \right)$.
Thus,
\begin{IEEEeqnarray}{rCl}
\IEEEeqnarraymulticol{3}{l}{
\Pr\left[ {{{\cal E} }_{\left( {\textbf{X}_{1}},{\textbf{X}_{2}} \right)}}\cap {\cal E}_{{\textbf{X}_{1}}}^{c}\cap
{\cal E}_{{\textbf{X}_{2}}}^{c}\cap {\cal E}_{\textbf{S}}^{c} \right] }\nonumber \\ \quad
& \, \le & \Pr\left[ {{{\cal A}}_{1}}\cap {\cal E}_{{\textbf{X}_{1}}}^{c}\cap {\cal E}_{{\textbf{X}_{2}}}^{c}\cap
{\cal E}_{\textbf{S}}^{c} \right]+\Pr\left[ {{{\cal A}}_{2}}\cap {\cal E}_{{\textbf{X}_{1}}}^{c}\cap {\cal E}_{{\textbf{X}_{2}}}^{c}\cap {\cal E}_{\textbf{S}}^{c} \right] \nonumber \\
&& +\Pr\left[ {{{\cal A}}_{3}}\cap {\cal E}_{{\textbf{X}_{1}}}^{c}\cap {\cal E}_{{\textbf{X}_{2}}}^{c}\cap
{\cal E}_{\textbf{S}}^{c} \right]+\Pr\left[ {{{\cal A}}_{4}}\cap {\cal E}_{{\textbf{X}_{1}}}^{c}\cap {\cal E}_{{\textbf{X}_{2}}}^{c}\cap {\cal E}_{\textbf{S}}^{c} \right] \nonumber \\
& \, \le & \Pr\left[ {{{\cal A}}_{1}}|{\cal E}_{{\textbf{X}_{1}}}^{c}\cap {\cal E}_{{\textbf{X}_{2}}}^{c}\cap
{\cal E}_{\textbf{S}}^{c} \right]+\Pr\left[ {{{\cal A}}_{2}}|{\cal E}_{\textbf{S}}^{c} \right]+\Pr\left[ {{{\cal A}}_{3}}|{\cal E}_{\textbf{S}}^{c} \right]+\Pr\left[ {{{\cal A}}_{4}}|
{\cal E}_{\textbf{S}}^{c} \right].
\label{eq:eps_x1,x2_proof}
\end{IEEEeqnarray}
The four terms on the r.h.s. of \eqref{eq:eps_x1,x2_proof} are now bounded in the following
two lemmas.

\medskip

\begin{lemma}\label{A1}
For $\epsilon<0.3$
\begin{equation*}
\Pr\left[ {{{\cal A}}_{1}}|{\cal E}_{\mathbf{S}}^{c} \cap {\cal E}_{{\mathbf{X}_{1}}}^{c}\cap {\cal E}_{{\mathbf{X}_{2}}}^{c} \right]=0.
\end{equation*}
\end{lemma}

\medskip

\begin{IEEEproof}
We first note that the term in the definition of ${\cal A}_1$ can be rewritten as
\begin{equation}\label{eq:A1_proof}
\frac{{{\nu}_{1}}{{\nu}_{2}} }{\left\| \textbf{u}_{1}^{*} \right\|\left\| \textbf{u}_{2}^{*} \right\|}\left\langle {\textbf{s}_{1}},{\textbf{s}_{2}} \right\rangle=\cos \sphericalangle \left( {\textbf{s}_{1}},\textbf{u}_{1}^{*} \right)\cos \sphericalangle \left( {\textbf{s}_{2}},\textbf{u}_{1}^{*} \right)\cos \sphericalangle \left( {\textbf{s}_{1}},{\textbf{s}_{2}} \right).
\end{equation}
We can now upper and lower bound the r.h.s. of \eqref{eq:A1_proof} for
$\left( {\textbf{s}_{1}},{\textbf{s}_{2}},{{\cal C}_{1}},{{\cal C}_{2}},{{\cal C}_{c}} \right)\in
{\cal E}_{\textbf{S}}^{c} \cap {\cal E}_{{\textbf{X}_{1}}}^{c}\cap {\cal E}_{{\textbf{X}_{2}}}^{c}$
by noticing that $\left( {\textbf{s}_{1}},{\textbf{s}_{2}},{{\cal C}_{1}},{{\cal C}_{2}},{{\cal C}_{c}} \right)\in
{\cal E}_{\textbf{S}}^{c}$ implies
\begin{equation*}
\left| \cos \sphericalangle \left( {\textbf{s}_{1}},{\textbf{s}_{2}} \right)-\rho  \right|<\rho \epsilon
,
\end{equation*}
that $\left( {\textbf{s}_{1}},{\textbf{s}_{2}},{{\cal C}_{1}},{{\cal C}_{2}},{{\cal C}_{c}} \right)\in
{\cal E}_{\textbf{X}_1}^{c}$ implies
\begin{equation*}
\left| \sqrt{1-{{2}^{-2{{R}_{1}}}}}-\cos \sphericalangle \left( {\textbf{s}_{1}},\textbf{u}_{1}^{*} \right) \right|<\epsilon \sqrt{1-{{2}^{-2{{R}_{1}}}}},
\end{equation*}
and that $\left( {\textbf{s}_{1}},{\textbf{s}_{2}},{{\cal C}_{1}},{{\cal C}_{2}},{{\cal C}_{c}} \right)\in
{\cal E}_{\textbf{X}_2}^{c}$ implies
\begin{equation*}
\left| \sqrt{1-{{2}^{-2{{R}_{2}}}}}-\cos \sphericalangle \left( {\textbf{s}_{2}},\textbf{u}_{2}^{*} \right) \right|<\epsilon \sqrt{1-{{2}^{-2{{R}_{2}}}}}.
\end{equation*}
Hence, combined with \eqref{eq:A1_proof} this gives
\begin{equation*}
\tilde{\rho }{{\left( 1-\epsilon  \right)}^{3}}\le \frac{{{\nu}_{1}}{{\nu}_{2}} }{\left\| \textbf{u}_{1}^{*} \right\|\left\| \textbf{u}_{2}^{*} \right\|}\left\langle {\textbf{s}_{1}},{\textbf{s}_{2}} \right\rangle\ \le \tilde{\rho }{{\left( 1+\epsilon  \right)}^{3}},
\end{equation*}
whenever
$\left( {\textbf{s}_{1}},{\textbf{s}_{2}},{{\cal C}_{1}},{{\cal C}_{2}},{{\cal C}_{c}} \right)\in
{\cal E}_{\textbf{S}}^{c} \cap {\cal E}_{{\textbf{X}_{1}}}^{c}\cap {\cal E}_{{\textbf{X}_{2}}}^{c}$.
The l.h.s. can be lower bounded by ${{\left( 1-3\epsilon  \right)}} \leq{{\left( 1-\epsilon  \right)}^{3}}$,
and the r.h.s. can be upper bounded by ${{\left( 1+\epsilon  \right)}^{3}}\leq {{\left( 1+4\epsilon  \right)}}$
whenever $\epsilon\leq 0.3$. Hence, for $\epsilon\leq 0.3$
\begin{equation*}
  \left| \tilde{\rho }-\frac{{{\nu}_{1}}{{\nu}_{2}} }{\left\| \textbf{u}_{1}^{*} \right\|\left\| \textbf{u}_{2}^{*} \right\|}\left\langle {\textbf{s}_{1}},{\textbf{s}_{2}} \right\rangle \right|\ \le 4\tilde{\rho }\epsilon \le 4\epsilon ,
\end{equation*}
and thus
\begin{flalign*}
&&
\Pr\left[ {{{\cal A}}_{1}}|{\cal E}_{\textbf{S}}^{c} \cap {\cal E}_{{\textbf{X}_{1}}}^{c}\cap {\cal E}_{{\textbf{X}_{2}}}^{c} \right]=0.
&&
\end{flalign*}
\end{IEEEproof}

\medskip

\begin{lemma}\label{A2,A3,A4}
For every $\delta>0$ and $\epsilon>0$ there exists an ${n'_{{\cal A}}}(\delta,\epsilon)$ such that for all
$n>{n'_{{\cal A}}}(\delta,\epsilon)$
\begin{equation*}
\Pr\left[ {{{\cal A}}_{2}}|{\cal E}_{\mathbf{S}}^{c}\right]<\delta, \qquad \qquad \Pr\left[ {{{\cal A}}_{3}}|{\cal E}_{\mathbf{S}}^{c}\right]<\delta, \qquad \qquad \Pr\left[ {{{\cal A}}_{4}}|{\cal E}_{\mathbf{S}}^{c}\right]<\delta.
\end{equation*}
\end{lemma}

\medskip

\begin{IEEEproof}
We start with the derivation of the bound on ${\cal A}_2$. To this end, we first
upper-bound the inner product between $\textbf{s}_1$ and $\textbf{w}_{2}$.
Let $\textbf{s}_{1,P}$ denote the projection of $\textbf{s}_1$ onto the subspace of $\mathbb{R}^{n}$
that is orthogonal to $\textbf{s}_2$, and that thus contains $\textbf{w}_{2}$. Hence,
\begin{IEEEeqnarray}{rCl}\label{eq:proof_A2}
\left| \frac{{{\nu }_{1}}}{\left\| \textbf{u}_{1}^{*} \right\|\left\| \textbf{u}_{2}^{*} \right\|}\left\langle {\textbf{s}_{1}},{\textbf{w}_{2}} \right\rangle  \right|\
& \overset{(a)}{\mathop{=}}&\,\left| \cos \sphericalangle \left( {\textbf{s}_{1}},\textbf{u}_{1}^{*} \right)\left\langle \frac{{\textbf{s}_{1}}}{\left\| {\textbf{s}_{1}} \right\|},\frac{{\textbf{w}_{2}}}{\left\| \textbf{u}_{2}^{*} \right\|} \right\rangle  \right| \nonumber \\
& \overset{(b)}{\mathop{\le }}&\,\left| \cos \sphericalangle \left( {\textbf{s}_{1}},\textbf{u}_{1}^{*} \right) \right|\left| \left\langle \frac{{\textbf{s}_{1}}}{\left\| {\textbf{s}_{1}} \right\|},\frac{{\textbf{w}_{2}}}{\left\| {\textbf{w}_{2}} \right\|} \right\rangle  \right| \nonumber \\
& \le &\left| \left\langle \frac{{\textbf{s}_{1}}}{\left\| {\textbf{s}_{1}} \right\|},\frac{{\textbf{w}_{2}}}{\left\| {\textbf{w}_{2}} \right\|} \right\rangle  \right| \nonumber \\
& = &\left| \left\langle \frac{{\textbf{s}_{1,P}}}{\left\| {\textbf{s}_{1}} \right\|},\frac{{\textbf{w}_{2}}}{\left\| {\textbf{w}_{2}} \right\|} \right\rangle  \right| \nonumber \\
& \le &\left| \left\langle \frac{{\textbf{s}_{1,P}}}{\left\| {\textbf{s}_{1,P}} \right\|},\frac{{\textbf{w}_{2}}}{\left\| {\textbf{w}_{2}} \right\|} \right\rangle  \right| \nonumber \\
& = &\left| \cos \sphericalangle \left( {\textbf{s}_{1,P}},{\textbf{w}_{2}} \right) \right|,
\end{IEEEeqnarray}
where (a) follows by the definition of $\nu_{1}$ and (b) follows since by the definition
of $\textbf{w}_{2}$ we have $\left\| \textbf{w}_2 \right\|\le \left\| \textbf{u}_{2}^{*} \right\|$.
By \eqref{eq:proof_A2} it now follows that
\begin{IEEEeqnarray*}{rCl}
\IEEEeqnarraymulticol{3}{l}{
\Pr\left[ {{{\cal A}}_{2}}|{\cal E}_{\textbf{S}}^{c} \right]
\le \Pr \left[ \left( {{\textbf{S}}_{1}},{{\textbf{S}}_{2}},{\mathscr{C}_{1}},{\mathscr{C}_{2}},{\mathscr{C}_{c}} \right): \left| \cos \sphericalangle \left( {{\textbf{S}}_{1,P}},{\textbf{W}_{2}} \right) \right|>\epsilon \ \mid {\cal E}_{\textbf{S}}^{c} \right]} \\ \quad
&=&{\sf{E}}_{\textbf{S}_1,\textbf{S}_2}\left[{\Pr}_{{\mathscr C}_1,{\mathscr C}_2,{\mathscr C}_c}
\left(\left| \cos \sphericalangle \left( {{\textbf{S}}_{1,P}},{\textbf{W}_{2}} \right) \right|>\epsilon \
\mid \left( {{\textbf{S}}_{1}},{{\textbf{S}}_{2}} \right)=\left( {{\textbf{s}}_{1}},{{\textbf{s}}_{2}} \right),{\cal E}_{\textbf{S}}^{c} \right)  \right],
\end{IEEEeqnarray*}
where in the last line we have denoted by ${{\Pr }_{{{\mathscr C}_{1}},{{\mathscr C}_{2}},{{\mathscr C}_{c}}}}\left( \cdot \mid \cdot \right)$
the conditional probability of the codebooks ${\mathscr C}_1, {\mathscr C}_2$ and ${\mathscr C}_3$ being such that
$ \left| \cos \sphericalangle \left( {{\textbf{S}}_{1,P}},{\textbf{W}_{2}} \right) \right|>\epsilon$,
given $\left( {{\textbf{S}}_{1}},{{\textbf{S}}_{2}} \right)=\left( {{\textbf{s}}_{1}},{{\textbf{s}}_{2}} \right)$
and ${\cal E}_{\textbf{S}}^{c}$.
To conclude our bound we now notice that conditioned on
$\left( {{\textbf{S}}_{1}},{{\textbf{S}}_{2}} \right)=\left( {{\textbf{s}}_{1}},{{\textbf{s}}_{2}} \right)$,
the random vector ${\textbf{W}_{2}}/\left\| {\textbf{W}_{2}} \right\|$ is distributed uniformly on the surface of
the centered $\mathbb{R}^{n-1}$-sphere of unit radius, that lies in the subspace that is
orthogonal to $\textbf{s}_2$. Hence, by \cite[Lemma B.1]{Bross}
\begin{IEEEeqnarray}{rCl}\label{eq:end_proof_A2}
\Pr\left[ {{{\cal A}}_{2}}|{\cal E}_{\textbf{S}}^{c} \right]
&\le &{{E}_{{{\textbf{S}}_{1}},{{\textbf{S}}_{2}}}}\left[ \left. \frac{2{{C}_{n-1}}\left(\Theta  \right)}{{{C}_{n-1}}\left( \pi  \right)} \right|{\cal E}_{\textbf{S}}^{c} \right] \nonumber\\
& \le &\frac{2{{C}_{n-1}}\left(\Theta  \right)}{{{C}_{n-1}}\left( \pi  \right)} \nonumber\\
& \overset{(a)} \le &\frac{2\Gamma \left( \frac{n+1}{2} \right)}{\left( n-1 \right)\Gamma \left( \frac{n}{2} \right)\sqrt{\pi }}\frac{{{\sin }^{\left( n-2 \right)}}\left( \Theta \right)}{\cos \left( \Theta  \right)}\nonumber \\
& \le &\frac{2\Gamma \left( \frac{n+1}{2} \right)}{\left( n-1 \right)\Gamma \left( \frac{n}{2} \right)\sqrt{\pi }\cos \left( \Theta  \right)} ,
\end{IEEEeqnarray}
where  $\Theta\triangleq \arccos(\epsilon)$, and where in (a) we have used Lemma~\ref{Cn}. \\
Upper bounding the ratio of Gamma functions by the asymptotic series of Lemma~\ref{gamma_ratio},
gives for every $\epsilon>0$ that $\Pr\left[ {{{\cal A}}_{2}}|{\cal E}_{\textbf{S}}^{c} \right]\to 0 $
as $n \to \infty$.
By similar arguments it also follows that $\Pr\left[ {{{\cal A}}_{3}}|{\cal E}_{\textbf{S}}^{c} \right]\to 0 $
as $n \to \infty$.

\medskip

To conclude the proof of Lemma~\ref{A2,A3,A4}, we derive the bound on ${\cal A}_4$.
The derivations are similar to those for ${\cal A}_2$.
First, let $\textbf{w}_{1,P}$ denote the projection of $\textbf{w}_1$ onto the subspace of $\mathbb{R}^{n}$
that is orthogonal to $\textbf{s}_2$, and that thus contains $\textbf{w}_{2}$.
As in \eqref{eq:proof_A2}, we can show that
\begin{equation}\label{eq:proof_A4}
\left| \frac{{{\nu }_{1}}}{\left\| \textbf{u}_{1}^{*} \right\|\left\| \textbf{u}_{2}^{*} \right\|}\left\langle {\textbf{w}_{1}},{\textbf{w}_{2}} \right\rangle  \right|\
\leq\,\left| \cos \sphericalangle \left( {\textbf{w}_{1,P}},{\textbf{w}_{2}} \right) \right| .
\end{equation}
Consequently,
\begin{IEEEeqnarray*}{rCl}
\IEEEeqnarraymulticol{3}{l}{
\Pr\left[ {{{\cal A}}_{4}}|{\cal E}_{\textbf{S}}^{c} \right]\le
\Pr \left[ \left( {{\textbf{S}}_{1}},{{\textbf{S}}_{2}},{\mathscr{C}_{1}},{\mathscr{C}_{2}},{\mathscr{C}_{c}} \right): \left| \cos \sphericalangle \left( {{\textbf{W}}_{1,P}},{\textbf{W}_{2}} \right) \right|>\epsilon \ \mid {\cal E}_{\textbf{S}}^{c} \right] }\\ \quad
&=&{\sf{E}}_{\textbf{S}_1,\textbf{S}_2,{\mathscr C}_1}\left[{\Pr}_{{\mathscr C}_2,{\mathscr C}_c}
\left(\left| \cos \sphericalangle \left( {{\textbf{W}}_{1,P}},{\textbf{W}_{2}} \right) \right|>\epsilon \
\mid \left( {{\textbf{S}}_{1}},{{\textbf{S}}_{2}},{{\textbf{U}}_{1}} \right)=
\left( {{\textbf{s}}_{1}},{{\textbf{s}}_{2}},{{\textbf{u}}_{1}} \right),{\cal E}_{\textbf{S}}^{c} \right)  \right],
\end{IEEEeqnarray*}
where in the last line we have denoted by ${\Pr }_{{{\mathscr C}_{2}},{{\mathscr C}_{c}}}\left( \cdot \mid \cdot \right)$ the conditional probability of the codebooks ${\mathscr C}_2$ and ${\mathscr C}_3$ being such that
$ \left| \cos \sphericalangle \left( {{\textbf{W}}_{1,P}},{\textbf{W}_{2}} \right) \right|>\epsilon$,
given $\left( {{\textbf{S}}_{1}},{{\textbf{S}}_{2}},{{\textbf{U}}_{1}} \right)=\left( {{\textbf{s}}_{1}},{{\textbf{s}}_{2}},{{\textbf{u}}_{1}} \right)$  (hence also given ${\textbf{W}}_{1,P}$)
and ${\cal E}_{\textbf{S}}^{c}$.\\
The desired upper bound now follows by noticing that conditioned on
$\left( {{\textbf{S}}_{1}},{{\textbf{S}}_{2}},{{\textbf{U}}_{1}} \right)=\left( {{\textbf{s}}_{1}},{{\textbf{s}}_{2}},{{\textbf{u}}_{1}} \right)$, and ${{\mathscr C}_{1}={\cal C}_1}$,
the random vector ${\textbf{W}_{2}}/\left\| {\textbf{W}_{2}} \right\|$ is distributed uniformly on the surface of the centered $\mathbb{R}^{n-1}$-sphere of unit radius, that lies in the subspace that is
orthogonal to $\textbf{s}_2$.
Hence, similarly as in \eqref{eq:end_proof_A2}
\begin{equation*}
\Pr\left[ {{{\cal A}}_{4}}|{\cal E}_{\textbf{S}}^{c} \right]
\le {{\sf{E}}_{{{\textbf{S}}_{1}},{{\textbf{S}}_{2}},{\mathscr{C}_{1}}}}\left[ \left. \frac{2{{C}_{n-1}}
\left( \Theta  \right)}{{{C}_{n-1}}\left( \pi  \right)} \right|{\cal E}_{\textbf{S}}^{c} \right] .
\end{equation*}
Therefore, for  every $\epsilon>0$, $\Pr\left[ {{{\cal A}}_{4}}|{\cal E}_{\textbf{S}}^{c} \right]\to 0 $
as $n \to \infty$.
\end{IEEEproof}

\medskip

Combining Lemma~\ref{A1} and Lemma~\ref{A2,A3,A4} with \eqref{eq:eps_x1,x2_proof} gives that for
every $\delta>0$ and $0<\epsilon<0.3$ there exists an $n'_{{\cal A}}(\delta,\epsilon)$ such that for all
$n>n'_{{\cal A}}(\delta,\epsilon)$
\begin{flalign*}
&&
\Pr\left[ {{\cal E}_{\left( {\textbf{X}_{1}},{\textbf{X}_{2}} \right)}}\cap {\cal E}_{{\textbf{X}_{1}}}^{c}\cap {\cal E}_{{\textbf{X}_{2}}}^{c}\cap {\cal E}_\textbf{S}^{c} \right]\le 3\delta.
&&
\qed
\end{flalign*}

\vskip.2truein

\textbf{Proof of inequality \eqref{eq:eps_x1,xv}:  }
By the notation in \eqref{eq:ortogonalization} we have
\begin{IEEEeqnarray}{rCl}\label{eq:cos(u1,v)}
\cos \sphericalangle \left( \textbf{u}_{1}^{*},\textbf{v}^{*} \right)
&=&\frac{\left\langle \textbf{u}_{1}^{*},\textbf{v}^{*} \right\rangle }{\left\| \textbf{u}_{1}^{*} \right\|\left\| \textbf{v}^{*} \right\|} \nonumber \\
&=&\frac{\left\langle \textbf{u}_{1}^{*},{{\nu}_{3}}{\textbf{z}_{\textnormal{Q}_1}}+{\textbf{w}_{3}} \right\rangle }{\left\| \textbf{u}_{1}^{*} \right\|\left\| \textbf{v}^{*} \right\|} \nonumber \\
&=&\frac{{{\nu}_{3}}\left\langle {\textbf{u}_{1}^{*}},{\textbf{z}_{\textnormal{Q}_1}} \right\rangle
+\left\langle {\textbf{u}_{1}^{*}},{\textbf{w}_{3}} \right\rangle }{\left\| \textbf{u}_{1}^{*} \right\|\left\| \textbf{v}^{*} \right\|},
\end{IEEEeqnarray}
where we recall that
${\nu}_{3}$ is a function of $\left\| \textbf{z}_{\textnormal{Q}_1} \right\|$ and
$\cos \sphericalangle \left( \textbf{z}_{\textnormal{Q}_1},\textbf{v}^{*} \right)$.
Now, define the two events
\begin{IEEEeqnarray*}{rCl}
{{{\cal A}}_{1}}&=&\left\{ \left( {\textbf{s}_{1}},{\textbf{s}_{2}},{{\cal C}_{1}},{{\cal C}_{2}},{{\cal C}_{c}} \right):
\left| \frac{{{\nu}_{3}} }{\left\| \textbf{u}_{1}^{*} \right\|\left\| \textbf{v}^{*} \right\|}\left\langle {\textbf{u}_{1}^{*}},{\textbf{z}_{\textnormal{Q}_1}} \right\rangle \right|>2\epsilon  \right\} \\
{{{\cal A}}_{2}}&=&\left\{ \left( {\textbf{s}_{1}},{\textbf{s}_{2}},{{\cal C}_{1}},{{\cal C}_{2}},{{\cal C}_{c}} \right):\left| \frac{{{\nu}_{1}}}{\left\| \textbf{u}_{1}^{*} \right\|\left\| \textbf{v}^{*} \right\|}\left\langle {\textbf{u}_{1}^{*}},{\textbf{w}_{3}} \right\rangle  \right|>\epsilon  \right\}  .
\end{IEEEeqnarray*}
Note that by \eqref{eq:cos(u1,v)}, ${{\cal E}_{\left( {\textbf{X}_{1}},{\textbf{X}_{v}} \right)}}=\left\{ \left( {\textbf{s}_{1}},{\textbf{s}_{2}},{{\cal C}_{1}},{{\cal C}_{2}},{{\cal C}_{c}} \right):\left| \cos \sphericalangle \left( \textbf{u}_{1}^{*},\textbf{v}^{*} \right) \right|>3\epsilon  \right\}\subset \left( {{{\cal A}}_{1}}\cup {{{\cal A}}_{2}} \right)$.
Thus,
\begin{IEEEeqnarray}{rCl}
\IEEEeqnarraymulticol{3}{l}{
\Pr\left[ {{{\cal E} }_{\left( {\textbf{X}_{1}},{\textbf{X}_{v}} \right)}}\cap {\cal E}_{{\textbf{X}_{1}}}^{c}\cap
{\cal E}_{{\textbf{X}_{v}}}^{c}\cap {\cal E}_{\textbf{S}}^{c} \right] }\nonumber \\ \quad
& \, \le & \Pr\left[ {{{\cal A}}_{1}}\cap {\cal E}_{{\textbf{X}_{1}}}^{c}\cap {\cal E}_{{\textbf{X}_{v}}}^{c}\cap
{\cal E}_{\textbf{S}}^{c} \right]+\Pr\left[ {{{\cal A}}_{2}}\cap {\cal E}_{{\textbf{X}_{1}}}^{c}\cap {\cal E}_{{\textbf{X}_{v}}}^{c}\cap {\cal E}_{\textbf{S}}^{c} \right] \nonumber \\
& \, \le &\Pr\left[ {{{\cal A}}_{1}}|{\cal E}_{{\textbf{X}_{1}}}^{c}\cap {\cal E}_{{\textbf{X}_{v}}}^{c}\cap
{\cal E}_{\textbf{S}}^{c} \right]+\Pr\left[ {{{\cal A}}_{2}}|{\cal E}_{\textbf{S}}^{c} \right].
\label{eq:eps_x1,xv_proof}
\end{IEEEeqnarray}
The two terms on the r.h.s. of \eqref{eq:eps_x1,xv_proof} are now bounded in the following
two lemmas.

\medskip

\begin{lemma}\label{A1_x1,xv}
For $0<\epsilon\leq 1$
\begin{equation*}
\Pr\left[ {{{\cal A}}_{1}}|{\cal E}_{\mathbf{S}}^{c} \cap {\cal E}_{{\mathbf{X}_{1}}}^{c}\cap {\cal E}_{{\mathbf{X}_{v}}}^{c} \right]=0.
\end{equation*}
\end{lemma}

\medskip

\begin{IEEEproof}
We first note that the term in the definition of ${\cal A}_1$ can be rewritten as
\begin{equation}\label{eq:A1_proof_x1,xv}
\frac{{{\nu}_{3}} }{\left\| \textbf{u}_{1}^{*} \right\|\left\| \textbf{v}^{*} \right\|}\left\langle {\textbf{u}_{1}^{*}},{\textbf{z}_{\textnormal{Q}_1}} \right\rangle=
\cos \sphericalangle \left( {\textbf{z}_{\textnormal{Q}_1}},\textbf{v}^{*} \right)
\cos \sphericalangle \left( {\textbf{u}_{1}^{*}},{\textbf{z}_{\textnormal{Q}_1}} \right).
\end{equation}
We can now upper and lower bound the r.h.s. of \eqref{eq:A1_proof_x1,xv} for
$\left( {\textbf{s}_{1}},{\textbf{s}_{2}},{{\cal C}_{1}},{{\cal C}_{2}},{{\cal C}_{c}} \right)\in
{\cal E}_{\textbf{S}}^{c} \cap {\cal E}_{{\textbf{X}_{2}}}^{c}\cap {\cal E}_{{\textbf{X}_{v}}}^{c}$
by noticing that
$\left( {\textbf{s}_{1}},{\textbf{s}_{2}},{{\cal C}_{1}},{{\cal C}_{2}},{{\cal C}_{c}} \right)\in
{\cal E}_{\textbf{X}_1}^{c}$ implies
\begin{equation*}
\left| \cos \sphericalangle \left( {\textbf{u}_{1}^{*}},\textbf{z}_{\textnormal{Q}_1} \right) \right|<\epsilon
,
\end{equation*}
and that $\left( {\textbf{s}_{1}},{\textbf{s}_{2}},{{\cal C}_{1}},{{\cal C}_{2}},{{\cal C}_{c}} \right)\in
{\cal E}_{\textbf{X}_v}^{c}$ implies
\begin{equation*}
\left| \sqrt{1-{{2}^{-2{{R}_\textnormal{c}}}}}-\cos \sphericalangle \left( {\textbf{z}_{\textnormal{Q}_1}},\textbf{v}^{*} \right) \right|<\epsilon \sqrt{1-{{2}^{-2{{R}_\textnormal{c}}}}}.
\end{equation*}
Hence, combined with \eqref{eq:A1_proof_x1,xv} this gives
\begin{equation*}
\left| \frac{{{\nu}_{3}} }{\left\| \textbf{u}_{1}^{*} \right\|\left\| \textbf{v}^{*} \right\|}\left\langle {\textbf{u}_{1}^{*}},{\textbf{z}_{\textnormal{Q}_1}} \right\rangle \right|
\le \sqrt{1-{{2}^{-2{{R}_\textnormal{c}}}}}\epsilon{{\left( 1+\epsilon  \right)}},
\end{equation*}
whenever
$\left( {\textbf{s}_{1}},{\textbf{s}_{2}},{{\cal C}_{1}},{{\cal C}_{2}},{{\cal C}_{c}} \right)\in
{\cal E}_{\textbf{S}}^{c} \cap {\cal E}_{{\textbf{X}_{1}}}^{c}\cap {\cal E}_{{\textbf{X}_{v}}}^{c}$.
The r.h.s. can be upper bounded by ${{ \epsilon(1+\epsilon)}}\leq 2\epsilon$
whenever $\epsilon\leq 1$. Hence, for $\epsilon\leq 1$
\begin{equation*}
\left| \frac{{{\nu}_{3}} }{\left\| \textbf{u}_{1}^{*} \right\|\left\| \textbf{v}^{*} \right\|}\left\langle {\textbf{u}_{1}^{*}},{\textbf{z}_{\textnormal{Q}_1}} \right\rangle \right|\ \le 2\sqrt{1-{{2}^{-2{{R}_\textnormal{c}}}}}\epsilon \le 2\epsilon,
\end{equation*}
and thus
\begin{flalign*}
&&
\Pr\left[ {{{\cal A}}_{1}}|{\cal E}_{\textbf{S}}^{c} \cap {\cal E}_{{\textbf{X}_{1}}}^{c}\cap {\cal E}_{{\textbf{X}_{v}}}^{c} \right]=0.
&&
\end{flalign*}
\end{IEEEproof}

\medskip

\begin{lemma}\label{A2_x1,xv}
For every $\delta>0$ and $\epsilon>0$ there exists an ${n'_{{\cal A}}}(\delta,\epsilon)$ such that for all
$n>{n'_{{\cal A}}}(\delta,\epsilon)$
\begin{equation*}
\Pr\left[ {{{\cal A}}_{2}}|{\cal E}_{\mathbf{S}}^{c}\right]<\delta.
\end{equation*}
\end{lemma}

\medskip

\begin{IEEEproof}
We first upper-bound the inner product between $\textbf{u}_{1}^{*}$ and $\textbf{w}_{3}$.
Let $\textbf{u}_{1,\textnormal{P}}$ denote the projection of $\textbf{u}_{1}^{*}$ onto the subspace of $\mathbb{R}^{n}$
that is orthogonal to $\textbf{z}_{\textnormal{Q}_1}$, and therfore contains $\textbf{w}_{3}$. Hence,
\begin{IEEEeqnarray}{rCl}\label{eq:proof_A2_x1,xv}
   \left| \frac{{{\nu }_{1}}}{\left\| \textbf{u}_{1}^{*} \right\|\left\| \textbf{v}^{*} \right\|}\left\langle {\textbf{u}_{1}^{*}},{\textbf{w}_{3}} \right\rangle  \right|\
& \overset{(a)}{\mathop{=}}& \,\left| \cos \sphericalangle \left( {\textbf{s}_{1}},\textbf{u}_{1}^{*} \right)\left\langle \frac{{\textbf{u}^{*}_{1}}}{\left\| {\textbf{s}_{1}} \right\|},\frac{{\textbf{w}_{3}}}{\left\| \textbf{v}^{*} \right\|} \right\rangle  \right| \nonumber \\
 & \overset{(b)}{\mathop{\le }}& \,\left| \cos \sphericalangle \left( {\textbf{s}_{1}},\textbf{u}_{1}^{*} \right) \right|\left| \left\langle \frac{{\textbf{u}_{1}^{*}}}{\left\| {\textbf{s}_{1}} \right\|},\frac{{\textbf{w}_{3}}}{\left\| {\textbf{w}_{3}} \right\|} \right\rangle  \right| \nonumber \\
 & \le & \left| \left\langle \frac{{\textbf{u}_{1}^{*}}}{\left\| {\textbf{s}_{1}} \right\|},\frac{{\textbf{w}_{3}}}{\left\| {\textbf{w}_{3}} \right\|} \right\rangle  \right| \nonumber \\
 & = & \left| \left\langle \frac{{\textbf{u}_{1,\textnormal{P}}}}{\left\| {\textbf{u}_{1}^{*}} \right\|},\frac{{\textbf{w}_{3}}}{\left\| {\textbf{w}_{3}} \right\|} \right\rangle  \right| \nonumber \\
 & \le & \left| \left\langle \frac{{\textbf{u}_{1,\textnormal{P}}}}{\left\| {\textbf{u}_{1,\textnormal{P}}} \right\|},\frac{{\textbf{w}_{3}}}{\left\| {\textbf{w}_{3}} \right\|} \right\rangle  \right| \nonumber \\
 & = & \left| \cos \sphericalangle \left( {\textbf{u}_{1,\textnormal{P}}},{\textbf{w}_{3}} \right) \right|,
\end{IEEEeqnarray}
where (a) follows by the definition of $\nu_{1}$ and (b) follows since by the definition
of $\textbf{w}_{3}$ we have $\left\| \textbf{w}_{3} \right\|\le \left\| \textbf{v}^{*} \right\|$.
By \eqref{eq:proof_A2_x1,xv} it now follows that
\begin{IEEEeqnarray*}{rCl}
\Pr  \left[ {{{\cal A}}_{2}}|{\cal E}_{\textbf{S}}^{c} \right]
& \le &
\Pr \left[ \left( {{\textbf{S}}_{1}},{{\textbf{S}}_{2}},{\mathscr{C}_{1}},{\mathscr{C}_{2}},{\mathscr{C}_{c}} \right)\colon \left| \cos \sphericalangle \left( {{\textbf{U}}_{1,\textnormal{P}}},{\textbf{W}_{3}} \right) \right|>\epsilon \mid {\cal E}_{\textbf{S}}^{c} \right] \\
&\overset{(a)}= &  {{\sf{E}}_{{{\textbf{S}}_{1}},{{\textbf{S}}_{2}},{\mathscr{C}_{1}}}}
\left[ {{\Pr }_{{\mathscr{C}_{2}},{\mathscr{C}_{c}}}}\left( \left| \cos \sphericalangle \left( {{\textbf{u}}_{1,P}},{\textbf{W}_{3}} \right) \right|>\epsilon \mid
\left( {{\textbf{S}}_{1}},{{\textbf{S}}_{2}},{{\textbf{U}}_{1}} \right)=\left( {{\textbf{s}}_{1}},{{\textbf{s}}_{2}},{{\textbf{u}}_{1}} \right),{\cal E}_{\textbf{S}}^{c} \right) \right],
\end{IEEEeqnarray*}
where $0<\epsilon \leq 1$, and where in (a) we have denoted by
${{\Pr }_{{{c}_{2}},{{c}_{c}}}}\left( \cdot \mid \cdot \right)$
the conditional probability of the codebooks $c_2$ and $c_c$ being such that
$\left| \cos \sphericalangle \left( {{\textbf{u}}_{1,P}},{\textbf{W}_{3}} \right) \right|>\epsilon $,
given $\left( {{\textbf{S}}_{1}},{{\textbf{S}}_{2}},{{\textbf{U}}_{1}} \right)=\left( {{\textbf{s}}_{1}},{{\textbf{s}}_{2}},{{\textbf{u}}_{1}} \right)$ and ${\cal E}_{\textbf{S}}^{c}$.
To conclude our bound, we now notice that conditioned on
$\left( {{\textbf{S}}_{1}},{{\textbf{S}}_{2}},{{\textbf{U}}_{1}} \right)=\left( {{\textbf{s}}_{1}},{{\textbf{s}}_{2}},{{\textbf{u}}_{1}} \right)$, the random vector ${\textbf{W}_{3}}/\left\| {\textbf{W}_{3}} \right\|$ is distributed uniformly on the surface of
the centered $\mathbb{R}^{n-1}$-sphere of unit radius, that lies in the subspace that is
orthogonal to $\textbf{z}_{\textnormal{Q}_1}$. Hence, according to  \cite[Lemma B.1]{Bross},
\begin{IEEEeqnarray}{rCl}\label{eq:A2_x1,v_proof_final}
\Pr\left[ {{{\cal A}}_{2}}|{\cal E}_{\textbf{S}}^{c} \right]
& \le & {{\sf{E}}_{{{\textbf{S}}_{1}},{{\textbf{S}}_{2}},{\mathscr{C}_{1}}}}\left[ \left. \frac{2{{C}_{n-1}}\left(\Theta  \right)}{{{C}_{n-1}}\left( \pi  \right)} \right|{\cal E}_{\textbf{S}}^{c} \right]\nonumber \\
& \le &\frac{2{{C}_{n-1}}\left( \Theta \right)}{{{C}_{n-1}}\left( \pi  \right)} ,
\end{IEEEeqnarray}
where $\Theta \triangleq \arccos(\epsilon)$.
Note that as $0<\epsilon \leq 1$, $\Theta\in (0,\frac{\pi}{2})$, and thus,
by \cite[Lemma B.4]{Bross}, the r.h.s. of \eqref{eq:A2_x1,v_proof_final} tends to 0 as $n \to \infty$,
and therefore $\Pr\left[ {{{\cal A}}_{2}}|{\cal E}_{\textbf{S}}^{c} \right]\to 0 $.
\end{IEEEproof}

\medskip

Combining Lemma~\ref{A1_x1,xv} and Lemma~\ref{A2_x1,xv} with \eqref{eq:eps_x1,xv_proof} gives that for
every $\delta>0$ and $0<\epsilon\leq1$ there exists an $n'_{{\cal A}}(\delta,\epsilon)$ such that for all
$n>n'_{{\cal A}}(\delta,\epsilon)$
\begin{flalign*}
&&
\Pr\left[ {{\cal E}_{\left( {\textbf{X}_{1}},{\textbf{X}_{v}} \right)}}\cap {\cal E}_{{\textbf{X}_{2}}}^{c}\cap {\cal E}_{{\textbf{X}_{v}}}^{c}\cap {\cal E}_\textbf{S}^{c} \right]\le \delta.
&&
\qed
\end{flalign*}

\vskip.2truein

\textbf{Proof of inequality \eqref{eq:eps_x2,xv}:  }
By the notation in \eqref{eq:ortogonalization} we have
\begin{IEEEeqnarray}{rCl}\label{eq:cos(u2,v)}
\cos \sphericalangle \left( \textbf{u}_{2}^{*},\textbf{v}^{*} \right)
&=&\frac{\left\langle \textbf{u}_{2}^{*},\textbf{v}^{*} \right\rangle }{\left\| \textbf{u}_{2}^{*} \right\|\left\| \textbf{v}^{*} \right\|} \nonumber \\
&=&\frac{\left\langle {{\nu}_{2}}{\textbf{s}_{2}}+{\textbf{w}_{2}},{{\nu}_{3}}{\textbf{z}_{\textnormal{Q}_1}}+{\textbf{w}_{3}} \right\rangle }{\left\| \textbf{u}_{2}^{*} \right\|\left\| \textbf{v}^{*} \right\|} \nonumber \\
&=&\frac{{{\nu}_{2}}{{\nu}_{3}}\left\langle {\textbf{s}_{2}},{\textbf{z}_{\textnormal{Q}_1}} \right\rangle
+{{\nu}_{2}}\left\langle {\textbf{s}_{2}},{\textbf{w}_{3}} \right\rangle
+{{\nu}_{3}}\left\langle {\textbf{z}_{\textnormal{Q}_1}},{\textbf{w}_{2}} \right\rangle
+\left\langle {\textbf{w}_{2}},{\textbf{w}_{3}} \right\rangle }{\left\| \textbf{u}_{2}^{*} \right\|\left\| \textbf{v}^{*} \right\|},
\end{IEEEeqnarray}
where we recall that ${\nu}_{2}$ is a function of $\left\| \textbf{s}_{2} \right\|$ and
$\cos \sphericalangle \left( \textbf{s}_{2},\textbf{u}_{2}^{*} \right)$ and similarly
${\nu}_{3}$ is a function of $\left\| \textbf{z}_{\textnormal{Q}_1} \right\|$ and
$\cos \sphericalangle \left( \textbf{z}_{\textnormal{Q}_1},\textbf{v}^{*} \right)$.
Now, define the four events
\begin{IEEEeqnarray*}{rCl}
{{{\cal A}}_{1}}&=&\left\{ \left( {\textbf{s}_{1}},{\textbf{s}_{2}},{{\cal C}_{1}},{{\cal C}_{2}},{{\cal C}_{c}} \right):\left| \bar{\rho }-\frac{{{\nu}_{2}}{{\nu}_{3}} }{\left\| \textbf{u}_{2}^{*} \right\|\left\| \textbf{v}^{*} \right\|}\left\langle {\textbf{s}_{2}},{\textbf{z}_{\textnormal{Q}_1}} \right\rangle \right|>4\epsilon  \right\} \\
{{{\cal A}}_{2}}&=&\left\{ \left( {\textbf{s}_{1}},{\textbf{s}_{2}},{{\cal C}_{1}},{{\cal C}_{2}},{{\cal C}_{c}} \right):\left| \frac{{{\nu}_{2}}}{\left\| \textbf{u}_{2}^{*} \right\|\left\| \textbf{v}^{*} \right\|}\left\langle {\textbf{s}_{2}},{\textbf{w}_{3}} \right\rangle  \right|>\epsilon  \right\} \\
{{{\cal A}}_{3}}&=&\left\{ \left( {\textbf{s}_{1}},{\textbf{s}_{2}},{{\cal C}_{1}},{{\cal C}_{2}},{{\cal C}_{c}} \right):\left| \frac{{{\nu}_{2}}}{\left\| \textbf{u}_{2}^{*} \right\|\left\| \textbf{v}^{*} \right\|}\left\langle {\textbf{z}_{\textnormal{Q}_1}},{\textbf{w}_{2}} \right\rangle  \right|>\epsilon  \right\} \\
{{{\cal A}}_{4}}&=&\left\{ \left( {\textbf{s}_{1}},{\textbf{s}_{2}},{{\cal C}_{1}},{{\cal C}_{2}},{{\cal C}_{c}} \right):\left| \frac{1}{\left\| \textbf{u}_{2}^{*} \right\|\left\| \textbf{v}^{*} \right\|}\left\langle {\textbf{w}_{2}},{\textbf{w}_{3}} \right\rangle  \right|>\epsilon  \right\} .
\end{IEEEeqnarray*}
Note that by \eqref{eq:cos(u2,v)}, \\
${{\cal E}_{\left( {\textbf{X}_{2}},{\textbf{X}_{v}} \right)}}=\left\{ \left( {\textbf{s}_{1}},{\textbf{s}_{2}},{{\cal C}_{1}},{{\cal C}_{2}},{{\cal C}_{c}} \right):\left| \bar{\rho }-\cos \sphericalangle \left( \textbf{u}_{2}^{*},\textbf{v}^{*} \right) \right|>7\epsilon  \right\}\subset \left( {{{\cal A}}_{1}}\cup {{{\cal A}}_{2}}\cup {{{\cal A}}_{3}}\cup {{{\cal A}}_{4}} \right)$. Thus,
\begin{IEEEeqnarray}{rCl}
\IEEEeqnarraymulticol{3}{l}{
\Pr\left[ {{{\cal E} }_{\left( {\textbf{X}_{2}},{\textbf{X}_{v}} \right)}}\cap {\cal E}_{{\textbf{X}_{2}}}^{c}\cap
{\cal E}_{{\textbf{X}_{v}}}^{c}\cap {\cal E}_{\textbf{S}}^{c} \right] }\nonumber \\ \quad
& \, \le &\Pr\left[ {{{\cal A}}_{1}}\cap {\cal E}_{{\textbf{X}_{2}}}^{c}\cap {\cal E}_{{\textbf{X}_{v}}}^{c}\cap
{\cal E}_{\textbf{S}}^{c} \right]+\Pr\left[ {{{\cal A}}_{2}}\cap {\cal E}_{{\textbf{X}_{2}}}^{c}\cap {\cal E}_{{\textbf{X}_{v}}}^{c}\cap {\cal E}_{\textbf{S}}^{c} \right] \nonumber \\
&& +\Pr\left[ {{{\cal A}}_{3}}\cap {\cal E}_{{\textbf{X}_{2}}}^{c}\cap {\cal E}_{{\textbf{X}_{v}}}^{c}\cap
{\cal E}_{\textbf{S}}^{c} \right]+\Pr\left[ {{{\cal A}}_{4}}\cap {\cal E}_{{\textbf{X}_{2}}}^{c}\cap {\cal E}_{{\textbf{X}_{v}}}^{c}\cap {\cal E}_{\textbf{S}}^{c} \right] \nonumber \\
& \, \le &\Pr\left[ {{{\cal A}}_{1}}|{\cal E}_{{\textbf{X}_{2}}}^{c}\cap {\cal E}_{{\textbf{X}_{v}}}^{c}\cap
{\cal E}_{\textbf{S}}^{c} \right]+\Pr\left[ {{{\cal A}}_{2}}|{\cal E}_{\textbf{S}}^{c} \right]+\Pr\left[ {{{\cal A}}_{3}}|{\cal E}_{\textbf{S}}^{c} \right]+\Pr\left[ {{{\cal A}}_{4}}|
{\cal E}_{\textbf{S}}^{c} \right].
\label{eq:eps_x2,xv_proof}
\end{IEEEeqnarray}
The four terms on the r.h.s. of \eqref{eq:eps_x2,xv_proof} are now bounded in the following
two lemmas.

\medskip

\begin{lemma}\label{A1_x2,xv}
For $\epsilon<0.3$
\begin{equation*}
\Pr\left[ {{{\cal A}}_{1}}|{\cal E}_{\mathbf{S}}^{c} \cap {\cal E}_{{\mathbf{X}_{2}}}^{c}\cap {\cal E}_{{\mathbf{X}_{v}}}^{c} \right]=0.
\end{equation*}
\end{lemma}

\medskip

\begin{IEEEproof}
We first note that the term in the definition of ${\cal A}_1$ can be rewritten as
\begin{equation}\label{eq:A1_proof_x2,xv}
\frac{{{\nu}_{2}}{{\nu}_{3}} }{\left\| \textbf{u}_{2}^{*} \right\|\left\| \textbf{v}^{*} \right\|}\left\langle {\textbf{s}_{2}},{\textbf{z}_{\textnormal{Q}_1}} \right\rangle=
\cos \sphericalangle \left( {\textbf{s}_{2}},\textbf{u}_{2}^{*} \right)
\cos \sphericalangle \left( {\textbf{z}_{\textnormal{Q}_1}},\textbf{v}^{*} \right)
\cos \sphericalangle \left( {\textbf{s}_{2}},{\textbf{z}_{\textnormal{Q}_1}} \right).
\end{equation}
We can now upper and lower bound the r.h.s. of \eqref{eq:A1_proof_x2,xv} for
$\left( {\textbf{s}_{1}},{\textbf{s}_{2}},{{\cal C}_{1}},{{\cal C}_{2}},{{\cal C}_{c}} \right)\in
{\cal E}_{\textbf{S}}^{c} \cap {\cal E}_{{\textbf{X}_{2}}}^{c}\cap {\cal E}_{{\textbf{X}_{v}}}^{c}$
by noticing that $\left( {\textbf{s}_{1}},{\textbf{s}_{2}},{{\cal C}_{1}},{{\cal C}_{2}},{{\cal C}_{c}} \right)\in
{\cal E}_{\textbf{S}}^{c}$ implies
\begin{equation}\label{eq:cos(s2,zq1)}
\left| \rho \sqrt{2^{-2R_1}}-\cos \sphericalangle \left( {\textbf{s}_{2}},{\textbf{z}_{\textnormal{Q}_1}} \right) \right|<\rho \sqrt{2^{-2R_1}} \epsilon \nonumber,
\end{equation}
that $\left( {\textbf{s}_{1}},{\textbf{s}_{2}},{{\cal C}_{1}},{{\cal C}_{2}},{{\cal C}_{c}} \right)\in
{\cal E}_{\textbf{X}_2}^{c}$ implies
\begin{equation*}
\left| \sqrt{1-{{2}^{-2{{R}_{2}}}}}-\cos \sphericalangle \left( {\textbf{s}_{2}},\textbf{u}_{2}^{*} \right) \right|<\epsilon \sqrt{1-{{2}^{-2{{R}_{2}}}}},
\end{equation*}
and that $\left( {\textbf{s}_{1}},{\textbf{s}_{2}},{{\cal C}_{1}},{{\cal C}_{2}},{{\cal C}_{c}} \right)\in
{\cal E}_{\textbf{X}_v}^{c}$ implies
\begin{equation*}
\left| \sqrt{1-{{2}^{-2{{R}_\textnormal{c}}}}}-\cos \sphericalangle \left( {\textbf{z}_{\textnormal{Q}_1}},\textbf{v}^{*} \right) \right|<\epsilon \sqrt{1-{{2}^{-2{{R}_\textnormal{c}}}}}.
\end{equation*}
Hence, combined with \eqref{eq:A1_proof_x2,xv} this gives
\begin{equation*}
\bar{\rho }{{\left( 1-\epsilon  \right)}^{3}}\le \frac{{{\nu}_{1}}{{\nu}_{2}} }{\left\| \textbf{u}_{1}^{*} \right\|\left\| \textbf{u}_{2}^{*} \right\|}\left\langle {\textbf{s}_{1}},{\textbf{s}_{2}} \right\rangle\ \le \bar{\rho }{{\left( 1+\epsilon  \right)}^{3}},
\end{equation*}
whenever
$\left( {\textbf{s}_{1}},{\textbf{s}_{2}},{{\cal C}_{1}},{{\cal C}_{2}},{{\cal C}_{c}} \right)\in
{\cal E}_{\textbf{S}}^{c} \cap {\cal E}_{{\textbf{X}_{2}}}^{c}\cap {\cal E}_{{\textbf{X}_{v}}}^{c}$.
The l.h.s. can be lower bounded by ${{\left( 1-3\epsilon  \right)}} \leq{{\left( 1-\epsilon  \right)}^{3}}$,
and the r.h.s. can be upper bounded by ${{\left( 1+\epsilon  \right)}^{3}}\leq {{\left( 1+4\epsilon  \right)}}$
whenever $\epsilon\leq 0.3$. Hence, for $\epsilon\leq 0.3$
\begin{equation*}
  \left| \bar{\rho }-\frac{{{\nu}_{1}}{{\nu}_{2}} }{\left\| \textbf{u}_{1}^{*} \right\|\left\| \textbf{u}_{2}^{*} \right\|}\left\langle {\textbf{s}_{1}},{\textbf{s}_{2}} \right\rangle \right|\ \le 4\bar{\rho }\epsilon \le 4\epsilon,
\end{equation*}
and thus
\begin{flalign*}
&&
\Pr\left[ {{{\cal A}}_{1}}|{\cal E}_{\textbf{S}}^{c} \cap {\cal E}_{{\textbf{X}_{2}}}^{c}\cap {\cal E}_{{\textbf{X}_{v}}}^{c} \right]=0.
&&
\end{flalign*}
\end{IEEEproof}

\medskip

\begin{remark}
To show \eqref{eq:cos(s2,zq1)}, note that
\begin{IEEEeqnarray*}{rCl}
\cos \sphericalangle \left( {\mathbf{s}_{2}},{\mathbf{z}_{{Q}_{1}}} \right)
&=&\frac{\left\langle {\mathbf{s}_{2}},\mathbf{z}_{\textnormal{Q}_{1}} \right\rangle }{\left\| {\mathbf{s}_{2}} \right\|\left\| \mathbf{z}_{\textnormal{Q}_{1}} \right\|}\ =\frac{\left\langle {\mathbf{s}_{2}},{\mathbf{s}_{1}}-\mathbf{u}_{1}^{*} \right\rangle }{\left\| {\mathbf{s}_{2}} \right\|\left\| \mathbf{z}_{\textnormal{Q}_{1}} \right\|}
=\frac{\left\langle {\mathbf{s}_{2}},{\mathbf{s}_{1}} \right\rangle -\left\langle {\mathbf{s}_{2}},\mathbf{u}_{1}^{*} \right\rangle }{\left\| {\mathbf{s}_{2}} \right\|\left\| \mathbf{z}_{\textnormal{Q}_{1}} \right\|}  \nonumber \\
& = & \frac{\left\langle {\mathbf{s}_{2}},{\mathbf{s}_{1}} \right\rangle -\left\langle {\mathbf{s}_{2}},{{\nu }_{1}}{\mathbf{s}_{1}}+{\mathbf{w}_{1}} \right\rangle }{\left\| {\mathbf{s}_{2}} \right\|\left\| \mathbf{z}_{\textnormal{Q}_{1}} \right\|} \\
& = & \frac{\left( 1-{{\nu}_{1}} \right)\left\langle {\mathbf{s}_{2}},{\mathbf{s}_{1}} \right\rangle -\left\langle {\mathbf{s}_{2}},{\mathbf{w}_{1}} \right\rangle }{\left\| {\mathbf{s}_{2}} \right\|\left\| \mathbf{z}_{\textnormal{Q}_{1}} \right\|}
=\frac{{{2}^{-2{{R}_{1}}}}\left\| {\mathbf{s}_{1}} \right\|\left\| {\mathbf{s}_{2}} \right\|\cos \sphericalangle \left( {\mathbf{s}_{1}},{\mathbf{s}_{2}} \right)}{\left\| {\mathbf{s}_{2}} \right\|\sqrt{{{n\sigma }^{2}}{{2}^{-2{{R}_{1}}}}}}-\frac{\left\langle {\mathbf{s}_{2}},{\mathbf{w}_{1}} \right\rangle }{\left\| {\mathbf{s}_{2}} \right\|\left\| \mathbf{z}_{\textnormal{Q}_{1}} \right\|} \\
& = & \sqrt{{{2}^{-2{{R}_{1}}}}}\cos \sphericalangle \left( {\mathbf{s}_{1}},{\mathbf{s}_{2}} \right)-\frac{\left\langle {\mathbf{s}_{2}},{\mathbf{w}_{1}} \right\rangle }{\left\| {\mathbf{s}_{2}} \right\|\left\| \mathbf{z}_{\textnormal{Q}_{1}} \right\|} .
\end{IEEEeqnarray*}
The second term vanishes when $n \to \infty$ as in proof of Lemma~\ref{A2,A3,A4}. \hfill \qed
\end{remark}

\medskip

\begin{lemma}\label{A2,A3,A4_x2,xv}
For every $\delta>0$ and $\epsilon>0$ there exists an ${n'_{{\cal A}}}(\delta,\epsilon)$ such that for all
$n>{n'_{{\cal A}}}(\delta,\epsilon)$
\begin{equation*}
\Pr\left[ {{{\cal A}}_{2}}|{\cal E}_{\mathbf{S}}^{c}\right]<\delta, \quad  \Pr\left[ {{{\cal A}}_{3}}|{\cal E}_{\mathbf{S}}^{c}\right]<\delta, \quad  \Pr\left[ {{{\cal A}}_{4}}|{\cal E}_{\mathbf{S}}^{c}\right]<\delta.
\end{equation*}
\end{lemma}

\medskip

\begin{IEEEproof}
We start with the derivation of the bound on ${\cal A}_2$. To this end, we first
upper-bound the inner product between $\textbf{s}_2$ and $\textbf{w}_{3}$.
Let $\textbf{s}_{2,P}$ denote the projection of $\textbf{s}_2$ onto the subspace of $\mathbb{R}^{n}$
that is orthogonal to $\textbf{z}_{\textnormal{Q}_1}$, and that thus contains $\textbf{w}_{2}$. Hence,
\begin{IEEEeqnarray}{rCl}
   \left| \frac{{{\nu }_{2}}}{\left\| \textbf{u}_{2}^{*} \right\|\left\| \textbf{v}^{*} \right\|}\left\langle {\textbf{s}_{2}},{\textbf{w}_{3}} \right\rangle  \right|\
& \overset{(a)}{\mathop{=}}& \,\left| \cos \sphericalangle \left( {\textbf{s}_{2}},\textbf{u}_{2}^{*} \right)\left\langle \frac{{\textbf{s}_{2}}}{\left\| {\textbf{s}_{2}} \right\|},\frac{{\textbf{w}_{3}}}{\left\| \textbf{v}^{*} \right\|} \right\rangle  \right| \nonumber \\
 & \overset{(b)}{\mathop{\le }}& \,\left| \cos \sphericalangle \left( {\textbf{s}_{2}},\textbf{u}_{2}^{*} \right) \right|\left| \left\langle \frac{{\textbf{s}_{2}}}{\left\| {\textbf{s}_{2}} \right\|},\frac{{\textbf{w}_{3}}}{\left\| {\textbf{w}_{3}} \right\|} \right\rangle  \right| \nonumber \\
 & \le &\left| \left\langle \frac{{\textbf{s}_{2}}}{\left\| {\textbf{s}_{2}} \right\|},\frac{{\textbf{w}_{3}}}{\left\| {\textbf{w}_{3}} \right\|} \right\rangle  \right| \nonumber \\
 & = & \left| \left\langle \frac{{\textbf{s}_{2,\textnormal{P}}}}{\left\| {\textbf{s}_{2}} \right\|},\frac{{\textbf{w}_{3}}}{\left\| {\textbf{w}_{3}} \right\|} \right\rangle  \right| \nonumber \\
 & \le & \left| \left\langle \frac{{\textbf{s}_{2,\textnormal{P}}}}{\left\| {\textbf{s}_{2,\textnormal{P}}} \right\|},\frac{{\textbf{w}_{3}}}{\left\| {\textbf{w}_{3}} \right\|} \right\rangle  \right| \nonumber \\
 & =& \left| \cos \sphericalangle \left( {\textbf{s}_{2,\textnormal{P}}},{\textbf{w}_{3}} \right) \right|,
\label{eq:proof_A2_x2,xv}
\end{IEEEeqnarray}
where (a) follows by the definition of $\nu_{2}$ and (b) follows since by the definition
of $\textbf{w}_{3}$ we have $\left\| \textbf{w}_{3} \right\|\le \left\| \textbf{v}^{*} \right\|$.
By \eqref{eq:proof_A2_x2,xv} it now follows that
\begin{IEEEeqnarray*}{rCl}
\Pr\left[ {{{\cal A}}_{2}}|{\cal E}_{\textbf{S}}^{c} \right]
&\le &\Pr \left[ \left( {{\textbf{S}}_{1}},{{\textbf{S}}_{2}},{\mathscr{C}_{1}},{\mathscr{C}_{2}},{\mathscr{C}_{c}} \right):\left| \cos \sphericalangle \left( {{\textbf{S}}_{2,\textnormal{P}}},{\textbf{W}_{3}} \right) \right|>\epsilon \ \mid {\cal E}_{\textbf{S}}^{c} \right] \\
&\overset{(a)}=&{{\sf{E}}_{{{\textbf{S}}_{1}},{{\textbf{S}}_{2}}}}\left[ {{\Pr }_{{\mathscr{C}_{1}},{\mathscr{C}_{2}},{\mathscr{C}_{c}}}}\left( \left| \cos \sphericalangle \left( {{\textbf{s}}_{2,P}},{\textbf{W}_{3}} \right) \right|>\epsilon \ \mid\left( {{\textbf{S}}_{1}},{{\textbf{S}}_{2}} \right)=\left( {{\textbf{s}}_{1}},{{\textbf{s}}_{2}} \right),{\cal E}_{\textbf{S}}^{c} \right) \right],
\end{IEEEeqnarray*}
where in (a) we have denoted by ${{\Pr }_{{\mathscr{C}_{1}},{\mathscr{C}_{2}},{\mathscr{C}_{c}}}}\left( \cdot \mid \cdot \right)$
the conditional probability of the codebooks $c_1, c_2$ and $c_3$ being such that
$\left| \cos \sphericalangle \left( {{\textbf{s}}_{2,P}},{\textbf{W}_{3}} \right) \right|>\epsilon $,
given $\left( {{\textbf{S}}_{1}},{{\textbf{S}}_{2}} \right)=\left( {{\textbf{s}}_{1}},{{\textbf{s}}_{2}} \right)$
and ${\cal E}_{\textbf{S}}^{c}$.
To conclude our bound we now notice that conditioned on
$\left( {{\textbf{S}}_{1}},{{\textbf{S}}_{2}} \right)=\left( {{\textbf{s}}_{1}},{{\textbf{s}}_{2}} \right)$,
the random vector ${\textbf{W}_{3}}/\left\| {\textbf{W}_{3}} \right\|$ is distributed uniformly on the surface of
the centered $\mathbb{R}^{n-1}$-sphere of unit radius, that lies in the subspace that is
orthogonal to $\textbf{s}_2$. Hence,
\begin{IEEEeqnarray}{rCl}\Pr\left[ {{{\cal A}}_{2}}|{\cal E}_{\textbf{S}}^{c} \right]
& \le & {{\sf{E}}_{{{\textbf{S}}_{1}},{{\textbf{S}}_{2}}}}\left[ \left. \frac{2{{C}_{n-1}}\left( \Theta  \right)}{{{C}_{n-1}}\left( \pi  \right)} \right|{\cal E}_{\textbf{S}}^{c} \right] \nonumber \\
& \le &\frac{2{{C}_{n-1}}\left( \Theta \right)}{{{C}_{n-1}}\left( \pi  \right)}  ,
\label{eq:A2_x2,v_proof_final}
\end{IEEEeqnarray}
where $\Theta \triangleq \arccos(\epsilon)$.
As $0<\epsilon \leq 1$, $\Theta\in (0,\frac{\pi}{2})$, and thus
by \cite[Lemma B.4]{Bross} the r.h.s. of \eqref{eq:A2_x2,v_proof_final} tends to 0 as $n \to \infty$,
and therefore $\Pr\left[ {{{\cal A}}_{2}}|{\cal E}_{\textbf{S}}^{c} \right]\to 0 $.
By similar arguments it also follows that $\Pr\left[ {{{\cal A}}_{3}}|{\cal E}_{\textbf{S}}^{c} \right]\to 0 $
as $n \to \infty$.

\medskip

To conclude the proof of Lemma~\ref{A2,A3,A4_x2,xv}, we derive the bound on ${\cal A}_4$.
The derivations are similar to those for ${\cal A}_2$.
First, let $\textbf{w}_{2,P}$ denote the projection of $\textbf{w}_2$ onto the subspace of $\mathbb{R}^{n}$
that is orthogonal to $\textbf{z}_{\textnormal{Q}_1}$, and that thus contains $\textbf{w}_{3}$.
As in \eqref{eq:proof_A2_x2,xv}, we can show that
\begin{equation}\label{eq:proof_A4_x2,xv}
\left| \frac{1}{\left\| \textbf{u}_{2}^{*} \right\|\left\| \textbf{v}^{*} \right\|}\left\langle {\textbf{w}_{2}},{\textbf{w}_{3}} \right\rangle  \right|\
\leq\,\left| \cos \sphericalangle \left( {\textbf{w}_{2,P}},{\textbf{w}_{3}} \right) \right|,
\end{equation}
from which it then follows that
\begin{equation*}
\Pr\left[ {{{\cal A}}_{4}}|{\cal E}_{\textbf{S}}^{c} \right]
\leq {{\sf{E}}_{{{\textbf{S}}_{1}},{{\textbf{S}}_{2}},{\mathscr{C}_{2}}}}\left[ {{\Pr }_{{\mathscr{C}_{1}},{\mathscr{C}_{c}}}}\left( \left| \cos \sphericalangle \left( {\textbf{w}_{2,P}},{\textbf{w}_{3}} \right) \right|>\epsilon \ \mid\left( {{\textbf{S}}_{1}},{{\textbf{S}}_{2}},{{\textbf{U}}_{2}} \right)=\left( {{\textbf{s}}_{1}},{{\textbf{s}}_{2}},{{\textbf{u}}_{2}} \right),{\cal E}_{\textbf{S}}^{c} \right) \right].
\end{equation*}
The desired upper bound now follows by noticing that conditioned on
$\left( {{\textbf{S}}_{1}},{{\textbf{S}}_{2}},{{\textbf{U}}_{2}} \right)=\left( {{\textbf{s}}_{1}},{{\textbf{s}}_{2}},{{\textbf{u}}_{2}} \right)$,
and ${{\mathscr C}_{2}={\cal C}_2}$, the random vector ${\textbf{W}_{3}}/\left\| {\textbf{W}_{3}} \right\|$ is distributed uniformly on the surface of the centered $\mathbb{R}^{n-1}$-sphere of unit radius, that lies in the subspace that is
orthogonal to $\textbf{z}_{\textnormal{Q}_1}$. Hence, similarly as in the derivation for ${\cal A}_2$
\begin{IEEEeqnarray}{l}
\Pr\left[ {{{\cal A}}_{4}}|{\cal E}_{\textbf{S}}^{c} \right]
\le {{\sf{E}}_{{{\textbf{S}}_{1}},{{\textbf{S}}_{2}},{\mathscr{C}_{2}}}}\left[ \left. \frac{2{{C}_{n-1}}\left(\Theta  \right)}{{{C}_{n-1}}\left( \pi  \right)} \right|{\cal E}_{\textbf{S}}^{c} \right],
\label{eq:Rnminusonesphere_proof_final}
\end{IEEEeqnarray}
where $\Theta \triangleq \arccos(\epsilon)$.
As $0<\epsilon \leq 1$, $\Theta\in (0,\frac{\pi}{2})$, and thus
by \cite[Lemma B.4]{Bross} the r.h.s. of \eqref{eq:Rnminusonesphere_proof_final} tends to 0 as $n \to \infty$,
and therefore $\Pr\left[ {{{\cal A}}_{4}}|{\cal E}_{\textbf{S}}^{c} \right]\to 0 $.
\end{IEEEproof}

\medskip

Combining Lemma~\ref{A1_x2,xv} and Lemma~\ref{A2,A3,A4_x2,xv} with \eqref{eq:eps_x2,xv_proof} gives that for
every $\delta>0$ and $0<\epsilon<0.3$ there exists an $n'_{{\cal A}}(\delta,\epsilon)$ such that for all
$n>n'_{{\cal A}}(\delta,\epsilon)$
\begin{flalign*}
&&
\Pr\left[ {{\cal E}_{\left( {\textbf{X}_{2}},{\textbf{X}_{v}} \right)}}\cap {\cal E}_{{\textbf{X}_{2}}}^{c}\cap {\cal E}_{{\textbf{X}_{v}}}^{c}\cap {\cal E}_\textbf{S}^{c} \right]\le 3\delta.
&&
\qed
\end{flalign*}


\textbf{Proof of inequality \eqref{eq:eps_xWZ}:}

The error probability analysis can be outlined as follows:
\begin{enumerate}
\item The pair $(\textbf{z}_{\textnormal{Q}_1},\textbf{s}_{2}) \notin {A_{\epsilon}^{*(n)}}$,
where $A^{*(n)}_{\epsilon}$ denotes the $\epsilon$- strongly jointly typical set of sequences (see \cite[Chapter~2]{EGYHKBook}).
The probability of this event is small for large enough $n$, by the weak low of large numbers.
\item The sequence $\textbf{z}_{\textnormal{Q}_1}$ is typical, but there does not exist a sequence $\textbf{v}\in {\cal C}_c$ such that
$(\textbf{z}_{\textnormal{Q}_1},\textbf{v}) \in {A_{\epsilon}^{*(n)}}$.
As in the proof of the rate distortion theorem, the probability of this event is small if for $\epsilon'<\epsilon$
\begin{equation*}
R_\textnormal{c}>I(V;Z_{\textnormal{Q}_1})+\delta(\epsilon'),
\end{equation*}
 where $\delta(\epsilon')\to 0$ as $\epsilon'\to 0$.
\item The pair $(\textbf{z}_{\textnormal{Q}_1},\textbf{v}) \in {A_{\epsilon}^{*(n)}}$,
but $(\textbf{v},\textbf{s}_2) \notin {A_{\epsilon}^{*(n)}}$, i.e. the codeword is not
jointly typical with the $\textbf{s}_{2}$ sequence. By the Markov lemma \cite[Lemma~12.1]{EGYHKBook}, the probability
of this event is small if $n$ is large enough, since $V\Markov Z_{Q_1}\Markov S_2$ forms a Markov chain.
\item There exists $\tilde{\textbf{v}}\in {\cal C}_c\setminus \textbf{v}^{*}$ within the same bin of $\textbf{v}^{*}$, such that
$(\tilde{\textbf{v}},\textbf{s}_{2}) \in {A_{\epsilon}^{*(n)}}$. Since the probability that a randomly chosen
$\tilde{\textbf{v}}$ is jointly typical with $\textbf{s}_{2}$ is $\approx 2^{-n[I(S_2;V)-\delta(\epsilon)]}$, the probability of the former event is upper bounded by
\begin{IEEEeqnarray*}{rCl}
\Pr (\exists \> \tilde{\textbf{v}}\in {\cal C}_c\setminus \textbf{v}^{*} : (\tilde{\textbf{v}},\textbf{s}_2)\in A_{\epsilon}^{*(n)})
& \leq &
 \frac{2^{nR_\textnormal{c}}}{2^{n[R_\textnormal{c}-I(S_2;V)-\frac{1}{2}\delta(\epsilon)]}}2^{-n[I(S_2;V)-\delta(\epsilon)]} \nonumber \\
& = & \frac{2^{n[I(V;Z_{Q_1})+\delta(\epsilon')]}}{2^{n[I(V;Z_{Q_1})+\delta(\epsilon')-I(S_2;V)-\frac{1}{2}\delta(\epsilon)]}}2^{-n[I(S_2;V)-\delta(\epsilon)]}
 \nonumber \\
& = & 2^{-\frac{n}{2}\delta(\epsilon)},
\end{IEEEeqnarray*}
which goes to zero as $n \rightarrow \infty $.
\end{enumerate}

\medskip

The formal detailed proof is as follows:
We start with a lemma that will be used to prove \eqref{eq:eps_xWZ}.

\begin{lemma}\label{eps_v,s2}
Define the event that the quantized sequence $\mathbf{v}^{*}$ and the source sequence $\mathbf{s}_2$
have an atypical angle to each other
\begin{equation*}
{{\cal E}_{\mathbf{v},\mathbf{s}_2}}=\Bigl\{ (\mathbf{s}_1,\mathbf{s}_2,{\cal C}_1,{\cal C}_2,{\cal C}_c):
\left|\rho_{{\mathbf{v}},\mathbf{s}_2}-\cos\sphericalangle({{\mathbf{v}}^{*}}( {{\mathbf{s}}_{1}},{{\cal C}_{1}}),\mathbf{s}_2)\right| > 5\epsilon \Bigr\}.
\end{equation*}
Then, for every $\delta > 0$ and $\epsilon >0$ there exists an $n'\left( \delta ,\epsilon  \right)\in \mathbb{N}$Â
such that for all $n>n'\left( \delta ,\epsilon  \right)$
\begin{equation*}
\Pr\left[ {{\cal E}_{\mathbf{v},\mathbf{s}_2}}\cap {\cal E}_{{\mathbf{X}_{v}}}^{c}\cap {\cal E}_\mathbf{S}^{c} \right]<\delta .
\end{equation*}
\end{lemma}

\medskip

\begin{IEEEproof}
We start with the following decomposition
\begin{IEEEeqnarray}{rCl}\label{eq:cos(v,s2)}
\cos \sphericalangle \left( \textbf{v}^{*},\textbf{s}_{2} \right)
&=&\frac{\left\langle \textbf{v}^{*},\textbf{s}_{2} \right\rangle }
{\left\| \textbf{v}^{*} \right\|\left\| \textbf{s}_{2} \right\|} \nonumber \\
&\overset{(a)}{\mathop{=}}&\,\frac{\left\langle {\textbf{v}^{*}},\rho{\textbf{s}_{1}}+{\textbf{z}_{\textnormal{G}_2}} \right\rangle }
{\left\| \textbf{v}^{*} \right\|\left\| \textbf{s}_{2} \right\|} \nonumber \\
& = & \frac{\rho\left\langle {\textbf{v}^{*}},{\textbf{s}_{1}} \right\rangle
+\left\langle {\textbf{v}^{*}},{\textbf{z}_{\textnormal{G}_2}} \right\rangle }
{\left\| \textbf{v}^{*} \right\|\left\| \textbf{s}_{2} \right\|},
\end{IEEEeqnarray}
where in (a) we represent $\textbf{s}_2$ as a scaled version
of $\textbf{s}_1$ corrupted by an additive gaussian noise $\textbf{z}_{G2}$.
More precisely,
\begin{equation}\label{s1,s2}
\textbf{s}_2=\rho\textbf{s}_1+\textbf{z}_{\textnormal{G}_2} \ \ \mbox{ where } \ \
\rho=\frac{\left\| \textbf{s}_{2} \right\|}{\left\| \textbf{s}_{1} \right\|}
\cos \sphericalangle \left( {\textbf{s}_{1}},\textbf{s}_{2} \right).
\end{equation}
With this choice of $\rho$, the vector $\textbf{z}_{\textnormal{G}2}$ is always orthogonal to $\textbf{s}_{1}$.

Now, define the two events
\begin{IEEEeqnarray*}{rCl}
{{{\cal A}}_{1}}&=&\left\{ \left( {\textbf{s}_{1}},{\textbf{s}_{2}},{{\cal C}_{1}},{{\cal C}_{2}},{{\cal C}_{c}} \right):\left| {\rho }_{\textbf{v},\textbf{s}_2}-\frac{\rho }{\left\| \textbf{v}^{*} \right\|\left\| \textbf{s}_{2} \right\|}
\left\langle {\textbf{v}^{*}},{\textbf{s}_{1}} \right\rangle \right|>4\epsilon  \right\} \\
{{{\cal A}}_{2}}&=&\left\{ \left( {\textbf{s}_{1}},{\textbf{s}_{2}},{{\cal C}_{1}},{{\cal C}_{2}},{{\cal C}_{c}} \right):\left| \frac{1}{\left\| \textbf{v}^{*} \right\|\left\| \textbf{s}_{2} \right\|}
\left\langle {\textbf{v}^{*}},{\textbf{z}_{\textnormal{G}_2}} \right\rangle  \right|>\epsilon  \right\} .
\end{IEEEeqnarray*}
Note that by \eqref{eq:cos(v,s2)}, ${{\cal E}_{\textbf{v},\textbf{s}_2}}=\left\{ \left( {\textbf{s}_{1}},{\textbf{s}_{2}},{{\cal C}_{1}},{{\cal C}_{2}},{{\cal C}_{c}} \right):
\left| \rho_{\textbf{v},\textbf{s}_2}-\cos \sphericalangle \left( \textbf{v}^{*},\textbf{s}_{2} \right) \right|>4\epsilon  \right\}
\subset \left( {{{\cal A}}_{1}}\cup {{{\cal A}}_{2}} \right)$.
Thus,
\begin{IEEEeqnarray}{rCl}\label{eq:WZ-eps_x_proof}
\Pr\left[ {{\cal E}_{\textbf{v},\textbf{s}_2}}\cap {\cal E}_{{\textbf{X}_{v}}}^{c}\cap {\cal E}_\textbf{S}^{c} \right]
& \, \le &\Pr\left[ {{{\cal A}}_{1}}\cap {\cal E}_{{\textbf{X}_{v}}}^{c}\cap
{\cal E}_{\textbf{S}}^{c} \right]
+\Pr\left[ {{{\cal A}}_{2}}\cap {\cal E}_{{\textbf{X}_{v}}}^{c}\cap {\cal E}_{\textbf{S}}^{c} \right] \nonumber \\
& \, \le &\Pr\left[ {{{\cal A}}_{1}}|{\cal E}_{{\textbf{X}_{v}}}^{c}\cap
{\cal E}_{\textbf{S}}^{c} \right]
+\Pr\left[ {{{\cal A}}_{2}}|{\cal E}_{\textbf{S}}^{c} \right].
\end{IEEEeqnarray}
The two terms on the r.h.s. of \eqref{eq:WZ-eps_x_proof} are now bounded in the following
two lemmas.

\medskip

\begin{lemma}\label{WZ-A1}
For $\epsilon<1$
\begin{equation*}
\Pr\left[ {{{\cal A}}_{1}}|{\cal E}_{\mathbf{S}}^{c} \cap {\cal E}_{{\mathbf{X}_{v}}}^{c} \right]=0.
\end{equation*}
\end{lemma}

\medskip

\begin{IEEEproof}
We first note that the term in the definition of ${\cal A}_1$ can be rewritten as
\begin{equation}\label{eq:WZ-A1_proof}
\frac{\rho }{\left\| \textbf{v}^{*} \right\|\left\| \textbf{s}_{2} \right\|}\left\langle {\textbf{v}^{*}},{\textbf{s}_{1}} \right\rangle
=\cos \sphericalangle \left( {\textbf{s}_{1}},\textbf{s}_{2} \right)
\cos \sphericalangle \left( {\textbf{v}^{*}},{\textbf{s}_{1}} \right).
\end{equation}
Note that the second term satisfies
\begin{equation}
\cos \sphericalangle \left( {\textbf{v}^{*}},{\textbf{s}_{1}} \right)
=\frac{\left\langle \textbf{v}^{*},\textbf{s}_{1} \right\rangle }
{\left\| \textbf{v}^{*} \right\|\left\| \textbf{s}_{1} \right\|}
=\frac{\left\langle \textbf{v}^{*},\textbf{u}^{*}_{1}+\textbf{z}_{\textnormal{Q}_1} \right\rangle }
{\left\| \textbf{v}^{*} \right\|\left\| \textbf{s}_{1} \right\|}
=\frac{\left\langle \textbf{v}^{*},\textbf{u}^{*}_{1} \right\rangle }
{\left\| \textbf{v}^{*} \right\|\left\| \textbf{s}_{1} \right\|}
+\frac{\left\langle \textbf{v}^{*},\textbf{z}_{\textnormal{Q}_1} \right\rangle}
{\left\| \textbf{v}^{*} \right\|\left\| \textbf{s}_{1} \right\|}.
\end{equation}
By \eqref{eq:u1,v_upper bound} and Lemma~\ref{u1,v}, the first term can be bounded
by
\begin{equation*}
\Bigl|\frac{\left\langle \textbf{v}^{*},\textbf{u}^{*}_{1} \right\rangle }
{\left\| \textbf{v}^{*} \right\|\left\| \textbf{s}_{1} \right\|}\Bigr|
\leq \frac{12\delta+3\epsilon}{\sqrt{2^{-2R_1}(1-2^{-2R_\textnormal{c}})}}=\epsilon_1.
\end{equation*}
The second term can be factorized
\begin{equation}
\frac{\left\langle \textbf{v}^{*},\textbf{z}_{\textnormal{Q}_1} \right\rangle}
{\left\| \textbf{v}^{*} \right\|\left\| \textbf{s}_{1} \right\|}
=\frac{\left\langle \textbf{v}^{*},\textbf{z}_{\textnormal{Q}_1} \right\rangle }
{\left\| \textbf{v}^{*} \right\|\left\| \textbf{z}_{\textnormal{Q}_1} \right\|}
\frac{\left\| \textbf{z}_{\textnormal{Q}_1} \right\|}{\left\| \textbf{s}_{1} \right\|}
=\cos \sphericalangle \left( {\textbf{z}_{\textnormal{Q}_1}},{\textbf{v}^{*}} \right)\sqrt{2^{-2R_1}}.
\end{equation}

We can now upper and lower bound the r.h.s. of \eqref{eq:WZ-A1_proof} for
$\left( {\textbf{s}_{1}},{\textbf{s}_{2}},{{\cal C}_{1}},{{\cal C}_{2}},{{\cal C}_{c}} \right)
\in {{\cal E}_{\textbf{S}}^{c} \cap {\cal E}_{{\textbf{X}_{v}}}^{c}}$
by noticing that
 $\left( {\textbf{s}_{1}},{\textbf{s}_{2}},{{\cal C}_{1}},{{\cal C}_{2}},{{\cal C}_{c}} \right)\in
{\cal E}_{\textbf{S}}^{c}$ implies
\begin{equation*}
\left| \cos \sphericalangle \left( {\textbf{s}_{1}},{\textbf{s}_{2}} \right)-\rho  \right|<\rho \epsilon
,
\end{equation*}
and that $\left( {\textbf{s}_{1}},{\textbf{s}_{2}},{{\cal C}_{1}},{{\cal C}_{2}},{{\cal C}_{c}} \right)\in
{\cal E}_{\textbf{X}_v}^{c}$ implies
\begin{equation*}
\left| \sqrt{1-{{2}^{-2{{R}_\textnormal{c}}}}}-\cos \sphericalangle \left( {\textbf{z}_{\textnormal{Q}_1}},\textbf{v}^{*} \right) \right|<\epsilon \sqrt{1-{{2}^{-2{{R}_\textnormal{c}}}}}.
\end{equation*}
Hence, combined with \eqref{eq:WZ-A1_proof} this gives
\begin{equation*}
{\rho}_{\textbf{v},\textbf{s}_2}{{\left( 1-\epsilon  \right)}^{2}}+\rho\epsilon_1(1-\epsilon)\le
\frac{\rho}{\left\| \textbf{v}^{*} \right\|\left\| \textbf{s}_{2} \right\|}\left\langle {\textbf{v}^{*}},{\textbf{s}_{1}} \right\rangle\
\le {\rho}_{\textbf{v},\textbf{s}_2}{{\left( 1+\epsilon  \right)}^{2}}+\rho\epsilon_1(1+\epsilon),
\end{equation*}
whenever
$\left( {\textbf{s}_{1}},{\textbf{s}_{2}},{{\cal C}_{1}},{{\cal C}_{2}},{{\cal C}_{c}} \right)\in
{\cal E}_{\textbf{S}}^{c} \cap {\cal E}_{{\textbf{X}_{v}}}^{c}$.
The l.h.s. can be lower bounded by ${{\left( 1-2\epsilon  \right)}} \leq{{\left( 1-\epsilon  \right)}^{2}}$,
and the r.h.s. can be upper bounded by ${{\left( 1+\epsilon  \right)}^{2}}\leq {{\left( 1+3\epsilon  \right)}}$
whenever $\epsilon\leq 1$. Hence, for $\epsilon\leq 1$
\begin{equation*}
\left| {\rho}_{\textbf{v},\textbf{s}_2}
-\frac{\rho }{\left\| \textbf{v}^{*} \right\|\left\| \textbf{s}_{1} \right\|}
\left\langle {\textbf{v}^{*}},{\textbf{s}_{2}} \right\rangle \right|\
\le 3{\rho}_{\textbf{v},\textbf{s}_2}\epsilon+\rho\epsilon_1(1+\epsilon) \le 4\epsilon,
\end{equation*}
and thus
\begin{flalign*}
&&
\Pr\left[ {{{\cal A}}_{1}}|{\cal E}_{\textbf{S}}^{c} \cap {\cal E}_{{\textbf{X}_{v}}}^{c} \right]=0.
&&
\end{flalign*}
\end{IEEEproof}

\medskip

\begin{lemma}\label{WZ-A2}
For every $\delta>0$ and $\epsilon>0$ there exists an ${n'_{{\cal A}}}(\delta,\epsilon)$ such that for all
$n>{n'_{{\cal A}}}(\delta,\epsilon)$
\begin{equation*}
\Pr\left[ {{{\cal A}}_{2}}|{\cal E}_{\mathbf{S}}^{c}\right]<\delta.
\end{equation*}
\end{lemma}

\medskip

\begin{IEEEproof}
By similar arguments as in proof of Lemma~\ref{A2,A3,A4}, it follows
that for every $\epsilon>0$, $\Pr\left[ {{{\cal A}}_{2}}|{\cal E}_{\textbf{S}}^{c} \right]\to 0 $
as $n \to \infty$.
\end{IEEEproof}

\medskip

Combining Lemma~\ref{WZ-A1} and Lemma~\ref{WZ-A2} with \eqref{eq:WZ-eps_x_proof} gives that for
every $\delta>0$ and $0<\epsilon<1$ there exists an $n'_{{\cal A}}(\delta,\epsilon)$ such that for all
$n>n'_{{\cal A}}(\delta,\epsilon)$
\begin{flalign*}
&&
\Pr\left[ {{\cal E}_{\textbf{v},\textbf{s}_2}}\cap {\cal E}_{{\textbf{X}_{v}}}^{c}\cap {\cal E}_\textbf{S}^{c} \right]\le \delta.
&&
\end{flalign*}
\end{IEEEproof}

\medskip


We now start with a definition that will be used to prove \eqref{eq:eps_xWZ}.
\begin{equation*}
{{\cal E}'_{\textbf{X}_\textnormal{WZ}}}\triangleq
\Bigl\{ (\textbf{s}_1,\textbf{s}_2,{\cal C}_1,{\cal C}_2,{\cal
C}_c):
\exists \> \tilde{\textbf{v}} \in {\cal C}_c\setminus \{\textbf{v}^{*}\}
\ \mbox{s.t.} \
\cos\sphericalangle(\tilde{\textbf{v}},\textbf{s}_2)\geq
\rho_{{\textbf{v}},\textbf{s}_2}-5\epsilon \Bigr\}.
\end{equation*}
Note that
\begin{equation}\label{eq:Rwz-proof1}
{{\cal E}_{\textbf{X}_\textnormal{WZ}}}
= \Bigl\{ (\textbf{s}_1,\textbf{s}_2,{\cal C}_1,{\cal C}_2,{\cal
C}_c):
\exists \> \tilde{\textbf{v}} \in {\cal C}_c\setminus \{\textbf{v}^{*}\}
\ \mbox{s.t.} \
\left|\rho_{{\textbf{v}},\textbf{s}_2}-\cos\sphericalangle(\tilde{\textbf{v}},\textbf{s}_2)\right|\leq
5\epsilon \Bigr\}
\subseteq
{{\cal E}'_{\textbf{X}_\textnormal{WZ}}}. \nonumber \\
\end{equation}

We now state one more lemma that will be used for the proof of \eqref{eq:eps_xWZ}:
\begin{lemma}\label{WZ_rate}
For every $\Delta \in (0,1]$, let the set ${\cal G}$ be given by
\begin{equation*}
{\cal G} =\left\{ \left( {\mathbf{s}_{1}},{\mathbf{s}_{2}},{{{\cal C}}_{1}},{{{\cal C}}_{2}},{{{\cal C}}_{c}} \right):\ \exists \> {{\mathbf{v}}}\in {{{\cal C}}_{c}}\backslash \left\{ \mathbf{v}^{*} \right\} \
\mbox{s.t.}  \ \cos \sphericalangle \left( \mathbf{s}_2,{{\mathbf{v}}} \right)\ge \Delta  \right\}.
\end{equation*}
Then,
\begin{equation*}
\frac{1}{n}\log M_b<-\frac{1}{2}\log \left( 1-{{\Delta }^{2}} \right)
\quad \implies \quad
\left( \underset{n\to \infty }{\mathop{\lim }}\,\ {\Pr}\left[ {\cal G} |{\cal E}_{{\mathbf{X}}_{v}}^{c} \right]=0, \ \epsilon >0 \right),
\end{equation*}
\end{lemma}
where $M_b$ denotes the bin size in the partitioned codebook ${\cal C}_c$, and ${\cal E}_{\textbf{X}_v}$ is defined in \eqref{eq:eps_xv_def}.

\medskip

\begin{IEEEproof}
The proof follows from upper-bounding in every point on ${\cal S}_c$ the
density of every ${{\textbf{v}}}\in {{{\cal C}}_{c}}\backslash \left\{ \textbf{v}^{*} \right\}$
and then using a standard argument from sphere-packing.
\end{IEEEproof}

\medskip

Next,
\begin{IEEEeqnarray}{l}
\Pr\left[ {{\cal E}_{\textbf{X}_\textnormal{WZ}}}\cap {\cal E}_{{\textbf{X}_{v}}}^{c}\cap {\cal E}_\textbf{S}^{c} \right]
\overset{(a)}{\mathop{\le }}\,\Pr\left[ {{\cal E}'_{\textbf{X}_\textnormal{WZ}}}\cap {\cal E}_{{\textbf{X}_{v}}}^{c}\cap {\cal E}_\textbf{S}^{c} \right]
\overset{(b)}{\mathop{\le }}\,\Pr\left[ {{\cal E}'_{\textbf{X}_\textnormal{WZ}}}
\middle|{\cal E}_{{\textbf{X}_{v}}}^{c} \right],
\label{eq:Rwz-proof2}
\end{IEEEeqnarray}
where (a) follows by \eqref{eq:Rwz-proof1} and (b) follows because ${\cal E}_{\textbf{X}}^{c}\subseteq {\cal E}_{{\textbf{X}_{v}}}^{c}$.

The proof of \eqref{eq:eps_xWZ} is now completed by combining \eqref{eq:Rwz-proof2} with Lemma~\ref{WZ_rate}.
This gives that for every $\delta >0$ and every $\epsilon >0$ there exists some
$n'(\delta ,\epsilon )$
such that for all $n>n'(\delta ,\epsilon )$,
we have
\begin{equation*}
\Pr\left[ {{\cal E}_{\textbf{X}_\textnormal{WZ}}}\cap {\cal E}_{{\textbf{X}_{v}}}^{c}\cap {\cal E}_\textbf{S}^{c} \right]<\delta ,
\end{equation*}
whenever
\begin{IEEEeqnarray}{l} \label{eq:Rwz-proof211}
\frac{1}{n}\log M_b<-\frac{1}{2}\log {\left(1-(\rho_{\textbf{v},\textbf{s}_2}-5\epsilon)^{2}\right)}.
\end{IEEEeqnarray}
The constraint  \eqref{eq:Rwz-proof211} yields the following bound on the bin size
\begin{equation}
M_b\leq
\left(1-\rho_{\textbf{v},\textbf{s}_2}^{2}\right)^{-\frac{n}{2}}2^{-n\delta(\epsilon)},
 \label{eq:Rwz-proof212}
\end{equation}
where $\delta(\epsilon)\to 0$ as $\epsilon\to 0$.

The desired result follows now by noticing that the bin size in our code construction
(defined in \eqref{eq:codebookbinsize}) satisfies  \eqref{eq:Rwz-proof212}. \hfill \qed



\medskip

\subsection{Upper bound on expected distortion --- {\bf  Proof of Proposition~\ref{proposition_genieaideddist12}}
}
\label{Upper Bound on Expected Distortion}

We derive an upper bound on the achievable distortion for the proposed
vector-quantizer scheme. By Corollary~\ref{corollary_rates}, it suffices to analyze
the genie-aided scheme. Since
$\hat{\textbf{S}}_1{\hspace{-.4em}}^{\textnormal{G}}={{\gamma }_{1,1}}{{\textbf{U}}_{1}}^{*}+{{\gamma }_{1,2}}{\textbf{U}_{2}}^{*}+{{\gamma }_{1,3}}{\textbf{V}^{*}}$,
we have
\begin{IEEEeqnarray}{rCl}
{{D}_{1}}& =& \frac{1}{n}{\sf{E}}\left[ {\| {\textbf{S}_{1}}-\hat{\textbf{S}}_{1}{\hspace{-.4em}}^\textnormal{G} \|^{2}} \right] \nonumber\\
& = &\frac{1}{n} {\sf{E}}\left[ {{\left\| {\textbf{S}_{1}}-\left( {{\gamma }_{1,1}}{\textbf{U}_{1}}^{*}
+{{\gamma }_{1,2}}{\textbf{U}_{2}}^{*}+{{\gamma }_{1,3}}{\textbf{V}^{*}} \right) \right\|}^{2}} \right]  \nonumber\\
 & = &  \frac{1}{n}\Bigl({\sf{E}}\left[ {{\left\| {\textbf{S}_{1}} \right\|}^{2}} \right]-2{{\gamma }_{1,1}}{\sf{E}}\left[ \left\langle {\textbf{S}_{1}},{\textbf{U}_{1}}^{*} \right\rangle  \right]
 -2{{\gamma }_{1,2}}{\sf{E}}\left[ \left\langle {\textbf{S}_{1}},{\textbf{U}_{2}}^{*} \right\rangle  \right]-2{{\gamma }_{1,3}}{\sf{E}}\left[ \left\langle {\textbf{S}_{1}},{\textbf{V}}^{*} \right\rangle  \right] \nonumber\\
 && \> +{{\gamma }_{1,1}}^{2}{\sf{E}}\left[ {{\left\| {\textbf{U}_{1}}^{*} \right\|}^{2}} \right]+2{{\gamma }_{1,1}}{{\gamma }_{1,2}}{\sf{E}}\left[ \left\langle {\textbf{U}_{1}}^{*},{\textbf{U}_{2}}^{*} \right\rangle  \right]+{{\gamma }_{1,2}}^{2}{\sf{E}}\left[ {{\left\| {\textbf{U}_{2}}^{*} \right\|}^{2}} \right]\nonumber \\
 &&  \> +2{{\gamma }_{1,1}}{{\gamma }_{1,3}}{\sf{E}}\left[ \left\langle {\textbf{U}_{1}}^{*},{\textbf{V}}^{*} \right\rangle  \right]+2{{\gamma }_{1,2}}{{\gamma }_{1,3}}{\sf{E}}\left[ \left\langle {\textbf{U}_{2}}^{*},{\textbf{V}}^{*} \right\rangle  \right]+{{\gamma }_{1,3}}^{2}{\sf{E}}\left[ {{\left\| {\textbf{V}}^{*} \right\|}^{2}} \right]\Bigr) \nonumber\\
 & =& {{\sigma }^{2}}-2{{\gamma }_{1,1}}\frac{1}{n}{\sf{E}}\left[ \left\langle {\textbf{S}_{1}},{\textbf{U}_{1}}^{*} \right\rangle  \right]-2{{\gamma }_{1,2}}\frac{1}{n}{\sf{E}}\left[ \left\langle {\textbf{S}_{1}},{\textbf{U}_{2}}^{*} \right\rangle  \right]-2{{\gamma }_{1,3}}\frac{1}{n}{\sf{E}}\left[ \left\langle {\textbf{S}_{1}},{\textbf{V}}^{*} \right\rangle  \right] \nonumber\\
 && \> +{{\gamma }_{1,1}}^{2}\left( {{\sigma }^{2}}\left( 1-{{2}^{-2{{R}_{1}}}} \right) \right)+2{{\gamma }_{1,1}}{{\gamma }_{1,2}}\frac{1}{n}{\sf{E}}\left[ \left\langle {\textbf{U}_{1}}^{*},{\textbf{U}_{2}}^{*} \right\rangle  \right]+{{\gamma }_{1,2}}^{2}\left( {{\sigma }^{2}}\left( 1-{{2}^{-2{{R}_{2}}}} \right) \right) \nonumber\\
 && \> +2{{\gamma }_{1,1}}{{\gamma }_{1,3}}\frac{1}{n}{\sf{E}}\left[ \left\langle {\textbf{U}_{1}}^{*},{\textbf{V}}^{*} \right\rangle  \right]+2{{\gamma }_{1,2}}{{\gamma }_{1,3}}\frac{1}{n}{\sf{E}}\left[ \left\langle {\textbf{U}_{2}}^{*},{\textbf{V}}^{*} \right\rangle  \right]+{{\gamma }_{1,3}}^{2}\left( {{\sigma }^{2}}{{2}^{-2{{R}_{1}}}}\left( 1-{{2}^{-2{{R}_\textnormal{c}}}} \right) \right), \nonumber\\*
\end{IEEEeqnarray}
where in the last equality all expected squared norms have been replaced
by their explicit values, i.e.
${\sf{E}}\bigl[ {{\left\| {\textbf{S}_{1}} \right\|}^{2}} \bigr]=n{{\sigma }^{2}}$,
and
${\sf{E}}\bigl[ {{\left\| {\textbf{U}_{i}} \right\|}^{2}} \bigr]=n{{\sigma }^{2}}\left( 1-{{2}^{-2{{R}_{i}}}} \right)$
for $i\in \left\{ 1,2 \right\}$
and
${\sf{E}}\bigl[ {{\left\| {\textbf{V}} \right\|}^{2}} \bigr]=n{{\sigma }^{2}}{2}^{-2{{R}_{1}}}\left( 1-{{2}^{-2{{R}_\textnormal{c}}}} \right)$.
The remaining expectations of the inner products are bounded in the following six lemmas.

\medskip

\begin{lemma}\label{s1,u1}
For every $\delta>0$ and $0.3>\epsilon>0$ and every positive integer $n$
\begin{equation*}
\frac{1}{n}{\sf{E}}\left[ \left\langle {\mathbf{S}_{1}},{\mathbf{U}_{1}}^{*} \right\rangle  \right]
\ge
\sigma^{2}(1-2^{-2R_{1}})\left( 1-2\epsilon  \right)\left( 1-13\delta  \right).
\end{equation*}
\end{lemma}

\medskip

\begin{IEEEproof}
\begin{IEEEeqnarray*}{rCl}
  \frac{1}{n}{\sf{E}}\left[ \left\langle {\textbf{S}_{1}},{\textbf{U}_{1}}^{*} \right\rangle  \right]
  & = & \frac{1}{n}\underbrace{{\sf{E}}\left[ \left. \left\| {\textbf{S}_{1}} \right\|\left\| {\textbf{U}_{1}}^{*} \right\|\cos \sphericalangle \left( {\textbf{S}_{1}},{\textbf{U}_{1}}^{*} \right) \right|{{\cal E}_{\textbf{S}}}\cup {{\cal E}_{\textbf{X}}} \right]}_{\ge 0} \Pr\left[ {{\cal E}_{\textbf{S}}}\cup {{\cal E}_{\textbf{X}}} \right]\\
  && + \frac{1}{n}{\sf{E}}\left[ \left. \left\| {\textbf{S}_{1}} \right\|\left\| {\textbf{U}_{1}}^{*} \right\|\cos \sphericalangle \left( {\textbf{S}_{1}},{\textbf{U}_{1}}^{*} \right) \right|
  {\cal E}^{c}_{\textbf{S}}\cap {\cal E}^{c}_{\textbf{X}} \right] \Pr\left[ {\cal E}^{c}_{\textbf{S}}  \cap
  {\cal E}^{c}_{\textbf{X}} \right] \\
 & \ge & \frac{1}{n}{\sf{E}}\left[ \left. \left\| {\textbf{S}_{1}} \right\|\left\| {\textbf{U}_{1}}^{*} \right\|\cos \sphericalangle \left( {\textbf{S}_{1}},{\textbf{U}_{1}}^{*} \right) \right|{\cal E}^{c}_{\textbf{S}}\cap
 {\cal E}^{c}_{\textbf{X}} \right] \Pr\left[ {\cal E}^{c}_{\textbf{S}}  \cap {\cal E}^{c}_{\textbf{X}} \right] \\
 & \ge & \sqrt{\sigma^{2}(1-\epsilon)\sigma^{2}(1-2^{-2R_{1}})(1-2^{-2R_{1}})}(1-\epsilon ) \Pr\left[ {\cal E}^{c}_{\textbf{S}}
 \cap {\cal E}^{c}_{\textbf{X}} \right] \\
 & \ge &  \sigma^{2}(1-2^{-2R_{1}}){{\left( 1-\epsilon  \right)}^{2}}
 \left(1-\Pr\left[ {{\cal E}_{\textbf{S}}}\cup {{\cal E}_{\textbf{X}}} \right] \right) \\
 & \ge & \sigma^{2}(1-2^{-2R_{1}})\left( 1-2\epsilon  \right)
 \left( 1-\Pr\left[ {{\cal E}_{\textbf{S}}} \right]-\Pr\left[ {{\cal E}_{\textbf{X}}} \right] \right),
\end{IEEEeqnarray*}
where in the first equality the first expectation term is non-negative
because if $\textbf{U}_{1}^{*}=0$, then it is equal to zero, and if $\textbf{U}_{1}^{*}\neq 0$,
then by the conditioning on ${{\cal E}_{\textbf{X}}}$ it follows that
$\cos \sphericalangle \left( {\textbf{S}_{1}},{\textbf{U}_{1}}^{*} \right)>0.$

\medskip

By Lemma~\ref{eps_s} and Lemma~\ref{eps_x} it now follows that for every
$\delta>0$ and $0.3>\epsilon>0$ there exists an
$n'\left( \delta ,\epsilon  \right)\in \mathbb{N}$
such that for all
$n>n'\left( \delta ,\epsilon  \right)$
\begin{flalign*}
&&
\frac{1}{n}{\sf{E}}\left[ \left\langle {\textbf{S}_{1}},{\textbf{U}_{1}}^{*} \right\rangle  \right]\ge
{{\sigma }^{2}}( 1-{{2}^{-2{{R}_{1}}}})\left( 1-2\epsilon  \right)\left( 1-13\delta  \right).
&&
\end{flalign*}
\end{IEEEproof}

\medskip

\begin{lemma}\label{u1,u2}
For every $\delta>0$ and $0.3>\epsilon>0$,
there exists an
$n'\left( \delta ,\epsilon  \right)\in \mathbb{N}$
such that for all
$n>n'\left( \delta ,\epsilon  \right)$
\begin{equation*}
\frac{1}{n}{\sf{E}}\left[ \left\langle {\mathbf{U}_{1}^{*}},{\mathbf{U}_{2}^{*}} \right\rangle  \right]\le
{{\sigma }^{2}}12\delta +\rho {{\sigma }^{2}}( 1-{{2}^{-2{{R}_{1}}}} )( 1-{{2}^{-2{{R}_{2}}}} )(1+7\epsilon ).
\end{equation*}
\end{lemma}

\begin{IEEEproof}
\begin{IEEEeqnarray*}{rCl}
  \frac{1}{n}{\sf{E}}\left[ \left\langle {\textbf{U}_{1}}^{*},{\textbf{U}_{2}}^{*} \right\rangle  \right]
  & = & \frac{1}{n}{\sf{E}}\left[ \left. \left\langle \textbf{U}_{1}^{*},\textbf{U}_{2}^{*} \right\rangle  \right|
  {{\cal E}_{\textbf{X}}} \right] \Pr\left[ {{\cal E}_{\textbf{X}}} \right]
  +\frac{1}{n}{\sf{E}}\left[ \left. \left\langle \textbf{U}_{1}^{*},\textbf{U}_{2}^{*} \right\rangle  \right|
  {{\cal E}^{c}_{\textbf{X}}} \right] \Pr\left[ {{\cal E}^{c}_{\textbf{X}}} \right] \\
 & \le & \frac{1}{n}{\sf{E}}\left[ \left. \left\| \textbf{U}_{1}^{*} \right\|\left\| \textbf{U}_{2}^{*} \right\| \right|{{\cal E}_{\textbf{X}}} \right] \Pr\left[ {{\cal E}_{\textbf{X}}} \right]
 +\frac{1}{n}{\sf{E}}\left[ \left. \left. \left\| \textbf{U}_{1}^{*} \right\|\left\| \textbf{U}_{2}^{*} \right\| \right|\cos \sphericalangle \left( \textbf{U}_{1}^{*},\textbf{U}_{2}^{*} \right) \right|
 {{\cal E}_{\textbf{X}}}^{c} \right] \\
 & \le & \frac{1}{n}\sqrt{n{{\sigma }^{2}}\left( 1-{{2}^{-2{{R}_{1}}}} \right)}\sqrt{n{{\sigma }^{2}}\left( 1-{{2}^{-2{{R}_{2}}}} \right)} \Pr\left[ {{\cal E}_{\textbf{X}}} \right] \\
 && +\frac{1}{n}  \sqrt{n{{\sigma }^{2}}\left( 1-{{2}^{-2{{R}_{1}}}} \right)}\sqrt{n{{\sigma }^{2}}\left( 1-{{2}^{-2{{R}_{2}}}} \right)}\tilde{\rho }(1+7\epsilon ) \\
 & \le & {{\sigma }^{2}}\Pr\left[ {{\cal E}_{\textbf{X}}} \right]+\rho {{\sigma }^{2}}\left( 1-{{2}^{-2{{R}_{1}}}} \right)\left( 1-{{2}^{-2{{R}_{2}}}} \right)(1+7\epsilon ).
 \end{IEEEeqnarray*}
Thus, it follows by Lemma~\ref{eps_x} that for every
$\delta>0$ and $0.3>\epsilon>0$
there exists an
$n'\left( \delta ,\epsilon  \right)\in \mathbb{N}$
such that for all
$n>n'\left( \delta ,\epsilon  \right)$
\begin{flalign*}
&&
\frac{1}{n}{\sf{E}}\left[ \left\langle {\textbf{U}_{1}^{*}},{\textbf{U}_{2}^{*}} \right\rangle  \right]\le
{{\sigma }^{2}}12\delta +\rho {{\sigma }^{2}}\left( 1-{{2}^{-2{{R}_{1}}}} \right)\left( 1-{{2}^{-2{{R}_{2}}}} \right)(1+7\epsilon ).
&&
\end{flalign*}
\end{IEEEproof}

\medskip

\begin{lemma}\label{s1,u2}
For every $\delta>0$ and $0.3>\epsilon>0$
there exists an
$n'\left( \delta ,\epsilon  \right)\in \mathbb{N}$
such that for all
$n>n'\left( \delta ,\epsilon  \right)$
\begin{equation*}
\frac{1}{n}{\sf{E}}\left[ \left\langle {\mathbf{S}_{1}^{*}},{\mathbf{U}_{2}^{*}} \right\rangle  \right]
\ge
\rho {{\sigma }^{2}}( 1-{{2}^{-2{{R}_{2}}}}){{\left( 1-\epsilon  \right)}^{3}}-{{\sigma }^{2}}\left( \epsilon +39\delta +12\delta \epsilon  \right).
\end{equation*}
\end{lemma}

\medskip

\begin{IEEEproof}
We begin with the following decomposition:
\begin{IEEEeqnarray}{rCl}\label{eq:s1,u2 decomposition}
\frac{1}{n}{\sf{E}}\left[ \left\langle {\textbf{S}_{1}},\textbf{U}_{2}^{*} \right\rangle  \right]
&=&\frac{1}{n}{\sf{E}}\left[ \left. \left\langle {\textbf{S}_{1}},\textbf{U}_{2}^{*} \right\rangle  \right|{{\cal E}_{\textbf{S}}}\cup
{{\cal E}_{{\textbf{X}_{2}}}} \right]\Pr\left[ {{\cal E}_{\textbf{S}}}\cup {{\cal E}_{{\textbf{X}_{2}}}} \right] \nonumber \\
&&+\frac{1}{n}{\sf{E}}\left[ \left. \left\langle {\textbf{S}_{1}},{\textbf{U}_{2}}^{*} \right\rangle  \right|{\cal E}^{c}_{\textbf{S}}  \cap
{\cal E}^{c}_{{\textbf{X}_{2}}} \right]\Pr\left[ {\cal E}^{c}_{\textbf{S}}  \cap
{\cal E}^{c}_{{\textbf{X}_{2}}} \right].
\end{IEEEeqnarray}
The first term on the r.h.s. of \eqref{eq:s1,u2 decomposition} is lower bounded as follows:
\begin{IEEEeqnarray}{l}
   \frac{1}{n}{\sf{E}}\left[ \left. \left\langle {\textbf{S}_{1}},{\textbf{U}_{2}}^{*} \right\rangle  \right|
  {{\cal E}_{\textbf{S}}}\cup {\cal E}_{{\textbf{X}_{2}}} \right]
  \Pr \left[ {{\cal E}_{\textbf{S}}}\cup {{\cal E}_{{\textbf{X}_{2}}}} \right] \nonumber \\
  \, \overset{(a)}{\mathop{\ge }}\,-\frac{1}{n}{\sf{E}}\left[ \left. {{\left\| {\textbf{S}_{1}} \right\|}^{2}}
  +{{\left\| {\textbf{U}_{2}}^{*} \right\|}^{2}} \right|{{\cal E}_{\textbf{S}}}\cup {{\cal E}_{{\textbf{X}_{2}}}} \right]
   \Pr \left( {{\cal E}_{\textbf{S}}}\cup {{\cal E}_{{\textbf{X}_{2}}}} \right) \nonumber \\
  \,\overset{(b)}{\mathop{\ge }}\,-\frac{1}{n}\Bigl( {\sf{E}}\left[ \left. {{\left\| {\textbf{S}_{1}} \right\|}^{2}} \right|
 {\cal E}_{{\textbf{S}}} \right]\Pr \left[ {{\cal E}_{\textbf{S}}} \right]
 +{\sf{E}}\left[ \left. {{\left\| {\textbf{S}_{1}} \right\|}^{2}} \right|{\cal E}^{c}_{\textbf{S}}\cap
 {{\cal E}_{{\textbf{X}_{2}}}} \right]
 \Pr\left[ {\cal E}^{c}_{\textbf{S}}\cap {{\cal E}_{\textbf{X}}} \right] \nonumber \\
   \, \qquad +{{\left\| {\textbf{U}_{2}}^{*} \right\|}^{2}}\left( \Pr \left[ {{\cal E}_{\textbf{S}}} \right]
 +\Pr \left[ {{\cal E}_{\textbf{X}}} \right] \right) \Bigr) \nonumber \\
  \,\overset{(c)}{\mathop{\ge }}\,-\left( {{\sigma }^{2}}\left( \epsilon
 +\Pr \left[ {{\cal E}_{\textbf{S}}} \right] \right)+{{\sigma }^{2}}\left( 1+\epsilon  \right)\Pr \left[ {{\cal E}_{\textbf{X}}} \right]
 +{{\sigma }^{2}}\left( 1-{{2}^{-2{{R}_{2}}}} \right)\left( \Pr \left[ {{\cal E}_{\textbf{S}}} \right]+\Pr \left[ {{\cal E}_{\textbf{X}}} \right] \right) \right) \nonumber \\
 \,  \ge -{{\sigma }^{2}}\left( \epsilon +2\Pr\left[ {\cal E}_{\textbf{S}} \right]+\left( 2+\epsilon  \right)\Pr
 \left[ {{\cal E}_{\textbf{X}}} \right] \right),
\label{eq:s1,u2|eps_s U eps_x2}
\end{IEEEeqnarray}
where in (a) we have used that for any two vectors $\textbf{v} \in \mathbb{R}^{n}$ and
$\textbf{w} \in \mathbb{R}^{n}$
\begin{IEEEeqnarray}{l}\label{eq:inner_product_bound}
|\langle{\textbf{v},\textbf{w}}\rangle\|
\leq \frac{1}{2}(\|\textbf{v}\|^2+\|\textbf{w}\|^2) 
 \leq \|\textbf{v}\|^2+\|\textbf{w}\|^2 ,
\end{IEEEeqnarray}
in (b) we have used that ${{\cal E}_{\textbf{X}}}\supseteq {{\cal E}_{{\textbf{X}_{2}}}}$,
and in (c) we have used Lemma~\ref{Genie_1}.

\medskip

We now turn to lower bounding the second term on the r.h.s. of \eqref{eq:s1,u2 decomposition}.
The probability term is lower bounded as follows:
\begin{IEEEeqnarray}{rCl}\label{eq:eps_s_c,eps_x2_c}
  \Pr\left[ {\cal E}^{c}_{\textbf{S}}\cap {\cal E}^{c}_{{\textbf{X}_{2}}} \right]
&=& 1-\Pr\left[ {\cal E}_{\textbf{S}}\cup {\cal E}_{{\textbf{X}_{2}}} \right] \nonumber \\
& \ge & 1-\left( \Pr\left[ {\cal E}_{\textbf{S}} \right]+\Pr\left[ {\cal E}_{{\textbf{X}}} \right] \right).
\end{IEEEeqnarray}
To lower bound the expectation term, we represent $\textbf{u}_{i}^{*}$ as a scaled version
of $\textbf{s}_{i}$ corrupted by an additive "quantization noise" $\textbf{w}_{i}^{*}$.
More precisely,
\begin{equation}\label{eq:ortogonalization}
\textbf{u}_{i}^{*}={{\nu }_{i}}{\textbf{s}_{i}}+{\textbf{w}_{i}} \ \ \mbox{ where } \ \
{{\nu }_{i}}=\frac{\left\| \textbf{u}_{i}^{*} \right\|}{\left\| \textbf{s}_{i}^{*} \right\|}\cos \sphericalangle \left( {\textbf{s}_{i}},\textbf{u}_{i}^{*} \right)\text{ },\text{   }i\in \left\{ 1,2 \right\}.
\end{equation}
With this choice of $\nu_{i}$, the vector $\textbf{w}_{i}$ is always orthogonal to $\textbf{s}_{i}$.
By \eqref{eq:ortogonalization}, the inner product $\left\langle {\textbf{S}_{1}},{\textbf{U}_{2}}^{*} \right\rangle $
can now be rewritten as ${{\nu }_{2}}\left\langle {\textbf{S}_{1}},{\textbf{S}_{2}} \right\rangle +\left\langle {\textbf{S}_{1}},{\textbf{W}_{2}} \right\rangle $.
Hence,
\begin{IEEEeqnarray}{rCl}
\IEEEeqnarraymulticol{3}{l}{
{\sf{E}}\left[ \left. \left\langle {\textbf{S}_{1}},{\textbf{U}_{2}}^{*} \right\rangle  \right|{\cal E}^{c}_{\textbf{S}}\cap {\cal E}^{c}_{{\textbf{X}_{2}}} \right]} \nonumber \\ \quad
 & \overset{(a)}{\mathop{=}} & \,{{\sf{E}}_{{\textbf{S}_{1}},{\textbf{S}_{2}}}}\left[ {{\sf{E}}_{{{\mathscr C}_{1}},{{\mathscr C}_{2}}}}\left[ \left. \left\langle {\textbf{s}_{1}},{\textbf{U}_{2}}^{*} \right\rangle  \right|\left( {\textbf{S}_{1}},{\textbf{S}_{2}} \right)=\left( {\textbf{s}_{1}},{\textbf{s}_{2}} \right),{\cal E}^{c}_{\textbf{S}}\cap {\cal E}^{c}_{{\textbf{X}_{2}}} \right] \right] \nonumber \\
& \overset{(b)}{\mathop{=}} & \,{{\sf{E}}_{{\textbf{S}_{1}},{\textbf{S}_{2}}}}\Biggl[ {{\sf{E}}_{{{\mathscr C}_{1}},{{\mathscr C}_{2}}}}\left[ \left. {{\nu }_{2}}\left\langle {\textbf{s}_{1}},{\textbf{s}_{2}} \right\rangle  \right|\left( {\textbf{S}_{1}},{\textbf{S}_{2}} \right)=\left( {\textbf{s}_{1}},{\textbf{s}_{2}} \right),{\cal E}^{c}_{\textbf{S}}\cap
 {\cal E}^{c}_{{\textbf{X}_{2}}} \right] \nonumber \\
 && +  \underbrace{{{\sf{E}}_{{{\mathscr C}_{1}},{{\mathscr C}_{2}}}}\left[ \left. \left\langle {\textbf{s}_{1}},{\textbf{W}_{2}} \right\rangle  \right|\left( {\textbf{S}_{1}},{\textbf{S}_{2}} \right)=\left( {\textbf{s}_{1}},{\textbf{s}_{2}} \right),{\cal E}^{c}_{\textbf{S}}\cap
 {\cal E}^{c}_{{\textbf{X}_{2}}} \right]}_{=0} \Biggr] \nonumber \\
 & = & {{\sf{E}}_{{\textbf{S}_{1}},{\textbf{S}_{2}}}}\left[ \frac{\left\| {\textbf{U}_{2}}^{*} \right\|}{\left\| {\textbf{S}_{2}} \right\|}\left\langle {\textbf{S}_{1}},{\textbf{S}_{2}} \right\rangle {{\sf{E}}_{{{\mathscr C}_{1}},{{\mathscr C}_{2}}}}\left[ \left. \cos \sphericalangle \left( {\textbf{s}_{2}},{\textbf{U}_{2}}^{*} \right) \right|\left( {\textbf{S}_{1}},{\textbf{S}_{2}} \right)=\left( {\textbf{s}_{1}},{\textbf{s}_{2}} \right),{\cal E}^{c}_{\textbf{S}}\cap {\cal E}^{c}_{{\textbf{X}_{2}}} \right] \right] \nonumber \\
 & \overset{(c)}{\mathop{\ge }}& \,{{\sf{E}}_{{\textbf{S}_{1}},{\textbf{S}_{2}}}}\left[ \left. \left\| {\textbf{U}_{2}^{*}} \right\|\left\| {\textbf{S}_{1}} \right\|\cos \sphericalangle \left( {\textbf{S}_{1}},{\textbf{S}_{2}} \right)\sqrt{\left( 1-{{2}^{-2{{R}_{2}}}} \right)}\left( 1-\epsilon  \right) \right|{\cal E}^{c}_{\textbf{S}}\cap
 {\cal E}^{c}_{{\textbf{X}_{2}}} \right] \nonumber \\
 & \overset{(d)}{\mathop{\ge }} & \,\sqrt{n{{\sigma }^{2}}\left( 1-{{2}^{-2{{R}_{2}}}} \right)}\sqrt{n{{\sigma }^{2}}\left( 1-\epsilon  \right)}\rho \left( 1-\epsilon  \right)\sqrt{\left( 1-{{2}^{-2{{R}_{2}}}} \right)}\left( 1-\epsilon  \right) \nonumber \\
 & \ge & n\rho {{\sigma }^{2}}\left( 1-{{2}^{-2{{R}_{2}}}} \right){{\left( 1-\epsilon  \right)}^{3}},
\end{IEEEeqnarray}\label{eq:s1,u2|eps_s_c,eps_x2_c}
where we have denoted by $\mathscr C_i$ the random codebook of user $\text{ }i\in \left\{ 1,2 \right\}$,
and where in (a) we have used law of total expectation,
in (b) the second expectation term is zero because for every
$\left( {\textbf{s}_{1}},{\textbf{s}_{2}} \right)\in {\cal E}^{c}_{\textbf{S}}$
\begin{equation*}
{{\sf{E}}_{{{{\mathscr C}_{2}}}}\left[ \left. \left\langle {\textbf{s}_{1}},{\textbf{W}_{2}} \right\rangle  \right|\left( {\textbf{S}_{1}},{\textbf{S}_{2}} \right)=\left( {\textbf{s}_{1}},{\textbf{s}_{2}} \right), {\cal E}^{c}_{{\textbf{X}_{2}}} \right]}=0.
\end{equation*}
This holds since in the expectation over the codebooks $\mathscr C_2$ with conditioning on
${\cal E}^{c}_{{\textbf{X}_{2}}}$, for every ${\textbf{w}_2} \in \mathbb{R}^{n}$, the sequences $\textbf{w}_2$ and
$-\textbf{w}_2$ are equiprobable, and thus their inner products with $\textbf{s}_1$ cancel off each other.
Inequality (c) follows from lower bounding $\cos \sphericalangle \left( {\textbf{s}_{2}},\textbf{U}_{2}^{*} \right)$
conditioned on ${\cal E}^{c}_{\textbf{X}}$ combined with the fact that conditioned on
${\cal E}^{c}_{\textbf{S}}$ the term $\cos \sphericalangle \left( {\textbf{S}_{1}},\textbf{S}_{2} \right)$
is positive.
Inequality (d) follows from lower bounding $\left\| {\textbf{S}_{1}} \right\|$ and
$\cos \sphericalangle \left( {\textbf{S}_{1}},\textbf{S}_{2} \right)$ conditioned
on ${\cal E}^{c}_{\textbf{S}}$.

\medskip

Combining \eqref{eq:s1,u2 decomposition} with \eqref{eq:s1,u2|eps_s U eps_x2}, \eqref{eq:eps_s_c,eps_x2_c}
and \eqref{eq:s1,u2|eps_s_c,eps_x2_c} gives
\begin{IEEEeqnarray*}{rCl}
   \frac{1}{n}{\sf{E}}\left[ \left\langle {\textbf{S}_{1}},{\textbf{U}_{2}}^{*} \right\rangle  \right]
  & \ge & -{{\sigma }^{2}}\left( \epsilon +2\Pr \left[ {{\cal E}_{\textbf{S}}} \right]+\left( 2+\epsilon  \right)
  \Pr \left[ {{\cal E}_{\textbf{X}}} \right] \right) \\
  && +\rho {{\sigma }^{2}}(1-2^{-2R_{2}}){{\left( 1-\epsilon  \right)}^{3}}
  \left( 1-\left( \Pr \left[ {{\cal E}_{\textbf{S}}} \right]+\Pr \left[ {{\cal E}_{\textbf{X}}} \right] \right) \right) \\
 & \ge & \rho {{\sigma }^{2}}(1-2^{-2R_{2}}){{\left( 1-\epsilon  \right)}^{3}}
 -{{\sigma }^{2}}\left( \epsilon +3\Pr\left[ {{\cal E}_{\textbf{S}}} \right]
 +\left( 3+\epsilon  \right)\Pr\left[ {{\cal E}_{\textbf{X}}} \right] \right) .
\end{IEEEeqnarray*}
Thus, by Lemma~\ref{eps_s} and Lemma~\ref{eps_x} it follows that for every
$\delta>0$ and $0.3>\epsilon>0$ there exists an $n'(\delta,\epsilon) \in \mathbb{N}$ such that
for all $n>n'(\delta,\epsilon)$
\newpage
\begin{flalign*}
&&
\frac{1}{n}{\sf{E}}\left[ \left\langle {\textbf{S}_{1}},{\textbf{U}_{2}}^{*} \right\rangle  \right]
\ge
\rho {{\sigma }^{2}}( 1-{{2}^{-2{{R}_{2}}}}){{\left( 1-\epsilon  \right)}^{3}}-{{\sigma }^{2}}\left( \epsilon +39\delta +12\delta \epsilon  \right).
&&
\end{flalign*}
\end{IEEEproof}

\medskip

\begin{lemma}\label{s1,v}
For every $\delta>0$ and $0.3>\epsilon>0$ and every positive integer n
\begin{equation*}
\frac{1}{n}{\sf{E}}\left[ \left\langle {\mathbf{S}_{1}},{\mathbf{V}^{*}} \right\rangle  \right]
\ge -{{\sigma }^{2}}\left( 12\delta +3\epsilon \right)+{{\sigma }^{2}}{2^{-2R_1}}( 1-{{2}^{-2{{R}_\textnormal{c}}}})\left( 1-2\epsilon  \right)\left( 1-13\delta  \right).
\end{equation*}
\end{lemma}

\medskip

\begin{IEEEproof}
We begin with the following decomposition:
\begin{equation}\label{eq:s1,v decomposition}
\frac{1}{n}{\sf{E}}\left[ \left\langle {\textbf{S}_{1}},{\textbf{V}}^{*} \right\rangle  \right]
=\frac{1}{n}{\sf{E}}\left[ \left\langle \textbf{U}_{1}^{*}+{\textbf{z}_{\textnormal{Q}_1}},\textbf{V}^{*} \right\rangle  \right]=\frac{1}{n}\left( {\sf{E}}\left[ \left\langle \textbf{U}_{1}^{*},\textbf{V}^{*} \right\rangle  \right]
+{\sf{E}}\left[ \left\langle \textbf{z}_{\textnormal{Q}_1},\textbf{V}^{*} \right\rangle  \right] \right).
\end{equation}

The first term on the r.h.s. of \eqref{eq:s1,v decomposition} is lower bounded as follows:
\begin{IEEEeqnarray}{rCl}\label{eq:u1,v_upper bound}
  \frac{1}{n}{\sf{E}}\left[ \left\langle \textbf{U}_{1}^{*},\textbf{V}^{*} \right\rangle  \right]
  &=&\frac{1}{n}{\sf{E}}\left[ \left. \left\langle \textbf{U}_{1}^{*},\textbf{V}^{*} \right\rangle  \right|
  {{\cal E}_{\textbf{X}}} \right] \Pr\left[ {{\cal E}_{\textbf{X}}} \right]
  +\frac{1}{n}{\sf{E}}\left[ \left. \left\langle \textbf{U}_{1}^{*},\textbf{V}^{*} \right\rangle  \right|
  {{\cal E}^{c}_{\textbf{X}}} \right] \Pr\left[ {{\cal E}^{c}_{\textbf{X}}} \right] \nonumber\\
 & \overset{(a)} {\mathop{\ge }}& - \frac{1}{n}{\sf{E}}\left[ \left. \left\| \textbf{U}_{1}^{*} \right\|^2
 +\left\| \textbf{V}^{*} \right\|^2 \right|
 {{\cal E}_{\textbf{X}}} \right] \Pr\left[ {{\cal E}_{\textbf{X}}} \right] \nonumber \\
&& +\frac{1}{n}{\sf{E}}\left[ \left. \left. \left\| \textbf{U}_{1}^{*} \right\|\left\| \textbf{V}^{*} \right\| \right|
 \cos \sphericalangle \left( \textbf{U}_{1}^{*},\textbf{V}^{*} \right) \right|
 {\cal E}^{c}_{\textbf{X}} \right]\Pr\left[ {{\cal E}^{c}_{\textbf{X}}} \right] \nonumber\\
 &\ge & -\frac{1}{n}\left(n{{\sigma }^{2}}\left( 1-{{2}^{-2{{R}_{1}}}} \right)+n{{\sigma }^{2}}2^{-2R_1}\left( 1-{{2}^{-2{{R}_\textnormal{c}}}} \right)\right) \Pr\left[ {{\cal E}_{\textbf{X}}} \right] \nonumber\\
 && + \frac{1}{n}  \sqrt{n{{\sigma }^{2}}\left( 1-{{2}^{-2{{R}_{1}}}} \right)}\sqrt{n{{\sigma }^{2}}2^{-2R_1}\left( 1-{{2}^{-2{{R}_\textnormal{c}}}} \right)}(-3\epsilon ) \nonumber\\
 & \ge & -{\sigma }^{2}\left(\Pr\left[ {{\cal E}_{\textbf{X}}}\right]+3\epsilon\right),
 \end{IEEEeqnarray}
where in (a) we have used \eqref{eq:inner_product_bound}.

We now turn to lower bounding the second term on the r.h.s. of \eqref{eq:s1,v decomposition}.
\begin{IEEEeqnarray}{rCl}
\IEEEeqnarraymulticol{3}{l}{
  \frac{1}{n}{\sf{E}}\left[ \left\langle {\textbf{z}_{\textnormal{Q}_1}},{\textbf{V}}^{*} \right\rangle  \right]}\nonumber \\ \qquad
  &=&\frac{1}{n}\underbrace{\mathbb E\left[ \left. \left\| {\textbf{z}_{\textnormal{Q}_1}} \right\|\left\| \textbf{V}^{*} \right\|\cos \sphericalangle \left( {\textbf{z}_{\textnormal{Q}_1}},{\textbf{V}}^{*} \right) \right|{{\cal E}_{\textbf{S}}}\cup
  {{\cal E}_{\textbf{X}}} \right]}_{\ge 0} \Pr\left[ {{\cal E}_{\textbf{S}}}\cup {{\cal E}_{\textbf{X}}} \right]\nonumber\\
  && +\frac{1}{n}{\sf{E}}\left[ \left. \left\| {\textbf{z}_{\textnormal{Q}_1}} \right\|\left\| {\textbf{V}}^{*} \right\|\cos \sphericalangle \left( {\textbf{z}_{\textnormal{Q}_1}},{\textbf{V}}^{*} \right) \right|
  {\cal E}^{c}_{\textbf{S}}\cap {\cal E}^{c}_{\textbf{X}} \right] \Pr\left[ {\cal E}^{c}_{\textbf{S}}  \cap
  {\cal E}^{c}_{\textbf{X}} \right] \nonumber\\
 & \ge & \frac{1}{n}{\sf{E}}\left[ \left. \left\| {\textbf{z}_{\textnormal{Q}_1}} \right\|\left\| {\textbf{V}}^{*} \right\|\cos \sphericalangle \left( {\textbf{z}_{\textnormal{Q}_1}},{\textbf{V}}^{*} \right) \right|{\cal E}^{c}_{\textbf{S}}\cap
 {\cal E}^{c}_{\textbf{X}} \right] \Pr\left[ {\cal E}^{c}_{\textbf{S}}  \cap {\cal E}^{c}_{\textbf{X}} \right] \nonumber\\
 & \ge & \sqrt{\sigma^{2}2^{-2R_1}(1-\epsilon)\sigma^{2}2^{-2R_1}(1-2^{-2R_\textnormal{c}})
 (1-2^{-2R_\textnormal{c}})}(1-\epsilon)
 \Pr\left[ {\cal E}^{c}_{\textbf{S}}  \cap {\cal E}^{c}_{\textbf{X}} \right] \nonumber\\
 & \ge &  \sigma^{2}2^{-2R_1}(1-2^{-2R_\textnormal{c}}){{\left( 1-\epsilon  \right)}^{2}}
 \left( 1-\Pr\left[ {{\cal E}_{\textbf{S}}}\cup {{\cal E}_{\textbf{X}}} \right] \right) \nonumber\\
 & \ge & \sigma^{2}2^{-2R_1}(1-2^{-2R_\textnormal{c}})\left( 1-2\epsilon  \right) \left( 1-\Pr\left[ {{\cal E}_{\textbf{S}}} \right]-\Pr\left[ {{\cal E}_{\textbf{X}}} \right] \right),
\end{IEEEeqnarray} \label{eq:zq1,v}
where in the first equality the first expectation term is non-negative
because if $\textbf{V}^{*}=0$, then it is equal to zero, and if $\textbf{V}^{*}\neq 0$,
then by the conditioning on ${{\cal E}_{\textbf{X}}}$ it follows that
$\cos \sphericalangle \left( {\textbf{z}_{\textnormal{Q}_1}},{\textbf{V}}^{*} \right)>0.$

\medskip

Combining \eqref{eq:s1,v decomposition}, with \eqref{eq:u1,v_upper bound} and \eqref{eq:zq1,v} gives
\begin{equation*}
\frac{1}{n}{\sf{E}}\left[ \left\langle {\textbf{S}_{1}},{\textbf{V}}^{*} \right\rangle  \right]
\ge -{{\sigma }^{2}}\left(\Pr\left[ {{\cal E}_{\textbf{X}}}\right]+3\epsilon\right)
+{{\sigma }^{2}}2^{-2R_1}(1-{{2}^{-2{{R}_\textnormal{c}}}})\left( 1-2\epsilon  \right)
\left( 1-\Pr\left[ {{\cal E}_{\textbf{S}}} \right]-\Pr\left[ {{\cal E}_{\textbf{X}}} \right] \right).
\end{equation*}

Thus, by Lemma~\ref{eps_s} and Lemma~\ref{eps_x} it now follows that for every
$\delta>0$ and $0.3>\epsilon>0$
there exists an
$n'\left( \delta ,\epsilon  \right)\in \mathbb{N}$
such that for all
$n>n'\left( \delta ,\epsilon  \right)$
\newpage
\begin{flalign*}
&&
\frac{1}{n}{\sf{E}}\left[ \left\langle {\textbf{S}_{1}},{\textbf{V}}^{*} \right\rangle  \right]
\ge -{{\sigma }^{2}}\left( 12\delta +3\epsilon \right)+{{\sigma }^{2}}{2^{-2R_1}}(1-{{2}^{-2{{R}_\textnormal{c}}}})\left( 1-2\epsilon  \right)\left( 1-13\delta  \right).
&&
\end{flalign*}
\end{IEEEproof}

\medskip

\begin{lemma}\label{u1,v}
For every $\delta>0$ and $0.3>\epsilon>0$,
there exists an
$n'\left( \delta ,\epsilon  \right)\in \mathbb{N}$
such that for all
$n>n'\left( \delta ,\epsilon  \right)$
\begin{equation*}
\frac{1}{n}{\sf{E}}\left[ \left\langle {\mathbf{U}^{*}_{1}},{\mathbf{V}^{*}} \right\rangle  \right]
\le {{\sigma }^{2}}\left( 12\delta +3\epsilon \right).
\end{equation*}
\end{lemma}

\medskip

\begin{IEEEproof}
\begin{IEEEeqnarray*}{rCl}
  \frac{1}{n}{\sf{E}}\left[ \left\langle \textbf{U}_{1}^{*},\textbf{V}^{*} \right\rangle  \right]
  & = & \frac{1}{n}{\sf{E}}\left[ \left. \left\langle \textbf{U}_{1}^{*},\textbf{V}^{*} \right\rangle  \right|
  {{\cal E}_{\textbf{X}}} \right] \Pr\left[ {{\cal E}_{\textbf{X}}} \right]
  +\frac{1}{n}{\sf{E}}\left[ \left. \left\langle \textbf{U}_{1}^{*},\textbf{V}^{*} \right\rangle  \right|
  {{\cal E}^{c}_{\textbf{X}}} \right] \Pr\left[ {{\cal E}^{c}_{\textbf{X}}} \right] \\
 & \le & \frac{1}{n}{\sf{E}}\left[ \left. \left\| \textbf{U}_{1}^{*} \right\|\left\| \textbf{V}^{*} \right\| \right|
 {{\cal E}_{\textbf{X}}} \right] \Pr\left[ {{\cal E}_{\textbf{X}}} \right]
 +\frac{1}{n}{\sf{E}}\left[ \left. \left. \left\| \textbf{U}_{1}^{*} \right\|\left\| \textbf{V}^{*} \right\| \right|
 \cos \sphericalangle \left( \textbf{U}_{1}^{*},\textbf{V}^{*} \right) \right|
 {\cal E}^{c}_{\textbf{X}} \right]\Pr\left[ {{\cal E}^{c}_{\textbf{X}}} \right] \\
 & \le & \frac{1}{n}\sqrt{n{{\sigma }^{2}}\left( 1-{{2}^{-2{{R}_{1}}}} \right)}\sqrt{n{{\sigma }^{2}}2^{-2R_1}\left( 1-{{2}^{-2{{R}_\textnormal{c}}}} \right)} \Pr\left[ {{\cal E}_{\textbf{X}}} \right] \\
 && +\frac{1}{n}  \sqrt{n{{\sigma }^{2}}\left( 1-{{2}^{-2{{R}_{1}}}} \right)}\sqrt{n{{\sigma }^{2}}2^{-2R_1}\left( 1-{{2}^{-2{{R}_\textnormal{c}}}} \right)}(3\epsilon ) \\
 & \le & {{\sigma }^{2}}\left(\Pr\left[ {{\cal E}_{\textbf{X}}}\right]+3\epsilon\right)  .
 \end{IEEEeqnarray*}
Thus, it follows by Lemma~\ref{eps_x} that for every
$\delta>0$ and $0.3>\epsilon>0$
there exists an
$n'\left( \delta ,\epsilon  \right)\in \mathbb{N}$
such that for all
$n>n'\left( \delta ,\epsilon  \right)$
\begin{flalign*}
&&
\frac{1}{n}{\sf{E}}\left[ \left\langle {\textbf{U}_{1}^{*}},{\textbf{V}^{*}} \right\rangle  \right]\le
{{\sigma }^{2}}\left( 12\delta +3\epsilon \right).
&&
\end{flalign*}
\end{IEEEproof}

\medskip

\begin{lemma}\label{u2,v}
For every $\delta>0$ and $0.3>\epsilon>0$,
there exists an
$n'\left( \delta ,\epsilon  \right)\in \mathbb{N}$
such that for all
$n>n'\left( \delta ,\epsilon  \right)$
\begin{equation*}
\frac{1}{n}{\sf{E}}\left[ \left\langle {\mathbf{U}^{*}_{2}},{\mathbf{V}^{*}} \right\rangle  \right]
\le {{\sigma }^{2}}11\delta+\rho\sigma^2 2^{-2R_1}(1-2^{-2R_2})(1-2^{-2R_\textnormal{c}})(1+7\epsilon).
\end{equation*}
\end{lemma}

\medskip

\begin{IEEEproof}
\begin{IEEEeqnarray*}{rCl}
  \frac{1}{n}{\sf{E}}\left[ \left\langle \textbf{U}_{2}^{*},\textbf{V}^{*} \right\rangle  \right]
  & = & \frac{1}{n}{\sf{E}}\left[ \left. \left\langle \textbf{U}_{2}^{*},\textbf{V}^{*} \right\rangle  \right|
  {{\cal E}_{\textbf{X}}} \right] \Pr\left[ {{\cal E}_{\textbf{X}}} \right]
  +\frac{1}{n}{\sf{E}}\left[ \left. \left\langle \textbf{U}_{2}^{*},\textbf{V}^{*} \right\rangle  \right|
  {{\cal E}^{c}_{\textbf{X}}} \right] \Pr\left[ {{\cal E}^{c}_{\textbf{X}}} \right] \\
 & \le & \frac{1}{n}{\sf{E}}\left[ \left. \left\| \textbf{U}_{2}^{*} \right\|\left\| \textbf{V}^{*} \right\| \right|
 {{\cal E}_{\textbf{X}}} \right] \Pr\left[ {{\cal E}_{\textbf{X}}} \right]
 +\frac{1}{n}{\sf{E}}\left[ \left. \left. \left\| \textbf{U}_{2}^{*} \right\|\left\| \textbf{V}^{*} \right\| \right|
 \cos \sphericalangle \left( \textbf{U}_{2}^{*},\textbf{V}^{*} \right) \right|
 {\cal E}^{c}_{\textbf{X}} \right]\Pr\left[ {{\cal E}^{c}_{\textbf{X}}} \right] \\
 & \le & \sqrt{\sigma^{2}\left(1-2^{-2R_{2}}\right)\sigma^{2}2^{-2R_1}\left(1-2^{-2R_\textnormal{c}} \right)}
 \Pr\left[ {{\cal E}_{\textbf{X}}} \right] \\
 && +\sqrt{\sigma^{2}\left(1-2^{-2R_{2}} \right)\sigma^{2}2^{-2R_1}\left(1-2^{-2R_\textnormal{c}} \right)}\bar{\rho}(1+7\epsilon ) \\
 & \le &{{\sigma }^{2}}\Pr\left[ {{\cal E}_{\textbf{X}}}\right]+\rho\sigma^2 2^{-2R_1}(1-2^{-2R_2})(1-2^{-2R_\textnormal{c}})(1+7\epsilon) .
 \end{IEEEeqnarray*}
Thus, it follows by Lemma~\ref{eps_x} that for every
$\delta>0$ and $0.3>\epsilon>0$
there exists an
$n'\left( \delta ,\epsilon  \right)\in \mathbb{N}$
such that for all
$n>n'\left( \delta ,\epsilon  \right)$
\begin{flalign*}
&&
\frac{1}{n}{\sf{E}}\left[ \left\langle {\textbf{U}_{2}^{*}},{\textbf{V}^{*}} \right\rangle  \right]\le
{{\sigma }^{2}}11\delta+\rho\sigma^2 2^{-2R_1}(1-2^{-2R_2})(1-2^{-2R_\textnormal{c}})(1+7\epsilon).
&&
\end{flalign*}
\end{IEEEproof}

\bigskip

The distortion $D_1$ of the genie-aided scheme is now upper bounded as follows:
\begin{IEEEeqnarray}{rCl}
{{D}_{1}}& = & \frac{1}{n}{\sf{E}}\left[ {\| {\textbf{S}_{1}}-\hat{\textbf{S}}_1{\hspace{-.4em}}^{\textnormal{G}} \|^{2}} \right] \nonumber \\
&=&{{\sigma }^{2}}-2{{\gamma }_{1,1}}\frac{1}{n}{\sf{E}}\left[ \left\langle {\textbf{S}_{1}},{\textbf{U}_{1}}^{*} \right\rangle  \right]-2{{\gamma }_{1,2}}\frac{1}{n}{\sf{E}}\left[ \left\langle {\textbf{S}_{1}},{\textbf{U}_{2}}^{*} \right\rangle  \right]-2{{\gamma }_{1,3}}\frac{1}{n}{\sf{E}}\left[ \left\langle {\textbf{S}_{1}},{\textbf{V}}^{*} \right\rangle  \right] \nonumber\\
&& +{{\gamma }_{1,1}}^{2}{{\sigma }^{2}}\left( 1-{{2}^{-2{{R}_{1}}}} \right)+2{{\gamma }_{1,1}}{{\gamma }_{1,2}}\frac{1}{n}{\sf{E}}\left[ \left\langle {\textbf{U}_{1}}^{*},{\textbf{U}_{2}}^{*} \right\rangle  \right]
+{{\gamma }_{1,2}}^{2}{{\sigma }^{2}}\left( 1-{{2}^{-2{{R}_{2}}}}\right) \nonumber\\
&& +2{{\gamma }_{1,1}}{{\gamma }_{1,3}}\frac{1}{n}{\sf{E}}\left[ \left\langle {\textbf{U}_{1}}^{*},{\textbf{V}}^{*} \right\rangle  \right]+2{{\gamma }_{1,2}}{{\gamma }_{1,3}}\frac{1}{n}{\sf{E}}\left[ \left\langle {\textbf{U}_{2}}^{*},{\textbf{V}}^{*} \right\rangle  \right]
+{{\gamma }_{1,3}}^{2}{{\sigma }^{2}}{{2}^{-2{{R}_{1}}}}\left( 1-{{2}^{-2{{R}_\textnormal{c}}}} \right) \nonumber \\
&\,\overset{(a)}{\mathop{\le }} & \,{{\sigma }^{2}}{{2}^{-2({{R}_{1}}+{{R}_\textnormal{c}})}}\frac{1-{{\rho }^{2}}\left( 1-{{2}^{-2{{R}_{2}}}} \right)}{1-{{\rho }^{2}}\left( 1-{{2}^{-2{{R}_{2}}}} \right)\left( 1-{{2}^{-2({{R}_{1}}+{{R}_\textnormal{c}})}} \right)}+\xi'(\delta,\epsilon),
\end{IEEEeqnarray}
where in (a) we have used Lemma~\ref{s1,u1}, Lemma~\ref{u1,u2}, Lemma~\ref{s1,u2}, Lemma~\ref{s1,v},
 Lemma~\ref{u1,v}, Lemma~\ref{u2,v} and Lemma~\ref{MMSE_coefficients_bounds}
and where $\underset{\delta ,\epsilon \to 0}{\mathop{\lim }}\,\xi '\left( \delta ,\epsilon  \right)=0$.

\bigskip

Now we upper-bound $D_2$.
By Corollary~\ref{corollary_rates}, it suffices to analyze the genie-aided scheme. Since
$\hat{\textbf{S}}_2{\hspace{-.4em}}^{\textnormal{G}}={{\gamma }_{2,1}}{{\textbf{U}}_{1}}^{*}+{{\gamma }_{2,2}}{\textbf{U}_{2}}^{*}
+{{\gamma }_{2,3}}{\textbf{V}^{*}}$,
we have
\begin{IEEEeqnarray}{rCl}
{{D}_{2}}& =& \frac{1}{n}{\sf{E}}\left[ {\| {\textbf{S}_{2}}-\hat{\textbf{S}}_2{\hspace{-.4em}}^\textnormal{G} \|^{2}} \right]
=\frac{1}{n}{\sf{E}}\left[ {{\left\| {\textbf{S}_{2}}-\left( {{\gamma }_{2,1}}{\textbf{U}_{1}}^{*}
+{{\gamma }_{2,2}}{\textbf{U}_{2}}^{*}+{{\gamma }_{2,3}}{\textbf{V}^{*}} \right) \right\|}^{2}} \right]  \nonumber\\
 & =& \frac{1}{n}\Bigl({\sf{E}}\left[ {{\left\| {\textbf{S}_{2}} \right\|}^{2}} \right]
 -2{{\gamma }_{2,1}}{\sf{E}}\left[ \left\langle {\textbf{S}_{2}},{\textbf{U}_{1}}^{*} \right\rangle  \right]
 -2{{\gamma }_{2,2}}{\sf{E}}\left[ \left\langle {\textbf{S}_{2}},{\textbf{U}_{2}}^{*} \right\rangle  \right]
 -2{{\gamma }_{2,3}}{\sf{E}}\left[ \left\langle {\textbf{S}_{2}},{\textbf{V}}^{*} \right\rangle  \right] \nonumber\\
 && \> +{{\gamma }_{2,1}}^{2}{\sf{E}}\left[ {{\left\| {\textbf{U}_{1}}^{*} \right\|}^{2}} \right]
 +2{{\gamma }_{2,1}}{{\gamma }_{2,2}}{\sf{E}}\left[ \left\langle {\textbf{U}_{1}}^{*},{\textbf{U}_{2}}^{*} \right\rangle  \right]+{{\gamma }_{2,2}}^{2}{\sf{E}}\left[ {{\left\| {\textbf{U}_{2}}^{*} \right\|}^{2}} \right]\nonumber \\
 &&  \> +2{{\gamma }_{2,1}}{{\gamma }_{2,3}}{\sf{E}}\left[ \left\langle {\textbf{U}_{1}}^{*},{\textbf{V}}^{*} \right\rangle  \right]+2{{\gamma }_{2,2}}{{\gamma }_{2,3}}{\sf{E}}\left[ \left\langle {\textbf{U}_{2}}^{*},{\textbf{V}}^{*} \right\rangle  \right]
 +{{\gamma }_{2,3}}^{2}{\sf{E}}\left[ {{\left\| {\textbf{V}}^{*} \right\|}^{2}} \right]\Bigr) \nonumber\\
 & =& {{\sigma }^{2}}-2{{\gamma }_{2,1}}\frac{1}{n}{\sf{E}}\left[ \left\langle {\textbf{S}_{2}},{\textbf{U}_{1}}^{*} \right\rangle  \right]-2{{\gamma }_{2,2}}\frac{1}{n}{\sf{E}}\left[ \left\langle {\textbf{S}_{2}},{\textbf{U}_{2}}^{*} \right\rangle  \right]-2{{\gamma }_{2,3}}\frac{1}{n}{\sf{E}}\left[ \left\langle {\textbf{S}_{2}},{\textbf{V}}^{*} \right\rangle  \right] \nonumber\\
 && \> +{{\gamma }_{2,1}}^{2}\left( {{\sigma }^{2}}\left( 1-{{2}^{-2{{R}_{1}}}} \right) \right)
 +2{{\gamma }_{2,1}}{{\gamma }_{2,2}}\frac{1}{n}{\sf{E}}\left[ \left\langle {\textbf{U}_{1}}^{*},{\textbf{U}_{2}}^{*} \right\rangle  \right]
 +{{\gamma }_{2,2}}^{2}\left( {{\sigma }^{2}}\left( 1-{{2}^{-2{{R}_{2}}}} \right) \right) \nonumber\\
 && \> +2{{\gamma }_{2,1}}{{\gamma }_{2,3}}\frac{1}{n}{\sf{E}}\left[ \left\langle {\textbf{U}_{1}}^{*},{\textbf{V}}^{*} \right\rangle  \right]
 +2{{\gamma }_{2,2}}{{\gamma }_{2,3}}\frac{1}{n}{\sf{E}}\left[ \left\langle {\textbf{U}_{2}}^{*},{\textbf{V}}^{*} \right\rangle  \right]
 +{{\gamma }_{2,3}}^{2}\left( {{\sigma }^{2}}{{2}^{-2{{R}_{1}}}}\left( 1-{{2}^{-2{{R}_\textnormal{c}}}} \right) \right), \nonumber\\*
\end{IEEEeqnarray}
where in the last equality all expected squared norms have been replaced
by their explicit values, i.e.
${\sf{E}}\bigl[ {{\left\| {\textbf{S}_{2}} \right\|}^{2}} \bigr]=n{{\sigma }^{2}}$,
and
${\sf{E}}\bigl[ {{\left\| {\textbf{U}_{i}} \right\|}^{2}} \bigr]=n{{\sigma }^{2}}\left( 1-{{2}^{-2{{R}_{i}}}} \right)$
for $i\in \left\{ 1,2 \right\}$
and
${\sf{E}}\bigl[ {{\left\| {\textbf{V}} \right\|}^{2}} \bigr]=n{{\sigma }^{2}}{2}^{-2{{R}_{1}}}\left( 1-{{2}^{-2{{R}_\textnormal{c}}}} \right)$.
The remaining expectations of the inner products are bounded in the following three lemmas.
\medskip

\begin{lemma}\label{s2,u2}
For every $\delta>0$ and $0.3>\epsilon>0$ and every positive integer n
\begin{equation*}
\frac{1}{n}{\sf{E}}\left[ \left\langle {\mathbf{S}_{2}},\mathbf{U}^{*}_{2} \right\rangle  \right]\ge
{{\sigma }^{2}}(1-{{2}^{-2{{R}_{2}}}})\left( 1-2\epsilon  \right)\left( 1-13\delta  \right).
\end{equation*}
\end{lemma}

\medskip

\begin{IEEEproof}
\begin{IEEEeqnarray*}{rCl}
  \frac{1}{n}{\sf{E}}\left[ \left\langle {\textbf{S}_{2}},{\textbf{U}_{2}}^{*} \right\rangle  \right]
  & = & \frac{1}{n}\underbrace{{\sf{E}}\left[ \left. \left\| {\textbf{S}_{2}} \right\|\left\| {\textbf{U}_{2}}^{*} \right\|\cos \sphericalangle \left( {\textbf{S}_{2}},{\textbf{U}_{2}}^{*} \right) \right|{{\cal E}_{\textbf{S}}}\cup {{\cal E}_{\textbf{X}}} \right]}_{\ge 0} \Pr\left[ {{\cal E}_{\textbf{S}}}\cup {{\cal E}_{\textbf{X}}} \right]\\
  && +\frac{1}{n}{\sf{E}}\left[ \left. \left\| {\textbf{S}_{2}} \right\|\left\| {\textbf{U}_{2}}^{*} \right\|\cos \sphericalangle \left( {\textbf{S}_{2}},{\textbf{U}_{2}}^{*} \right) \right|
  {\cal E}^{c}_{\textbf{S}}\cap {\cal E}^{c}_{\textbf{X}} \right] \Pr\left[ {\cal E}^{c}_{\textbf{S}}  \cap
  {\cal E}^{c}_{\textbf{X}} \right] \\
 & \ge & \frac{1}{n}{\sf{E}}\left[ \left. \left\| {\textbf{S}_{2}} \right\|\left\| {\textbf{U}_{2}}^{*} \right\|\cos \sphericalangle \left( {\textbf{S}_{2}},{\textbf{U}_{2}}^{*} \right) \right|{\cal E}^{c}_{\textbf{S}}\cap
 {\cal E}^{c}_{\textbf{X}} \right] \Pr\left[ {\cal E}^{c}_{\textbf{S}}  \cap {\cal E}^{c}_{\textbf{X}} \right] \\
 & \ge & \sqrt{\sigma^{2}(1-\epsilon)\sigma^{2}(1-2^{-2R_{2}})(1-2^{-2R_{2}})}(1-\epsilon ) \Pr\left[ {\cal E}^{c}_{\textbf{S}}
 \cap {\cal E}^{c}_{\textbf{X}} \right] \\
 & \ge &  {{\sigma }^{2}}(1-{{2}^{-2{{R}_{2}}}}){{\left( 1-\epsilon  \right)}^{2}}
 \left( 1-\Pr\left[ {{\cal E}_{\textbf{S}}}\cup {{\cal E}_{\textbf{X}}} \right] \right) \\
 & \ge & {{\sigma }^{2}}(1-{{2}^{-2{{R}_{2}}}})\left( 1-2\epsilon  \right)
 \left( 1-\Pr\left[ {{\cal E}_{\textbf{S}}} \right]-\Pr\left[ {{\cal E}_{\textbf{X}}} \right] \right),
\end{IEEEeqnarray*}
where in the first equality the first expectation term is non-negative
because if $\textbf{U}_{2}^{*}=0$, then it is equal to zero, and if $\textbf{U}_{2}^{*}\neq 0$,
then by the conditioning on ${{\cal E}_{\textbf{X}}}$ it follows that
$\cos \sphericalangle \left( {\textbf{S}_{2}},{\textbf{U}_{2}}^{*} \right)>0.$

\medskip

By Lemma~\ref{eps_s} and Lemma~\ref{eps_x} it now follows that for every
$\delta>0$ and $0.3>\epsilon>0$ there exists an
$n'\left( \delta ,\epsilon  \right)\in \mathbb{N}$
such that for all
$n>n'\left( \delta ,\epsilon  \right)$
\begin{flalign*}
&&
\frac{1}{n}{\sf{E}}\left[ \left\langle {\textbf{S}_{2}},{\textbf{U}_{2}}^{*} \right\rangle  \right]
\ge \sigma^{2}(1-2^{-2R_{2}})\left( 1-2\epsilon  \right)\left( 1-13\delta  \right).
&&
\end{flalign*}
\end{IEEEproof}

\begin{lemma}\label{s2,u1}
For every $\delta>0$ and $0.3>\epsilon>0$
there exists an
$n'\left( \delta ,\epsilon  \right)\in \mathbb{N}$
such that for all
$n>n'\left( \delta ,\epsilon  \right)$
\begin{equation*}
\frac{1}{n}{\sf{E}}\left[ \left\langle {\mathbf{S}_{2}^{*}},{\mathbf{U}_{1}^{*}} \right\rangle  \right]\ge
\rho {{\sigma }^{2}}(1-2^{-2R_{1}}){{\left( 1-\epsilon  \right)}^{3}}-{{\sigma }^{2}}\left( \epsilon +39\delta +12\delta \epsilon  \right).
\end{equation*}
\end{lemma}

\medskip

\begin{IEEEproof}
The proof is following in a similar manner as the proof of Lemma~\ref{s1,u2}.
\end{IEEEproof}

\medskip

\begin{lemma}\label{s2,v}
For every $\delta>0$ and $0.3>\epsilon>0$ and every positive integer n
\begin{equation*}
\frac{1}{n}{\sf{E}}\left[ \left\langle {\mathbf{S}_{2}},{\mathbf{V}^{*}} \right\rangle  \right]
\ge -\rho{{\sigma }^{2}}\left( 12\delta +3\epsilon \right)+\rho{{\sigma }^{2}}{2^{-2R_1}}\left( 1-{{2}^{-2{{R}_\textnormal{c}}}} \right)\left( 1-2\epsilon  \right)\left( 1-13\delta  \right).
\end{equation*}
\end{lemma}

\medskip

\begin{IEEEproof}
We begin with the following decomposition.
\begin{IEEEeqnarray}{rCl} \label{eq:s2,v decomposition}
\frac{1}{n}{\sf{E}}\left[ \left\langle {\textbf{S}_{2}},{\textbf{V}}^{*} \right\rangle  \right]
&=&\frac{1}{n}{\sf{E}}\left[ \left\langle \rho \textbf{S}_{1}+{\textbf{Z}_{G_2}},\textbf{V}^{*} \right\rangle  \right] \nonumber \\
&=&\frac{1}{n}\left( \rho{\sf{E}}\left[ \left\langle \textbf{S}_{1},\textbf{V}^{*} \right\rangle  \right]
+{\sf{E}}\left[ \left\langle \textbf{Z}_{G_2},\textbf{V}^{*} \right\rangle  \right] \right).
\end{IEEEeqnarray}

The second term on the r.h.s. of \eqref{eq:s2,v decomposition} vanishes as follows:
\begin{equation*}
{\sf{E}}\left[ \left\langle \textbf{Z}_{G_2},\textbf{V}^{*} \right\rangle  \right]
={\sf{E}}_{\textbf{S}_1,{\mathscr C}_1,{\mathscr C}_c}\Bigl[{\sf{E}}_{\textbf{S}_2}\left[ \left\langle \textbf{Z}_{G_2},\textbf{v}^{*} \right\rangle
\mid \textbf{S}_1=\textbf{s}_1,{\mathscr C}_1={\cal C}_1,{\mathscr C}_c={\cal C}_c \right]\Bigr]=0.
\end{equation*}
This holds, since conditionally on $\textbf{S}_1$ the random variable $\textbf{Z}_{G_2}$ is independent of $(\textbf{S}_1,\textbf{V}^{*})$, and therefore in the expectation over $\textbf{S}_2$,
for every $\textbf{z}_{\textnormal{G}_2}\in \mathbb{R}$, the sequences
$\textbf{z}_{\textnormal{G}_2}$ and $-\textbf{z}_{\textnormal{G}_2}$ are equiprobable and thus their inner products with $\textbf{v}^{*}$
cancel off each other.
\end{IEEEproof}

\bigskip

The distortion $D_2$ of the genie-aided scheme is now upper bounded as follows:
\begin{IEEEeqnarray}{rCl}
D_2 & =& \frac{1}{n}{\sf{E}}\left[ {\| {\textbf{S}_{2}}-\hat{\textbf{S}}_2{\hspace{-.4em}}^{\textnormal{G}} \|^{2}} \right] \nonumber \\
&=&{{\sigma }^{2}}-2{{\gamma }_{2,1}}\frac{1}{n}{\sf{E}}\left[ \left\langle {\textbf{S}_{2}},{\textbf{U}_{1}}^{*} \right\rangle  \right]-2{{\gamma }_{2,2}}\frac{1}{n}{\sf{E}}\left[ \left\langle {\textbf{S}_{2}},{\textbf{U}_{2}}^{*} \right\rangle  \right]-2{{\gamma }_{2,3}}\frac{1}{n}{\sf{E}}\left[ \left\langle {\textbf{S}_{2}},{\textbf{V}}^{*} \right\rangle  \right] \nonumber\\
&& +{{\gamma }_{2,1}}^{2}\left( {{\sigma }^{2}}\left( 1-{{2}^{-2{{R}_{1}}}} \right) \right)
+2{{\gamma }_{2,1}}{{\gamma }_{2,2}}\frac{1}{n}{\sf{E}}\left[ \left\langle {\textbf{U}_{1}}^{*},{\textbf{U}_{2}}^{*} \right\rangle  \right]+{{\gamma }_{2,2}}^{2}\left( {{\sigma }^{2}}\left( 1-{{2}^{-2{{R}_{2}}}} \right) \right) \nonumber\\
&& +2{{\gamma }_{2,1}}{{\gamma }_{2,3}}\frac{1}{n}{\sf{E}}\left[ \left\langle {\textbf{U}_{1}}^{*},{\textbf{V}}^{*} \right\rangle  \right]+2{{\gamma }_{2,2}}{{\gamma }_{2,3}}\frac{1}{n}{\sf{E}}\left[ \left\langle {\textbf{U}_{2}}^{*},{\textbf{V}}^{*} \right\rangle  \right]
+{{\gamma }_{2,3}}^{2}\left( {{\sigma }^{2}}{{2}^{-2{{R}_{1}}}}\left( 1-{{2}^{-2{{R}_\textnormal{c}}}} \right) \right) \nonumber \\
&\,\overset{(a)}{\mathop{\le }}&\,{{\sigma }^{2}}2^{-2{R}_{2}}\frac{1-{{\rho }^{2}}\left( 1-{{2}^{-2({{R}_{1}}+{{R}_\textnormal{c}})}} \right)}{1-{{\rho }^{2}}\left( 1-{{2}^{-2{{R}_{2}}}} \right)\left( 1-{{2}^{-2({{R}_{1}}+{{R}_\textnormal{c}})}} \right)}+\xi'(\delta,\epsilon),
\end{IEEEeqnarray}
where in (a) we have used Lemma~\ref{u1,u2}, Lemma~\ref{u1,v}, Lemma~\ref{u2,v},
Lemma~\ref{s2,u2}, Lemma~\ref{s2,u1}, Lemma~\ref{s2,v} and Lemma~\ref{MMSE_coefficients_bounds}
and where $\underset{\delta ,\epsilon \to 0}{\mathop{\lim }}\,\xi '\left( \delta ,\epsilon  \right)=0$.

\end{appendix}

\medskip



\begin{thebibliography}{99}

\bibitem{Stephan} A. Lapidoth and S. Tinguely, ``Sending a bivariate Gaussian over a Gaussian MAC,''
{\it IEEE Trans. Inform. Theory,} vol. IT-56, no. 6, pp. 2714-2752, June 2010.

\bibitem{Stephan feedback} A. Lapidoth and S. Tinguely, ``Sending a bivariate Gaussian source over a Gaussian MAC with feedback,''
{\it IEEE Trans. Inform. Theory,} vol. IT-56, no. 4, pp. 1852-1864, April 2010.

\bibitem{CovEl} T. M. Cover, A. El Gamal and M. Salehi, ``Multiple access channels with arbitrarily correlated
 sources,'' {\it IEEE Trans. Inform. Theory,} vol. IT-26, no. 6, pp. 648-657, Nov. 1980.

 \bibitem{SpWolf} D. Slepian and J. K. Wolf, ``Noiseless coding of correlated information sources'',
{\it IEEE Trans. Inform. Theory,} vol. IT-19, no. 4, pp. 471-480, July 1973.

\bibitem{DeBru} K. De Bruyn, V. V. Perlov and E. C. Van der Meulen, ``Reliable transmission of correlated
sources over an asymmetric multiple-access channel,''
 {\it IEEE Trans. Inform. Theory,} vol. IT-33, no. 5, pp. 716-718, Sep. 1987.

\bibitem{Deniz} D. Gunduz and E. Erkip, ``Correlated sources over an asymmetric MAC with one distortion criterion,''
CISS, Baltimore, MD, March 2007.

\bibitem{Xiao} J. Xiao and Z. Luo, ``Compression of Gaussian correlated sources under individual distortion criteria,''
{\it Proceedings 43rd Allerton Conference,} Illinois, Sep. 2005.

\bibitem{WZ} A. D. Wyner and J. Ziv, ``The rate-distortion function for source coding with
side information at the decoder,'' {\it IEEE Trans. Inform. Theory,}
vol. IT-22, no. 1, pp. 1-10, Jan. 1976.
\
\bibitem{Kaspi} A. H. Kaspi and T. Berger, ``Rate-distortion for correlated sources with partially separated encoders",
 {\it IEEE Trans. Inform. Theory,} vol. 28, no. 6, pp. 828-840, Sep. 1982.

\bibitem{Michelle} S.~I. Bross, A. Lapidoth and M.~A. Wigger, ``The Gaussian MAC with conferencing encoders,''
in {\it Proceedings IEEE International Symposium on Information
Theory,} pp. 2702-2706, July 6-11 2008.

\bibitem{Bross} S. I. Bross, A. Lapidoth, and M. Wigger, ``Dirty-paper coding for the Gaussian multiaccess
channel with conferencing",
 {\it IEEE Trans. Inform. Theory,} vol. 58, no. 9, Sep. 2012.

\bibitem{Willems} F.~M.~J. Willems, ``The discrete memoryless multiple access channel with partially cooperating encoders,''
{\it IEEE Trans. Inform. Theory,} vol. IT-29, no. 3, pp. 441-445,
May 1983.

\bibitem{Oohama} Y. Oohama, ``Gaussian multiterminal source coding,''
  {\it IEEE Trans. Inform. Theory,} vol. IT-43, No. 6, pp. 1912-1923, Nov. 1997.

\bibitem{Wagner} A. B. Wagner, S. Tavildar and P. Viswanath, ``The rate region of the quadratic Gaussian two-terminal source-coding problem,''
{\it IEEE Trans. Inform. Theory,} vol. IT-54, no. 5, pp. 1938-1961, May 2008.

\bibitem{cov} T. M. Cover and J. A. Thomas, ``{\it Elements of Information Theory,}'' Wiley, 1991.

\bibitem{Ozar} L. H. Ozarow, ``The capacity of the white Gaussian MAC with feedback,''
{\it IEEE Trans. Inform. Theory,} vol. IT-30, no. 4, pp. 623-629, July 1985.

\bibitem{Michelle2} M.~A. Wigger, ``Cooperation on the multiple-access channel,''
Ph.D. Thesis, ETH, Z\"{u}rich, Switzerland, 2008.

\bibitem{WIT} H.~S. Witsenhausen, ``On sequences of pairs of dependent random variables,''
{\it SIAM Journal on Applied Mathematics,} vol. 28, no.~1, pp. 100-113,
January 1975.

\bibitem{ROZ} Y.~A. Rozanov (translated by A. Feinstein), ``Stationary Random Processes,''
Holden-Day, 1967.

\bibitem{Dembo} W. Bryc, A. Dembo and A. Kagan, ``On the maximum correlation coefficient,''
{\it SIAM Journal on Theory of Probability and its Applications,} vol. 49, no.~1, pp. 132-138,
2005.


\bibitem{Shannon} C. E. Shannon â``Probability of error for optimal codes
 in a Gaussian channel",
 {\it Bell System Technical Journal,} vol. 38, pp. 611-656, May 1959.

\bibitem{Wyner}A. D. Wyner, ``Random packings and coverings of the unit $n$-sphere",
Bell System Technical Journal, vol. 46, pp. 2111-2118,
November 1967.

\bibitem{Graham}R. L. Graham, D. E. Knuth, and O. Patashnik, Concrete Mathematics:
``A Foundation for Computer Science", 2nd ed. Addison-Wesley, 1994.


\bibitem{EGYHKBook}  A. El Gamal and Y. H. Kim,
\emph{Network Information Theory,} Cambridge University Press, 2012.

\end{thebibliography}
\end{document}